\pdfoutput=1
\documentclass[10pt,twoside,openright]{memoir}
\usepackage[paperwidth=8.267in, paperheight=11.692in,bindingoffset=.75in]{geometry}
\usepackage[latin1]{inputenc}
\usepackage[T1]{fontenc}
\usepackage{amsmath}
\usepackage{amsfonts}
\usepackage{amssymb}
\usepackage{lipsum}
\usepackage{comment}
\usepackage{braket}
\usepackage[square,numbers]{natbib}
\usepackage{mathrsfs}
\usepackage{stmaryrd}
\usepackage[stable]{footmisc}
\usepackage{slashed}
\usepackage{graphicx}
\usepackage{subfig}
\usepackage{amsthm}
\usepackage{pbox}
\usepackage{slashbox}
\usepackage{color}
\usepackage{tikz-cd}
\usepackage{afterpage}
\usepackage{tgtermes}
\usepackage[protrusion=true,expansion=true]{microtype}
\usepackage[backref]{hyperref}
\hypersetup{
    colorlinks=true,
    citecolor=blue,
    filecolor=black,
    linkcolor=blue,
    urlcolor=gray
}

\newcommand\blankpage{%
    \null
    \thispagestyle{empty}%
    \addtocounter{page}{-1}%
    \newpage}

\theoremstyle{plain}
\newtheorem{thm}{Theorem}[chapter]
\newtheorem{defn}{Definition}[chapter]
\newtheorem{lem}{Lemma}[chapter]

\newtheorem{prop}{Proposition}[chapter]
\newtheorem{cor}{Corollary}[chapter]

\theoremstyle{definition}
\newtheorem*{demo}{Proof}

\newcommand*\modulo{\mathrm{\; mod \;}}

\hypersetup{pageanchor=false}
\makeatletter
\def\maketitle{
  \null
  \thispagestyle{empty}
  \vfill
  
  \begin{center}\leavevmode
    \normalfont
    {\LARGE\raggedleft \@author\par}
    \hrulefill\par
    {\huge\raggedright \@title\par}
    \vskip 1cm
  \end{center}
  
  \begin{flushright} \LARGE
	\emph{Supervisors:} Christian Brouder \\
	Fabien Besnard \\
	Thierry Masson
  \end{flushright}
  
  \vskip 1cm
  
  \begin{flushright} \LARGE
	\emph{Jury:} Benjamin Fuks \\
	Fedele Lizzi \\
	John Barrett \\
	Sylvie Paycha \\
	Jean-Christophe Wallet \\
	Christian Brouder \\
	Fabien Besnard
  \end{flushright}
  
  \vskip 3cm
  
  \begin{center}
	\includegraphics[height=2cm]{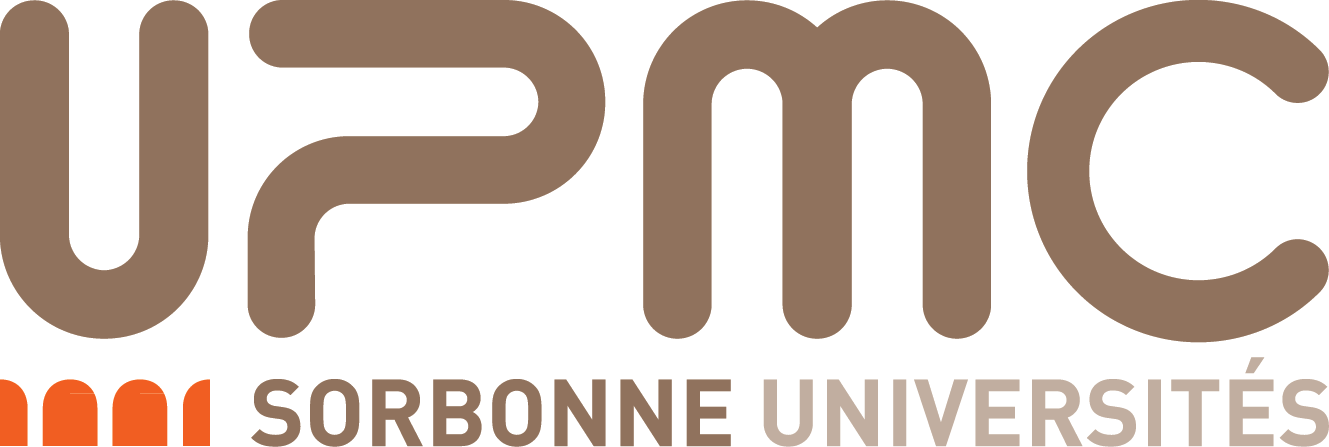}
	\hskip 1cm
	\includegraphics[height=2cm]{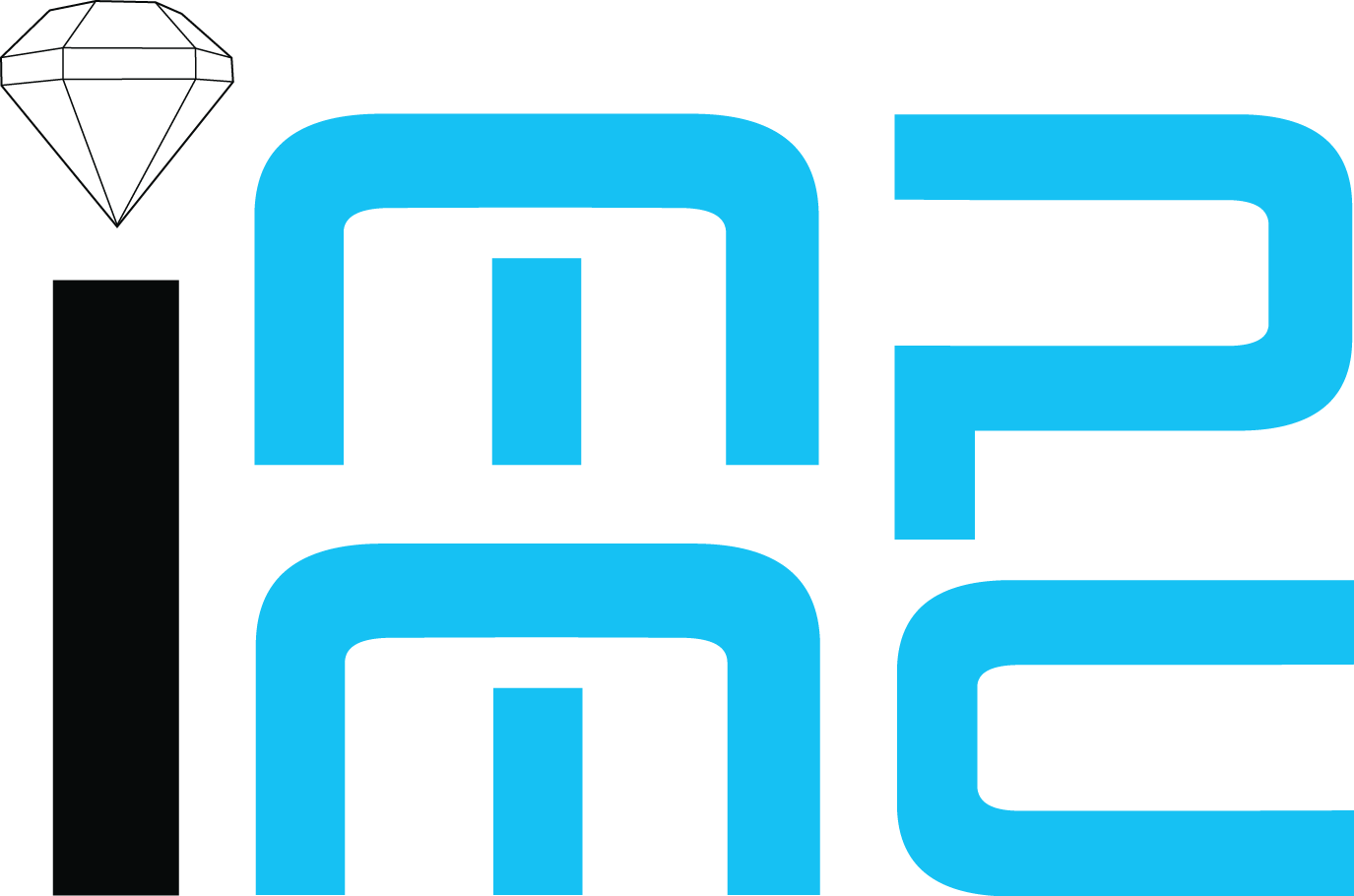}
  \end{center}
  
  \vfill
  \null
  \cleardoublepage
  }
\makeatother
\author{NADIR BIZI\footnote{E-mail contact: nadirxyz@hotmail.com}}
\title{SEMI-RIEMANNIAN NONCOMMUTATIVE GEOMETRY, GAUGE THEORY, AND THE STANDARD MODEL OF PARTICLE PHYSICS}
\date{}

\begin{document}

\let\cleardoublepage\clearpage

\maketitle

\frontmatter

\blankpage

\vspace*{0.2\textheight}
\begin{center}

\textit{To my Loved Ones: }

\textit{my Parents and Family, whose unconditional love never ceases to amaze me.}

\textit{My Friends, who have always believed (way too much) in me.}

\textit{My Muse, without whom my life would be dull.}

\textit{And of course, my Two Cats.}

\end{center}

\chapter*{Acknowledgments}

I would like to thank my thesis supervisors Christian Brouder and Fabien Besnard who, 4 years ago, took me on a unforgettable, life-changing trip down the stream of Noncommutative Geometry, and helped me stay afloat whenever the stream got rocky. On one bank was physics, on the other mathematics, and they helped me keep both in sight.

I would also like to thank my thesis supervisor Thierry Masson for the warm welcome at Marseille and the very interesting discussions, as well as the work we started together, work that I wish I had the time and opportunity to complete.

Finally, during my thesis, I had many enriching exchanges with fellow noncommutative geometers. I would like to thank Latham Boyle and Shane Farnsworth for their warm welcome at the Perimeter Institute and all our other encounters. I would also like to thank John Barrett, Michal Eckstein, and many others whose names I regret never asking. Whoever you are, wherever you are, thank you.

\newpage\blankpage\newpage

\tableofcontents

\newpage\blankpage\newpage\blankpage

\mainmatter

\chapter{Introduction} \label{ChapterIntro}

Our current understanding of particles and forces is based on two theories which are geometric in nature: the Standard Model of particle physics, which is a gauge theory, and General Relativity.

Gauge theory, which is based on connections and Lie groups, and general relativity, which is based on metric and the infinite dimensional diffeomorphism group, may seem quite different. However, it is worth mentioning that the former grew out of the later through the works of Weyl, Kaluza and Klein, to name only the early contributors \cite{ORaifeartaigh}. 
Indeed, dimensional reduction allows to break the full diffeomorphism invariance of a Kaluza-Klein theory into gauge invariance $\times$ space-time diffeomorphism invariance of coupled general relativity + gauge theory. 
What Connes' noncommutative geometry (NCG) provides is a  framework 
--almost-commutative manifolds-- in which  the full symetry group is already the (semidirect) product of the gauge group and the spacetime diffeomorphism group, with no need for any additional \emph{ad hoc} principle to reduce it. 
In its full glory, with the spectral action principle, it also provides a metric theory, which fulfills the dream of a (classical) unified theory of all interactions. At least it should, were it free of problems... 
Let us recall now the main stages of its development, with its successes but without hiding some of its troublesome issues. We stick to the interconnection of the theory with particle physics, and do not try to give an account of its mathematical origin or applications, which are deep and beyond our scope.

There are several distinct theories named NCG. Connes' one is distinguished by: i) an algebraic aparatus which plays the role of a spin structure (Hilbert space, chirality, real structure); ii) a Dirac operator which plays a major r\^ole. In its most recent formulation, the theory is based on the notion of a \emph{real spectral triple}, which will be defined in the body of this thesis. Its connection to geometry stems from the fact that a commutative spectral triple is (up to some technical assumptions) a Riemannian spin manifold \cite{Connes-13}.

The relation between NCG and particle physics starts in 1988 when Connes
Connes rewrites for the first time the Yang-Mills action in the terms of NCG \cite{Connes-88}. Using this time an almost-commutative manifold, Connes and Lott show that the Higgs phenomenon naturally emerges from NCG without additional input \cite{Connes-90}. Dubois-Violette, Kerner and Madore obtain the same conclusion from a derivation-based NCG \cite{Dubois-90}.

In 1995 Connes introduces the real structure of NCG in the ground-breaking paper \cite{Connes-95-reality}. It is a crucial ingredient: no $J$, no spin structure. No spin structure, no particles. In the same paper the algebra of the standard model is found to be $A=\mathbb{C}\oplus\mathbb{H}\oplus M_3(\mathbb{C})$, and the fluctuation of the metric $D_\omega=D+\omega+J \omega J^{-1}$ is introduced. Note that, in this paper the fluctuation is only in the fermionic action. The bosonic action is still written in terms of the curvature of $\omega$.

The following year the Spectral Action is introduced \cite{Connes-96}. The bosonic variable is now the fluctuated metric. The Lagrangian of the standard model coupled to gravity is derived, and the renormalization group is applied (assuming the big desert hypothesis) to predict a Higgs mass
of $170\pm 10$ GeV (see \cite{Chamseddine-96,Chamseddine-97}).

Let us make a little pause at this point to give a few more details on the pros and cons of the approach. On the good side:

\begin{enumerate}
\item We have the Higgs for free.
\item In the version with the spectral action, all the forces are unified with gravity. This is the fulfilling of Einstein's dream of a unified theory.
\item Contrarily to String Theory, where the Kaluza-Klein idea is also used, there is no moduli stabilization problem. The extra dimensions are stable because they are discrete.
\item There is not a plethora of different particle physics models compatible with this picture. For one thing the symmetry group must be one of a finite-dimensional real algebra, so not all Lie groups are allowed. There are also arguments constraining the algebra further \cite{Chamseddine-08}. Moreover the Dirac operator must fulfill the first-order condition, and this puts some constraints on the possible terms in the Lagragian, independently of any consideration of renormalizability. An even more constraining second-order condition, which is satisfied by the Standard Model has recently emerged \cite{Boyle,Farnsworth-15,3B-DGA}. In the other direction, even if we relax the first-order condition, the NCG route towards GUT does not seem
to go beyond the Pati-Salam model \cite{Chamseddine-13}.
\end{enumerate}
 
On the bad side:

\begin{enumerate}
\item This is not necessarily a problem, but it should be stressed that NCG is a classical theory. It yields a Lagragian, and from this point on, usual QFT takes over. This is the mainstream approach and the one we will follow in this thesis.  However, would it be possible to devise a Quantum NCG? Only very scarce and preliminary works are available on this 
question \cite{Rovelli-99,Besnard-07}. 
\item\label{fd} There is a doubling (even quadrupling in fact) of the fermionic degrees of freedom, as first spotted in \cite{Lizzi-97}. Indeed, the Hilbert space of spectral triple is the tensor product of a term $L^2(S)$, where $S$ is the spinor bundle, and a term $H_F$ which is a finite-dimensional Hilbert space with a basis made of all elementary particles, including antiparticles as well as their right/left versions. 
\item As we said in the little list of virtues of the approach, not all Lie groups are available as gauge groups, but only those which arise as unitary groups of finite-dimensional algebras. However this virtue turns out to be an pitfall, since $SU(3)$ is not among those! Hence the theory must be supplemented with a rather \emph{ad hoc} ``unimodularity condition'', in order to accomodate for the gauge group of the Standard Model. This condition can be justified when the representation of the algebra is complex \cite{Lazzarini-01}, which is unfortunately not the case in the Standard Model. This is certainly one of the most troubling problem of the theory.
\item\label{no} In the original model of Connes and Chamseddine, there was no room for neutrino oscillations.
\item\label{lor} Everything is defined on a Riemannian ``spacetime'', so that a Euclidean version of the Standard Model Lagragian is delivered. 
\item As is now clear, the prediction of  Higgs mass was off target. 
\end{enumerate}

Some of this concerns have been adressed in the subsequent development of the theory. First, there was a breakthrough in 2006 when Barrett \cite{Barrett} noticed that problems \ref{fd}, \ref{no} and \ref{lor} were linked. More precisely, using the fact that the KO-dimension of a Lorentzian manifold is different from the one of a Riemannian manifold, Barrett showed that not only the fermion doubling can be solved, using  
\begin{equation}
J\Psi=\Psi,\qquad \chi\Psi=\Psi\quad\mbox{ (Barrett's conditions)}\nonumber
\end{equation}
but also that a right-handed neutrino and see-saw mechanism can be added to the formalism. Connes came to the same conclusion in a purely Euclidean framework by also allowing the internal space to have a KO-dimension different from $0$ \cite{Connes-06}. Note that Barrett's work does not 
address the question of the possible formulation of a Lorentzian, or more generally semi-Riemannian Noncommutative Geometry. Works towards that goal started, on the mathematical side, with \cite{Strohmaier} and include \cite{Paschke-06,Franco-12,Franco-15,VPR}. On the physics side, relatively few works have appeared so far, with the notable exception of \cite{Elsner}, where the Connes-Lott model has been promoted to the Lorentzian signature, and \cite{VdD}, where the consequences of a Lorentzian spacetime on the fermionic action has been explored. Finally, in order to solve the discrepancy between the observed and predicted mass of the Higgs boson, Chamseddine and Connes introduced a new scalar field in the model \cite{Chamseddine-12}. This boson had already been considered in the physics literature as a way to solve the metastability issue of the electroweak vacuum. However, in NCG it does not emerge from a fluctuation of the metric, as the gauge bosons and the Higgs do, which can be viewed as unnatural. In order to give a better conceptual foundation of this important addition, it has been proposed to modify the model in order to incorporate a ``Grand Symmetry'' \cite{Devastato-14}, or even to change the way symmetries are represented in the theory \cite{Besnard-09}. Note however that it is also possible to just central extend $A$ by a second $\mathbb{C}$ term, and interpret the new boson as a Majoron associated to B-L symmetry \cite{Besnard-18-2}.

In this thesis we will mainly adress the question of the formulation of Semi-Riemannian Noncommutative Geometry, i.e. the definition of \emph{indefinite spectral triples}. The aim is the application to the Standard Model, hence we will limit the mathematical developments to those which are needed to handle almost-commutative manifolds. Obviously we will have to deal with the manifold (commutative) case, which is a template for the whole theory, and also with tensor products. Though these matters seem to be well-known, they have some surprises in store. In the Riemannian case, one can associate canonically a spectral triple to a spin manifold. In the non-Riemannian case, this cannot be done \emph{stricto sensu}. The reason is that the completion of the space of compactly supported spinors is not unique, a phenomenon that does not seem to have been noticed before. The definition of the tensor product of indefinite spectral triple is not so obvious either. First we have to handle the case of two Clifford algebras $\mathrm{Cl}(V_1,g_1)$ and $\mathrm{Cl}(V_2,g_2)$,  represented irreducibly on  spinor spaces $S_1$ and $S_2$, respectively. On $S_1$ (resp. $S_2$) there exists a non-degenerate sesquilinear form, unique up to rescaling, such that the vectors of $V_1$ (resp. $V_2$) act in a self-adjoint manner. We call it the \emph{Robinson form}. It turns out that the Robinson form on $\mathrm{Cl}(V_1,g_1)\hat\otimes \mathrm{Cl}(V_2,g_2)\simeq \mathrm{Cl}(V_1\oplus V_2,g_1\oplus g_2)$ is given by a kind of non-intuitive formula, that has to be taken into account in the definition of the tensor product of two indefinite spectral triples. Once this is done, a new invariant in the form of an ordered pair $(m,n)\in (\mathbb{Z}/8\times \mathbb{Z}/8)/(4,4)$, the \emph{KO-metric pair}, can be defined, which generalizes the sole KO-dimension $n$. This invariant is additive for the tensor product \cite{3B-IST}. All these subtleties will play a role in the formulation of the Standard Model. However, it must be said that the Spectral Action Principle will have to be abandoned, since no semi-Riemannian replacement for it is available to date (see chapter 6). Hence, we will fall back to the earlier Connes-Lott model, extended to spaces equipped with an indefinite product by Elsner \cite{Elsner}, with the important nuance that the KO-dimension of the finite spectral triple is not zero. We will also solve the fermion doubling problem using conditions slightly more general than Barrett's. However we will show that Barrett's condition are singled out by the Standard Model. In the text we have paid a particular attention to the different sign conventions existing in spin geometry. In particular we have not feared to write down some results twice, using two different combinations of sign conventions (to which we have given names), for the ease of the reader. In the Standard Model part, we have also handled simultaneously the case of commuting and Grassmann variables.

The thesis is organized as follows: in chapter 2 we recall the general theory of Krein spaces. The only small novelty here with respect to \cite{Bognar} is the proof that any two fundamental symmetries are unitarily equivalent. Chapter 3 introduces the necessary background in the theory of Clifford algebras. We give here the formula for the Robinson form on a tensor product of spinor spaces. Chapter 4 contains many classical material, due in particular to Baum \cite{Baum}. We will add a few results to her work. In particular we will show that the sufficient condition given by Baum for the existence of a local Robinson form is also necessary. In the same spirit, we will show that in order for the Dirac operator to be symmetric on the sections of the spinor bundle equipped with a metric $B$, it is necessary that the connection on spinors preserve $B$ and that the tangent vectors act self-adjointly for $B$ (or anti-self-adjointly, depending on the convention). In this chapter we will distinguish 3 different kinds of structures on a manifold: Clifford structures, spin-c structures and spin structures (the difference between the two first is invisible in the Riemannian case). Each structure corresponds to a particular class of connections: Clifford connections, spin-c connections and spin connections. We  will prove the existence and uniqueness of a spin connection compatible with a given spin structure, a well-known result in the Riemannian case. We will also give a formula for the difference between  the  Dirac operators associated with two different spin structures compatible with a given Clifford structure. In this chapter we will also observe that two different fundamental symmetries on the space of compactly supported spinor fields generally yield different Krein spaces, a problem we have alluded to before. We will end this chapter with a discussion of the tensor product of indefinite spectral triples associated with manifolds. Finally, chapter 5 builds on these results to arrive at a natural definition for what we like to call Indefinite Spectral Triples, that generalize spectral triples to the indefinite case.

The second part of the thesis derives the indefinite spectral triple corresponding to the Standard Model of particle physics. In chapter 7, the most general noncommutative gauge theory is constructed, and its action defined. The theory is constructed carefully, and the important role of fermion doubling is emphasized. In chapter 8, this noncommutative gauge theory is applied to the Standard Model. The seesaw mechanism and the antilinear charge conjugation are discussed.

\chapter{Krein Spaces} \label{ChapterKrein}

The main object of this chapter is the presentation of Krein spaces. In our study of spinors on pseudo-Riemannian spaces, we will see that they are naturally equipped with a canonical indefinite hermitian form that represents better the symmetries of spinors. But we would also like to have at our disposal a positive definite inner product, that enables us to use results from functional analysis. This leads naturally to the use of Krein spaces. Note that the use of Krein spaces in NCG has been advocated in \cite{VdD, PS, Strohmaier, VPR, VS2} among others. Let us define these Krein spaces.

\section{Definitions and Examples}

For our exposition of Krein spaces, we will rely on Reference \cite{Bognar}. Throughout this chapter, we will use the following notations. Let $\mathcal{K}$ be a - possibly infinite-dimensional - complex vector space, and let $(\cdot, \cdot)$ be a \emph{nondegenerate} hermitian form\footnote{We remind the reader that a hermitian form $h$ on a complex vector space $E$ is a map $E \times E \rightarrow \mathbb{C}$ that is linear in the second argument, anti-linear in the first, and that satisfies: $\overline{h(\psi, \varphi)} = h(\varphi,\psi)$.} on $\mathcal{K}$. If the form is positive definite and the space complete with respect to the induced norm, then $\mathcal{K}$ is said to be a Hilbert space. But what if the form is not definite? How can the space be given an adequate topology? Let us answer these questions.
The idea here is to turn the indefinite form into a definite one. This can be achieved through the use of fundamental decompositions and fundamental symmetries. 

\begin{defn}\textbf{(\cite{Bognar}, P.49, P.52)}

A decomposition:
\begin{equation}
\mathcal{K} = \mathcal{K}_{+} \oplus \mathcal{K}_{-}
\label{FundamentalDecomposition}
\end{equation}
is said to be a fundamental decomposition of $\mathcal{K}$ if and only if:
\begin{itemize}
	\item The form $(\cdot, \cdot)$ is positive definite on $\mathcal{K}_{+}$
	\item The form $(\cdot, \cdot)$ is negative definite on $\mathcal{K}_{-}$
	\item The subspaces $\mathcal{K}_{+}$ and $\mathcal{K}_{-}$ are mutually orthogonal with respect to $(\cdot, \cdot)$. In other words: $(\mathcal{K}_{+}, \mathcal{K}_{-}) = 0$.
\end{itemize}

An operator $\eta$ on $\mathcal{K}$ is said to be a fundamental symmetry if there exists a fundamental decomposition \eqref{FundamentalDecomposition} such that:
\begin{itemize}
	\item $\eta = 1$ on $\mathcal{K}_{+}$
	\item $\eta = -1$ on $\mathcal{K}_{-}$
\end{itemize}

\end{defn}

Note that a fundamental decomposition is characterized uniquely by its fundamental symmetry. It is in fact easy to see that the projectors on the subspaces $\mathcal{K}_{\pm}$ are given by:
\begin{equation}
P_{\pm} = \frac{1 \pm \eta}{2}.
\label{FundamentalProjectors}
\end{equation}
We will, from now on, use fundamental symmetries to parametrize fundamental decompositions.

The importance of fundamental symmetries is illustrated by the following result:

\begin{prop} Let $\eta$ be a fundamental symmetry. Then the sesquilinear form:
\begin{equation}
\langle \cdot, \cdot \rangle_\eta = (\cdot, \eta \cdot)
\label{EtaForm}
\end{equation}
is a positive definite, hermitian form. In other words, it is an inner product on $\mathcal{K}$.
\end{prop}

\begin{demo}
Let $\eta$ be a fundamental symmetry, and $\mathcal{K} = \mathcal{K}_{+} \oplus \mathcal{K}_{-}$ be the corresponding decomposition. Let $\psi = \psi_{+} + \psi_{-}, \varphi = \varphi_{+} + \varphi_{-}$ be elements of $\mathcal{K} = \mathcal{K}_{+} \oplus \mathcal{K}_{-}$. We have: $\eta \varphi = \varphi_+ - \varphi_-$, which implies that:
\begin{equation}
\begin{aligned}
\langle \psi, \varphi \rangle_\eta &= (\psi, \eta \varphi) \\
&= (\psi_{+} + \psi_{-}, \varphi_+ - \varphi_-) \\
\langle \psi, \varphi \rangle_\eta &= (\psi_{+},\varphi_{+}) - (\psi_{-},\varphi_{-}) \text{   (the subspaces are orthogonal)} .
\end{aligned}
\end{equation}
We now see that:
\begin{equation*}
\begin{aligned}
\langle \varphi, \psi \rangle_\eta &= (\varphi_{+},\psi_{+}) - (\varphi_{-},\psi_{-}) \\
&= \overline{(\psi_{+},\varphi_{+})} - \overline{(\psi_{-},\varphi_{-})} \\
&= \overline{(\psi_{+},\varphi_{+}) - (\psi_{-},\varphi_{-})} \\
\langle \varphi, \psi \rangle_\eta &= \overline{\langle \psi, \varphi \rangle_\eta},
\end{aligned}
\end{equation*}
which proves that $\langle \cdot, \cdot \rangle_\eta$ is a hermitian form. Moreover, for $\varphi = \psi$, we have:
\begin{equation}
\langle \psi, \psi \rangle_\eta = (\psi_{+},\psi_{+}) - (\psi_{-},\psi_{-}).
\end{equation}
Since $\psi_\pm \in \mathcal{K}_\pm$, we know that $(\psi_{+},\psi_{+})$ is positive, and $(\psi_{-},\psi_{-})$ is negative. We infer that $\langle \psi, \psi \rangle_\eta$ is positive. The hermitian form $\langle \cdot, \cdot \rangle_\eta$ is thus positive. Finally, let us prove that it is definite. Let us assume that $\langle \psi, \psi \rangle_\eta = 0$. Since both summands $(\psi_{+},\psi_{+})$ and $- (\psi_{-},\psi_{-})$ are positive, they both must vanish. But we know that $(\cdot, \cdot)$ is \emph{definite} on the subspaces $\mathcal{K}_\pm$. We therefore infer that $\psi_+ = \psi_- = 0$, and thus that $\psi = 0$.

\qed
\end{demo}

With this construction of inner products in hand, we can define Krein spaces and state their general properties. We follow \cite{Bognar}, P.100-102,  more or less to the letter. 

\begin{defn}

The pair $(\mathcal{K}, (\cdot, \cdot))$ is said to be a Krein space if and only if there exists a fundamental symmetry $\eta$ for which $(\mathcal{K} , \langle \cdot, \cdot \rangle_\eta)$ is a Hilbert space.
\end{defn}

The norm $\|\cdot\|_\eta$ associated to the scalar product $\langle \cdot, \cdot \rangle_\eta$ is called the \emph{$\eta$-norm}. We remark that $\eta$ is a bounded operator for the $\eta$-norm, since $\|\eta\psi\|_\eta^2=(\eta\psi,\eta^2\psi)=(\psi,\eta\psi)=\|\psi\|_\eta^2$.

What is asked in the definition, besides the mere existence of a fundamental symmetry, is that ${\mathcal K}$ be complete for the $\eta$-norm. Equivalently, we can ask the eigenspaces ${\mathcal K}_+$ and ${\mathcal K}_-$ of $\eta$ to be complete for their \emph{intrinsic topology}, which is the topology defined on ${\mathcal K}_\pm$ by the norm $\psi\mapsto \sqrt{\pm(\psi,\psi)}$. Obviously, this requirement is always satisfied if ${\mathcal K}$ is finite-dimensional.

If $\eta$ is a fundamental symmetry which turns ${\mathcal K}$ into a Krein space, what about the others? The following theorem is a key result.  

\begin{thm}
If $(\mathcal{K}, (\cdot, \cdot))$ is a Krein space, then:
\begin{enumerate}
\item Every fundamental symmetry turns ${\mathcal K}$ into a Hilbert space, 
\item if $\eta$ and $\eta'$ are two fundamental symmetries, then the $\eta$-norm and the $\eta'$-norm are equivalent.
\end{enumerate}
\end{thm}
\begin{demo}
The proof of these two claims are obtained in \cite{Bognar} as simple consequences of a long excursion in topology that we cannot reproduce here. Let us give nonetheless  a shorter proof of the second claim following the method of \cite{Langer}. We first define the norm $\|\cdot\|:=\|\cdot\|_\eta+\|\cdot\|_{\eta'}$. Let us show that ${\mathcal K}$ is complete for this norm. We let $(\psi_n)$ be a Cauchy sequence for $\|\cdot\|$. It is thus a Cauchy sequence for the $\eta$-norm and $\eta'$-norm. Since ${\mathcal K}$ is a Banach space for these two norms, there exists an $\eta$-limit $\psi_\eta$ and an $\eta'$-limit $\psi_{\eta'}$ to the sequence $(\psi_n)$. We then have, for all $\varphi\in{\mathcal K}$:
\begin{equation*}
\begin{aligned}
(\varphi,\psi_n-\psi_\eta)&= \langle\varphi,\eta(\psi_n-\psi_\eta)\rangle_\eta\longrightarrow 0, 
\end{aligned}
\end{equation*}
since $\eta$ is bounded for the $\eta$-norm. Similarly, $(\varphi,\psi_n-\psi_{\eta'})\longrightarrow 0$. Substracting we obtain $(\varphi,\psi_{\eta'}-\psi_\eta)=0$ for all $\varphi$, which yields $\psi_{\eta}=\psi_{\eta'}$ by the non-degeneracy of $(\cdot,\cdot)$. Thus $\|\psi_n-\psi_\eta\|\rightarrow 0$. Hence ${\mathcal K}$ is complete for $\|\cdot\|$. Now we have $\|\cdot\|_\eta\le\|\cdot\|$, which proves that the identity map from $({\mathcal K},\|\cdot\|)$ to $({\mathcal K},\|\cdot\|_\eta)$ is bounded. By Banach's bounded inverse theorem,  the identity map from $({\mathcal K},\|\cdot\|_\eta)$ to $({\mathcal K},\|\cdot\|)$ is also bounded, which proves that $\|\cdot\|$ and $\|\cdot\|_\eta$ are equivalent. Since we can do the same with the $\eta'$-norm, the $\eta$-norm and $\eta'$-norm are equivalent.
\qed 
\end{demo}
Hence, while the fundamental symmetry $\eta$ is far from unique, the topology defined by the $\eta$-norm is. This topology is called by Bognar the \emph{strong topology}. Since all $\eta$-norms are equivalent, an operator is bounded with respect to one if and only if it is bounded with respect to all of them. It thus makes sense to define $B(\mathcal{K})$, the algebra of bounded operators on a Krein space.

Note that one can define real Krein spaces, just as one can define real Hilbert spaces. All definitions above generalize to the real case, with hermitian forms becoming bilinear forms.

Let us give a few examples of Krein spaces:
\begin{itemize}
	\item Any Hilbert space is a Krein space, with the following trivial decomposition and fundamental symmetry:
	\begin{equation*}
\begin{aligned}
\mathcal{K}_+ &= \mathcal{K} \\
\mathcal{K}_- &= \{0\} \\
\eta &= 1.
\end{aligned}
\end{equation*}
	
	\item The most useful example for our purposes is the finite-dimensional Krein-space. Note that finite Krein spaces are studied extensively in chapter 2 of \cite{Neretin}. Since any finite-dimensional vector space equipped with an inner product is a Hilbert space, we deduce the following:
\begin{prop} If $\mathcal{K}$ is finite-dimensional, then it is a Krein space. 
\end{prop}
Let us explore this example a bit. Since the form $(\cdot, \cdot)$ is non-degenerate, one can always construct a pseudo-orthonormal basis by "diagonalizing" the hermitian form $(\cdot, \cdot)$. Let $(e_i)_i$ be this basis. It is such that:
\begin{equation*}
(e_i, e_j) = \begin{cases}
-\delta_{ij} &\text{ if $i,j = 1, \dots, q$} \\
\delta_{ij} &\text{ if $i,j = p+1, \dots, n$} \\
0 &\text{ otherwise.} \end{cases}
\end{equation*}
Here $n$ is the total dimension of the vector space. We denote $p = n-q$ the number of positive signs. The ordered pair $(q,p)$ is called the \emph{signature} of the indefinite product $(\cdot, \cdot)$. Our convention is to always put the number of negative signs at the beginning. The vector space $\mathcal{K}$ is often denoted $\mathbb{C}^{q,p}$.

One can associate to this basis the fundamental decomposition given by
\begin{equation*}
\begin{aligned}
\mathcal{K}_- &= \text{Span}(e_1,\dots,e_q) \\
\mathcal{K}_+ &= \text{Span}(e_{q+1},\dots,e_n).
\end{aligned}
\end{equation*}
In this basis, the corresponding fundamental symmetry takes the simple form:
\begin{equation*}
\eta = \begin{pmatrix} -I_q & 0 \\ 0 & I_p \end{pmatrix}.
\end{equation*}
We will see later that any fundamental symmetry can be obtained this way.

	\item Perhaps the most known example of a real Krein space is Minkowski space-time, often denoted $\mathbb{R}^{1,3}$ or $\mathbb{R}^{3,1}$. It is a real vector space of signature $(3,1)$ or $(1,3)$ (see later chapters). In both cases, a fundamental decomposition splits space-time into: (i) a one-dimensional subspace that corresponds to time, spanned by a timelike vector field, and (ii) a three-dimensional subspace that corresponds to space. By orthogonality, the latter is specified uniquely by the former. Any fundamental decomposition is thus specified by a timelike vector field. This corresponds (up to rotations) to a choice of inertial reference frame. This justifies the intuition that \emph{a choice of fundamental symmetry corresponds to a choice of reference frame or observer}. We will explore and justify this further below.

\end{itemize}

From now on, we will always consider $\mathcal{K}$ to be a Krein space.

Before we conclude this section, let us introduce a few more definitions and notations. We define adjunction similarly to how it is defined for Hilbert spaces (here we follow \cite{Bognar} again, chap. VI).

\begin{defn}
Let $T$ be an operator on $\mathcal{K}$ with domain ${\mathcal D}(T)$. The Krein-adjoint, or adjoint of $T$, denoted $T^\times$ is the operator defined on those $\varphi$ for which there exists $\varphi'$ such that
\begin{equation}
\forall \psi\in{\mathcal D}(T): \; (\psi,  \varphi') = (T \psi, \varphi), 
\label{KreinAdjoint}
\end{equation}
by $T^\times\varphi:=\varphi'$.
\end{defn}
In other words, $T^\times$ is the operator with largest possible domain which satisfies

\begin{equation*}
\forall \psi\in{\mathcal D}(T),\varphi\in{\mathcal D}(T^\times): \; (T\psi,\varphi)=(\psi,T^\times \varphi).
\end{equation*}

An operator $T$ is said to be Krein-self-adjoint, or self-adjoint, if and only if $T^\times = T$: that is, if $T$ and $T^\times$ have the same domain and coincide on said domain. Similarly to the Hilbert space case, one can define \emph{symmetric} operators (sometimes also called \emph{formally self-adjoint operators}), to be the operators satisfying $(T\psi,\varphi)=(\psi,T\varphi)$ for all $\psi,\varphi$ in the domain of $T$. The notions of self-adjoint and symmetric operators obviously coincide for bounded operators.

An operator $T$ is said to be Krein-unitary, or unitary, if it satisfies: $T^\times T = T T^\times = 1$ over some domain. That is, if it is both an isometry (\emph{i.e.} if it preserves the indefinite product $(\cdot,\cdot)$) and a co-isometry (\emph{i.e.} if its adjoint $T^\times$ is an isometry). The set of all Krein-unitaries will be denoted $U(\mathcal{K})$. Note that a Krein-unitary operator is not necessarily bounded or everywhere defined, unlike a unitary operator on a Hilbert space. Let us provide a counterexample. Let $\mathcal{K}$ be the space of square-integrable sequences of vectors that take values in $\mathbb{C}^2$:
\begin{equation*}
\mathcal{K} = \{ \varphi = (\varphi_n)_n \in (\mathbb{C}^2)^{\mathbb{N}^\ast} | \sum_n \psi_n^\dagger \psi_n < \infty \}.
\end{equation*}
We equip it with the indefinite form:
\begin{equation*}
(\psi,\varphi) = \sum_n \psi_n^\dagger \begin{pmatrix} 0 & 1 \\ 1 & 0 \end{pmatrix} \varphi_n,
\end{equation*}
which turns it into a Krein space (that this hermitian form is defined everywhere on $\mathcal{K}$ is a simple consequence of the constant bound of the square matrix that appears above, with respect to the index $n$). We leave it to the reader to check that the operator:
\begin{equation*}
U = \bigoplus_n \begin{pmatrix} 0 & n \\ 1/n & 0 \end{pmatrix}
\end{equation*}
is a Krein-unitary operator, whilst not being bounded or everywhere defined.

Note that one can expand this notation to elements of the Krein space by writing:
\begin{equation*}
\psi^\times = (\psi, \cdot).
\end{equation*}
for $\psi \in \mathcal{K}$.

Let us illustrate with examples:
\begin{itemize}
	\item For the finite dimensional Krein space $\mathbb{C}^{q,p}$, the unitary group is often denoted $U(q,p)$. It is sometimes called a pseudo-unitary group.
	
	\item For Minkowski space-time, the group of unitaries is simply the full Lorentz group $O(1,3)$.
\end{itemize}

The notions defined above clearly reduce to the usual ones when the Krein space is a Hilbert space. Moreover, using fundamental symmetries, one can see $\mathcal{K}$ as a Hilbert space, and use the corresponding adjunctions. Hence the following quite self-explanatory definitions, which we gather here for future reference:

\begin{defn}

Let $\eta$ be a fundamental symmetry on $\mathcal{K}$. Let $T$ be an operator on $\mathcal{K}$. The  $\eta$-adjoint of $T$, denoted $T^{\dagger\eta}$ is the adjoint of $T$ for the scalar product $\langle\cdot,\cdot\rangle_\eta$.  When no confusion is possible (\text{i.e.} when only one fundamental symmetry is relevant), the $\eta$-adjoint will be called a Hilbert-adjoint, and simply denoted $T^\dagger$.

$T$ is said to be $\eta$-self-adjoint, or Hilbert-self-adjoint, if and only if $T^{\dagger\eta}=T$.

$T$ is said to be $\eta$-unitary, or Hilbert-unitary, if it satisfies: $T^{\dagger\eta} T = T T^{\dagger\eta} = 1$.
\end{defn}

\section{Properties of Fundamental Symmetries} \label{propsym}

We start this section with two straightforward properties of fundamental symmetries:

\begin{prop}
Let $\eta$ be a fundamental symmetry of $\mathcal{K}$. Then $\eta$ is a Krein-self-adjoint involution:
\begin{equation*}
\begin{aligned}
\eta^2 &= 1 \\
\eta^\times &= \eta.
\end{aligned}
\end{equation*}
Moreover $\eta$ is bounded for every $\eta'$-norm (i.e. it is continuous for the strong topology).

\end{prop}

\begin{demo}
Let $\mathcal{K} = \mathcal{K}_{+} \oplus \mathcal{K}_{-}$ be the fundamental decomposition corresponding to $\eta$. From the definition of $\eta$: $\eta = \pm 1$ on $\mathcal{K}_\pm$, we easily infer that $\eta^2 = 1$.

We have already proved above that $\langle \cdot, \cdot \rangle_\eta = (\cdot, \eta \cdot)$ is a hermitian form. This is equivalent to the Krein-self-adjointness of $\eta$.  Clearly $\eta$ is everywhere defined, and we have already remarked that it is bounded for the $\eta$-norm, hence for every $\eta'$-norm.

\qed
\end{demo}

From $\eta^2 = 1$, one can deduce that:
\begin{prop}
Let $\eta$ be a fundamental symmetry on $\mathcal{K}$. Let $T$ be an operator on $\mathcal{K}$ with domain ${\mathcal D}(T)$. Then:
\begin{equation}
T^{\dagger\eta} = \eta T^\times \eta
\end{equation}
on ${\mathcal D}(T^{\dagger\eta})=\eta {\mathcal D}(T^\times)$.

\end{prop}

\begin{demo}
Let $\psi\in{\mathcal D}(T)$ and $\varphi \in \eta{\mathcal D}(T^\times)$. We have:
\begin{equation*}
\begin{aligned}
\langle \psi, T^{\dagger\eta} \varphi \rangle_\eta &= \langle T \psi, \varphi \rangle_\eta \\
&= (T \psi, \eta \varphi) \\
&= (\psi, T^\times \eta \varphi) \\
&= (\psi, \eta^2 T^\times \eta \varphi) \\
&= \langle \psi, (\eta T^\times \eta) \varphi \rangle_\eta.
\end{aligned}
\end{equation*}
This shows that $\eta{\mathcal D}(T^\times)={\mathcal D}(\eta T^\times \eta)\subset{\mathcal D}(T^{\dagger\eta})$. But conversely, using exactly the same steps with the Krein and scalar products exchanged, we can prove that $\eta{\mathcal D}(T^{\dagger})={\mathcal D}(\eta T^{\dagger\eta}\eta)\subset{\mathcal D}(T^\times)$. Since $\eta^2=1$, we obtain ${\mathcal D}(T^{\dagger\eta})={\mathcal D}(\eta T^\times \eta)$, hence the two operators are equal.

\qed
\end{demo}

Let $\eta, \nu$ be two fundamental symmetries on $\mathcal{K}$, with fundamental decompositions $\mathcal{K} = \mathcal{K}_{+}^\eta \oplus \mathcal{K}_{-}^\eta$ and $\mathcal{K} = \mathcal{K}_{+}^\nu \oplus \mathcal{K}_{-}^\nu$ respectively. It seems natural to look for an operator $U$ that would map $\mathcal{K}_\pm^\eta$ to $\mathcal{K}_\pm^\nu$, and then relate $\eta$ to $\nu$ through the operator $U$. Since it maps a positive subspace to a positive one, and a negative subspace to a negative one, we will look for an operator that preserves the indefinite product - that is, a Krein-unitary operator.

\begin{thm} \label{FundamentalSymmetriesKreinUnitary}
Let $\eta, \nu$ be two fundamental symmetries on $\mathcal{K}$. There exists a Krein-unitary operator $U$ such that:
\begin{equation}
\nu  = U^\times \eta U.
\label{FundamentalSymmetriesKreinUnitaryFormula}
\end{equation}

Moreover, $U$ can be chosen bounded, with bounded inverse, and positive definite with respect to $\langle \cdot, \cdot \rangle_\eta$.

\end{thm}

From the relation $\nu  = U^\times \eta U$, it is easy to see that $U^\times = U^{-1}$ maps $\mathcal{K}_\pm^\eta$ isometrically to $\mathcal{K}_\pm^\nu$, since they are the subspaces of $\eta$ and $\nu$ respectively, of eigenvalue $\pm 1$. Moreover, the $\eta$ and $\nu$-inner products are related by a simple insertion of the operator $U$. Indeed, for any $\psi, \varphi \in \mathcal{K}$, we have:
\begin{equation*}
\begin{aligned}
\langle \psi, \varphi \rangle_\nu &= (\psi, \nu \varphi) \\
&= (\psi, U^\times \eta U \varphi) \\
&= (U \psi, \eta U \varphi) \\
\langle \psi, \varphi \rangle_\nu &= \langle U\psi, U\varphi \rangle_\eta.
\end{aligned}
\end{equation*}

\begin{demo}
To prove the lemma, we will be working in the Hilbert space $(\mathcal{K} , \langle \cdot, \cdot \rangle_\eta)$. In other words, we will work with the $\eta$ "frame of reference". Let us denote $H = \eta \nu$. Since $\eta$ and $\nu$ are both bounded invertible operators, then so is $H$. We have:
\begin{equation*}
\langle \cdot, \cdot\rangle_\nu = (\cdot, \nu \cdot) = \langle \cdot, \eta \nu \cdot\rangle_\eta = \langle \cdot, H \cdot\rangle_\eta.
\end{equation*}
The product $\langle \cdot, \cdot\rangle_\nu$ is a positive definite inner product. We deduce that $H$ is a bounded invertible positive definite operator. Note that $\nu = \eta H$. From $\nu^2 = (\eta H)^2 = 1$, we deduce that $\eta H \eta H = 1$. Multiplying by $\eta$ on both the left and right then gives us: $H \eta H \eta = 1$. We deduce that $H$ is the left and right inverse of $\eta H \eta$:
\begin{equation*}
H^{-1} = \eta H \eta.
\end{equation*}
Note that $\eta$ is Hilbert-self-adjoint: $\eta^{\dagger\eta} = \eta$, and thus Hilbert-unitary. We can thus apply self-adjoint functional calculus to the formula above. For any well-behaved enough function $f$ on $\mathbb{R}$, and using $\eta^2 = 1$, we have:
\begin{equation*}
f(H^{-1}) = \eta f(H) \eta.
\end{equation*}
In particular, for $f(h) = |h|^{1/2}$, have:
\begin{equation}
U^{-1} = \eta U \eta,
\end{equation}
where:
\begin{equation}
U = f(H) = H^{1/2}
\end{equation}
is a bounded positive definite operator. The relation $U^{-1} = \eta U \eta$ implies that $U^{-1}$ is bounded as well\footnote{This is also simply a consequence of the bounded inverse theorem}. Since $U$ is Hilbert self-adjoint, we have:
\begin{equation*}
U^{-1} = \eta U^{\dagger\eta} \eta = U^\times,
\end{equation*}
proving that $U$ is Krein-unitary. Finally, we have:
\begin{equation*}
\begin{aligned}
\nu &= \eta H \\
&= \eta U^2 \\
&= \eta U^{\dagger\eta} U \\
\nu &= U^\times \eta U.
\end{aligned}
\end{equation*}

\qed
\end{demo}

Since the fundamental symmetries $\eta$ and $\nu$ are arbitrary, it is easy to arrive at the following corollary:
\begin{cor}
For any fundamental symmetry $\eta$, the set of all fundamental symmetries is:
\begin{equation*}
\mathcal{F} = \{U^\times \eta U | U \in U(\mathcal{K}) \cap B(\mathcal{K}) \}.
\end{equation*}
\end{cor}
Note that the set $\mathcal{F}$ of all fundamental symmetries is isomorphic to the set of all fundamental decompositions.

\begin{demo}
We have seen in the previous theorem that any fundamental symmetry is of the form $U^\times \eta U$, with $U$ bounded. Let us now prove the converse. Let $U$ be a bounded Krein-unitary operator. Since $U$ is invertible, we can define the following decomposition of $\mathcal{K}$:
\begin{equation*}
\begin{aligned}
\mathcal{K} &= \mathcal{K}'_{+} \oplus \mathcal{K}'_{-} \\
\mathcal{K}'_\pm &= U \mathcal{K}_\pm.
\end{aligned}
\end{equation*}
Let $\psi = U \varphi \in \mathcal{K}'_\pm$. We have:
\begin{equation*}
\pm (\psi, \psi) = \pm (U \varphi, U \varphi) = \pm (\varphi, \varphi) \geq 0.
\end{equation*}
Moreover, if $\pm (\psi, \psi) = 0$, then $(\varphi, \varphi)=0$. Since $(\cdot, \cdot)$ is definite on $\mathcal{K}_\pm$, $\varphi$ must vanish. And so must $\psi$. This proves that $\pm (\cdot, \cdot)$ is positive definite on $\mathcal{K}'_\pm$, implying that $\mathcal{K} = \mathcal{K}'_{+} \oplus \mathcal{K}'_{-}$ is a fundamental decomposition. We leave it to the reader to check that the corresponding fundamental symmetry is $\nu = U^\times \eta U$.

\qed
\end{demo}

Let us illustrate with a few examples:
\begin{itemize}
	\item For the finite Krein space $\mathbb{C}^{q,p}$, we saw that a fundamental symmetry $\eta$ can be constructed out of a pseudo-orthonormal basis $(e_i)_i$. One can construct $\eta$ explicitly:
\begin{equation*}
\eta = - \sum_{i=1}^{q} e_i \otimes e_i^\times + \sum_{i=q+1}^{n} e_i \otimes e_i^\times.
\end{equation*}
Any other fundamental symmetry $\nu$ can be obtained from $\eta$ by a Krein-unitary $U$ as in theorem \ref{FundamentalSymmetriesKreinUnitary}. It is then easy to see that $\nu$ is the fundamental symmetry built from the pseudo-orthonormal basis $(f_i = U^\times e_i)_i$:
\begin{equation*}
\nu = - \sum_{i=1}^{q} f_i \otimes f_i^\times + \sum_{i=q+1}^{n} f_i \otimes f_i^\times.
\end{equation*}
Thus, any fundamental symmetry can be constructed this way. In \cite{Neretin}, it is proven (P.76) that $\mathcal{F}$ is isomorphic to the homogeneous space $U(q,p)/(U(q) \times U(p))$ of dimension $2qp$.

	\item For Minkowski space-time, the physical content of the theorem is that one can go from any frame to any other frame by a Lorentz transformation. The fact that this Lorentz transformation can be chosen positive means that it is a pure boost, as one can check explicitly, using a representation such as the one used in \cite{Jackson}.
	
\end{itemize}

\section{Tensor Product of Krein Spaces}

We conclude this chapter with a definition of the topological tensor product of two Krein spaces.

\begin{defn} \label{TensorKrein}
The topological tensor product of two Krein spaces $(\mathcal{K}_1, (\cdot, \cdot)_1)$ and $(\mathcal{K}_2, (\cdot, \cdot)_2)$ is the Krein space $(\mathcal{K}, (\cdot, \cdot))$, where 
\begin{itemize}
	\item $\mathcal{K} = \overline{\mathcal{K}_1 \otimes \mathcal{K}_2}$ is the completion of $\mathcal{K}_1 \otimes \mathcal{K}_2$ with respect to the inner product $\langle \cdot, \cdot \rangle_{\eta_1} \otimes \langle \cdot, \cdot \rangle_{\eta_2}$, with $\eta_1$ and $\eta_2$ any two fundamental symmetries of $(\mathcal{K}_1, (\cdot, \cdot)_1)$ and $(\mathcal{K}_2, (\cdot, \cdot)_2)$ respectively,
	\item $(\cdot,\cdot) = (\cdot, \cdot)_1 \otimes (\cdot, \cdot)_2$ is the hermitian form defined uniquely by its action:
\begin{equation}
(\psi_1 \otimes \psi_2, \varphi_1 \otimes \varphi_2) = (\psi_1, \varphi_1)_1 (\psi_2, \varphi_2)_2.
\end{equation}
on $\mathcal{K}_1 \otimes \mathcal{K}_2$.
\end{itemize}

\end{defn}

To prove that $(\mathcal{K}, (\cdot, \cdot))$ is a Krein space, it suffices to notice that $\eta =\eta_1 \otimes \eta_2$ is a fundamental symmetry for any $\eta_1$ and $\eta_2$ fundamental symmetries of $(\mathcal{K}_1, (\cdot, \cdot)_1)$ and $(\mathcal{K}_2, (\cdot, \cdot)_2)$ respectively, and that the inner product:
\begin{equation*}
\langle \cdot, \cdot \rangle_\eta = \langle \cdot, \cdot \rangle_{\eta_1} \otimes \langle \cdot, \cdot \rangle_{\eta_2}
\end{equation*}
induces the appropriate topology.

\chapter{Clifford Algebras} \label{ChapterClifford}

In this chapter, we study an important class of algebras: Clifford algebras. Their importance lies in the necessity of using spinors (see next chapter) to describe matter in the Standard Model of particle physics. But, as we will see, they can also be used to describe the geometry of space-time, hence their importance in NCG. We will start with a definition of Clifford algebras, and we will briefly present their classification and irreducible representations. For this we rely on references \cite{LawsonMichelsohn, Crum}. Then, we define canonical objects on these representations: the Robinson product and the charge conjugation operator, and study their properties and interactions. We will rely for this on references \cite{Robinson, GBVF, VS}. Next, we study tensor products of Clifford algebras as a preliminary step to the study of tensor products of spectral triples. Finally, as a check, we will use an explicit representation of Clifford algebras to derive explicitly most results derived in previous sections. Throughout this chapter, we will focus on \emph{even} Clifford algebras.

\section{Definition and Classification of Even Clifford Algebras}

Let $V$ be a real vector space of even dimension $d \geq 2$, and $g$ a real non-degenerate bilinear form of signature $(q,p)$ on $V$ (with $q+p=d$ and $(q,p) \neq (0,0)$). That is, $V \cong \mathbb{R}^{q,p}$. The Clifford algebra $\mathrm{Cl}(V)$ is the \emph{real} algebra generated by the elements of $V$ quotiented by the equivalence relation:
\begin{equation*}
u v + v u \sim 2 g(u,v).
\end{equation*}
This relation induces a $\mathbb{Z}_2$ grading on the Clifford algebra, where the grading of an element is given by the parity of the number of its generators. Let $\omega \in \mathrm{Cl}(V)$. We denote its grading $|\omega | \in \{0,1\}$. We denote $\gamma: V \hookrightarrow \mathrm{Cl}(V)$ the canonical embedding of $V$ into $\mathrm{Cl}(V)$. We thus have the following relation:
\begin{equation}
\{\gamma(u), \gamma(v)\} = 2 g(u,v).
\label{CliffordDefinition}
\end{equation}
To find a convenient set of generators of the algebra, one can choose a pseudo-orthonormal basis $(e^a)_a$ of $V$:
\begin{equation*}
g(e^a, e^b) = \begin{cases}
-\delta_{ab} &\text{ if $a,b = 1, \dots, q$} \\
\delta_{ab} &\text{ if $a,b = p+1, \dots, d$} \\
0 &\text{ otherwise.} \end{cases}
\end{equation*}
The corresponding elements of $\mathrm{Cl}(V)$ are then simply denoted $\gamma^a = \gamma(e^a)$, and they satisfy:
\begin{equation*}
\{ \gamma^a, \gamma^b \} = \begin{cases}
-\delta_{ab} &\text{ if $a,b = 1, \dots, q$} \\
\delta_{ab} &\text{ if $a,b = p+1, \dots, d$} \\
0 &\text{ otherwise.} \end{cases}
\end{equation*}
This set of generators makes it clear that the algebra $\mathrm{Cl}(V)$ depends only on the signature $(q,p)$. When no confusion is possible, we will simply denote it: $\mathrm{Cl}(V) \cong \mathrm{Cl}(q,p)$. A basis for the algebra $\mathrm{Cl}(q,p)$ is formed of the elements $(\gamma^I)_{I \subset \llbracket 1, d \rrbracket}$ where $\gamma^I = \gamma^{i_1}...\gamma^{i_{{|I|}}}$ for $I=\{i_1 < ... < i_{|I|}\} \subset \llbracket 1, d \rrbracket$ (the symbol $|I|$ denotes the cardinality of the set $I$). The real dimension of the algebra is the number of its basis elements: $2^d$.

On a Clifford algebra, one can define the main anti-automorphism $\alpha$ which leaves generators invariant and switches the order of elements. That it, it is defined by:
\begin{equation}
\alpha(\gamma(u_1) \dots \gamma(u_i)) = \gamma(u_i) \dots \gamma(u_1).
\end{equation}
The action of $\alpha$ on the basis elements is given by:
\begin{equation*}
\begin{aligned}
\alpha(\gamma^I) &= \alpha(\gamma^{i_1}...\gamma^{i_{|I|}}) \\
&= \gamma^{i_{|I|}}...\gamma^{i_1} \\
&= (-1)^{{|I|}({|I|}-1)/2} \gamma^{i_1}...\gamma^{i_{|I|}} \\
\alpha(\gamma^I) &= (-1)^{{|I|}({|I|}-1)/2} \gamma^I.
\end{aligned}
\end{equation*}
One can also define the graded anti-automorphism $\alpha'$ that flips all generators: $\alpha'(\gamma(u)) = -\gamma(u)$. It is related to $\alpha$ by the relation:
\begin{equation}
\alpha'(\omega) = (-1)^{|\omega|} \alpha(\omega),
\end{equation}
for $\omega \in \mathrm{Cl}(V)$.

We will often be interested in Clifford algebras of direct sums of vector spaces. We will thus often consider the sum $V \oplus W$, where $V$ is the space of signature $(q,p)$ described above, and $W$ a vector space of even dimensions $d'$, with a bilinear $g'$ of signature $(q',p')$. The space $V \oplus W$ is then naturally equipped with the bilinear $g \oplus g'$. The corresponding Clifford algebra $\mathrm{Cl}(V \oplus W)$ can be described using a real tensor product of $\mathrm{Cl}(V)$ and $\mathrm{Cl}(W)$. More accurately, the following is an isomorphism of graded algebras:
\begin{equation}
\begin{aligned}
\mathrm{Cl}(V \oplus W) &\longrightarrow &&\mathrm{Cl}(V) \hat{\otimes} \mathrm{Cl}(W) \\
\gamma(v \oplus w) &\longmapsto &&\gamma(v) \hat{\otimes} 1 + 1 \hat{\otimes} \gamma(w).
\end{aligned}
\label{TensorCliffordDef}
\end{equation}
Here $\hat{\otimes}$ is a \emph{graded} tensor product. The multiplication rule is the following:
\begin{equation}
(\omega_1 \hat{\otimes} \theta_1) (\omega_2 \hat{\otimes} \theta_2) = (-1)^{|\theta_1||\omega_2|}(\omega_1 \omega_2 \hat{\otimes} \theta_1 \theta_2)
\end{equation}
for $\omega_i \in \mathrm{Cl}(V)$ and $\theta_i \in \mathrm{Cl}(W)$. The grading on the factors extends consistently to the tensor product according to:
\begin{equation}
|\omega \hat{\otimes} \theta| \equiv |\omega| + |\theta|.
\end{equation}
When no confusion is possible, we will simply write: $\gamma(v \oplus w) = \gamma(v) \hat{\otimes} 1 + 1 \hat{\otimes} \gamma(w)$. Let us compute the tensor form of the main anti-automorphism $\alpha$:
\begin{equation*}
\begin{aligned}
\alpha(\omega \hat{\otimes} \theta) &= \alpha((\omega \hat{\otimes} 1)(1 \hat{\otimes} \theta)) \\
&= \alpha(1 \hat{\otimes} \theta) \alpha(\omega \hat{\otimes} 1) \\
&= (1 \hat{\otimes} \alpha(\theta))(\alpha(\omega) \hat{\otimes} 1) \\
\alpha(\omega \hat{\otimes} \theta) &= (-1)^{|\theta||\omega|} \alpha(\omega) \hat{\otimes} \alpha(\theta).
\end{aligned}
\end{equation*}
A similar relation holds for $\alpha'$:
\begin{equation*}
\alpha'(\omega \hat{\otimes} \theta) = (-1)^{|\theta||\omega|} \alpha'(\omega) \hat{\otimes} \alpha'(\theta).
\end{equation*}
As we construct more and more objects related to Clifford algebras, we will also determine how they combine in tensor products of Clifford algebras, and update the rules of this tensor product.

Using this tensor product, any Clifford algebra can be decomposed into a tensor product of smaller Clifford algebras. The smallest Clifford algebras can be explicitly described in terms of algebras of matrices, and bigger Clifford algebras can then be reconstructed in terms of algebras of matrices. As a result, even Clifford algebras have been classified. Let $D=2^{d/2}$. One finds that:
\begin{itemize}
	\item if $p-q \modulo 8 \in \{0,2\}$, then $\mathrm{Cl}(q,p) \cong \mathrm{M}_D(\mathbb{R})$,
	\item if $p-q \modulo 8 \in \{4,6\}$, then $\mathrm{Cl}(q,p) \cong \mathrm{M}_{D/2}(\mathbb{H})$.
\end{itemize}

One can also define the complexification of a Clifford algebra. The complexification of $\mathrm{Cl}(V) \cong \mathrm{Cl}(q,p)$ will be denoted $\mathbb{C}\mathrm{l}(V) \equiv \mathbb{C} \otimes_\mathbb{R} \mathrm{Cl}(V)$, or simply $\mathbb{C}\mathrm{l}(q,p)$. Note that the monomials $(\gamma^I)_{I \subset \llbracket 1, d \rrbracket}$ form a complex linear basis of $\mathbb{C}\mathrm{l}(V)$. This algebra naturally inherits the $\mathbb{Z}_2$ grading of $\mathrm{Cl}(V)$. The classification above for real Clifford algebras tells us that complexified even Clifford algebras are necessarily of the form:
\begin{equation*}
\mathbb{C}\mathrm{l}(q,p) \cong \mathrm{M}_D(\mathbb{C}).
\end{equation*}
On this algebra, one can define an antilinear automorphism called \emph{charge conjugation} by applying complex conjugation on the first factor of the tensor product:
\begin{equation*}
\overline{\lambda \otimes \omega} = \overline{\lambda} \otimes \omega,
\end{equation*}
for any $\lambda \in \mathbb{C}$ and $\omega \in \mathrm{Cl}(V)$. This induces a real structure on $\mathbb{C}\mathrm{l}(V)$, and $\mathrm{Cl}(V)$ can be recovered as the real part of $\mathbb{C}\mathrm{l}(V)$:
\begin{equation*}
\mathrm{Cl}(q,p) = \{\Omega \in \mathbb{C}\mathrm{l}(V) | \overline{\Omega} = \Omega \}.
\end{equation*}
The anti-automorphisms $\alpha$ and $\alpha'$ can be naturally extended to $\mathbb{C}\mathrm{l}(V)$ as well, by defining: $\alpha(\lambda \otimes \omega) = \lambda \otimes \alpha(\omega)$. The tensor product built above for real Clifford algebras extends naturally to their complexifications, with complex numbers acting "diagonally": $\lambda \equiv \lambda \hat{\otimes} 1 + 1 \hat{\otimes} \lambda$. The extended $\alpha$ and $\alpha'$ obey the same tensor product rules as above. Charge conjugation extends trivially: $\overline{\Omega \hat{\otimes} \Theta} = \overline{\Omega} \hat{\otimes} \overline{\Theta}$.

In this algebra, one can define an element called a \emph{chirality} operator:
\begin{equation}
\chi = \pm i^{(p-q)/2} \gamma^1 ... \gamma^d.
\label{ChiralityDef}
\end{equation}
This element has the same form in all pseudo-orthonormal bases, and is thus well-defined. The sign depends on the orientation of the basis with respect to some chosen orientation for the vector space $V$. Its main property is that it anticommutes with all generators of the Clifford algebra. As a result, for all homogeneous $\Omega \in \mathbb{C}\mathrm{l}(V)$, we have:
\begin{equation}
\chi \Omega = (-1)^{|\Omega|} \Omega \chi.
\end{equation}
The $i^{(p-q)/2}$ factor has been chosen so that:
\begin{equation}
\chi^2 = 1
\end{equation}
would hold. Finally, for a tensor product of Clifford algebras, the total chirality operator is given by:
\begin{equation}
\chi_{V \oplus W} = \pm \chi_V \hat{\otimes} \chi_W.
\end{equation}
where $\chi_V$ and $\chi_W$ are the chirality operators for $\mathrm{Cl}(V)$ and $\mathrm{Cl}(W)$ respectively. Indeed, let $(v^a)_a$ be a pseudo-orthonormal basis of $V$, and $(w^b)_b$ be a pseudo-orthonormal basis of $W$. Then $(v^1 \oplus 0, \dots, v^d \oplus 0, 0 \oplus w^1, \dots, 0 \oplus w^{d'})$ is a pseudo-orthonormal basis of $V \oplus W$. The chirality operator is given by:
\begin{equation*}
\begin{aligned}
\chi_{V \oplus W} &= \pm i^{(p+p'-q-q')/2} \gamma(v^1 \oplus 0) \dots \gamma(v^d \oplus 0) \gamma(0 \oplus w^1) \dots \gamma(0 \oplus w^{d'}) \\
&= \pm i^{(p+p'-q-q')/2} (\gamma(v^1) \hat{\otimes} 1) \dots (\gamma(v^d) \hat{\otimes} 1) (1 \hat{\otimes} \gamma(w^1)) \dots (1 \hat{\otimes} \gamma(w^{d'})) \\
&= \pm i^{(p+p'-q-q')/2} (\gamma(v^1) \dots \gamma(v^d)) \hat{\otimes} (\gamma(w^1) \dots \gamma(w^{d'})) \\
&= \pm (i^{(p-q)/2} \gamma(v^1) \dots \gamma(v^d)) \hat{\otimes} (i^{(p'-q')/2} \gamma(w^1) \dots \gamma(w^{d'})) \\
\chi_{V \oplus W} &= \pm \chi_V \hat{\otimes} \chi_W.
\end{aligned}
\end{equation*}

\section{Spinor Spaces and the Robinson Product} \label{SectionSpin}

As a result of the previous classification, the algebras $\mathrm{Cl}(q,p)$ and $\mathbb{C}\mathrm{l}(q,p)$ are central simple algebras (CSA), and their irreducible complex representations are necessarily on vector spaces of the form $S \cong \mathbb{C}^D$, called spaces of (Dirac) spinors. Because these representations are faithful, we will often omit them when acting with Clifford algebra elements on spinors. Since $\mathrm{M}_D(\mathbb{C})$ was constructed as a complexification of either $\mathrm{M}_D(\mathbb{R})$ or $\mathrm{M}_{D/2}(\mathbb{H})$ (to which $\mathrm{Cl}(q,p)$ can be identified), there exists a representation $\pi_0 : \mathbb{C}\mathrm{l}(q,p) \longrightarrow \mathrm{End}_{\mathbb{C}}(S)$ and a basis of $S$, such that $\mathrm{Cl}(q,p)$ is mapped to either $\mathrm{M}_D(\mathbb{R})$ or $\mathrm{M}_{D/2}(\mathbb{H})$, depending on the value of $p-q$. In general, a representation $\pi$ of $\mathbb{C}\mathrm{l}(q,p)$ on $S$ does not satisfy such a condition, and most representations used in physics (Weyl, Dirac, imaginary Majorana) do not. For future use, we indicate that $\mathbb{H}$ is embedded into its complexification $\mathrm{M}_2(\mathbb{C})$ through the following algebra morphism:
\begin{equation*}
q=\alpha+\beta j \longmapsto
\begin{pmatrix}
\alpha & \beta \\
-\overline{\beta} & \overline{\alpha}	
\end{pmatrix}
\in \mathrm{M}_2(\mathbb{C}).
\end{equation*}

Let $\mathrm{tr}$ be the usual trace of operators on $S$. It is well-known that:
\begin{equation*}
\begin{aligned}
\mathrm{tr}(1) &= D \\
\mathrm{tr}(\gamma^I) &= 0 \text{ if $I \neq \emptyset$}.
\end{aligned}
\end{equation*}
In particular, we have: $\mathrm{tr}(\chi) = 0$. 

Since the chirality operator $\chi$ implements grading on the Clifford algebra through the relation $\chi \Omega \chi = (-1)^{|\Omega|} \Omega$, we can use it to lift the grading to spinors. Indeed, $\chi$ is an involution: $\chi^2=1$. As a result, $S$ splits into two eigenspaces of $\chi$ of eigenvalues $\pm 1$:
\begin{equation}
\begin{aligned}
S &= S^- \oplus S^+ \\
S^\pm &= \mathrm{Ker}(\chi \mp 1).
\end{aligned}
\end{equation}
The elements of $S^\pm$ are called chiral spinors, or Weyl spinors. We call the elements of $S^+$ (resp. $S^-$) right (resp. left) spinors. From $\mathrm{tr}(\chi) = 0$, we deduce that $S^+$ and $S^-$ have both the same dimension $D/2$. The grading of a spinor $\psi$ will be denoted $|\psi | \in \{0,1\}$. We thus have:
\begin{equation}
\chi \psi = (-1)^{|\psi|} \psi.
\end{equation}

From the tensor product rule \eqref{TensorCliffordDef} for Clifford algebras, we see that we can represent $\mathrm{Cl}(V \oplus W)$ on $S_V \otimes S_W$, where $S_V$ and $S_W$ are spinor spaces for $\mathrm{Cl}(V)$ and $\mathrm{Cl}(W)$ respectively. For this representation to be consistent with the graded tensor product, it must be itself a graded tensor product of representations. This uses the lift of the gradings of $\mathrm{Cl}(V)$ and $\mathrm{Cl}(W)$ to $S_V$ and $S_W$, and the rule is the following:
\begin{equation}
(\omega \hat{\otimes} \theta) (\psi \hat{\otimes} \varphi) = (-1)^{|\theta||\psi|}(\omega \psi \hat{\otimes} \theta \varphi)
\end{equation}
for any $\omega \hat{\otimes} \theta \in \mathrm{Cl}(V \oplus W)$ and $\psi \hat{\otimes} \varphi \in S_V \hat{\otimes} S_W$. Note that the tensor product of the spinor spaces is denoted as a graded tensor product, as a reminder to use the graded product of representations above. The total representation space is thus:
\begin{equation}
S_{V \oplus W} \cong S_V \hat{\otimes} S_W,
\end{equation}
Dimension counting then proves that this is indeed a spinor space. Its chiral subspaces are:
\begin{equation*}
\begin{aligned}
S_{V \oplus W}^{+} &\cong S_{V}^{+} \otimes S_{W}^{+} \oplus S_{V}^{-} \otimes S_{W}^{-} \\
S_{V \oplus W}^{-} &\cong S_{V}^{+} \otimes S_{W}^{-} \oplus S_{V}^{-} \otimes S_{W}^{+}
\end{aligned}
\end{equation*}
The graded tensor product of representations extends of course to complexified Clifford algebras.

\subsection{A Canonical Adjunction}

Charge conjugation and the main anti-automorphism $\alpha$ commute, and can be combined to define the following hermitian adjunction (\textit{i.e.} involutive antilinear anti-automorphism) on $\mathbb{C}\mathrm{l}(V)$:
\begin{equation*}
\Omega \longmapsto \Omega^\times = \alpha(\overline{\Omega}),
\end{equation*}
where $\Omega \in \mathbb{C}\mathrm{l}(V)$. It is the unique hermitian adjunction on $\mathbb{C}\mathrm{l}(V)$ that leaves its generators invariant:
\begin{equation}
\gamma(u)^\times = \gamma(u).
\end{equation}
Its action extends to the basis elements as:
\begin{equation*}
(\gamma^I)^\times = (-1)^{{|I|}({|I|}-1)/2} \gamma^I.
\end{equation*}
In particular, its action on chirality is given by:
\begin{equation}
\chi^\times = (-1)^q \chi.
\end{equation}
Can this adjunction be lifted to spinors, the same way the hermitian adjunction of matrices can be extended to vectors? The answer is yes, following Robinson \cite{Robinson}:

\begin{thm} P.264, \cite{Robinson}

There exists a unique (up to a non-vanishing real factor) non-degenerate hermitian form $(\cdot, \cdot)$ on $S$ such that, for all $\psi, \varphi \in S$ and $\Omega \in \mathbb{C}\mathrm{l}(V)$:
\begin{equation}
(\psi, \Omega \varphi) = (\Omega^\times \psi, \varphi).
\label{RobinsonDefinition}
\end{equation}
Equivalently, there exists a unique non-degenerate hermitian form on $S$ such that, for all $u \in V$:
\begin{equation*}
(\psi, \gamma(u) \varphi) = (\gamma(u) \psi, \varphi).
\end{equation*}

\end{thm}

Let us list a few properties of this Robinson product (see \cite{Robinson} for proof):
\begin{itemize}
	\item If $g$ is positive definite (that is, if $q=0$), then the Robinson product is definite. It can then be chosen positive definite by a redefinition.
	\item If $q>0$, then the Robinson product is indefinite of signature $(D/2, D/2)$.
	\item The space $S$ equipped with $(\cdot, \cdot)$ is thus a finite-dimensional Krein space.
	\item In particular, if $p=0$, then the Robinson product is definite with opposite signs on $S^+$ and $S^-$. As a result $S = S^- \oplus S^+$ is a fundamental decomposition, and the corresponding fundamental symmetry is $\chi$ if $(\cdot, \cdot)$ is positive on $S^+$, and $-\chi$ if $(\cdot, \cdot)$ is negative on $S^+$.
\end{itemize}
Additionally:	
\begin{itemize}	
	\item If $p,q$ are even, then $S^+$ and $S^-$ are mutually orthogonal.
	\item If $p,q$ are odd, then $S^+$ and $S^-$ are self-orthogonal.
	\item As a result, the product $(\psi, \varphi)$ is nonvanishing only if $|\psi|+|\varphi| = q \modulo 2$.
\end{itemize}

Let us now construct fundamental symmetries for the Robinson product. Let $(e^a)_a$ be a pseudo-orthonormal basis of $V$ as in the beginning of this chapter, and $\gamma^a = \gamma(e^a)$.
\begin{thm} \label{RobinsonFundamentalSymmetryTheorem}
There is a sign for which the operator:
\begin{equation}
\eta_+ = 
\begin{cases}
\pm \chi_- &\text{ if $p,q$ are even,} \\
\pm i\chi_+ &\text{ if $p,q$ are odd,}
\end{cases}
\label{RobinsonFundamentalSymmetry}
\end{equation}
is a fundamental symmetry for $(\cdot, \cdot)$, where we use the "partial chirality" operators:
\begin{equation}
\begin{aligned}
\chi_- &= i^{q/2} \gamma^1...\gamma^q \\
\chi_+ &= i^{p/2} \gamma^{q+1}...\gamma^d.
\end{aligned}
\end{equation}

\end{thm}
Note that the partial chirality operators are such that:
\begin{equation*}
\begin{aligned}
\chi_+^2 &= \chi_-^2 = \pm 1 \\
\chi &= \pm \chi_+ \chi_-.
\end{aligned}
\end{equation*}
These partial chirality operators will be very important in the next chapter, where they will be used to measure space- and time-orientability of manifolds.

\begin{demo}
Consider a representation of $\mathbb{C}\mathrm{l}(V)$ on $S$ and an inner product $(\psi, \varphi) \mapsto \psi^\dagger \cdot \varphi$, such that:
\begin{equation*}
(\gamma^a)^\dagger = \epsilon_a \gamma^a,
\end{equation*}
where $\epsilon_a = (\gamma^a)^2$. Notice that this means that the $\gamma^a$ are unitary. Such a representation can always be built (see the last section of this chapter for example). Let $(\cdot,\cdot)$ be a Robinson product on $S$. Then there exists an invertible self-adjoint operator $F=F^\dagger$ acting on $S$ such that:
\begin{equation*}
(\psi, \varphi) = \psi^\dagger F \varphi.
\end{equation*}  
Let $T \in \mathrm{End}(S)$. Then:
\begin{equation*}
\begin{aligned}
(\psi, T \varphi) &= \psi^\dagger F T \varphi \\
&= \psi^\dagger FTF^{-1} F \varphi \\
&= (F^{-1} T^\dagger F\psi)^\dagger F \varphi \\
(\psi, T \varphi) &= (F^{-1} T^\dagger F \psi, \varphi).
\end{aligned}
\end{equation*}
The Robinson adjoint of $T$ is thus $T^\times = F^{-1} T^\dagger F$. We know that $\gamma^a$ is self-adjoint. This can be now written: $\gamma^a  = F^{-1} (\gamma^a)^\dagger
 F$. Using the unitarity of the $\gamma^a$, we rewrite the equation as:
\begin{equation}
\gamma^a F \gamma^a  = F,
\label{JEq}
\end{equation}
for all $a$ (no summation). The operator $F$ is a linear operator acting on $S$, and thus an element of $\mathbb{C}\mathrm{l}(q,p)$. As such, it can be written as:
\begin{equation}
F = \sum_{I \subset \llbracket 1,d \rrbracket} \alpha_I \gamma^I.
\label{JBasis}
\end{equation}
Let $a \in \llbracket 1,d \rrbracket, I \subset \llbracket 1,d \rrbracket$. If $a \notin I$, then $\gamma^a$ anti-commutes with all the $\gamma^b$ that appear in $\gamma^I$. We thus have that $\gamma^a \gamma^I \gamma^a = (-1)^{|I|} (\gamma^a)^2 \gamma^I = (-1)^{|I|} \epsilon_a \gamma^I$. If $a \in I$, then it anti-commutes with all $\gamma^b$ but itself, and there is one less minus sign: $\gamma^a \gamma^I \gamma^a = (-1)^{|I|-1} \epsilon_a \gamma^I$. We summarize this in the following way:
\begin{equation*}
\gamma^a \gamma^I \gamma^a = \sigma(a,I) \gamma^I,
\end{equation*}
where:
\begin{equation*}
\sigma(a,I) = \begin{cases}
(-1)^{|I|+1} \epsilon_a &\text{  if  } a \in I, \\
(-1)^{|I|} \epsilon_a &\text{  otherwise}.
\end{cases}
\end{equation*}
We thus have:
\begin{equation*}
\gamma^a F \gamma^a = \sum_{I \in \llbracket 1,n \rrbracket} \sigma(a,I) \alpha_I \gamma^I,
\end{equation*}
Replacing this in (\ref{JEq}), and comparing with (\ref{JBasis}), it is clear that the only authorized $\gamma^I$ are those for which $\sigma(a,I)=+1$ for all $a$. 

Let us at first assume that $|I|$ is odd. Then we have:
\begin{equation*}
\sigma(a,I) = \begin{cases}
+\epsilon_a &\text{  if  } a \in I, \\
-\epsilon_a &\text{  otherwise}.
\end{cases}
\end{equation*}
The values of $\sigma(a,I)$ are given in table \ref{sigma}. According to the table and the requirement $\sigma(a,I)=+1$, $a \in I$ if and only if $a \in \llbracket q+1,d \rrbracket$. We conclude that $I=\llbracket q+1,d \rrbracket$. The set $I$ being of odd cardinality, this solution is only valid when $q$ and $p$ are odd.

\begin{table}[!h]
	\centering
		\begin{tabular}{|c|c|c|}
			\hline
 			& $a \in I$ & $a \notin I$ \\ \hline
 			$a \in \llbracket 1,q \rrbracket$ & -1 & +1 \\ \hline
 			$a \in \llbracket q+1,d \rrbracket$ & +1 & -1 \\ \hline
		\end{tabular}
		\caption{Values of $\sigma(a,I)$}
		\label{sigma}
\end{table}

If one assumes $|I|$ to be even, one can follow a similar reasoning, and one finds that $I=\llbracket 1,q \rrbracket$. This is of course possible only when $q$ is even. There are thus two possible cases:

\begin{itemize}

\item Odd $q,p$: Then $F$ can only be of the form:
\begin{equation*}
F = \alpha \gamma^{q+1}...\gamma^d,
\end{equation*}
with $\alpha \in \mathbb{C}$. One can prove that:
\begin{equation*}
F^\dagger = (-1)^{p(p-1)/2} \overline{\alpha} \gamma^{t+1}...\gamma^d.
\end{equation*}
Asking that $F$ be self-adjoint gives us the solution:
\begin{equation}
F = \lambda i^{p(p-1)/2} \gamma^{q+1}...\gamma^d,
\label{JEvenT}
\end{equation}
where $\lambda$ is a nonzero real number.

\item Even $q,p$: A similar reasoning gives us the solution:
\begin{equation}
F = \lambda i^{q(q+1)/2} \gamma^1...\gamma^q.
\label{JOddT}
\end{equation}

\end{itemize}

To summarize:
\begin{equation*}
F =
\begin{cases}
\lambda i^{(p-1)/2} \gamma^{q+1}...\gamma^d &\text{for odd  } q,p \\
\lambda i^{q/2} \gamma^1...\gamma^q &\text{for even  } q,p. \\
\end{cases}
\end{equation*}
For both cases, it is straightforward to prove that $F^2 = \lambda^2$. We define:
\begin{equation*}
\eta_+ = \frac{F}{|\lambda|} = 
\begin{cases}
\mathrm{sgn}(\lambda) i^{(p-1)/2} \gamma^{q+1}...\gamma^d &\text{for odd  } q,p \\
\mathrm{sgn}(\lambda) i^{q/2} \gamma^1...\gamma^q &\text{for even  } q,p. \\
\end{cases}
\end{equation*}
so that $\eta_+^2 = 1$. Moreover, one can prove that:
\begin{equation*}
\langle \psi, \varphi \rangle_{\eta_+} = (\psi, \eta_+ \varphi) = |\lambda| \psi^\dagger \varphi.
\end{equation*} 
This proves that $\langle \cdot, \cdot \rangle_{\eta_+}$ is a positive definite hermitian form. Using $\eta_+^2 = 1$, one can construct an eigendecomposition of $S$. This decomposition is clearly a fundamental decomposition of $S$, whose fundamental symmetry is $\eta_+$.

To conclude, let us note that $(\cdot,\cdot)$ depends on the representation used, and let us prove that $\eta_+$ is a fundamental symmetry in all representations. To this end, we will make the representations explicit. Let $\pi$ be the representation used above, and $\pi'$ be any other one. Since the two representations are irreducible, there exists an invertible linear operator $T$ on $S$ such that:
\begin{equation*}
\pi'(\Omega) = T \pi(\Omega) T^{-1}.
\end{equation*}
One can also prove that $(\cdot, \cdot)$ is a Robinson product for $\pi$ if and only if $(\cdot, \cdot)'=(T^{-1} \cdot, T^{-1} \cdot)$ is a Robinson product for $\pi'$. Indeed, let us assume that $(\cdot, \cdot)$ is a Robinson product for $\pi$. Since $T$ is invertible, it is easy to see that $(T^{-1} \cdot, T^{-1} \cdot)$ is a non-degenerate Hermitian form. Moreover, for any $\Omega \in \mathbb{C}\mathrm{l}(V)$ and $\psi,\varphi \in S$:
\begin{equation*}
\begin{aligned}
(\psi, \pi'(\Omega) \varphi)' &= (T^{-1} \psi, T^{-1} (T \pi(\Omega) T^{-1}) \varphi) \\
&= (T^{-1} \psi, \pi(\Omega) T^{-1} \varphi) \\
&= (\pi(\Omega^\times) T^{-1} \psi, T^{-1} \varphi) \\
&= (T^{-1} (T \pi(\Omega^\times) T^{-1}) \psi, T^{-1} \varphi) \\
(\psi, \pi'(\Omega) \varphi)' &= (\pi'(\Omega^\times) \psi, \varphi)'.
\end{aligned}
\end{equation*}
The converse proof is identical. Now, we have:
\begin{equation*}
\begin{aligned}
(\psi, \pi'(\eta^+) \psi)' &= (T^{-1} \psi, T^{-1} (T \pi(\eta^+) T^{-1})\psi) \\
&= (T^{-1} \psi, \pi(\eta^+) (T^{-1}\psi)) \\
(\psi, \pi'(\eta^+) \psi)' &= \langle T^{-1}\psi, T^{-1}\psi \rangle_{\eta_+}.
\end{aligned}
\end{equation*}
Since the latter is positive definite, then so is the former.

\qed
\end{demo}

Another important result of this section is the following theorem:
\begin{thm}
The Robinson product on $S_{V \oplus W}$ is given in terms of $(\cdot,\cdot)_V$ on $S_V$ and $(\cdot,\cdot)_W$ on $S_W$ by:
\begin{equation}
(\psi_1 \hat{\otimes} \varphi_1, \psi_2 \hat{\otimes} \varphi_2)_{V \oplus W} = (\psi_1, \psi_2)_V (\varphi_1, \beta \varphi_2)_W
\label{RobinsonTensorProduct}
\end{equation}
up to a real factor, where:
\begin{equation}
\beta = (i^{q'} \chi_W)^{q} = \begin{cases}
1 &\text{if $q$ is even,} \\
\chi_W &\text{if $q$ is odd and $q'$ is even,} \\
i \chi_W &\text{if $q$ and $q'$ are both odd,}
\end{cases}
\end{equation}
and $\psi_1 \hat{\otimes} \varphi_1, \psi_2 \hat{\otimes} \varphi_2 \in S_{V \oplus W}$.

\end{thm}

Note that the case where $q$ is even has already been studied in \cite{Robinson}. To prove the theorem, we will first need the following lemma:
\begin{lem}
The Krein-adjunction for the hermitian form on the right-hand side of equation \eqref{RobinsonTensorProduct} is given by:
\begin{equation}
(\Omega \hat{\otimes} \Theta)^\times = (-1)^{|\Omega||\Theta|} \Omega^\times \hat{\otimes} \Theta^\times
\end{equation}
where $\Omega \in \mathbb{C}\mathrm{l}(V)$ and $\Theta \in \mathbb{C}\mathrm{l}(W)$.

\end{lem}

\begin{demo}
Let $(\cdot,\cdot)$ be the tensor sesquilinear form on the right-hand side of equation \eqref{RobinsonTensorProduct}. First, note that $\chi_W^\times = (-1)^{q'} \chi_W$ implies that $i^{q'} \chi_W$ is (Krein-)self-adjoint. It is also invertible. So is $\beta$ as a result, and $(\cdot, \cdot)$ is a non-degenerate Hermitian form for which one can define a Krein-adjoint. Let $\Omega \in \mathbb{C}\mathrm{l}(V), \Theta \in \mathbb{C}\mathrm{l}(W)$, and $\psi_i \in S_V, \varphi_i \in S_W$. We have:
\begin{equation*}
\begin{aligned}
((\Omega \hat{\otimes} \Theta)(\psi_1 \hat{\otimes} \varphi_1), \psi_2 \hat{\otimes} \varphi_2) &= (-1)^{|\Theta||\psi_1|} (\Omega \psi_1 \hat{\otimes} \Theta \varphi_1, \psi_2 \hat{\otimes} \varphi_2)_{V \oplus W} \\
&= (-1)^{|\Theta||\psi_1|} (\Omega \psi_1, \psi_2)_V (\Theta \varphi_1, \beta \varphi_2)_W \\
&= (-1)^{|\Theta||\psi_1|} (\psi_1, \Omega^\times \psi_2)_V (\varphi_1, \Theta^\times \beta \varphi_2)_W. \\
\end{aligned}
\end{equation*}
From the definition of $\beta$, one can can see that $\Theta \beta = (-1)^{q |\Theta|} \beta \Theta$. The same holds for $\Theta^\times$, thanks to the self-adjointness of $\beta$. We now have:
\begin{equation*}
\begin{aligned}
((\Omega \hat{\otimes} \Theta)(\psi_1 \hat{\otimes} \varphi_1), \psi_2 \hat{\otimes} \varphi_2)_{V \oplus W} &= (-1)^{|\Theta|(|\psi_1|+q)} (\psi_1, \Omega^\times \psi_2)_V (\varphi_1, \beta \Theta^\times \varphi_2)_W \\
&= (-1)^{|\Theta|(|\psi_1|+q)} (\psi_1 \hat{\otimes} \varphi_1, \Omega^\times\psi_2 \hat{\otimes} \Theta^\times \varphi_2) \\
&= (-1)^{|\Theta|(|\psi_1|+|\psi_2|+q)} (\psi_1 \hat{\otimes} \varphi_1, (\Omega^\times \hat{\otimes} \Theta^\times) (\psi_2 \hat{\otimes} \varphi_2)).
\end{aligned}
\end{equation*}
The product $(\Omega \psi_1, \psi_2)_V$ is non vanishing only if $|\Omega|+|\psi_1|+|\psi_2| = q \modulo 2$. This implies that $(-1)^{|\Theta|(|\psi_1|+|\psi_2|+q)} = (-1)^{|\Omega||\Theta|}$, and the result follows.

\qed
\end{demo}

We can now prove the theorem:
\begin{demo}
Let $v \oplus w \in V \oplus W$. We have:
\begin{equation*}
\begin{aligned}
\gamma(v \oplus w)^\times &= (\gamma(v) \hat{\otimes} 1 + 1 \hat{\otimes} \gamma(w))^\times \\
&= (-1)^{|\gamma(v)||1|} \gamma(v)^\times \hat{\otimes} 1 + (-1)^{|\gamma(w)||1|} 1 \hat{\otimes} \gamma(w)^\times \\
&= \gamma(v) \hat{\otimes} 1 + 1 \hat{\otimes} \gamma(w) \\
\gamma(v \oplus w)^\times &= \gamma(v \oplus w)
\end{aligned}
\end{equation*}
which proves that it is indeed the Robinson product.

\qed
\end{demo}

\subsection{Graded Adjunction}

Similarly to what was done above with $\alpha$, one can define another adjunction by combining charge conjugation with $\alpha'$. The result is the following adjunction:
\begin{equation*}
\Omega \longmapsto \Omega^+ = \alpha'(\overline{\Omega}),
\end{equation*}
where $\Omega \in \mathbb{C}\mathrm{l}(V)$. It is the unique hermitian adjunction on $\mathbb{C}\mathrm{l}(V)$ that flips its generators:
\begin{equation}
\gamma(u)^+ = -\gamma(u).
\end{equation}
It is related to the previous adjunction by:
\begin{equation*}
\Omega^+ = (-1)^{|\Omega |} \Omega^\times.
\end{equation*}
Its action on the basis elements is:
\begin{equation*}
(\gamma^I)^\times = (-1)^{{|I|}({|I|}+1)/2} \gamma^I.
\end{equation*}
In particular, its action on chirality is given by:
\begin{equation}
\chi^+ = (-1)^q \chi.
\end{equation}

This "graded" adjunction lifts to spinors to define what we call the anti-Robinson product\footnote{This is actually the product Robinson studies in his paper \cite{Robinson}. But the methods and results are identical for both graded and non-graded adjunctions.}:

\begin{thm} P.264, \cite{Robinson}

There exists a unique (up to a non-vanishing real factor) non-degenerate hermitian form $[\cdot, \cdot]$ on $S$ such that, for all $\psi, \varphi \in S$ and $\Omega \in \mathbb{C}\mathrm{l}(V)$:
\begin{equation}
[\psi, \Omega \varphi] = [\Omega^+ \psi, \varphi].
\label{AntiRobinsonDefinition}
\end{equation}
Equivalently, there exists a unique non-degenerate hermitian form on $S$ such that, for all $u \in V$:
\begin{equation*}
[\psi, \gamma(u) \varphi] = -[\gamma(u) \psi, \varphi].
\end{equation*}

\end{thm}

Robinson and anti-Robinson product are related by the following result:
\begin{thm}
The hermitian form $(\cdot, \cdot)$ is a Robinson product if and only if $(\cdot, i^q \chi \cdot)$ is an anti-Robinson product.
\end{thm}

\begin{demo}
Let $(\cdot, \cdot)$ be a Robinson product. Since $i^q \chi$ is Krein-adjoint and invertible, $(\cdot, i^q \chi \cdot)$ is a non-degenerate hermitian form. Let us compute the adjoint of $\Omega \in \mathbb{C}\mathrm{l}(q,p)$ with respect to it:
\begin{equation*}
\begin{aligned}
(\psi, i^q \chi \Omega \varphi) &= (-1)^{|\Omega|} (\psi, \Omega i^q \chi \varphi) \\
&= (-1)^{|\Omega|} (\Omega^\times \psi, i^q \chi \varphi) \\
(\psi, i^q \chi \Omega \varphi) &= (\Omega^+ \psi, i^q \chi \varphi),
\end{aligned}
\end{equation*}
proving that $(\cdot, i^q \chi \cdot)$ is an anti-Robinson product. The converse proof is identical.

\qed
\end{demo}

This means that up to a real non-vanishing factor, Robinson and anti-Robinson product are related by:
\begin{equation}
[\cdot, \cdot] = (\cdot, i^q \chi \cdot).
\label{RobinsonAntiRobinson}
\end{equation}
All properties of the anti-Robinson product can thus be deduced from the properties of the Robinson product (and conversely):
\begin{itemize}
	\item If $g$ is negative definite (that is, if $p=0$), then the anti-Robinson product is definite. It can then be chosen positive definite by a redefinition.
	\item If $p>0$, then the anti-Robinson product is indefinite of signature $(D/2, D/2)$.
	\item The space $S$ equipped with $[\cdot, \cdot]$ is a finite-dimensional Krein space.
	\item In particular, if $q=0$, then the anti-Robinson product is definite with opposite signs on $S^+$ and $S^-$.
\end{itemize}
Additionally:	
\begin{itemize}	
	\item If $p,q$ are even, then $S^+$ and $S^-$ are mutually orthogonal.
	\item If $p,q$ are odd, then $S^+$ and $S^-$ are self-orthogonal.
	\item As a result, the product $[\psi, \varphi]$ is nonvanishing only if $|\psi|+|\varphi| = q \modulo 2$.
\end{itemize}

The fundamental symmetries of $[\cdot, \cdot]$ can be deduced from those of $(\cdot, \cdot)$ by a multiplication by $i^q \chi$. Indeed, let $\eta_+$ be a fundamental symmetry of $(\cdot, \cdot)$. Then the hermitian form:
\begin{equation*}
[\cdot, (-i)^q \chi \eta_+ \cdot] = (\cdot, \eta_+ \cdot)
\end{equation*}
is positive definite. We infer that $(-i)^q \chi \eta_+$ is a fundamental symmetry for $[\cdot, \cdot]$. This results in the following theorem:
\begin{thm}
Let $(e^a)_a$ be a pseudo-orthonormal basis of $V$ as in the beginning of this chapter, and $\gamma^a = \gamma(e^a)$. There is a sign for which the operator:
\begin{equation}
\eta_- = (-i)^q \chi \eta_+ =
\begin{cases}
\pm \chi_+ &\text{ if $p,q$ are even,} \\
\pm \chi_- &\text{ if $p,q$ are odd,}
\end{cases}
\label{AntiRobinsonFundamentalSymmetry}
\end{equation}
is a fundamental symmetry for $[\cdot, \cdot]$

\end{thm}

Finally, let us construct a tensor product rule for the anti-Robinson product and adjunction:
\begin{thm}
The anti-Robinson product on $S_{V \oplus W}$ is given in terms of $[\cdot,\cdot]_V$ on $S_V$ and $[\cdot,\cdot]_W$ on $S_W$ by:
\begin{equation}
[\psi_1 \hat{\otimes} \varphi_1, \psi_2 \hat{\otimes} \varphi_2]_{V \oplus W} = [\psi_1, \psi_2]_V [\varphi_1, \beta \varphi_2]_W
\label{AntiRobinsonTensorProduct}
\end{equation}
up to a real factor, where $\beta = (i^{q'} \chi_W)^{q}$ and $\psi_1 \hat{\otimes} \varphi_1, \psi_2 \hat{\otimes} \varphi_2 \in S_{V \oplus W}$.

\end{thm}

\begin{demo}
Using \eqref{RobinsonTensorProduct}, we find:
\begin{equation*}
\begin{aligned}
\left[\psi_1 \hat{\otimes} \varphi_1, \psi_2 \hat{\otimes} \varphi_2\right]_{V \oplus W} &= (\psi_1 \hat{\otimes} \varphi_1, i^{q+q'}(\chi_V \hat{\otimes} \chi_W)(\psi_2 \hat{\otimes} \varphi_2))_{V \oplus W} \\
&= (\psi_1 \hat{\otimes} \varphi_1, i^q \chi_V \psi_2 \hat{\otimes} i^{q'} \chi_W \varphi_2)_{V \oplus W} \\
&= (\psi_1, i^q \chi_V \psi_2)_V (\varphi_1, \beta i^{q'} \chi_W \varphi_2)_W \\
[\psi_1 \hat{\otimes} \varphi_1, \psi_2 \hat{\otimes} \varphi_2]_{V \oplus W}  &= [\psi_1, \psi_2]_V [\varphi_1, \beta \varphi_2]_W
\end{aligned}
\end{equation*}

\qed
\end{demo}

As for the adjunction, one can prove that:
\begin{prop}
The Krein-adjunction for $[\cdot, \cdot]_{V \oplus W}$ is given by:
\begin{equation}
(\Omega \hat{\otimes} \Theta)^+ = (-1)^{|\Omega||\Theta|} \Omega^+ \hat{\otimes} \Theta^+
\end{equation}
where $\Omega \in \mathbb{C}\mathrm{l}(V)$ and $\Theta \in \mathbb{C}\mathrm{l}(W)$.

\end{prop}

\section{Pin and Spin Groups}

In this section, we present the so-called Pin and Spin groups. See \cite{LawsonMichelsohn} for a detailed study. Let us define them:

\begin{defn}
The Pin group $\mathrm{Pin}(V)$, also denoted $\mathrm{Pin}(q,p)$, is the group generated by all pseudo-normalized generators of $\mathrm{Cl}(V)$:
\begin{equation}
\mathrm{Pin}(V) = \{ \gamma(u_1) \dots \gamma(u_n) | g(u_i, u_i) = \pm 1 \}
\label{PinGroup}
\end{equation}
The Spin group $\mathrm{Spin}(V)$, also denoted $\mathrm{Spin}(q,p)$, is the even part of the Pin group:
\begin{equation}
\mathrm{Spin}(V) = \{ \gamma(u_1) \dots \gamma(u_{2n}) | n \in \mathbb{N}, g(u_i, u_i) = \pm 1 \}
\label{SpinGroup}
\end{equation}

\end{defn}

These two groups are sub-groups of $\mathrm{Cl}(V)^\ast$, the group of invertible elements of the Clifford algebra $\mathrm{Cl}(V)$:
\begin{equation*}
\mathrm{Spin}(V) \subset \mathrm{Pin}(V) \subset \mathrm{Cl}(V)^\ast.
\end{equation*}
They can thus be represented on a spinor space $S$. One can prove that $S^+$ are $S^-$ are irreducible representations of $\mathrm{Spin}(V)$, while $S$ is an irreducible representation of $\mathrm{Pin}(V)$.

\subsection{Adjoint Action and the Orthogonal Group}

The Pin group can be constructed as a lift of orthogonal endomorphisms of $V$ to spinors. Indeed, let $u \in V$ such that $g(u,u) = \pm 1$. One can compute the adjoint action on $\gamma(u)$ on a generator of the Clifford algebra, and one finds that, for all $v \in V$:
\begin{equation*}
\mathrm{Ad}_{\gamma(u)} \gamma(v) = -\gamma(R_u v),
\end{equation*}
where $R_u$ is the reflection with respect to the subspace orthogonal to $u$:
\begin{equation*}
R_u v = v - 2 \frac{g(u,v)}{g(u,u)} u.
\end{equation*}
The endomorphism $R_u$ preserves the product $g$, and we thus have that $R_u \in \mathrm{O}(V)$. Since the Pin group is generated by pseudo-normalized vectors such as $\gamma(u)$, this adjoint action can be extended to all its elements, and to any $\omega \in \mathrm{Pin}(V)$, one can associate a unique $R_\omega \in \mathrm{O}(V)$ such that, for all $v \in V$:
\begin{equation*}
\mathrm{Ad}_{\omega} \gamma(v) = (-1)^{|\omega|}\gamma(R_\omega v).
\end{equation*}
Conversely, the orthogonal group is generated by reflections of the form $R_u$. As a result, the map $\omega \mapsto R_\omega$ is surjective. This is what we mean when we say that the Pin groups lifts the orthogonal groups to spinors.

\subsection{Topology}

The topology of the Pin and Spin groups depends on the signature $(p,q)$:

\textbf{If $g$ is indefinite:} then $\mathrm{Pin}(V)$ has four connected components. The group is split according to the parity of the number of positive and negative generators $\gamma(u)$:
\begin{itemize}

	\item The connected component of the identity, which is generated by an even number of positive generators, and an even number of negative generators: 
	\begin{equation}
	\mathrm{Pin}_0(V) = \{ \gamma(u_1) \dots \gamma(v_1) \dots \gamma(u_{2m}) \dots \gamma(v_{2n}) | m,n \in \mathbb{N}, g(u_i, u_i) = 1, g(v_i, v_i) = -1 \}.
	\end{equation}
	Note that all orders are allowed for the product.
	
	\item The negative-space-reversing\footnote{This alludes to the fact that the corresponding orthogonal transformation reverses the orientation of a negative subspace of $V$ in some appropriate fundamental decomposition.} component, which is generated by an even number of positive generators, and an odd number of negative generators:
	\begin{equation}
	\mathrm{Pin}_-(V) = \{ \gamma(u_1) \dots \gamma(v_1) \dots \gamma(u_{2m}) \dots \gamma(v_{2n+1}) | m,n \in \mathbb{N}, g(u_i, u_i) = 1, g(v_i, v_i) = -1 \}.
	\end{equation}
	
	\item The positive-space-reversing component, which is generated by an odd number of positive generators, and an even number of negative generators: 
	\begin{equation}
	\mathrm{Pin}_+(V) = \{ \gamma(u_1) \dots \gamma(v_1) \dots \gamma(u_{2m+1}) \dots \gamma(v_{2n}) | m,n \in \mathbb{N}, g(u_i, u_i) = 1, g(v_i, v_i) = -1 \}.
	\end{equation}
	
	\item The second orientation-preserving component, which is generated by an odd number of positive generators, and an odd number of negative generators:
	\begin{equation}
	\mathrm{Pin}_{-+}(V) = \{ \gamma(u_1) \dots \gamma(v_1) \dots \gamma(u_{2m+1}) \dots \gamma(v_{2n+1}) | m,n \in \mathbb{N}, g(u_i, u_i) = 1, g(v_i, v_i) = -1 \}.
	\end{equation}
	
\end{itemize}

The Spin group has two connected components:
\begin{itemize}
	
	\item The connected component of the identity: $\mathrm{Spin}_0(V) = \mathrm{Pin}_0(V)$.
	
	\item The orientation-reversing component: $\mathrm{Spin}_{-+}(V) = \mathrm{Pin}_{-+}(V)$.

\end{itemize}

\textbf{If $g$ is positive definite ($q=0$):} then $\mathrm{Pin}_-(V)$ and $\mathrm{Pin}_{-+}(V)$ are empty. As a result, $\mathrm{Spin}(V)$ is connected:
\begin{equation*}
\mathrm{Spin}(V) = \mathrm{Pin}_0(V);
\end{equation*}
while $\mathrm{Pin}(V)$ has two connected components:
\begin{itemize}

	\item The connected component of the identity, which is the Spin group itself: $\mathrm{Pin}_0(V) = \mathrm{Spin}(V)$.
	
	\item The odd, orientation-reversing part of the Pin group: $\mathrm{Pin}_+(V)$.
	
\end{itemize}

\textbf{If $g$ is negative definite ($p=0$):} then $\mathrm{Pin}_+(V)$ and $\mathrm{Pin}_{-+}(V)$ are empty. As a result, $\mathrm{Spin}(V)$ is connected:
\begin{equation*}
\mathrm{Spin}(V) = \mathrm{Pin}_0(V);
\end{equation*}
while $\mathrm{Pin}(V)$ has two connected components:
\begin{itemize}

	\item The connected component of the identity, which is the Spin group itself: $\mathrm{Pin}_0(V) = \mathrm{Spin}(V)$.
	
	\item The odd, orientation-reversing part of the Pin group: $\mathrm{Pin}_-(V)$.
	
\end{itemize}

In all cases above, the most distinguished part of the Pin and Spin groups is the connected component of the identity: $\mathrm{Spin}_0(V) = \mathrm{Pin}_0(V)$. One can prove that $S^+$ and $S^-$ are irreducible representations of $\mathrm{Spin}_0(V)$. Its Lie algebra is the space of "two-forms":
\begin{equation}
\mathrm{spin}(V) = \mathrm{Span}_\mathbb{R} ([\gamma(u), \gamma(v)])
\end{equation}
for which a basis is $(\gamma^a \gamma^b)_{a \leq b}$. The exponential map to $\mathrm{Spin}_0(V)$ is surjective, and as a result we can write that:
\begin{equation*}
\mathrm{Spin}_0(V) = \{e^{t_{ab} \gamma^a \gamma^b} | t \in \mathrm{M}_d(\mathbb{R}), \; t^T = -t \}.
\end{equation*}

\subsection{The Robinson product and the Pin group}

As we have seen above, the Pin group is the lift of the orthogonal group of $V$ to spinors. It is thus natural to ask whether the elements of the Pin group preserve the Robinson and anti-Robinson products. One arrives at the following proposition:

\begin{prop}
Let $\omega \in \mathrm{Pin}(V)$. Then:
\begin{equation*}
\omega^\times \omega = \begin{cases}
+1 &\text{if $\omega \in \mathrm{Pin}_0(V)$ or $\mathrm{Pin}_+(V)$} \\
-1 &\text{if $\omega \in \mathrm{Pin}_-(V)$ or $\mathrm{Pin}_{-+}(V)$},
\end{cases}
\end{equation*}
and:
\begin{equation*}
\omega^+ \omega = \begin{cases}
+1 &\text{if $\omega \in \mathrm{Pin}_0(V)$ or $\mathrm{Pin}_-(V)$} \\
-1 &\text{if $\omega \in \mathrm{Pin}_+(V)$ or $\mathrm{Pin}_{-+}(V)$},
\end{cases}
\end{equation*}

\end{prop}

\begin{demo}
Let $\omega = \gamma(u_1) \dots \gamma(u_n) \in \mathrm{Pin}(V)$, with $u_i \in V$ such that: $g(u_i, u_i) = \pm 1$. We have:
\begin{equation*}
\begin{aligned}
\omega^\times \omega &= (\gamma(u_1) \dots \gamma(u_n))^\times (\gamma(u_1) \dots \gamma(u_n)) \\
&= \gamma(u_n) \dots \gamma(u_1) \gamma(u_1) \dots \gamma(u_n) \\
\omega^\times \omega &= g(u_1, u_1) \dots g(u_n,u_n).
\end{aligned}
\end{equation*}
We infer from this that $\omega^\times \omega = +1$ if it is generated by an even number of negative vectors, and $\omega^\times \omega = -1$ if it is generated by an odd number of negative vectors.

Similarly,we have:
\begin{equation*}
\begin{aligned}
\omega^+ \omega &= (\gamma(u_1) \dots \gamma(u_n))^\times (\gamma(u_1) \dots \gamma(u_n)) \\
&= (-\gamma(u_n)) \dots (-\gamma(u_1)) \gamma(u_1) \dots \gamma(u_n) \\
\omega^\times \omega &= (-g(u_1, u_1)) \dots (-g(u_n,u_n)).
\end{aligned}
\end{equation*}
We infer that $\omega^+ \omega = +1$ if it is generated by an even number of positive vectors, and $\omega^+ \omega = -1$ if it is generated by an odd number of positive vectors.

\qed
\end{demo}

The following interesting corollary results from the previous proposition:
\begin{cor}
An element of the Pin group preserves both Robinson and anti-Robinson products if and only if it is in the connected component of the identity.
\end{cor}

\section{Charge Conjugation on Spinors} \label{SectionCC}

We now prove that charge conjugation can be lifted to spinors, and study the resulting operator. We refer the reader to \cite{GBVF} (chapters 5 and 9 in particular), from which we adapted a certain number of proofs.

An invertible antilinear operator $K$ acting on spinors is said to implement charge conjugation if it satisfies:
\begin{equation*}
\forall \Omega \in \mathbb{C}\mathrm{l}(V): \overline{\Omega} = K \Omega K^{-1}.
\end{equation*}
The real Clifford algebra $\mathrm{Cl}(V)$ can now be extracted from its complexification in the following way. 
\begin{equation*}
\forall \Omega \in \mathbb{C}\mathrm{l}(V): \Omega \in \mathrm{Cl}(V) \Leftrightarrow [K, \Omega] = 0.
\end{equation*}
Conversely, any antilinear operator that satisfies
\begin{equation*}
\forall \omega \in \mathrm{Cl}(V):  [K, \omega] = 0
\end{equation*}
implements charge conjugation. Indeed:
\begin{equation*}
K (\lambda \omega) K^{-1} = \overline{\lambda} K \omega K^{-1} = \overline{\lambda} \omega = \overline{\lambda \omega},
\end{equation*}
for all $\lambda \in \mathbb{C}$, $\omega \in \mathrm{Cl}(V)$. We obviously identified $\lambda \otimes \omega$ with $\lambda\omega$ using the trivial embedding of $\mathrm{Cl}(V)$ into its complexification.

\begin{thm} \label{CCThm}
There exists an invertible antilinear operator $J_+$ that implements charge conjugation on spinors:
\begin{equation}
\forall \Omega \in \mathbb{C}\mathrm{l}(V): \overline{\Omega} = J_+ \Omega J_+^{-1},
\label{CCJ}
\end{equation}
and such that:
\begin{equation}
J_+^2 = \epsilon
\end{equation}
where $\epsilon \in \{-1, 1\}$. The charge conjugation operator $J_+$ is unique up to a phase, and the sign $\epsilon$ is determined uniquely by $(q,p)$.
\end{thm}

\begin{demo}
Let $K'$ be any invertible antilinear operator on $S$. Then $\phi: \Omega \mapsto K'^{-1} \overline{\Omega} K'$ is an automorphism of $\mathbb{C}\mathrm{l}(V)$. Because it is a simple algebra, the automorphism $\phi$ has to be inner: there exists an invertible $T \in \mathbb{C}\mathrm{l}(V)$ such that: $\phi(\Omega) = T \Omega T^{-1}$. Let $K = K' T$. Then $K$ is an invertible antilinear operator such that:
\begin{equation*}
\overline{\Omega} = K \Omega K^{-1}.
\end{equation*}
We have proven that charge conjugation can be implemented on spinors with an invertible antilinear operator. We will now prove that this charge conjugation operator can be conveniently normalized.

From $\Omega = \overline{\overline{\Omega}}$, we infer that $\Omega = K^2 \Omega K^{-2}$. This implies that $K^2$ commutes with all elements of the complexified Clifford algebra. Since it is a linear operator, it is an element of the algebra. It is clearly in its center. We conclude that there exists a complex number $\mu \in \mathbb{C}$ such that: $K^2 = \mu$. We now define our final charge conjugation operator:
\begin{equation*}
J_+ = \frac{1}{|\mu|^{1/2}} K.
\end{equation*} 
We then have:
\begin{equation*}
\overline{\Omega} = J_+ \Omega J_+^{-1},
\end{equation*}
as well as:
\begin{equation*}
J_+^2 = \epsilon
\end{equation*}
where $\epsilon = \mathrm{sgn}(\mu)$.

Let us prove that $J_+$ is unique up to a phase. Let $J_+,J_+'$ be two charge conjugation operators, normalized so that: $J_+^2 = \epsilon$ and $J_+^{\prime 2} = \epsilon'$. Then, for all $\Omega \in \mathbb{C}\mathrm{l}(V)$:
\begin{equation*}
\overline{\Omega} = J_+ \Omega J_+^{-1} = J'_+ \Omega J_+^{\prime -1}.
\end{equation*}
This yields the following identity:
\begin{equation*}
\Omega J_+^{-1} J_+' = J_+^{-1} J_+' \Omega.
\end{equation*}
The linear operator (and thus Clifford algebra element) $J_+^{-1} J_+'$ is in the center of the algebra: there exists a complex number $\eta$ such that $J_+^{-1} J_+' = \eta$. This implies that $J_+' = \eta J_+$. This in turn yields:
\begin{equation*}
J_+'^2 = |\eta|^2 J_+^2
\end{equation*}
from which we infer that:
\begin{equation*}
\epsilon' = |\eta|^2 \epsilon
\end{equation*}
It is then easy to see that $\eta$ is a phase: $|\eta|=1$, and that $\epsilon' = \epsilon$. The normalized charge conjugation operator is thus unique up to a phase, and the sign $\epsilon$ is the same for all charge conjugation operators (that differ by a phase).

Finally, let us note that $J$ depends on the representation on the spinor space, and let us prove that $\epsilon$ is the same for all spinor representations. Let $\pi, \pi'$ be two representations of $\mathbb{C}\mathrm{l}(V)$ on $S$. There exists an invertible linear operator $T$ on $S$ such that:
\begin{equation*}
\pi'(\Omega) = T \pi(\Omega) T^{-1}.
\end{equation*}
It is then easy to check that $J_+$ is a normalized charge conjugation operator for $\pi$ if and only if $J'_+=TJ_+T^{-1}$ is a normalized charge conjugation operator for $\pi'$. Indeed, let us assume that $J_+$ is a normalized charge conjugation operator. We have, for any $\Omega \in \mathbb{C}\mathrm{l}(V)$:
\begin{equation*}
\begin{aligned}
J'_+ \pi'(\Omega) J_+^{\prime -1} &= (TJ_+T^{-1}) (T \pi(\Omega) T^{-1}) (TJ_+T^{-1})^{-1} \\
&= T J_+ \pi(\Omega) J_+^{-1} T^{-1}\\
&= T \pi(\overline{\Omega}) T^{-1} \\
J'_+ \pi'(\Omega) J_+^{\prime -1} &= \pi'(\overline{\Omega}).
\end{aligned}
\end{equation*}
We also have that $J_+^{\prime 2}=(TJ_+T^{-1})^2 = TJ_+^2T^{-1} = \pm 1$. The converse proof is identical. From $J_+^{\prime 2}= TJ_+^2T^{-1} = \epsilon$, we infer that $\epsilon$ can only depend on $q$ and $p$.
\qed
\end{demo}

In the following sections, we will prove that $\epsilon$ depends only on $p-q$.

An interesting property that follows straightforwardly from the definition of $\chi$ and $J_+$ is the following:
\begin{prop}
The charge conjugation and chirality operators always commute or anticommute:
\begin{equation}
J_+ \chi = (-1)^{(p-q)/2} \chi J_+.
\end{equation}
In other words, $J_+$'s parity is $|J_+| = (p-q)/2 \modulo 2$.

\end{prop}

This is immediate to prove from the commutation of $J_+$ with elements of the Clifford algebra, taking into account the phase in equation \eqref{ChiralityDef}. We also have the following result:

\begin{thm} \label{CCThm2}
Let $J$ be a (normalized) charge conjugation operator. Then it is either self-adjoint or anti-self-adjoint:
\begin{equation}
\begin{aligned}
J_+^\times &= \kappa J_+ \\
J_+^+ &= (-1)^{(p+q)/2} \kappa J_+,
\end{aligned}
\end{equation}
where $\kappa \in \{-1, 1\}$ is determined uniquely by $(q,p)$.
\end{thm}

We remind the reader that the adjoint of an antilinear operator $K$ is given by:
\begin{equation*}
(\psi, K \phi) = \overline{(K^\times \psi, \phi)} = (\phi, K^\times \psi),
\end{equation*}
and, similarly for the anti-Robinson product:
\begin{equation*}
[\psi, K \phi] = \overline{[K^+ \psi, \phi]} = [\phi, K^+ \psi],
\end{equation*}

\begin{demo}
Let $\Omega \in \mathbb{C}\mathrm{l}(V)$. We have:
\begin{equation}
\overline{\Omega} J_+ = J_+ \Omega.
\label{J1}
\end{equation}
Taking the adjoint, we find:
\begin{equation*}
J_+^\times \alpha(\Omega) = \Omega^\times J_+^\times.
\end{equation*}
We replace $\Omega$ with $\Omega^\times$:
\begin{equation}
J_+^\times \overline{\Omega} = \Omega J_+^\times.
\label{J2}
\end{equation}
Combining equations \eqref{J1} and \eqref{J2}, one finds:
\begin{equation*}
J_+^\times J_+ \Omega = J_+^\times \overline{\Omega} J_+ = \Omega J_+^\times J_+.
\end{equation*}
The operator $J_+^\times J_+$ is a linear operator acting on $S$. It is thus an element of the complex Clifford algebra. It is clearly in the center of this algebra, as the equation $J_+^\times J_+ \Omega = \Omega J_+^\times J_+$ holds for all $\Omega \in \mathbb{C}\mathrm{l}(V)$. We conclude that there exists $\lambda \in \mathbb{C}$ such that $J_+^\times J_+=\lambda$. Let $\psi \in S$ such that $(\psi, \psi) \ne 0$ (such a vector always exists because the Robinson product is nondegenerate). We have:
\begin{equation*}
(J_+ \psi, J_+ \psi) = (\psi, J_+^\times J_+ \psi) = \lambda (\psi, \psi),
\end{equation*}
from which we conclude that $\lambda$ is a real number. Since $J_+$ is invertible, $J_+^\times J_+ = \lambda$ is invertible as well. This implies that $\lambda$ is nonzero. It is either strictly positive or strictly negative. We have:
\begin{equation*}
\begin{aligned}
1 &= \epsilon^2 \\
&= (J_+^2)^\times J_+^2 \\
&= J_+^\times(J_+^\times J_+)J_+ \\
&= \lambda J_+^\times J_+ \\
1 &= \lambda^2.
\end{aligned}
\end{equation*}
We thus have $\lambda \in \{-1, 1\}$. Let $\kappa = \lambda \epsilon$. We can see that:
\begin{equation*}
\begin{aligned}
J_+^\times J_+ &= \lambda \\
&= \kappa \epsilon \\
J_+^\times J_+ &= \kappa J_+^2
\end{aligned}
\end{equation*}
from which we conclude that $J_+^\times = \kappa J_+$.

Let us prove that $\kappa$ does not depend on $J_+$. Let $J_+, J_+'$ be two charge conjugation operators, and $\kappa, \kappa'$ signs such that $J_+^\times = \kappa J_+$ and $J_+^{\prime\times} = \kappa' J_+'$. There exists a phase $\eta$ such that $J_+' = \eta J_+$. Thus:
\begin{equation*}
\begin{aligned}
\kappa' &= J_+^{\prime\times} J_+^{\prime -1} \\
&= (\eta J_+)^\times (\eta J_+)^{-1} \\
&= J_+^\times \overline{\eta} J_+^{-1} \overline{\eta} \\
&= J_+^\times J_+^{-1} \eta \overline{\eta} \\
\kappa' &= \kappa.
\end{aligned}
\end{equation*}

Now, let us note that $J_+$ and $(\cdot, \cdot)$ depend on the representation on the spinor space, and let us prove that $\kappa$ is the same for all spinor representations. Let $\pi, \pi'$ be two representations of $\mathbb{C}\mathrm{l}(V)$ on $S$. There exists an invertible linear operator $T$ on $S$ such that:
\begin{equation*}
\pi'(\Omega) = T \pi(\Omega) T^{-1}.
\end{equation*}
We saw in the proof of theorem \ref{CCThm} that $J_+$ is a normalized charge conjugation operator for $\pi$ if and only if $J'_+=TJ_+T^{-1}$ is a normalized charge conjugation operator for $\pi'$, and in theorem \ref{RobinsonFundamentalSymmetryTheorem} that $(\cdot, \cdot)$ is a Robinson product for $\pi$ if and only if $(\cdot, \cdot)'=(T^{-1} \cdot, T^{-1} \cdot)$ is a Robinson product for $\pi'$. Let us compute $\kappa$ for $\pi'$:
\begin{equation*}
\begin{aligned}
(\psi, J'_+ \varphi)' &= (T^{-1} \psi, T^{-1} (TJ_+T^{-1}) \varphi) \\
&= (T^{-1} \psi, J_+ T^{-1} \varphi) \\
&= \kappa \overline{(J_+ T^{-1} \psi, T^{-1} \varphi)} \\
&= \kappa \overline{(T^{-1} (T J_+ T^{-1}) \psi, T^{-1} \varphi)} \\
(\psi, J'_+ \varphi)' &= \kappa \overline{(J'_+ \psi, \varphi)'}.
\end{aligned}
\end{equation*}
This proves that $\kappa$ does not depends on the choice of representation. It can thus only depend on $(q,p)$.

Finally, let us prove that $J_+^+ = (-1)^{(p+q)/2} \kappa J_+$. Using \eqref{RobinsonAntiRobinson}, we can compute $J_+^+$: 
\begin{equation*}
\begin{aligned}
\left[\psi, J_+ \varphi\right] &= (\psi, i^q \chi J_+ \varphi) \\
&= (-1)^{(p+q)/2} (\psi, J_+ i^q \chi \varphi) \\
&= (-1)^{(p+q)/2} \kappa \overline{(J_+ \psi, i^q \chi \varphi)} \\
[\psi, J_+ \varphi] &= (-1)^{(p+q)/2} \kappa \overline{[J_+ \psi, \varphi]}.
\end{aligned}
\end{equation*}

\qed
\end{demo}

We will prove later that $\kappa$ depends only on $d=q+p$.

Finally, let us formulate the rule for tensor products of charge conjugation operators:
\begin{thm}
Let $J_{+V}, J_{+W}$ be normalized charge conjugation operators for $\mathbb{C}\mathrm{l}(V)$ and $\mathbb{C}\mathrm{l}(W)$ respectively. Then the charge conjugation operator for $\mathbb{C}\mathrm{l}(V \oplus W)$ is given (up to a phase) by:
\begin{equation}
J_{+,V \oplus W} = \chi_V^{(p'-q')/2} J_{+V} \hat{\otimes} \chi_W^{(p-q)/2} J_{+W}.
\label{CCTensor}
\end{equation}

\end{thm}

\begin{demo}
Let $J$ be the anti-linear operator defined on the right-hand side of equation \eqref{CCTensor}. We will begin by proving that $J$ commutes with the generators of $\mathrm{Cl}(V \oplus W)$. Let $v \oplus w \in V \oplus W$. We have:
\begin{equation*}
\begin{aligned}
J\gamma(v \oplus w) =& (\chi_V^{(p'-q')/2} J_{+V} \hat{\otimes} \chi_W^{(p-q)/2} J_{+W}) (\gamma(v) \hat{\otimes} 1 + 1 \hat{\otimes} \gamma(w)) \\
=& (-1)^{(p'-q')/2} \chi_V^{(p'-q')/2} J_{+V} \gamma(v) \hat{\otimes} \chi_W^{(p-q)/2} J_{+W} \\ &+ \chi_V^{(p'-q')/2} J_{+V} \hat{\otimes} \chi_W^{(p-q)/2} J_{+W} \gamma(w),
\end{aligned}
\end{equation*}
where we have used that the parity of $J_{+W}$ is $(p'-q')/2 \modulo 2$. Next, we take all charge conjugation operators to the right:
\begin{equation*}
\begin{aligned}
J\gamma(v \oplus w) =& (-1)^{(p'-q')/2} \chi_V^{(p'-q')/2} \gamma(v) J_{+V} \hat{\otimes} \chi_W^{(p-q)/2} J_{+W} \\ &+ \chi_V^{(p'-q')/2} J_{+V} \hat{\otimes} \chi_W^{(p-q)/2} \gamma(w) J_{+W} \\
\end{aligned}
\end{equation*}
Finally, we take $\gamma(v)$ and $\gamma(w)$ to the left, and factorize using the rules of graded tensor products:
\begin{equation*}
\begin{aligned}
J\gamma(v \oplus w) =& \gamma(v) (\chi_V^{(p'-q')/2} J_{+V}) \hat{\otimes} \chi_W^{(p-q)/2} J_{+W} \\ &+ (-1)^{(p-q)/2} \chi_V^{(p'-q')/2} J_{+V} \hat{\otimes} \gamma(w) (\chi_W^{(p-q)/2} J_{+W}) \\
=& (\gamma(v) \hat{\otimes} 1 + 1 \hat{\otimes} \gamma(w)) (\chi_V^{(p'-q')/2} J_{+V} \hat{\otimes} \chi_W^{(p-q)/2} J_{+W}). \\
J\gamma(v \oplus w) =& \gamma(v \oplus w)J.
\end{aligned}
\end{equation*}
Since $\mathrm{Cl}(V \oplus W)$ is generated by all vectors of the form $\gamma(v \oplus w)$, we deduce that $J$ commutes with all elements of $\mathrm{Cl}(V \oplus W)$. It is also an anti-linear operator. It is thus a charge conjugation operator. It remains to prove that it is normalized. We have:
\begin{equation*}
\begin{aligned}
J^2 &= (\chi_V^{(p'-q')/2} J_{+V} \hat{\otimes} \chi_W^{(p-q)/2} J_{+W})(\chi_V^{(p'-q')/2} J_{+V} \hat{\otimes} \chi_W^{(p-q)/2} J_{+W}) \\
&= \pm (\chi_V^{(p'-q')/2} J_{+V})^2 \hat{\otimes} (\chi_W^{(p-q)/2} J_{+W})^2
\end{aligned}
\end{equation*}
(we do not care about the exact value of the sign). We have, thanks to the homogeneity of $J_{+V}$:
\begin{equation*}
(\chi_V^{(p'-q')/2} J_{+V})^2 = \pm (\chi_V^{(p'-q')/2})^2 (J_{+V})^2 = \pm 1,
\end{equation*}
and similarly for $\chi_W^{(p-q)/2} J_{+W}$. We thus find:
\begin{equation*}
J^2 = \pm 1,
\end{equation*}
which proves that it is a normalized charge conjugation operator.

\qed
\end{demo}

\subsection{Graded Charge Conjugation}

Additionally to charge conjugation, one can define graded charge conjugation, and a corresponding graded charge conjugation operator. Graded charge conjugation is the following antilinear operation:
\begin{equation*}
\Omega \longmapsto (-1)^{|\Omega|} \overline{\Omega}.
\end{equation*}
A graded charge conjugation operator is an anti-linear operator $K$ that lifts graded charge conjugation to spinors:
\begin{equation*}
\forall \Omega \in \mathbb{C}\mathrm{l}(V): (-1)^{|\Omega|} \overline{\Omega} = K \Omega K^{-1}.
\end{equation*}
Graded and ungraded normalized charge conjugation operators are related by the following:
\begin{thm} \label{GradedUngradedCC}
The operator $J_+$ is a (normalized) charge conjugation operator if and only if $J_- = \chi J_+$ is a (normalized) graded charge conjugation operator.
\end{thm}

\begin{demo}
Let $J_+$ be a (normalized) charge conjugation operator, and let $J_- = \chi J_+$. For any $\Omega \in \mathbb{C}\mathrm{l}(V)$:
\begin{equation*}
\begin{aligned}
J_- \Omega J_-^{-1} &= \chi J_+ \Omega J_+^{-1} \chi \\
&= \chi \overline{\Omega} \chi \\
J_- \Omega J_-^{-1} &= (-1)^{|\Omega|} \overline{\Omega},
\end{aligned}
\end{equation*}
which proves that $J_-$ is a graded charge conjugation operator. We also have:
\begin{equation*}
J_-^2 = (\chi J_+)^2 = \pm \chi^2 J_+^2 = \pm 1,
\end{equation*}
thanks to the homogeneity of $J_+$. We infer that $J_-$ is a normalized graded charge conjugation operator.

The converse proof is identical.

\qed
\end{demo}

As a consequence of this theorem, we have:
\begin{thm} \label{CCThmGraded}
There exists a unique (up to a phase) normalized graded charge conjugation operator $J_-$ on spinors:
\begin{equation}
\forall \Omega \in \mathbb{C}\mathrm{l}(V): (-1)^{|\Omega|} \overline{\Omega} = J_- \Omega J_-^{-1}.
\label{CCJGraded}
\end{equation}
It is such that:
\begin{equation}
\begin{aligned}
J_-^2 &= (-1)^{(p-q)/2} \epsilon \\
J_-^\times &= (-1)^{(p+q)/2} \kappa J_- \\
J_-^+ &= \kappa J_-.
\end{aligned}
\end{equation}
Moreover, it satisfies:
\begin{equation}
J_- \chi = (-1)^{(p-q)/2} \chi J_-.
\end{equation}
In other words, $J_-$'s parity is $|J_-| = (p-q)/2 \modulo 2$.

\end{thm}

\begin{demo}
The existence and uniqueness are direct consequences of theorem \ref{CCThm}, and the isomorphism between graded and ungraded charge conjugations constructed in theorem \ref{GradedUngradedCC}. Now, let $J_-$ be a graded charge conjugation operator. There exists a charge conjugation operator $J_+$ such that $J_- = \chi J_+$. We have, using the parity of $J_+$:
\begin{equation*}
J_-^2 = (\chi J_+)^2 = (-1)^{(p-q)/2} \chi^2 J_+^2 = (-1)^{(p-q)/2} \epsilon.
\end{equation*}
We also have:
\begin{equation*}
\begin{aligned}
J_-^\times &= (\chi J_+)^\times \\
&= J_+^\times \chi^\times \\
&= \kappa (-1)^q J_+ \chi \\
&= \kappa (-1)^q (-1)^{(p-q)/2} \chi J_+ \\
J_-^\times &= (-1)^{(p+q)/2} \kappa J_-.
\end{aligned}
\end{equation*}
A similar computation gives: $J_-^+ = \kappa J_-$. Finally, since $\chi$ is even, $J_+$ and $J_-$ have the same parity: $|J_-| = |J_+| = (p-q)/2 \modulo 2$.

\qed
\end{demo}

We now give the tensor product rule for graded charge conjugation:
\begin{thm}
Let $J_{-V}, J_{-W}$ be normalized graded charge conjugation operators for $\mathbb{C}\mathrm{l}(V)$ and $\mathbb{C}\mathrm{l}(W)$ respectively. Then the graded charge conjugation operator for $\mathbb{C}\mathrm{l}(V \oplus W)$ is given (up to a phase) by:
\begin{equation}
J_{-,V \oplus W} = \chi_V^{(p'-q')/2} J_{-V} \hat{\otimes} \chi_W^{(p-q)/2} J_{-W}.
\end{equation}

\end{thm}

\begin{demo}
We have:
\begin{equation*}
\begin{aligned}
J_{-,V \oplus W} &= \chi_{V \oplus W} J_{V \oplus W} \\
&= (\chi_V \hat{\otimes} \chi_W)(\chi_V^{(p'-q')/2} J_V \hat{\otimes} \chi_W^{(p-q)/2} J_W) \\
&= \pm \chi_V^{(p'-q')/2+1} J_V \hat{\otimes} \chi_W^{(p-q)/2+1} J_W \\
J_{-,V \oplus W} &= \pm \chi_V^{(p'-q')/2} J_{-V} \hat{\otimes} \chi_W^{(p-q)/2} J_{-W}
\end{aligned}
\end{equation*}

\qed
\end{demo}

\subsection{Charge Conjugation for Physicists}

We want to relate the charge conjugation operators described here with the charge conjugation \emph{matrix} operator used in the Dirac theory of a spin $1/2$ particle. Particle physicists often construct a matrix $C$ (in some specific representation of the Clifford algebra) such that:
\begin{equation*}
\overline{\gamma^\mu} = C^{-1} \gamma^\mu C,
\end{equation*}
where $\overline{\gamma^\mu}$ is the complex conjugate matrix of $\gamma^\mu$ in the chosen representation. It has to satisfy $C\overline{C} = \pm 1$. Charge conjugation on spinors is then defined as the anti-linear operation:
\begin{equation*}
K :\psi \longmapsto \psi^c = \eta C \overline{\psi},
\end{equation*}
Where $\eta$ is a phase. Let us now prove that this is indeed a charge conjugation operator. We have:
\begin{equation*}
\begin{aligned}
K \gamma^\mu \psi &= \eta C \overline{\gamma^\mu \psi} \\
&= \eta C \overline{\gamma^\mu} \overline{\psi} \\
&= \eta CC^{-1} \gamma^\mu C \overline{\psi} \\
K \gamma^\mu \psi &= \gamma^\mu K \psi.
\end{aligned}
\end{equation*}
The antilinear operator $K$ commutes with all $\gamma^\mu$, and thus with all elements of the Clifford algebra. It consequently implements charge conjugation. Moreover:
\begin{equation*}
\begin{aligned}
K^2 \psi &= \eta C^{-1} \overline{\eta C^{-1} \overline{\psi}} \\
&= |\eta|^2 C^{-1}\overline{C}^{-1} \psi \\
K^2 \psi &= \pm \psi.
\end{aligned}
\end{equation*}
We conclude that $K^2 = \pm 1$, and $K^\times = \pm K$ as well, thanks to theorem \ref{CCThm2}. It is thus a properly normalized charge conjugation operator.

Most of the times, the matrix $C$ is chosen so that $\overline{\gamma^\mu} = - C^{-1} \gamma^\mu C$. It is then easy to prove that $K$ implements graded charge conjugation.

As an example, the matrix $C$ implementing \emph{graded} charge conjugation is defined in \cite{Weinberg} as $C=\gamma^2$, where the Clifford algebra $\mathrm{Cl}(1,3)$ is represented using the Weyl (chiral) representation. We have explicitly:
\begin{equation}
\gamma^0 = -i\begin{pmatrix} 0 & 1 \\ 1 & 0 \end{pmatrix}, \;\;\; \gamma^k = -i\begin{pmatrix} 0 & \sigma^k \\ -\sigma^k & 0 \end{pmatrix}, \;\;\; \chi=\gamma^5 = \begin{pmatrix} 1 & 0 \\ 0 & -1 \end{pmatrix}
\end{equation}
where the $\sigma^k$ are the Pauli matrices. It is easy to check that $\overline{\gamma^\mu} = - C^{-1} \gamma^\mu C$ and that $C\overline{C} = -1$ hold. To define a non-graded charge conjugation operator, one could choose $C' = \gamma^5 C = \gamma^5 \gamma^2$ as a charge conjugation matrix. It is easy to check that $C'\overline{C'} = +1$.

\subsection{Computation of $\epsilon$ and $\kappa$}

We will now compute the $\epsilon$ and $\kappa$ signs. To compute $\epsilon$, we use the $\pi_0$ representation of $\mathbb{C}\mathrm{l}(p,q)$ described at the beginning of section \ref{SectionSpin}. We then construct in this representation a ``nicely behaving'' charge conjugation operator whose square is easy to compute. This is inspired from chapter 4 in \cite{VS}. There are two cases that need to be dealt with separately:

\begin{itemize}
\item $p-q \modulo 8 \in \{0,2\}$:
In this case, the representation $\pi_0$ maps the real Clifford algebra $\mathrm{Cl}(p,q)$ to $\mathrm{M}_D(\mathbb{R})$. Consider the operator $J_+$ that maps a spinor $\psi \in S$ to its complex conjugate $\overline{\psi}$ in the basis associated to the representation $\pi_0$. It is clear that this antilinear operator commutes with - the representation of - $\mathrm{Cl}(p,q)$. It is also properly normalized, since $J_+^2 = 1$. We conclude that it is the charge conjugation operator, and that $\epsilon = 1$ for $p-q \modulo 8 \in \{0,2\}$.

\item $p-q \modulo 8 \in \{4,6\}$:
We now have a representation $\pi_0$ that maps the real Clifford algebra $\mathrm{Cl}(p,q)$ to $\mathrm{M}_{D/2}(\mathbb{H})$. We now construct the operator $J_+: \psi \mapsto Q \overline{\psi}$. The operator $Q$ is the following block-diagonal matrix:
\begin{equation*}
Q=\begin{pmatrix}
\begin{matrix} 0 & -1 \\ 1 & 0 \end{matrix} & & \\
& \ddots & \\
& & \begin{matrix} 0 & -1 \\ 1 & 0 \end{matrix}
\end{pmatrix}
\end{equation*}
We leave to the reader to check that $J_+$ commutes with all elements of $\mathrm{M}_{D/2}(\mathbb{H})$, and thus $\mathrm{Cl}(p,q)$. It is also properly normalized, since $J_+^2 = -1$. It is thus the charge conjugation operator, and we conclude that $\epsilon  = -1$ for $p-q \modulo 8 \in \{4,6\}$.
\end{itemize}

To put these results in a more useful form, we define the following sign function on even integers:
\begin{equation}
\begin{aligned}
a \colon 2 \mathbb{Z} &\longrightarrow \{ -1, +1 \} \\
n &\longmapsto (-1)^{n(n+2)/8} = (-1)^{[-n/4]}
\end{aligned}
\label{defa}
\end{equation}
This function can be shown to be periodic of period 8. It has the following symmetries:
\begin{equation}
\begin{aligned}
a(n+2) &= -a(-n) \\
a(n+4) &= -a(n) \\
a(n+6) &= a(-n) \\
a(n+8) &= a(n).
\end{aligned}
\label{ASym}
\end{equation}
Another important property is the following:
\begin{equation}
a(m+n) a(m-n) = (-1)^{(m+1)n/2},
\label{a(m,n)}
\end{equation}
it is a result of the identity:
\begin{equation*}
\frac{(m+n)(m+n+2)}{8} = \frac{(m-n)(m-n+2)}{8} + \frac{(m+1)n}{2},
\end{equation*}
which is straightforward to prove. Substituting $m=0$, one finds the following symmetry:
\begin{equation*}
a(-n) = (-1)^{n/2} a(n).
\end{equation*}
The values of $a(n), a(-n), (-1)^{n/2}$ are summarized in table \ref{avsn}.
\begin{table}[!h]
\centering
\setlength{\tabcolsep}{2em}
\begin{tabular}{c|cccc}
$n \;\mathrm{mod}\; 8$ & 0 & 2 & 4 & 6 \\
\hline
$a(n)$ & 1 & -1 & -1 & 1 \\
$a(-n)$ & 1 & 1 & -1 & -1 \\
$(-1)^{n/2}$ & 1 & -1 & 1 & -1
\end{tabular}
\caption{Values of $a(n),a(-n),(-1)^{n/2}$}
\label{avsn}
\end{table}

Using table \ref{avsn}, we see that:
\begin{equation*}
J_+^2 = \epsilon = a(q-p).
\end{equation*}
We thus have:
\begin{equation*}
J_-^2 = (-1)^{(p-q)/2} \epsilon = (-1)^{(p-q)/2} a(q-p) = a(p-q).
\end{equation*}

Let us compute $\kappa$. To this end, we will use the fundamental symmetry constructed in theorem \ref{RobinsonFundamentalSymmetryTheorem}:
\begin{equation*}
\eta_+ = \begin{cases}
\pm i^{(p-1)/2} \gamma^{q+1}...\gamma^d &\text{for odd  } q,p \\
\pm i^{q/2} \gamma^1...\gamma^q &\text{for even  } q,p. \\
\end{cases}
\end{equation*}
From the commutation of $J_+$ with the $\gamma^a$, we find:
\begin{equation*}
J_+ \eta_+ = \begin{cases}
(-1)^{(p-1)/2} \eta_+ J_+ &\text{for odd  } q,p \\
(-1)^{q/2} \eta_+ J_+ &\text{for even  } q,p. \\
\end{cases}
\end{equation*}
This can be put in the more succinct form:
\begin{equation*}
J_+ \eta_+ = (-1)^{q(p-1)/2} \eta_+ J_+.
\end{equation*}
Indeed, if $q$ and $p$ are odd, then $(q-1)(p-1)/2$ is en even integer, from which we deduce that $(-1)^{q(p-1)/2} = (-1)^{(q-1)(p-1)/2+(p-1)/2} = (-1)^{(p-1)/2}$. If they are both even, then $qp/2$ is even, which implies that  $(-1)^{q(p-1)/2} = (-1)^{qp/2-q/2} = (-1)^{q/2}$. Now, we have:
\begin{equation*}
\begin{aligned}
J_+^{\dagger \eta_+} J_+ &= \eta_+ J_+^\times \eta_+ J_+ \\
&= \kappa \eta_+ J_+ \eta_+ J_+ \\
&= \kappa (-1)^{q(p-1)/2} \eta_+^2 J_+^2 \\
&= \kappa (-1)^{q(p-1)/2} a(q-p) \\
J_+^{\dagger \eta_+} J_+ &= a(-(q+p)) \kappa,
\end{aligned}
\end{equation*}
where we used equations \eqref{a(m,n)} and \eqref{ASym} to get the last equality. We know that $J_+^{\dagger \eta_+} J_+$ must be a positive definite linear operator. We deduce that:
\begin{equation*}
\kappa = a(-(q+p)).
\end{equation*}

All other signs can be deduced from the properties of the $a,b$ functions and the results of theorems \ref{CCThm2} and \ref{CCThmGraded}. We summarize these results in table \ref{CCSummaryTable}.

\begin{table}[!h]
\centering
\setlength{\tabcolsep}{2em}
\renewcommand{\arraystretch}{3}
\begin{tabular}{|c|c|c|}
\hline
& \pbox{20cm}{$(\cdot, \cdot)$ \\} & \pbox{20cm}{$[\cdot, \cdot]$ \\}\\
\hline
\pbox{20cm}{$J_+$ \\} & \pbox{20cm}{$J_+^2 = a(q-p)$ \\ $J_+^\times = a(-(p+q)) J_+$ \\} & \pbox{20cm}{$J_+^2 = a(q-p)$ \\ $J_+^+ = a(p+q) J_+$ \\} \\
\hline
\pbox{20cm}{$J_-$ \\} & \pbox{20cm}{$J_-^2 = a(p-q)$ \\ $J_-^\times = a(p+q) J_-$ \\} & \pbox{20cm}{$J_-^2 = a(p-q)$ \\ $J_-^+ = a(-(p+q)) J_-$ \\} \\
\hline
\end{tabular}
\caption{Summary of Signs}
\label{CCSummaryTable}
\end{table}

\section{Summary of Tensor Product, Non-Graded Form} \label{SectionTensorSummary}

Let us summarize the rules for the tensor products of Clifford algebras:
\begin{equation}
\begin{aligned}
\gamma(v \oplus w) &= \gamma(v) \hat{\otimes} 1 + 1 \hat{\otimes} \gamma(w) \\
S_{V \oplus W} &= S_V \hat{\otimes} S_W \\
\chi_{V \oplus W} &= \chi_V \hat{\otimes} \chi_W \\
J_{+,V \oplus W} &= \chi_V^{(q'-p')/2} J_{+V} \hat{\otimes} \chi_W^{(q-p)/2} J_{+W} \\
J_{-,V \oplus W} &= \chi_V^{(q'-p')/2} J_{-V} \hat{\otimes} \chi_W^{(q-p)/2} J_{-W} \\
(\omega_1 \hat{\otimes} \theta_1, \omega_2 \hat{\otimes} \theta_2)_{V \oplus W} &= (\omega_1, \omega_2)_V (\theta_1, \beta \theta_2)_W \\
[\omega_1 \hat{\otimes} \theta_1, \omega_2 \hat{\otimes} \theta_2]_{V \oplus W} &= [\omega_1, \omega_2]_V [\theta_1, \beta \theta_2]_W \\
\beta &= (i^{q'} \chi_W)^{q}
\end{aligned}
\label{CliffordTensorSummary}
\end{equation}

Although the graded nature of the tensor product makes it more symmetric and esthetically appealing, it is more useful for applications to use a non-graded version of the same tensor product. To this end, one can rewrite a tensor product of operators (whether linear or antilinear) $R \hat{\otimes} T$ on $S_V \hat{\otimes} S_W$ in the following form:
\begin{equation}
R \hat{\otimes} T \cong R \chi_V^{|T|} \otimes T,
\label{NonGradedRep}
\end{equation}
whereas vectors stay the same:
\begin{equation*}
\psi \hat{\otimes} \varphi \cong \psi \otimes \varphi
\end{equation*}
One can indeed check that $R \hat{\otimes} T$ and $R \chi_V^{|T|} \otimes T$ act the same way on $S_V \hat{\otimes} S_W$:
\begin{equation*}
(R \chi_V^{|T|} \otimes T)(\psi \otimes \varphi) = (-1)^{|T||\psi|} (R\psi \otimes T\varphi),
\end{equation*}
and that operators multiply according to the graded rule:
\begin{equation*}
(R \chi_V^{|T|} \otimes T)(R' \chi_V^{|T'|} \otimes T') = (-1)^{|T||R'|} (RR'\chi_V^{|TT'|} \otimes TT').
\end{equation*}
The tensor product rules become:
\begin{equation}
\begin{aligned}
\gamma(v \oplus w) &= \gamma(v) \otimes 1 + \chi_V \otimes \gamma(w) \\
S_{V \oplus W} &= S_V \otimes S_W \\
\chi_{V \oplus W} &= \chi_V \otimes \chi_W \\
J_{+,V \oplus W} &= \begin{cases}
J_{+V} \otimes J_{+W} \text{    if $(p-q)/2$ is even} \\
J_{+V} \otimes J_{-W} \text{    if $(p-q)/2$ is odd}
\end{cases}\\
J_{-,V \oplus W} &= \begin{cases}
J_{-V} \otimes J_{-W} \text{    if $(p-q)/2$ is even} \\
J_{-V} \otimes J_{+W} \text{    if $(p-q)/2$ is odd}
\end{cases}\\
(\omega_1 \otimes \theta_1, \omega_2 \otimes \theta_2)_{V \oplus W} &= \begin{cases}
(\omega_1, \omega_2)_V (\theta_1, \theta_2)_W  \text{    if $q$ is even} \\
(\omega_1, \omega_2)_V [\theta_1, \theta_2]_W  \text{    if $q$ is odd}.
\end{cases} \\
[\omega_1 \otimes \theta_1, \omega_2 \otimes \theta_2]_{V \oplus W} &= \begin{cases}
[\omega_1, \omega_2]_V [\theta_1, \theta_2]_W  \text{    if $q$ is even} \\
[\omega_1, \omega_2]_V (\theta_1, \theta_2)_W  \text{    if $q$ is odd}.
\end{cases}
\end{aligned}
\label{CliffordNonGraded}
\end{equation}
We used theorem \ref{GradedUngradedCC} and equation \eqref{RobinsonAntiRobinson} to simplify the tensor product, making it less symmetric.

\section{An Explicit Spinor Representation with Fock Spaces} \label{Fock}

In this section we construct an explicit representation of Clifford algebras on Fock spaces. This representation will enable us to check all the properties of Clifford algebras studied above, as well as the existence of a representation where all generators are either self-adjoint or anti-self-adjoint. The results presented here are a generalization of the existing results in the literature (see \cite{GBVF,PlymenRobinson} and references therein) from the positive definite case $q=0$, to the indefinite case. In addition, we construct in this representation the Robinson product and charge conjugation operators explicitly. We start first with a reminder on Fock spaces.

\subsection{Antisymmetric Fock Spaces}

Let $E$ be a finite complex Krein space of dimension $d$, with indefinite product $(\cdot,\cdot)$ of signature $(q,p)$. Quite similarly to what is done to construct Clifford algebras, one can construct the so-called \emph{exterior algebra} of $E$, denoted $\bigwedge E$ as the algebra generated by the elements of $E$, quotiented by the equivalence relation:
\begin{equation*}
u v + v u \sim 0,
\end{equation*}
for all $u, \in E$. In other words, the elements of $E$ anticommute in $\bigwedge E$. The resulting product, called exterior product, is denoted with the symbol $\wedge$. We can thus write that:
\begin{equation}
u \wedge v + v \wedge u = 0.
\end{equation}
The underlying vector space is often called a fermionic Fock space, or antisymmetric Fock space. Note that exterior algebras are a limiting case of Clifford algebras, when the bilinears $g$ vanish.

This algebra is equipped with an $\mathbb{N}$-grading. Let $n \in \mathbb{N}$. The $n$-th component of $\bigwedge E$ is denoted $\bigwedge^n E$, and is spanned by elements that are products of $n$ vectors:
\begin{equation*}
\bigwedge^n E = \mathrm{Span}(u_1 \wedge \dots \wedge u_n).
\end{equation*}
The elements of $\bigwedge^n E$ are called $n$-vectors. Due to the antisymmetry of the exterior product, a monomial of the form $u_1 \wedge \dots \wedge u_n$ is nonvanishing if and only if all factors are linearly independent vectors. As a result, all components for which $n > d$ are zero.

To study the non-empty components, one can pick a basis $(E^a)_a$ of $E$. One can then prove that the monomials $E^I = E^{i_1} \wedge \dots \wedge E^{i_{|I|}}$, with $I=\{i_1 < ... < i_{|I|}\} \subset \llbracket 1, d \rrbracket$, are a basis of $\bigwedge E$. In particular, the monomials $E^I$ such that $|I| = n$ are a basis of $\bigwedge^n E$. From some simple combinatorics, one can infer that the dimension of $\bigwedge^n E$ is $\binom{n}{k}$. In particular, $\bigwedge^0 E$ is unidimensional, and is spanned by the identity $1$. We will later use the exterior algebra as a vector space on which to act with operators. To avoid any confusion, the identity of the exterior algebra will be denoted:
\begin{equation}
\Omega = 1 \in \bigwedge^0 E.
\end{equation}
This vector is sometimes called a vacuum. The top component $\bigwedge^d E$ is also unidimensional, and is spanned by the vector:
\begin{equation}
\tilde{\Omega} = E^1 \wedge \dots \wedge E^d \in \bigwedge^d E.
\end{equation}

The underlying Fock space can be equipped with a non-degenerate Hermitian form (\emph{i.e.} a Krein product), called the Hodge product, defined by its action on monomials:
\begin{equation}
(u_1 \wedge \dots \wedge u_k, v_1 \wedge \dots \wedge v_l) = \delta_{kl} \mathrm{det}(((u_i, v_j))_{1 \leq i,j \leq k}).
\end{equation}
Let us choose the basis $(E^a)_a$ so that it is pseudo-orthonormal:
\begin{equation*}
(E^a, E^b) = \begin{cases}
-\delta_{ab} &\text{ if $a,b = 1, \dots, q$} \\
\delta_{ab} &\text{ if $a,b = p+1, \dots, d$} \\
0 &\text{ otherwise.} \end{cases}
\end{equation*}
Then one can prove that the monomials $E^I$ are a pseudo-orthonormal basis of $\bigwedge E$, and that their "squared-norm" is $\pm 1$ depending on the number of factors of negative "squared-norm". As a result, $(\cdot,\cdot)$ is positive definite if $q=0$, and indefinite of signature $(2^{d-1}, 2^{d-1})$ if $q > 0$. Note that the vacuum vector $\Omega$ is normalized:
\begin{equation*}
(\Omega, \Omega) = 1,
\end{equation*}
whereas:
\begin{equation*}
(\tilde{\Omega}, \tilde{\Omega}) = (-1)^q,
\end{equation*}
since it contains all negative vectors.

On this Fock space, one can define so-called \emph{creation operators}, defined by:
\begin{equation}
a^\times(v)(v_1 \wedge \dots \wedge v_l) = v \wedge v_1 \wedge \dots \wedge v_l.
\end{equation}
The map $a^\times: E \rightarrow \mathrm{End}(\bigwedge E)$ is linear. Creation operators are simply left multiplication operators on the exterior algebra. The adjoint of $a^\times(v)$ with respect to the Hodge product, called an annihilation operator, is given by (see theorem 2.3.1 in \cite{PlymenRobinson}):
\begin{equation}
a(v) (v_1 \wedge \dots \wedge v_l) = \sum_{i} (-1)^{i-1} (v,v_i) v_1 \wedge \dots \wedge \hat{v_i} \wedge \dots \wedge v_l
\end{equation}
where $\hat{v_i}$ means that $v_i$ is omitted. The map $a: E \rightarrow \mathrm{End}(\bigwedge E)$ is anti-linear. One can prove that annihilation operators are graded derivations of the exterior algebra. The most important properties of annihilation and creation operators are their anti-commutation relations (see \cite{GBVF,PlymenRobinson} for proof of this standard result):
\begin{equation}
\begin{aligned}
\{a(u), a(v) \} &= 0 \\
\{a(u)^\times, a(v)^\times \} &= 0 \\
\{a(u), a(v)^\times\} &= (u,v).
\end{aligned}
\end{equation}
These relations will be used to construct a representation of Clifford algebras in the next subsections. One can prove that the vacuum $\Omega$ is the unique vector (up to a real factor) that is annihilated by all annihilation operators:
\begin{equation*}
\forall v \in E: a(v) \Omega = 0,
\end{equation*}
whereas $\tilde{\Omega}$ is the unique vector that is annihilated by all creation operators:
\begin{equation*}
\forall v \in E: a^\times(v) \tilde{\Omega} = 0.
\end{equation*}
We now have almost everything we need to construct complex representations of Clifford algebras. It remains to choose the complex vector space $E$. This is the subject of the next subsection.

\subsection{Complex Structures and the Complexification of Real Vector Spaces}

We now go back to our real vector space $V$ with its bilinear $g$ of signature $(q,p)$. We wish to construct a complex vector space from $V$ in order to build a complex Fock space. This can be done with complex structures:
\begin{defn}
An endomorphism $C$ of $V$ is said to be a complex structure operator if it squares to minus the identity:
\begin{equation*}
C^2 = -1.
\end{equation*}
Moreover, it is said to be an orthogonal complex structure if it is an isometry:
\begin{equation*}
g(Cu, Cv) = g(u,v).
\end{equation*}

\end{defn}

Equivalently, a complex structure is orthogonal if it is skew-symmetric:
\begin{equation*}
g(Cu, v) = -g(u,Cv)
\end{equation*}
(this can be seen by replacing $v$ with $Cv$). We have the following existence theorem for orthogonal complex structures:
\begin{thm}
The real vector space $(V,g)$ admits an orthogonal complex structure $C$ if and only if $q$ and $p$ are both even.
\end{thm}

\begin{demo}
From the fact $C^2 = -1$, we deduce that: $\mathrm{det}(C)^2 = \mathrm{det}(-1) = (-1)^d$. Since $\mathrm{det}(C)^2$ is positive, we conclude that $d$ must be even. Because the operator squares to $-1$, its eigenvalues (as a complex operator) are $\pm i$. Let $m$ and $n$ be the multiplicities of $i$ and $-i$ respectively. We have: $\mathrm{tr}(C) = (m-n)i$. Since $C$ is a real operator, its trace is real, from which we conclude that $m=n$. We can now infer the determinant of $C$: $\mathrm{det}(C) = i^m (-i)^n = i^n (-i)^n = 1$. The operator $C$ is unimodular.

Now, let us consider a pseudo-orthonormal basis for $V$ in which the bilinear $g$ takes the form:
\begin{equation*}
g(u,v) = u^T \eta v
\end{equation*}
with $\eta = \mathrm{diag}(-, \dots -, +, \dots, +)$ a signature matrix for $g$. The statement that $C$ is orthogonal now takes the form:
\begin{equation*}
C^T \eta C = \eta,
\end{equation*}
from which we infer that $\eta C$ is skew-symmetric and real (we used here $C$ to denote the matrix representation of $C$ itself). It is a well-known fact of linear algebra that such a matrix has a positive determinant. Since it is invertible, its determinant is strictly positive. This means that $\mathrm{det}(C)$ and $\mathrm{det}(\eta) = (-1)^q$ have the same sign. This implies that $q$ must be even. Because $d$ is even, $p$ must be as well.

Conversely, let us assume that $q$ and $q$ are even. In the same pseudo-orthonormal basis described above, we construct the operator $C$ whose matrix is:
\begin{equation}
C=\begin{pmatrix}
\begin{matrix} 0 & -1 \\ 1 & 0 \end{matrix} & & \\
& \ddots & \\
& & \begin{matrix} 0 & -1 \\ 1 & 0 \end{matrix}
\end{pmatrix}
\label{blockC}
\end{equation}
This is clearly a skew-symmetric matrix that squares to -1. Moreover, because $q$ and $p$ are even, it commutes with $\eta$. This means that $\eta C$ is skew-symmetric. The operator $C$ is thus an orthogonal complex structure.

\qed
\end{demo}

From now on, we will assume that $q$ and $p$ are always even. One can prove that all complex structures are of the form described above:

\begin{prop} \label{Cbasis}
Let $C$ be an orthogonal complex structure on $V$. There exists a pseudo-orthonormal basis of $V$ in which $C$ takes the form (\ref{blockC}).
\end{prop}

\begin{demo}
Let $e^1$ be a negative normalized vector: $g(e^1, e^1) = -1$. Consider the vector $Ce^1$. It is negative and normalized as well: $g(Ce^1, Ce^1) = g(e^1, e^1) = -1$. It is also orthogonal to $e^1$:
\begin{equation*}
\begin{aligned}
g(e^1, Ce^1) &= -g(Ce^1, e^1) \\
&= -g(e^1, Ce^1) \\
&= 0.
\end{aligned}
\end{equation*}
Let $V_1 = \mathrm{Span}(e^1, Ce^1)$ be the subspace spanned by the two vectors, and let $V_1'$ be its orthogonal in $V = V_1 \oplus V_1'$. The subspace $V_1$ is clearly an invariant subspace for $C$. Since $C$ is an isometry, $V_1'$ is invariant as well. This means that (the restriction of) $C$ is a complex structure on $V_1'$, whose signature is $(q-2,p)$. One repeats the procedure above until one is left the with the invariant subspace $V_{q/2}'$, of signature $(0,p)$. One then repeats the procedure using positive normalized vectors, until the space is depleted of all vectors. The result is a pseudo-orthonormal basis $(e^1, Ce^1, \dots, e^{d/2}, Ce^{d/2})$ where the product $g$ takes the form:
\begin{equation*}
g(u,v) = u^T \eta v
\end{equation*}
with $\eta = \mathrm{diag}(-, \dots -, +, \dots, +)$, and where $C$ takes the form
\begin{equation*}
C=\begin{pmatrix}
\begin{matrix} 0 & -1 \\ 1 & 0 \end{matrix} & & \\
& \ddots & \\
& & \begin{matrix} 0 & -1 \\ 1 & 0 \end{matrix}
\end{pmatrix}
\end{equation*}

\qed
\end{demo}

Since all orthonormal bases are related by orthogonal transformations, one can easily infer the following corollary:
\begin{cor}
\textit{(a)} Any two orthogonal complex structures $C$ and $C'$ are related by an orthogonal transformation $O$ of $(V,g)$:
\begin{equation*}
C' = O C O^{-1}.
\end{equation*}
\textit{(b)} For any orthogonal complex structure $C$, the set of all orthogonal complex structures is:
\begin{equation*}
\{O C O^{-1} | O \in \mathrm{O}(V) \}. 
\end{equation*}
\end{cor}

A complex structure $C$ enables one to define multiplication by complex numbers on $V$ using the rule $iv \equiv Cv$, as $C^2=-1$. This is rigorously done by extending $C$ to the complexification $\mathbb{C} \otimes V$ of $V$, and then building the vector space:
\begin{equation}
V_C = \mathrm{Ker}(C-i) = \mathrm{Im}(C+i) \subset \mathbb{C} \otimes V.
\end{equation}
This is a complex vector space in which all elements satisfy: $iv=Cv$. It is also isomorphic, as a real vector space, to $V$ itself. The isomorphism is given by the $\mathbb{R}$-linear map:
\begin{equation}
\begin{aligned}
c: V &\longrightarrow V_C \\
v &\longmapsto \frac{(1-iC)}{2} v,
\end{aligned}
\end{equation}
which satisfies the relation: $c(Cv) = i c(v)$, as required. In $V_C$, the vectors $v$ and $Cv$ become co-linear. Let $(e^i, Ce^i)_i$ be the basis constructed above for $C$. Then the image of the basis in $V_C$ is the basis $(e^i = c(e^i))_i$, as $c(Ce^i)=ic(e^i)$ is co-linear to $c(e^i)$. The space $V_C$ is thus a $d/2$-dimensional complex vector space.

A canonical hermitian form for $V_C$ can be constructed:
\begin{equation}
(c(u), c(v))_C = g(u,v) - i g(u, Cv),
\end{equation}
by extending $g$ to $\mathbb{C} \otimes V$, and then restricting the vector space to $V_C$. We leave it to the reader to check that it is indeed a hermitian form, and that:
\begin{equation*}
(e^i, e^j)_C = g(e^i, e^j) = \epsilon_i \delta_{ij},
\end{equation*}
where $\epsilon_i = g(e^i, e^i) = \pm 1$, from which we conclude that $(\cdot, \cdot)$ has the signature $(q/2, p/2)$. The corresponding matrix $\eta_C = \mathrm{diag}(- \dots - + \dots +) = \mathrm{diag}(\epsilon_i)$ is a fundamental symmetry for $(\cdot, \cdot)_C$.

\subsection{The Even-Even Case}

In this subsection, we consider the case where $q$ and $p$ are both even. Let $C$ be an orthogonal complex structure on $V$. We will represent the Clifford algebra $\mathrm{Cl}(V)$ on the complex, indefinite Fock space: $S_C = \bigwedge V_C$ built on the Krein space $(V_C, (\cdot,\cdot)_C)$. The corresponding Hodge product will be denoted $(\cdot,\cdot)_C$.

Let us now define the representation $\pi_C$ of $\mathrm{Cl}(V)$ on $S_C = \bigwedge V_C$. Let $v \in V$. The representation is defined by the following action for the generators:
\begin{equation}
\pi_C (\gamma(v)) = a(c(v)) + a(c(v))^\times.
\end{equation}
When no confusion is possible, and only one complex structure is being used, we will omit $\pi_C$ and write $\gamma(v) = a(c(v)) + a(c(v))^\times$. This will be the case for the rest of this section. This is indeed a representation of the Clifford algebra:
\begin{equation*}
\begin{aligned}
\{\gamma(u), \gamma(v)\} &= \{ a(c(u)) + a(c(u))^\times, a(c(v)) + a(c(v))^\times \} \\
&= \{ a(c(u)), a(c(v))^\times \} + \{ a(c(u))^\times, a(c(v)) \} \\
&= (c(u), c(v))_C + (c(v), c(u))_C \\
&= 2 \mathrm{Re}(g(u,v) - i g(u, Cv)) \\
\{\gamma(u), \gamma(v)\} &= 2 g(u,v).
\end{aligned}
\end{equation*}
Moreover, $S_C$ is $D=2^{d/2}$-dimensional. It is thus a spinor space for $\mathrm{Cl}(V)$. From now on we will simply call it $S \equiv S_C$. 

The annihilation and creation operators can be expressed using the Clifford generators. Indeed, let $v \in V$. We have:
\begin{equation*}
\begin{aligned}
\gamma(v) &= a(c(v)) + a(c(v))^\times \\
\gamma(Cv) &= i(a(c(v))^\times -a(c(v)))
\end{aligned}
\end{equation*}
which can be inverted to yield:
\begin{equation}
\begin{aligned}
a(c(v)) &= \frac{\gamma(v) + i \gamma(Cv)}{2} \\
a(c(v))^\times &= \frac{\gamma(v) - i \gamma(Cv)}{2}.
\end{aligned}
\end{equation}
A basis of generators of $\mathrm{Cl}(V)$ is $(\gamma(e^i), \gamma(Ce^i))_i$, which can be rewritten:
\begin{equation}
\begin{aligned}
\Gamma_R^i &= \gamma(e^i) = a(e^i) + a(e^i)^\times \\
\Gamma_I^i &= \gamma(Ce^i) = i(a(e^i)^\times - a(e^i))
\end{aligned}
\end{equation}
(the subscripts $R$ and $I$ stand for "real" and "imaginary").

Let us now construct the canonical objects associated to Clifford algebras. We start with the chirality operators:
\begin{thm} \label{FockChiralityThm}
There exists a value of the sign in the definition of the chirality operator $\chi = \pm i^{(p-q)/2} \gamma^1 \dots \gamma^d$ for which it is a parity operator for the Fock space $S_C = \bigwedge V_C$: it takes the value $+1$ on $\bigwedge^{\mathrm{ev}} V_C$, the even part of $\bigwedge V_C$, and $-1$ on $\bigwedge^{\mathrm{odd}} V_C$, its odd part.
\end{thm}

\begin{demo}
For any $v \in V$, $\chi$ anti-commutes with $\gamma(v)$. This implies that $\chi$ anti-commutes with all annihilation and creation operators. Since $c$ is an isomorphism, we conclude that for any $v \in V_C$:
\begin{equation*}
\begin{aligned}
\chi a(v)^\times \chi &= -a(v)^\times \\
\chi a(v) \chi &= -a(v)
\end{aligned}
\end{equation*}
We thus have: $a(v) \chi \Omega = -\chi a(v) \Omega = 0$. There thus exists a complex number $\sigma$ such that $\chi \Omega = \sigma \Omega$. Since $\chi^2 = 1$, we must have: $\sigma = \pm 1$. From all of this, we can deduce the action of $\chi$ on $S_V$:
\begin{equation*}
\begin{aligned}
\chi (v_1 \wedge \dots \wedge v_l) &= \chi a(v_1)^\times \dots a(v_l)^\times \Omega \\
&= (\chi a(v_1)^\times \chi) \dots (\chi a(v_l)^\times \chi) \chi \Omega \\
&= (-a(v_1)^\times) \dots (-a(v_l)^\times) \sigma \Omega \\
\chi (v_1 \wedge \dots \wedge v_l) &=  \sigma (-1)^l (v_1 \wedge \dots \wedge v_l).
\end{aligned}
\end{equation*}
The chirality operator thus takes the value $\sigma$ on $\bigwedge^{\mathrm{ev}} V_C$, and $-\sigma$ on $\bigwedge^{\mathrm{odd}} V_C$. The sign $\sigma$ depends on the arbitrary sign that appears in the definition of $\chi$. It is easy to see that there exists a choice for which $\sigma = 1$, hence the theorem.

\qed
\end{demo}

We can thus make the identifications:
\begin{equation*}
\begin{aligned}
S^+ &= \bigwedge\nolimits^{\!\mathrm{ev}} V_C \\
S^- &= \bigwedge\nolimits^{\!\mathrm{odd}} V_C,
\end{aligned}
\end{equation*}

We now turn our attention towards the Robinson and anti-Robinson products:
\begin{thm} \label{FockRobinsonThm}
The Hodge product:
\begin{equation}
(u_1 \wedge \dots \wedge u_k, v_1 \wedge \dots \wedge v_l)_C = \delta_{kl} \mathrm{det}(((u_i, v_j)_C)_{1 \leq i,j \leq k})
\end{equation}
is a Robinson product on $\bigwedge V_C$. The graded Hodge product:
\begin{equation}
\left[u_1 \wedge \dots \wedge u_k, v_1 \wedge \dots \wedge v_l\right]_C = (-1)^k \delta_{kl} \mathrm{det}((u_i, v_j)_{1 \leq i,j \leq k})
\end{equation}
is an anti-Robinson product on $\bigwedge V_C$.

\end{thm}

\begin{demo}
It is clear that $\gamma(v) = a(c(v)) + a^\times(c(v))$ is self-adjoint for the Hodge product. The Hodge product is thus the canonical Robinson product of this representation. To construct an anti-Robinson product, one simply ought to insert the chirality operator, according to equation \eqref{RobinsonAntiRobinson}. Thanks to the previous theorem, we know that this is simply the parity operator of $\bigwedge V_C$, hence the graded Hodge product defined above.

\qed
\end{demo}

Finally, we construct the graded and non-graded charge conjugation operators $J_\pm$:
\begin{thm} \label{FockCCThm}
The charge conjugation operator $J_+$ is defined completely and uniquely by the following property:
\begin{equation}
\omega \wedge J_+ \varphi = \lambda (-1)^{l(l-1)/2} (\varphi, \omega)_C \tilde{\Omega},
\label{JFock}
\end{equation}
for all $l$-vectors $\omega, \varphi$, with $\lambda$ an arbitrary phase. Similarly, we have:
\begin{equation}
\omega \wedge J_- \varphi = \lambda (-1)^{l(l+1)/2+d/2} (\varphi, \omega)_C \tilde{\Omega},
\end{equation}
for all $l$-vectors $\omega, \varphi$.

\end{thm}

\begin{demo}
Given that $J_+$ and $J_-$ are related by the chirality operator $\chi$, the two formulas above can be easily deduced from each other. We will thus only prove the first one. That these two formulas define the action of $J_\pm$ completely and uniquely is a standard result of the theory of Hodge duality. We will give a short proof of this for the readers convenience. Let us prove the first equation.

The defining property of $J_+$ is that is an anti-linear operator that commutes with the $\gamma(v)$'s. We thus have:
\begin{equation*}
\begin{aligned}
J_+ a(c(v)) &= J_+ \frac{\gamma(v) + i \gamma(Cv)}{2} \\
&= \frac{\gamma(v) - i \gamma(Cv)}{2} J_+ \\
J_+ a(c(v)) &= a(c(v))^\times J_+,
\end{aligned}
\end{equation*}
and similarly for $a(c(v))^\times$:
\begin{equation*}
J_+ a(c(v))^\times = a(c(v)) J_+
\end{equation*}
for all $v \in V$. Using the fact that $c$ is an isomorphism gives:
\begin{equation}
\begin{aligned}
J_+ a(v) &= a(v)^\times J_+ \\
J_+ a(v)^\times &= a(v) J_+
\end{aligned}
\end{equation}
for all $v \in V_C$, which shows that there is a relation between $J_+$ and Hodge duality on $S$. From: $a(v)^\times J_+ \Omega = J_+ a(v) \Omega = 0$, we infer the existence of a complex number $\lambda$ such that:
\begin{equation*}
J_+ \Omega = \lambda \tilde{\Omega}.
\end{equation*}
Let $\varphi = u_1 \wedge \dots \wedge u_l \in S$ be an $l$-vector. We have:
\begin{equation*}
\begin{aligned}
J_+ \varphi &= J_+ a(u_1)^\times \dots a(u_l)^\times \Omega \\
&= a(u_1) \dots a(u_l) J_+ \Omega \\
J_+ \varphi &= \lambda a(u_1) \dots a(u_l) \tilde{\Omega} \in \bigwedge\nolimits^{\! d/2 - l} V_C.
\end{aligned}
\end{equation*}
The operator $J_+$ maps $\bigwedge^l V_C$ to $\bigwedge^{d/2-l} V_C$. Let $\omega = v_1 \wedge \dots \wedge v_l \in S$ be another $l$-vector. We have $\omega \wedge J_+ \varphi \in \bigwedge^{d/2} V_C$ which makes it co-linear to $\tilde{\Omega}$. The linear coefficient between the two multivectors specifies $J_+ \varphi$ completely, so let us compute it. We have:
\begin{equation*}
\begin{aligned}
\omega \wedge J_+ \varphi &= v_1 \wedge \dots \wedge v_l \wedge J_+ \varphi \\
&= a(v_1)^\times \dots a(v_l)^\times J_+ \varphi \\
&= J_+ a(v_1) \dots a(v_l) \varphi.
\end{aligned}
\end{equation*}
The multivector $a(v_1) \dots a(v_l) \varphi$ is a $0$-vector, and is thus proportional to $\Omega$:
\begin{equation*}
\begin{aligned}
\omega \wedge J_+ \varphi &= J_+ a(v_1) \dots a(v_l) \varphi \\
&= J_+ (\Omega, a(v_1) \dots a(v_l) \varphi)_C \Omega \\
&= \lambda \overline{(\Omega, a(v_1) \dots a(v_l) \varphi)_C} \tilde{\Omega} \\
&= \lambda (a(v_1) \dots a(v_l) \varphi, \Omega)_C \tilde{\Omega} \\
&= \lambda (\varphi, a(v_l)^\times \dots a(v_1)^\times \Omega)_C \tilde{\Omega}
\end{aligned}
\end{equation*}
Reordering the annihilation operators gives:
\begin{equation*}
\begin{aligned}
\omega \wedge J_+ \varphi &= \lambda (-1)^{l(l-1)/2} (\varphi, a(v_1)^\times \dots a(v_l)^\times \Omega)_C \tilde{\Omega} \\
&= \lambda (-1)^{l(l-1)/2} (\varphi, \omega)_C \tilde{\Omega}.
\end{aligned}
\end{equation*}
By linearity, we conclude that for all $l$-vectors $\omega, \varphi$:
\begin{equation*}
\omega \wedge J_+ \varphi = \lambda (-1)^{l(l-1)/2} (\varphi, \omega)_C \tilde{\Omega},
\end{equation*}
Applying this to $\omega = \varphi = \tilde{\Omega}$, one finds:
\begin{equation*}
\tilde{\Omega} \wedge J_+ \tilde{\Omega} = \lambda (-1)^{d(d-2)/8+q/2} \tilde{\Omega}
\end{equation*}
which implies that
\begin{equation*}
J_+ \tilde{\Omega} = \lambda (-1)^{d(d-2)/8+q/2} \Omega.
\end{equation*}
Applying $J$ twice to $\Omega$ yields:
\begin{equation*}
\begin{aligned}
J_+^2 \Omega &= J_+ \lambda \tilde{\Omega} \\
&= \bar{\lambda} J_+ \tilde{\Omega} \\
J_+^2 \Omega &= |\lambda|^2 (-1)^{d(d-2)/8+q/2} \Omega,
\end{aligned}
\end{equation*}
from which we deduce that $\lambda$ must be a phase. This is the arbitrary phase in the definition of charge conjugation. To see that equation \eqref{JFock} defines the action of $J_+$, consider a given $d/2 - l$-vector $\theta$. We have:
\begin{equation*}
\omega \wedge \theta = \lambda (-1)^{l(l-1)/2} (J_+^{-1} \theta, \omega)_C \tilde{\Omega}.
\end{equation*}
The product $(J_+^{-1} \theta, \omega)_C$ is uniquely specified by $\omega \wedge \theta$. We thus have a linear form on $l$-vectors: $\omega \mapsto (J_+^{-1} \theta, \omega)_C$. The $l$-vector $J_+^{-1} \theta$ is then the dual of this linear form with respect to the Hodge product. This specifies completely and uniquely the action of $J_+^{-1}$, and thus that of $J_+$.

\qed
\end{demo}

For simplicity, we will choose $\lambda = 1$. Let us compute the action of $J_+$ on the basis multivectors $E^{i_1} \wedge \dots \wedge E^{i_l}$. Let $\sigma \in S_{d/2}$ be a permutation of $d/2$ integers, and $\epsilon(\sigma)$ its parity. We have:
\begin{equation*}
\begin{aligned}
J_+ \bigwedge_{i=1}^{l} E^{\sigma(i)} &= J_+ a(E^{\sigma(1)})^\times \dots a(E^{\sigma(l)})^\times \Omega \\
&= a(E^{\sigma(1)}) \dots a(E^{\sigma(l)}) \tilde{\Omega} \\
&= a(E^{\sigma(1)}) \dots a(E^{\sigma(l)}) E^1 \wedge \dots \wedge E^{d/2}.
\end{aligned}
\end{equation*}
By reordering the annihilation operators and the basis vectors, we pick up two additional signs:
\begin{equation*}
\begin{aligned}
J_+ \bigwedge_{i=1}^{l} E^{\sigma(i)} &= \epsilon(\sigma) (-1)^{l(l-1)/2} a(E^{\sigma(l)}) \dots a(E^{\sigma(1)}) E^{\sigma(1)} \wedge \dots \wedge E^{\sigma(d/2)} \\
&= \epsilon(\sigma) (-1)^{l(l-1)/2} (\prod_{1 \leq i \leq l} \epsilon_i) E^{\sigma(l+1)} \wedge \dots \wedge E^{\sigma(d/2)}.
\end{aligned}
\end{equation*}
Using $\prod_{1 \leq i \leq d/2} \epsilon_i = (-1)^{q/2}$, one finds:
\begin{equation}
J_+ \bigwedge_{i=1}^{l} E^{\sigma(i)} = \epsilon(\sigma) (-1)^{l(l-1)/2+q/2} \bigwedge_{i=l+1}^{d/2} (\epsilon_{\sigma(i)} E^{\sigma(i)}).
\label{FockCCBasis}
\end{equation}
A similar formula can be found for the graded conjugation operator $J_- = \chi J$:
\begin{equation}
J_- \bigwedge_{i=1}^{l} E^{\sigma(i)} = \epsilon(\sigma) (-1)^{l(l+1)/2+p/2} \bigwedge_{i=l+1}^{d/2} (\epsilon_{\sigma(i)} E^{\sigma(i)}).
\end{equation}

We now have a complete Fock space representation of Clifford algebras and their associated canonical objects. Let us use it to prove some of the results of the previous subsections. We start with the claim that for a given pseudo-orthonormal basis $(e^a)_a$ of $V$, there exists a representation of $\mathrm{Cl}(V)$ and an inner product on $S$ such that $\gamma(e^a)^\dagger = \epsilon_a \gamma(e^a)$, with $\epsilon_a = g(e^a, e^a)$. For this purpose, we will start by proving that $\eta_+ = \pm i^{q/2} \gamma(e^1)...\gamma(e^q)$ is a fundamental symmetry for the Robinson product\footnote{Note that this can be used to shorten the proof of theorem \ref{RobinsonFundamentalSymmetryTheorem}: we prove here that $i^{q/2} \gamma(e^1)...\gamma(e^q)$ is a fundamental symmetry in some representation. Then, according to the last part of the proof of the theorem, it has to be a fundamental symmetry in all representations.}. Let $C$ be the complex structure associated to the basis $(e^a)_a$ as in \eqref{blockC}. Using the notations previously defined, the operator $\eta_+$ takes the form: $\eta_+ = \pm i^{q/2} \prod_{i=1}^{q/2} (\Gamma_R^i \Gamma_I^i)$. It is easy to see that $\eta_+$ anti-commutes with $\gamma(e^i)$ and $\gamma(Ce^i)$ for $i \leq q/2$, and commutes with them for $i > q/2$. From this one infers that it anticommutes with $a(E^i)^\times$ and $a(E^i)$ for $i \leq q/2$, and commutes with them for $i > q/2$. Using linearity, we put this in the following form:
\begin{equation*}
\begin{aligned}
\eta_+ a(v)^\times \eta_+ &= a(\eta_C v)^\times \\
\eta_+ a(v) \eta_+ &= a(\eta_C v),
\end{aligned}
\end{equation*}
where $\eta_C$ is the fundamental symmetry of $V$ associated with the basis $(e^a)_a$. We have: $a(v) \eta_+ \Omega = \eta_+ a(\eta_C v) \Omega = 0$. Hence the existence of a complex number $\sigma'$ such that $\eta_+ \Omega = \sigma' \Omega$. We can now compute the action of $\eta_+$:
\begin{equation*}
\begin{aligned}
\eta_+ (v_1 \wedge \dots \wedge v_l) &= \eta_+ a(v_1)^\times \dots a(v_l)^\times \Omega \\
&= (\eta_+ a(v_1)^\times \eta_+) \dots (\eta_+ a(v_l)^\times \eta_+) \eta_+ \Omega \\
&= a(\eta_C v_1)^\times \dots a(\eta_C v_l)^\times \sigma' \Omega \\
\eta_+ (v_1 \wedge \dots \wedge v_l) &= \sigma' \eta_C v_1 \wedge \dots \wedge \eta_C v_l.
\end{aligned}
\end{equation*}
We thus have:
\begin{equation*}
\begin{aligned}
(u_1 \wedge \dots \wedge u_k, \eta_+ (v_1 \wedge \dots \wedge v_l))_C &= \sigma' (u_1 \wedge \dots \wedge u_k, \eta_C v_1 \wedge \dots \wedge \eta_C v_l)_C \\
(u_1 \wedge \dots \wedge u_k, \eta_+ (v_1 \wedge \dots \wedge v_l))_C &= \sigma' \delta_{kl} \mathrm{det}(((u_i, \eta_C v_j)_C)_{1 \leq i,j \leq k}).
\end{aligned}
\end{equation*}
The product $(\cdot, \eta_+ \cdot)_C$ is thus the Hodge product of $\sigma' (\cdot, \eta_C \cdot)_C$. Since the latter is positive definite, the former is positive definite if and only if $\sigma'=1$. This fixes the sign in the definition $\eta_+ = \pm i^{q/2} \prod_{i=1}^{q/2} (\Gamma_R^i \Gamma_I^i)$ of the fundamental symmetry. Let us compute this sign explicitly. We have:
\begin{equation*}
\begin{aligned}
\Gamma_R^i \Gamma_I^i \Omega &= (a(E^i) + a(E^i)^\times) i E^i \\
&= i \epsilon_i \Omega
\end{aligned}
\end{equation*}
from which we deduce that:
\begin{equation*}
\eta_+ \Omega = \pm i^{q/2} \prod_{i=1}^{q/2}(i \epsilon_i) \Omega = \pm \Omega
\end{equation*}
and thus that $\eta_+ = i^{q/2} \prod_{i=1}^{q/2} (\Gamma_R^i \Gamma_I^i)$ is a fundamental symmetry for the Hodge/Robinson product. It is then straightforward to prove that:
\begin{equation*}
\gamma(e^a)^{\dagger \eta_+} = \epsilon_a \gamma(e^a).
\end{equation*}

We can also use this representation to compute the signs $\epsilon$ and $\kappa$. Indeed, we saw above that:
\begin{equation*}
J_+^2 \Omega = (-1)^{d(d-2)/8+q/2} \Omega,
\end{equation*}
from which we deduce that $J_+^2 = (-1)^{d(d-2)/8+q/2} = (-1)^{q/2} a(-q-p)$. Using the properties of the $a$ function, one can see that $J_+^2 = (-1)^{q/2} (-1)^{q(p-1)/2} a(q-p) = a(q-p) (-1)^{qp/2} = a(q-p)$, as expected. Let us compute $\kappa$ now. From its definition:
\begin{equation*}
(\omega, J_+ \varphi)_C = \kappa (\varphi, J_+ \omega)_C.
\end{equation*}
In particular:
\begin{equation*}
\begin{aligned}
(\Omega, J_+ \tilde{\Omega})_C = \kappa (\tilde{\Omega}, J_+ \Omega)_C &\Rightarrow \lambda (-1)^{d(d-2)/8+q/2} (\Omega, \Omega)_C = \kappa \lambda (\tilde{\Omega}, \tilde{\Omega})_C \\
&\Rightarrow \lambda (-1)^{d(d-2)/8+q/2} = \kappa \lambda (-1)^{q/2} \\
&\Rightarrow \kappa = (-1)^{d(d-2)/8} = a(-(q+p)),
\end{aligned}
\end{equation*}
as expected.

\subsection{The Odd-Odd Case}

We conclude this section with the case where $q$ and $p$ are odd. A Fock representation cannot be built due to the nonexistence of a complex structure on $V$. One can nonetheless use Fock spaces to represent $\mathrm{Cl}(V)$. To this end, one decomposes $V$ into the sum of two orthogonal subspaces $V = V_1 \oplus V_2$ of signatures $(1,1)$ and $(q-1, p-1)$ respectively. Since $q-1$ and $p-1$ are even, a Fock representation of $\mathrm{Cl}(V_2)$ can be built using a complex structure $C$. As for $\mathrm{Cl}(V_1)$, we use a simple representation based on Pauli matrices. One can then build the tensor product of the Clifford algebras to obtain $\mathrm{Cl}(V)$.

Let us start with $\mathrm{Cl}(V_1)$. Its spinor space is $S_1 \equiv \mathbb{C}^2$. Let $(f_-, f_+)$ be a pseudo-orthonormal basis of $V_1$ such that $g(f_+,f_+) = -g(f_-,f_-) = 1$. Any vector $v$ in $V_1$ can be written: $v = v_+ f_+ + v_- f_-$. The representation is defined by the representation of the generators $\gamma(f_\pm)$:
\begin{equation*}
\begin{aligned}
\gamma_- = \gamma(f_-) &= \begin{pmatrix} 0 & -1 \\ 1 & 0 \end{pmatrix} \\
\gamma_+ = \gamma(f_+) &= \begin{pmatrix} 0 & 1 \\ 1 & 0 \end{pmatrix}
\end{aligned}
\end{equation*}
From which we deduce that
\begin{equation*}
\gamma(v) = v_+ \gamma_+ + v_- \gamma_- =\begin{pmatrix} 0 & v_+ - v_- \\ v_+ + v_- & 0 \end{pmatrix}.
\end{equation*}
The chirality is:
\begin{equation*}
\chi_1 = \gamma_+ \gamma_- = \begin{pmatrix} 1 & 0 \\ 0 & -1 \end{pmatrix}.
\end{equation*}
Since the representation is real, charge conjugation is simply complex conjugation:
\begin{equation*}
J_{1+} \varphi = \overline{\varphi}
\end{equation*}
for any $\varphi \in S_1 \equiv \mathbb{C}^2$. Finally, the canonical product on $S_1$ is simply:
\begin{equation*}
(\omega, \varphi)_1 = \omega^\dagger \gamma_+ \varphi = \omega^\dagger \begin{pmatrix} 0 & 1 \\ 1 & 0 \end{pmatrix} \varphi.
\end{equation*}

We can now build the tensor product of the algebras using the rules given in section \ref{SectionTensorSummary}. We will put this product in a non-graded form. The total spinor space is $S \cong \mathbb{C}^2 \otimes \bigwedge V_{2,C}$. We choose to see it as two copies of the Fock space $\bigwedge V_{2,C}$:
\begin{equation}
S \equiv \bigwedge V_{2,C} \oplus \bigwedge V_{2,C}.
\end{equation}
The Robinson product on spinors is:
\begin{equation}
(\omega \oplus \omega', \varphi \oplus \varphi') = (\omega, \chi_2 \varphi')_C + (\omega', \chi_2 \varphi)_C = \left[\omega, \varphi'\right]_C + \left[\omega', \varphi\right]_C.
\label{OddOddRobinson}
\end{equation}
The representation of the Clifford algebra is given by:
\begin{equation}
\gamma(v_1 \oplus v_2) = \begin{pmatrix} \gamma(v_2) & v_{1+} - v_{1-} \\ v_{1+} + v_{1-} & - \gamma(v_2) \end{pmatrix}.
\end{equation}
This is a $2\times 2$ block matrix, with the blocks in $\mathrm{End}(\bigwedge V_{2,C})$. The total chirality is:
\begin{equation}
\chi= \begin{pmatrix} \chi_2 & 0 \\ 0 & -\chi_2 \end{pmatrix},
\end{equation}
while the total charge conjugation is:
\begin{equation}
J_+ = \begin{pmatrix} J_{2+} & 0 \\ 0 & J_{2+} \end{pmatrix}.
\end{equation}

We can now check that the results proven using the Fock representation in the even-even case can be extended to the odd-odd case. For example, The square of the charge conjugation operator is given by:
\begin{equation*}
J_+^2 = J_{2+}^2 = a((q-1)-(p-1)) = a(q-p),
\end{equation*}
as expected. Its adjoint can also be computed using \eqref{OddOddRobinson}:
\begin{equation*}
\begin{aligned}
(\omega \oplus \omega', J_+ (\varphi \oplus \varphi')) =& \left[\omega, J_{2+} \varphi' \right]_C + \left[\omega', J_{2+} \varphi \right]_C \\
=& \left[J_{2+}^+ \omega, \varphi' \right]_C + \left[J_{2+}^+ \omega', \varphi \right]_C \\
=& a((p-1)+(q-1)) \left[J_{2+} \omega, \varphi' \right]_C + \left[J_{2+} \omega', \varphi \right]_C \\
(\omega \oplus \omega', J_+ (\varphi \oplus \varphi')) =& a(p+q-2) (J_+ (\omega \oplus \omega'), \varphi \oplus \varphi'),
\end{aligned}
\end{equation*}
from which we deduce that $J_+^\times = a(p+q-2) J_+ = a(-(p+q)) J_+$. Finally, let us find a fundamental symmetry for the product \eqref{OddOddRobinson}. If $\eta_{2,-}$ is a fundamental symmetry for $[\cdot,\cdot]_C$, then one can easily check that the following block operator:
\begin{equation*}
\eta_+ = \begin{pmatrix} 0 & \eta_{2,-} \\ \eta_{2,-} & 0 \end{pmatrix} = \gamma(f_+) \otimes \eta_{2,-}.
\end{equation*}
is a fundamental symmetry for the Robinson product. We will now choose a specific $\eta_{2,-}$. Let $(e^a)_a$ be the basis associated to $C$. We saw in the even-even case that a fundamental symmetry for the Robinson product $(\cdot,\cdot)_C$ of $\mathrm{Cl}(V_2)$ is given by: $\eta_{2,+} = i^{(q-1)/2} \prod_{i=1}^{(q-1)/2} (\Gamma_R^i \Gamma_I^i)$. One then finds a fundamental symmetry for the anti-Robinson product by multiplying by chirality: $\eta_{2,-} = \pm i^{(p-1)/2} \prod_{i=(q+1)/2}^{(d-2)/2} (\Gamma_R^i \Gamma_I^i)$. The total fundamental symmetry is thus given by:
\begin{equation*}
\eta_+ = \pm i^{(p-1)/2} (\gamma(f_+) \otimes \prod_{i=(q+1)/2}^{(d-2)/2} (\Gamma_R^i \Gamma_I^i)) = \pm i^{(p-1)/2} \gamma(f_+ \oplus 0) \prod_{i=q}^{d-2} \gamma(0 \oplus e^a). 
\end{equation*}
Notice that the basis $(f_+ \oplus 0,0 \oplus e^1,\dots,0 \oplus e^{q-1},f_+\oplus 0,0 \oplus e^q,\dots,0 \oplus e^{d-2})$ is a pseudo-orthonormal basis of $V$.

As an example, we use this double-Fock representation to construct explicit representations of the Clifford algebra $\mathrm{Cl}(1,3)$. For this algebra, the space $V_2$ has the signature $(0,2)$. If $(e^1, e^2)$ is an orthonormal basis of this space, we can choose the following complex structure:
\begin{equation*}
C = \begin{pmatrix} 0 & -1 \\ 1 & 0 \end{pmatrix}.
\end{equation*}
In this case, $V_{2,C}$ is a one-dimensional complex vector space, with one basis vector $E_1$. The corresponding Fock space $\bigwedge V_{2,C}$ admits the basis $(\Omega,E_1)$. In this space, the annihilation and creation operators take the form:
\begin{equation*}
\begin{aligned}
a(E_1) &= \begin{pmatrix} 0 & 1 \\ 0 & 0 \end{pmatrix} \\
a(E_1)^\times &= \begin{pmatrix} 1 & 0 \\ 0 & 0 \end{pmatrix}.
\end{aligned}
\end{equation*}
The corresponding Clifford algebra generators are:
\begin{equation*}
\begin{aligned}
\gamma(e^1) = a(E_1) + a(E_1)^\times &= \begin{pmatrix} 0 & 1 \\ 1 & 0 \end{pmatrix} \\
\gamma(e^2) = i(a(E_1) - a(E_1)^\times) &= \begin{pmatrix} 0 & i \\ -i & 0 \end{pmatrix}.
\end{aligned}
\end{equation*}
We know (see theorem \ref{FockRobinsonThm}) that the Robinson product coincides with the Hodge product. It is easy to see that $(\Omega,E_1)$ is an orthonormal basis for the Hodge product. As a result, the Robinson product is given by:
\begin{equation*}
(\omega, \varphi)_C = \omega^\dagger \varphi
\end{equation*}
in this basis. Since chirality corresponds to grading, it is given by:
\begin{equation*}
\chi_2 = \begin{pmatrix} 1 & 0 \\ 0 & -1 \end{pmatrix}.
\end{equation*}
The anti-Robinson product is thus given by:
\begin{equation*}
\left[ \omega, \varphi \right]_C = \omega^\dagger \begin{pmatrix} 1 & 0 \\ 0 & -1 \end{pmatrix} \varphi
\end{equation*}
Finally, using equation \eqref{FockCCBasis}, we find:
\begin{equation*}
\begin{aligned}
J_{2+} \Omega &= E_1
J_{2+} E_1 &= \Omega,
\end{aligned}
\end{equation*}
and thus that:
\begin{equation*}
J_{2+} = \begin{pmatrix} 0 & 1 \\ 1 & 0 \end{pmatrix} \circ CC = \gamma(e^1) \circ CC,
\end{equation*}
where $CC$ is complex conjugation in the basis $(\Omega,E_1)$. We now take a look at the total vector space $V$. It has the pseudo-orthonormal basis: $(f_- \oplus 0, f_+ \oplus 0, 0 \oplus e^1, 0 \oplus e^2)$. These vectors are represented by the matrices:
\begin{equation*}
\begin{aligned}
\gamma^0 =& \gamma(f_- \oplus 0) = \begin{pmatrix} 0 & 0 & -1 & 0 \\ 0 & 0 & 0 & -1 \\ 1 & 0 & 0 & 0 \\ 0 & 1 & 0 & 0 \end{pmatrix} \\
\gamma^1 =& \gamma(f_+ \oplus 0) = \begin{pmatrix} 0 & 0 & 1 & 0 \\ 0 & 0 & 0 & 1 \\ 1 & 0 & 0 & 0 \\ 0 & 1 & 0 & 0 \end{pmatrix} \\
\gamma^2 =& \gamma(0 \oplus e^1) = \begin{pmatrix} 0 & 1 & 0 & 0 \\ 1 & 0 & 0 & 0 \\ 0 & 0 & 0 & -1 \\ 0 & 0 & -1 & 0 \end{pmatrix} \\
\gamma^3 =& \gamma(0 \oplus e^2) = \begin{pmatrix} 0 & i & 0 & 0 \\ -i & 0 & 0 & 0 \\ 0 & 0 & 0 & -i \\ 0 & 0 & i & 0 \end{pmatrix}.
\end{aligned}
\end{equation*}
The total chirality operator is given by:
\begin{equation*}
\chi = \begin{pmatrix} 1 & 0 & 0 & 0 \\ 0 & -1 & 0 & 0 \\ 0 & 0 & -1 & 0 \\ 0 & 0 & 0 & 1 \end{pmatrix}.
\end{equation*}
The total charge conjugation operator is given by:
\begin{equation*}
J_+ = \begin{pmatrix} 0 & 1 & 0 & 0 \\ 1 & 0 & 0 & 0 \\ 0 & 0 & 0 & 1 \\ 0 & 0 & 1 & 0 \end{pmatrix} \circ CC = \gamma^0 \gamma^2 \circ CC.
\end{equation*}
The total Robinson product is:
\begin{equation*}
(\omega, \varphi) = \omega^\dagger \begin{pmatrix} 0 & 0 & 1 & 0 \\ 0 & 0 & 0 & -1 \\ 1 & 0 & 0 & 0 \\ 0 & -1 & 0 & 0 \end{pmatrix} \varphi,
\end{equation*}
for $\omega,\varphi \in \mathbb{C}^4$. By inserting the chirality operator, one finds the anti-Robinson product:
\begin{equation*}
\left[ \omega, \varphi \right] = \omega^\dagger \begin{pmatrix} 0 & 0 & -1 & 0 \\ 0 & 0 & 0 & -1 \\ 1 & 0 & 0 & 0 \\ 0 & 1 & 0 & 0 \end{pmatrix} \varphi = \omega^\dagger \gamma^0 \varphi.
\end{equation*}

\chapter{Semi-Riemannian Spin Geometry} \label{ChapterSpin}

In this chapter, we study Spinor bundles on Semi-Riemannian manifolds, and construct the associated canonical objects. The aim of this chapter is to generalize the components of commutative spectral triples on Riemannian manifolds to Semi-Riemannian manifolds, with the hope to arrive at a Semi-Riemannian generalization of spectral triples. We will call those generalized triples \emph{indefinite spectral triples}. To this end, we will mirror the construction of Riemannian commutative spectral triples as presented in \cite{GBVF}. We will also use classic results on Riemannian Spin geometry presented in \cite{LawsonMichelsohn}. Finally, we will use results on Semi-Riemannian Spin geometry presented in \cite{Baum}. Some novel results here are presented in \cite{BB}.

\section{Vector Bundles}

We start with a quick reminder on vector bundles on manifolds, in order to set a few notations. We refer the reader to \cite{Husemoller} for the theory of vector bundles. Let $M$ be a smooth manifold of dimension $d \geq 1$, and $F$ a vector space (real or complex). A (smooth) vector bundle $E$ on the manifold $M$, with fiber $F$, is a manifold such that:
\begin{itemize}
	\item There exists a smooth surjective map $\pi: E \rightarrow M$ called the bundle projection, such that for all $x \in M$, its pre-image $\pi^{-1}(x)$ is a vector space isomorphic to the fiber $F$. We denote this pre-image $E_x \equiv \pi^{-1}(x)$, and call it the local fiber over $x$.
	\item For every open neighborhood $U \subset M$, there exists a diffeomorphism $\Phi_U$ that maps the pre-image of $U$, denoted $E|_U \equiv \pi^{-1}(U)$, to $U \times F$. This diffeomorphism takes the form: 
	\begin{equation}
	\begin{aligned}
	\Phi_U : E|_U &\longrightarrow U \times F \\
	e &\longmapsto (\pi(e), \Pi_U (e)),
	\end{aligned}
	\end{equation}
	where $\Pi_U$ is an isomorphism of vector spaces between $E_x$ and $F$, for any $x \in U$. The pair $(U, \Phi_U)$ is called a local trivialization of the vector bundle.
\end{itemize}

A smooth section of the bundle $E$ is any smooth map $f: M \rightarrow E$ such that the image of $x \in M$ is in the local fiber $E_x$:
\begin{equation*}
\begin{aligned}
f: M &\longrightarrow E \\
x &\longmapsto f(x) \in E_x.
\end{aligned}
\end{equation*}
In other words, it has to satisfy: $\pi \circ f = \mathrm{id}_M$. We denote the set of smooth sections $\Gamma(M,E)$. This set is a vector space, with linear combinations of elements being built locally in each local fiber $E_x$. If $F$ is also an algebra, then $\Gamma(M,E)$ is an algebra as well, with multiplication defined locally.

We denote $\Gamma(\mathrm{End}(E))$ the algebra of endomorphisms of $E$. That is, the set of maps $A : E \rightarrow E$ that map each local fiber $E_x$ linearly to itself, while preserving the base point $x$ (in other words: $\pi \circ A = \pi$). An element of $\Gamma(\mathrm{End}(E))$ can also be seen as a function on $M$ that takes values in the space of endomorphisms of the local fiber:
\begin{equation*}
\begin{aligned}
A: M &\longrightarrow \mathrm{End}(E) \\
x &\longmapsto A(x) \in \mathrm{End}(E_x),
\end{aligned}
\end{equation*}
where $\mathrm{End}(E)$ is a vector bundle over $M$ of fiber $\mathrm{End}(E_x)$. The algebra $\Gamma(\mathrm{End}(E))$ is then simply the space of smooth sections of this bundle.

Let $(U, \Phi_U), (V, \Phi_V)$ be two local trivializations of $E$ such that $U \cap V$ is non-empty. From the linearity of the maps $\Pi_U, \Pi_V$, one can conclude that there must exist a smooth map:
\begin{equation}
g_{UV}: U \cap V \longrightarrow \mathrm{GL}(F)
\end{equation}
such that:
\begin{equation}
\Pi_U(e) = g_{UV}(x) \Pi_V(e),
\end{equation}
for any $x = \pi(e) \in U \cap V$. The map $g_{UV}$ is called a transition function. Transition functions have to form a Cech cocycle. That is, they have to obey the following conditions:
\begin{equation}
\begin{aligned}
g_{UU}(x) &= \mathrm{id}_F \\
g_{UV}(x) &= g_{VU}^{-1}(x) \\
g_{UV}(x) g_{VW}(x) g_{WU}(x) &= \mathrm{id}_F.
\end{aligned}
\end{equation}
If all transition functions take values in some subgroup $G$ of $\mathrm{GL}(F)$, then $G$ is said to be a structure group of the bundle $E$.

Finally, we will denote $TM$ and $T^\ast M$ respectively the tangent and cotangent bundles of $M$, and $T_x M$ and $T^\ast_x M$ their respective local fibers over $x$. We will denote $\Lambda T^\ast M$ the bundle of differential forms, and $\Lambda T^\ast_x M$ its local fibers.

A construction similar to the bundle of differential forms can be done for a general vector bundle $E$. The exterior bundle $\Lambda E$ is the bundle of local fiber $\Lambda E_x$ over the base point $x \in M$. We can use this bundle to define and measure the orientability of the bundle $E$, just as one uses differential forms to measure the orientability of the (co-)tangent bundle. The bundle $E$ is said to be orientable if there exists a nowhere vanishing smooth section of $\Lambda^{\mathrm{dim}F} E$, the bundle of multivectors of maximal degree.

\section{Clifford Bundles and Spin Structures}

Let $M$ be a manifold of even dimension $d$. Let $g$ be a metric (a real symmetric bilinear form):
\begin{equation}
g^{-1} : x \longrightarrow (g_x : T^\ast_x M \times T^\ast_x M \rightarrow \mathbb{R})
\end{equation}
on its cotangent bundle, of signature $(q,p)$. We denote $g$ its extension to the tangent bundle. For each point $x \in M$, the cotangent space $T^\ast_x M$ is equipped with a product $g^{-1}_x$. We can thus construct a Clifford algebra $\mathrm{Cl}(T^\ast_x M)$. Since each of these Clifford algebras is generated by the elements of the local fibers of $T^\ast M$, which has the structure of a bundle, the local algebras $\mathrm{Cl}(T^\ast_x M)$ can be "stitched up" together to form what is called the \emph{Clifford bundle}, denoted $\mathrm{Cl}(T^\ast M)$, whose local fiber over $x$ is $\mathrm{Cl}(T^\ast_x M) \cong \mathrm{Cl}(q,p)$. Similarly, their complexifications can be used to construct the bundle $\mathbb{C}\mathrm{l}(T^\ast M)$, whose local fiber bundle is $\mathbb{C}\mathrm{l}(T^\ast_x M) \cong \mathbb{C}\mathrm{l}(q,p)$.

The canonical embedding $\gamma$ of $T^\ast_x M$ in $\mathrm{Cl}(T^\ast_x M)$ can be extended to an embedding of bundles, and more importantly, of smooth sections:
\begin{equation*}
\gamma: \Gamma(M, T^\ast M) \longrightarrow \Gamma(M, \mathrm{Cl}(T^\ast M)),
\end{equation*}
which enables us to embed 1-forms in smooth sections of the Clifford bundle. One can extend this construction to all differential forms, and obtain the so-called Chevalley-Riesz isomorphism:
\begin{equation}
\begin{aligned}
\gamma : \Gamma(M, \Lambda T^\ast M) &\longrightarrow \Gamma(M, \mathrm{Cl}(T^\ast M)) \\
\omega^1 \wedge \dots \wedge \omega^n &\longmapsto \mathrm{Alt}(\gamma(\omega^1), \dots, \gamma(\omega^n)),
\end{aligned}
\end{equation}
with $\omega^1,...,\omega^n$ 1-forms. Here the Alt symbol is the antisymmetrizer:
\begin{equation}
\mathrm{Alt}[\gamma(\omega^1),...,\gamma(\omega^n)] = \frac{1}{n!} \sum_{\sigma \in S_n} \epsilon(\sigma) \gamma(\omega^{\sigma(1)})...\gamma(\omega^{\sigma(n)}).
\end{equation}
Note that the tangent bundle can be embedded in the Clifford bundle through the musical isomorphism:
\begin{equation*}
\gamma(X) \equiv \gamma(g(X,\cdot)),
\end{equation*}
for any $X \in \Gamma(M, TM)$.

For each $x \in M$, the local fiber $\mathbb{C}\mathrm{l}(T^\ast_x M) \cong \mathbb{C}\mathrm{l}(q,p)$ of the (complexified) vector bundle is the algebra of endomorphisms of some spinor space $S_0$ of $\mathbb{C}\mathrm{l}(q,p)$. One is tempted to consider a vector bundle of fiber $S_0$, on which the Clifford bundle could act. This leads us to the following definition:
\begin{defn}
The manifold $M$ is said to possess a Clifford structure if there exists a bundle $S_M \rightarrow M$, called a \emph{spinor bundle}, that satisfies: $\Gamma(\mathbb{C}\mathrm{l}(TM)) \cong \Gamma(\mathrm{End}(S_M))$. The typical fiber of $S_M$ is then $S_0$.
\end{defn}

To construct a spectral triple, we need the manifold to be orientable. In that case, there exists a nowhere vanishing volume differential form $\omega \in \Gamma(M, \Lambda^d T^\ast M)$. Let $(e^a)_a$ be a pseudo-orthonormal basis of the cotangent bundle over some sufficiently small neighborhood $U$, such that:
\begin{equation*}
g^{-1}(e^a, e^b) = \begin{cases}
-\delta_{ab} &\text{ if $a,b = 1, \dots, q$} \\
\delta_{ab} &\text{ if $a,b = p+1, \dots, d$} \\
0 &\text{ otherwise.} \end{cases}
\end{equation*}
Then there exists a smooth non-zero real function $\lambda$ over $U$ such that $\omega = \lambda e^1 \wedge \dots \wedge e^d$. The Hodge square of $\omega$ (see section \ref{Fock}) is given by: $g^{-1}(\omega,\omega) = (-1)^q \lambda^2$. We are now led to define the normalized differential form:
\begin{equation}
\hat{\omega} = \frac{\omega}{\sqrt{(-1)^q g^{-1}(\omega, \omega)}},
\end{equation}
which has the form: $\hat{\omega} = \pm e^1 \wedge \dots \wedge e^d$ over $U$. Its image in the Clifford bundle is the section: $\gamma(\hat{\omega}) = \pm \gamma(e^1) \dots \gamma(e^d)$, which can be used to define a chirality operator over $U$. We can thus define a global chirality operator:
\begin{equation}
\chi_M = i^{(p-q)/2} \gamma(\hat{\omega}) \in \Gamma(M, \mathrm{Cl}(T^\ast M)).
\end{equation}
over all $M$. The chirality operator is usually defined up to a sign. In the formula above, this sign is determined by the orientation form $\omega$, and hence by the orientation chosen for the manifold $M$.

\section{The Global Robinson Product}

In this section, we assume that the manifold $M$ has a spinor bundle $S_M$. We will determine a necessary and sufficient condition for the existence of globally defined Robinson and anti-Robinson products on smooth sections of the spinor bundles. To this end, we will need to define partial orientability requirements for the cotangent bundle. 

At any point $x \in M$, the tangent bundle $T_x M$ has a fundamental decomposition into a negative subspace $E_{-x}$ and positive subspace $E_{+x}$ that are mutually orthogonal. This decompositions can always be chosen so as to form two smooth vector bundles $E_-$ and $E_+$ such that $TM = E_- \oplus E_+$ (see \cite{FN,Baum} for example). We can now define the partial orientability notions we will need later:
\begin{defn}
When $q>0$ (resp. $p>0$), the manifold $(M,g)$ is said to be negative-orientable (resp. positive-orientable) if and only if there exists a splitting of the tangent bundle $TM = E_- \oplus E_+$ such that the subbundle $E_-$ (resp. $E_+$) is orientable.
\end{defn}

It is easy to see, using local orthogonal transformations, that $(M,g)$ is negative-orientable (resp. positive-orientable) if and only if the subbundle $E_-$ (resp. $E_+$) is orientable for \emph{all} splittings of the tangent bundle $TM = E_- \oplus E_+$.

Just like the overall orientability of a manifold, positive- and negative-orientability can be characterized using differential forms:

\begin{lem}
The manifold $(M,g)$ is negative-orientable if and only if there exists a nowhere vanishing differential $q$-form $\tau$ such that its kernel $\mathrm{Ker} \; \tau_x$ is at all points $x \in M$ a $p$-dimensional positive subspace of $T_x M$. The manifold is positive-orientable if and only if there exists a nowhere vanishing differential $p$-form $\sigma$ such that its kernel $\mathrm{Ker} \; \sigma_x$ is at all points $x \in M$ a $q$-dimensional timelike subspace of $T_x M$.
\end{lem}

\begin{demo}
We will prove the lemma for negative-orientability only. The proof for positive-orientability is very similar. 

Let us prove the direct statement first. Let us assume that $(M,g)$ is negative-orientable. Then there exists a splitting of the tangent bundle $TM = E_- \oplus E_+$ such that $E_-$ is orientable. This means that there exists a nowhere vanishing volume element on $E_-$. That is, a nowhere vanishing form $\tau_- \in \Gamma(M,\bigwedge^q E_-^\ast)$. This form can be extended to a differential $q$-form $\tau \in \Gamma(M,\bigwedge^q T^\ast M)$. Indeed, let $P: TM \rightarrow E_-$ be the smooth projection on $E_-$. Then we define:
\begin{equation*}
\forall X_1,...,X_q \in \Gamma(M,TM): \tau(X_1,...,X_q) = \tau_-(P(X_1),...,P(X_q))
\end{equation*}
This is obviously a smooth differential form which coincides with $\tau_-$ when $X_1,...,X_q \in \Gamma(M,E_-)$. We infer from this that it is nowhere vanishing. Let us prove that the kernel of $\tau_x$ is $(E_+)_x$ for all $x \in M$. Let $X \in T_x M$. If $X \in (E_+)_x$, then $P(X)=0$, which implies that $\tau_x(X,X_2,...,X_q)=0$ for any $X_2,...,X_q \in T_x M$. This means that $X \in \mathrm{Ker} \; \tau_x$. If $X \notin (E_+)_x$, then $P(X) \neq 0$. There thus exist vectors $X_2,...X_q \in (E_-)_x$ such that the family $(P(X),X_2,...,X_q)$ is linearly independent. Since $\tau_-$ is a volume form, we have: $\tau(X,X_2,...,X_q) = \tau_-(P(X),X_2,...,X_q) \neq 0$. This implies that $X \notin \mathrm{Ker} \; \tau_x$. We conclude that $\mathrm{Ker} \; \tau_x = (E_+)_x$.

Conversely, let $\tau$ be a nowhere vanishing differential $q$-form such that its kernel $\mathrm{Ker} \; \tau_x$ is at all points $x \in M$ an $q$-dimensional positive subspace of $T_x M$. We define $E_+ = \bigcup_x \mathrm{Ker} \; \tau_x$. This is a smooth $q$-dimensional positive subbundle of the tangent bundle \cite{Curtis}. Let $E_-$ be the orthogonal smooth subbundle. It is necessarily a negative subbundle for dimensional reasons. We thus have a splitting $TM = E_- \oplus E_+$ of the tangent bundle. Let $\tau_- \in \Gamma(\bigwedge^q E_-^\ast)$ be the restriction of $\tau$ to $E_-$. For any $x \in M$, $(\tau_-)_x$ cannot vanish, otherwise the kernel of $\tau_x$ would be all of $T_x M$. This proves that the subbundle $E_-$ is orientable.

\qed
\end{demo}

Notice that the differential form $\sigma$ or $\tau$ contains more information than just the positive or negative-orientation of the manifold, since it also gives a preferred splitting of the tangent bundle. This means that a positive- or negative-orientation is represented by a class of differential forms whose kernels might differ, but are "similarly oriented".

For a Lorentzian manifold ($q=1$ or $p=1$), this definition of orientability coincides with the usual notions of space- and time-orientability. Indeed, according to this lemma, for $q=1$, negative-orientability is equivalent to the existence of a nowhere vanishing 1-form, whose kernel is positive. The associated vector field through the metric $g$ (by raising indices) is thus timelike and nowhere vanishing, and the manifold is time-orientable. For $p=1$, time-orientability is equivalent to positive-orientability.

We now wish to construct a \emph{local Robinson product} $H$. That is, a map:
\begin{equation}
H : x \longrightarrow (H_x : S_x \times S_x \rightarrow \mathbb{C})
\end{equation}
such that $H_x$ is a Robinson product for the Clifford algebra $\mathbb{C}\mathrm{l}(T^\ast_x M)$ and the spinor space $S_x$. The necessary and sufficient conditions for the existence of such a product are given by the following theorem:

\begin{thm} \label{LocalRobinsonThm}
There exists a nowhere degenerate local Robinson product on the spinor bundle $S_M$ if and only if the manifold $M$ is negative-orientable when $q,p$ are even, and if and only if it is positive-orientable when $q,p$ are odd. 

\end{thm}

This theorem extends lemma 3.4 in \cite{Baum} (see P.141).

\begin{demo}
In the following, $(U_\alpha)_{\alpha \in I}$ will be a covering of $M$ with open subsets sufficiently small so that the tangent and spinor bundle are trivial, and $(f_\alpha)_\alpha$ will be a smooth partition of unity subordinate to the covering $(U_\alpha)_\alpha$. For all $x \in M$ we define the set $I_x = \{\alpha | x \in U_\alpha \} \neq \emptyset$, and the neighborhood $U_x = \bigcap_{\alpha \in I_x} U_\alpha$. We will use a splitting of the tangent bundle $TM = E_- \oplus E_+$. We consider the $q,p$ even case only. The odd case can be proven similarly.

Let $H$ be a local Robinson product. On each $U_\alpha$, the spinor bundle is trivial $S|_{U_\alpha} \simeq U_\alpha \times S_0$. There thus exists a constant nonvanishing section $\psi_\alpha$. Let $TM = E_- \oplus E_+$ be a splitting of the tangent bundle. We define the following complex-valued smooth $q$-form on $E_-$:
\begin{equation}
\tau_-(X_1,...,X_q)(x) = \sum_{\alpha \in I_x} f_\alpha(x) H_x(\psi_\alpha, i^{q/2} \mathrm{Alt}[\gamma(X_1),...,\gamma(X_q)] \psi_\alpha),
\label{tform}
\end{equation}
for all $X_1,...,X_q \in \Gamma(M,E_-)$. Let $x \in M$. Let $(e_1,...,e_d)$ be a local pseudo-orthonormal basis of the tangent space over $U_x$, such that $(e_1,...,e_q)$ is a basis of $E_-$, and $(e_{q+1},...,e_d)$ is a basis of $E_+$. Because the $\gamma(e_a)$ anti-commute, we have that $\mathrm{Alt}[\gamma(e_1),...,\gamma(e_q)] = \gamma(e_1)...\gamma(e_q)$, which gives us:
\begin{equation*}
\tau_-(e_1,...,e_q)(x) = \sum_{\alpha \in I_x} f_\alpha(x) H_x(\psi_\alpha, i^{q/2} \gamma(e_1)...\gamma(e_q) \psi_\alpha).
\end{equation*}
According to theorem \ref{RobinsonFundamentalSymmetryTheorem}, there exists a sign $\epsilon  = \pm 1$ such that $ \epsilon i^{q/2} \gamma(e_1)...\gamma(e_q)$ is a fundamental symmetry for the indefinite product $H_x(\cdot, \cdot)$. This means that $\epsilon H_x(\cdot, i^{q/2} \gamma(e_1)...\gamma(e_q) \cdot)$ is positive definite, and thus that $\epsilon H_x(\psi_\alpha, i^{q/2} \gamma(e_1)...\gamma(e_q) \psi_\alpha) > 0$ for all $\alpha \in I_x$. Since there is at least one $\alpha \in I_x$ such that $f_\alpha(x)>0$, we conclude that $\epsilon \tau_-(e_1,...,e_q)(x) > 0$. The form $\tau_-$ is thus nowhere vanishing. It remains to prove that it is real-valued. To do this, it suffices to notice that due to the anti-symmetry of $\tau_-$ we have:
\begin{equation*}
\tau_-(X_1,...,X_q) = \mathrm{det}(X^a_i) \tau_-(e_1,...,e_q),
\end{equation*} 
where the $X^a_i$ are the components of the $X_i$ in the local basis $(e_1,...,e_q)$ of $E_-$. This proves that $(M,g)$ is negative-orientable. Note that the sign $\epsilon$ depends only on $H_x$, and not on the $U_\alpha$ or $\psi_\alpha$. It is indeed the sign that makes $\epsilon i^{q/2} \gamma(e_1)...\gamma(e_q)$ a fundamental symmetry for $H_x$. This means that the orientation defined by $H$ does not depend on the choice of trivialization or constant vectors, but only on $H$ itself.

Conversely, let us assume that $(M,g)$ is negative-orientable. Then there exists a negative-volume differential form $\tau$, according to the previous theorem. Just as the total volume form can be normalized, the form $\tau$ can be normalized too. Over the open set $U_\alpha$, the spinor bundle becomes trivial: $S|_{U_\alpha} \simeq U_\alpha \times S_0$. Let $(\cdot, \cdot)_0$ be a particular realization of the Robinson product on $S_0$. This induces a constant hermitian form $(\cdot,\cdot)_\alpha$ over $U_\alpha$. Let $(e_1,...,e_d)$ be a local pseudo-orthonormal basis of the tangent space over $U_\alpha$, such that $(e_1,...,e_q)$ is a basis of $E_- = \mathrm{Ker}(\tau)$. Then $\tau$ takes the form: $\tau = \lambda e^1 \wedge ... \wedge e^q$, with $\lambda$ a nowhere vanishing real function over $U_\alpha$, and $(e^a)_a$ the dual basis of $(e_a)_a$. One finds that: $g(\tau,\tau) = (-1)^q \lambda^2$, which leads us to construct the nowhere vanishing, normalized differential form:
\begin{equation}
\hat{\tau} = \frac{\tau}{\sqrt{(-1)^q g^{-1}(\tau, \tau)}},
\end{equation}
which takes the form: $\hat{\tau} = \pm e^1 \wedge ... \wedge e^q$ over $U_\alpha$. We now define the Clifford smooth section:
\begin{equation}
\eta_+ = i^{q/2} \gamma(\hat{\tau}) = \pm i^{q/2} \gamma(e^1)...\gamma(e^q).
\end{equation}
There exists a sign $\epsilon_\alpha = \pm 1$ such that $\epsilon_\alpha \eta_+$ is a fundamental symmetry for $(\cdot,\cdot)_\alpha$. This means that the form $\epsilon_\alpha (\cdot, \eta_+ \cdot)_\alpha$ is positive definite for all $\alpha$. Next, we construct the hermitian form:
\begin{equation*}
\tilde{H}_x(\cdot,\cdot) = \sum_{\alpha \in I_x} f_\alpha(x) \epsilon_\alpha (\cdot, \eta_+ \cdot)_\alpha.
\end{equation*}
This is clearly a positive definite form. Finally, we define:
\begin{equation*}
H_x(\cdot,\cdot) = \tilde{H}_x(\cdot,\eta_+ \cdot) = \sum_{\alpha \in I_x} f_\alpha(x) \epsilon_\alpha (\cdot, \cdot)_\alpha.
\end{equation*}
where we have used $\eta_+^2=1$. This hermitian form is nowhere degenerate, because, $\tilde{H}$ is everywhere positive definite, and $\eta_+$ is invertible. It is smooth, because the $(\cdot, \cdot)_\alpha$ are locally constant (and thus smooth). And last but not least, it is a Robinson product, because the $(\cdot, \cdot)_\alpha$ are. This concludes our proof.

\qed
\end{demo}

In the Riemannian case ($q=0$), The Robinson product always exists thanks to its positiveness, which allows one to use directly a partition of unity. This is not possible for other signatures (including an anti-Riemannian signature $p=0$), as the hermitian form $H$ is then indefinite. Hence the need for a globally defined fundamental symmetry to turn the indefinite product into a positive definite one. This explains the need for a positive or negative orientation.

A few remarks are in order. Notice that when the local Robinson product exists, the proof of the theorem provides a globally defined fundamental symmetry for $H$. This is simply $\eta_+ = i^{q/2} \gamma(\hat{\tau})$ in the even case, and $\eta_+ = i^{(p-1)/2} \gamma(\hat{\sigma})$ in the odd case, where:
\begin{equation}
\hat{\sigma} = \frac{\sigma}{\sqrt{g^{-1}(\sigma, \sigma)}}
\end{equation}
is the normalized form of $\sigma$, a positive-orientation form.

For a given Robinson product $H$ satisfying the requirements of the theorem, the Robinson uniqueness theorem implies that all possible products are of the form $\lambda H$, where $\lambda \in \mathscr{C}^\infty (M, \mathbb{R}^\ast)$. This implies that there are two classes of Robinson products. In each class, the products are related by \emph{strictly positive} smooth functions. One goes from one class to the other by multiplying by a negative smooth function. These two classes induce two classes of differential forms, which correspond to the two possible choices of negative- or positive-orientation.

A similar theorem (with a similar proof) can be formulated for anti-Robinson products. A \emph{local anti-Robinson product} $G$ is a map:
\begin{equation}
G : x \longrightarrow (G_x : S_x \times S_x \rightarrow \mathbb{C})
\end{equation}
such that $G_x$ is an anti-Robinson product for the Clifford algebra $\mathbb{C}\mathrm{l}(T^\ast_x M)$ and the spinor space $S_x$. The theorem is the following:
\begin{thm} \label{LocalAntiRobinsonThm}
There exists a nowhere degenerate local anti-Robinson product on the spinor bundle $S_M$ if and only if the manifold $M$ is positive-orientable when $q,p$ are even, and if and only if it is negative-orientable when $q,p$ are odd. 

\end{thm}

The orientability requirements are now reversed with respect to the previous theorem. In particular, an anti-Robinson product always exists for an anti-Riemannian signature ($p=0$). Just as for Robinson products, anti-Robinson products come naturally with their fundamental symmetries, and are divided in two classes that each correspond to a choice of negative- or positive-orientation.

For spectral triples, we need the manifold to be orientable. In that case, negative- and positive-orientability are equivalent, and we simply say that the manifold is Space- and Time-Orientable (STO). We have the following corollary:
\begin{cor} \label{RobinsonCor}
Let $(M,g)$ be an orientable manifold with a Clifford structure. The following statements are equivalent:
\begin{enumerate}
	\item The manifold is STO
	\item There exists a nowhere degenerate local Robinson product on the spinor bundle
	\item There exists a nowhere degenerate local anti-Robinson product on the spinor bundle.
\end{enumerate}
\end{cor}
In the case of an orientable manifold, a globally defined local chirality operator $\chi_M$ always exists, and can be used to relate Robinson and anti-Robinson products when they both exist:
\begin{equation}
G(\cdot, \cdot) = \lambda H(\cdot, i^q \chi_M \cdot),
\end{equation}
with $\lambda$ a nowhere vanishing real function.

This leads us to the following Semi-Riemannian generalization of Spin$^c$ structures:
\begin{defn}
A manifold $(M,g)$ is said to possess a Spin$^c$ structure if and only if it is STO, and has a Clifford structure.
\end{defn}

Local Robinson and anti-Robinson products can be used to construct a global product through integration. We thus define the associated global Robinson product:
\begin{equation}
\begin{aligned}
(\cdot, \cdot)_M : \Gamma_c(M,S_M) \times \Gamma_c(M,S_M) &\longrightarrow \mathbb{C} \\
(\psi,\varphi) &\longmapsto (\psi,\varphi)_M = \int_{x \in M} H_x (\psi,\varphi) \sqrt{|g|} d^d x,
\end{aligned}
\end{equation}
where $g$ is the determinant of the metric, and $\Gamma_c(M,S_M)$ is the space of smooth sections of the spinor bundle with \emph{compact support}. We also define the associated global anti-Robinson product:
\begin{equation}
\begin{aligned}
\left[\cdot, \cdot \right]_M : \Gamma_c(M,S_M) \times \Gamma_c(M,S_M) &\longrightarrow \mathbb{C} \\
(\psi,\varphi) &\longmapsto [\psi,\varphi]_M = \int_{x \in M} G_x (\psi,\varphi) \sqrt{|g|} d^d x.
\end{aligned}
\end{equation}
The restriction to spinors with compact support is to ensure the convergence of the global products. For any $\Omega \in \Gamma(M,\mathbb{C}\mathrm{l}(T^\ast M))$, the local and global adjoints coincide in the sense that $(\Omega^\times)(x) = \Omega(x)^\times$ and $(\Omega^+)(x) = \Omega(x)^+$

\section{The Global Charge Conjugation Operator}

We now move on to charge conjugation, and look for a globally defined \emph{local charge conjugation operator} for Spin manifolds. That is, an operator:
\begin{equation}
J_{M+} : x \longrightarrow (J_{+,x} : S_x \rightarrow S_x)
\end{equation}
such that $J_{+,x}$ is a charge conjugation operator for the Clifford algebra $\mathbb{C}\mathrm{l}(T^\ast_x M)$ and the spinor space $S_x$. To construct this operator, we need an additional requirement on the spinor bundle:
\begin{defn}
The manifold $M$ is said to possess a Spin structure if it is STO (and thus has a $\mathrm{Spin}^c$ structure), and the structure group of the spinor bundle can be reduced to $\mathrm{Spin}_0(q,p)$.
\end{defn}
Note that not all manifolds possess a Clifford, Spin$^c$, or Spin structure. The topological requirements for the existence of closely related structures (notably for orientable manifolds) have been studied in \cite{GBVF, LawsonMichelsohn, Trautman, Baum}.

We have the following:
\begin{thm} \label{CCSpinManifoldThm}
Let $M$ be a Spin manifold. Then there exists a local charge conjugation operator $J_{M+}$.
\end{thm}

\begin{demo}
Let $\pi: S_M \rightarrow M$ be the projection of the spinor bundle $S_M$ on its base $M$. Let $(U_\alpha, \Phi_\alpha)_{\alpha \in I}$ be a family of trivializations of $S_M$ with $\mathrm{Spin}_0 (q,p)$-valued transition functions\footnote{Here $\mathrm{Spin}_0 (q,p)$ is seen as a subgroup of the invertible elements of $\mathbb{C}\mathrm{l}(q,p) = \mathrm{End}(S_0)$.}, and $(U_\alpha)_\alpha$ an open cover of $M$. We can assume, without any loss of generality, that every $U_\alpha$ is simply connected. Such a family of trivializations always exists thanks to the Spin structure of $S_M$. The map $\Phi_\alpha$ is the diffeomorphism that trivializes $S_M$ over $U_\alpha$:
\begin{equation*}
\begin{aligned}
\Phi_\alpha : S_M|_{U_\alpha} &\longrightarrow U_\alpha \times S_0 \\
s &\longmapsto (\pi(s), \Pi_\alpha (s)).
\end{aligned}
\end{equation*}
For $\alpha,\beta \in I$ such that $U_\alpha \cap U_\beta \neq \emptyset$, there exists a smooth transition function $g_{\alpha \beta}: U_\alpha \cap U_\beta \rightarrow \mathrm{Spin}_0 (q,p)$ such that, for $\pi(s) = x \in U_\alpha \cap U_\beta$ we have $\Pi_\alpha (s) = g_{\alpha \beta}(x) \Pi_\beta (s)$.

Let $J_0$ be a charge conjugation operator on $S_0$. We define a local charge conjugation operator $J_\alpha$ on each $U_\alpha$ whose local form is $J_0$:
\begin{equation*}
\Pi_\alpha(J_\alpha s) = J_0 \Pi_\alpha(s), 
\end{equation*}
for $s \in S|_{U_\alpha}$. Now, let $s \in S|_{U_\alpha \cap U_\beta}$, and $x=\pi(s) \in U_\alpha \cap U_\beta$. We thus have:
\begin{equation*}
\begin{aligned}
\Pi_\alpha(J_\alpha s) &= J_0 \Pi_\alpha(s) \\
\Pi_\beta(J_\beta s) &= J_0 \Pi_\beta(s).
\end{aligned}
\end{equation*}
Using the fact that $J_0$ commutes with all elements of $\mathrm{Spin}_0 (q,p)$, we find:
\begin{equation*}
\begin{aligned}
\Pi_\alpha(J_\alpha s) &= J_0 \Pi_\alpha(s) \\
&= J_0 g_{\alpha \beta}(x) \Pi_\beta (s) \\
&= g_{\alpha \beta}(x) J_0 \Pi_\beta (s) \\
&= g_{\alpha \beta}(x) \Pi_\beta(J_\beta s) \\
\Pi_\alpha(J_\alpha s) &= \Pi_\alpha(J_\beta s),
\end{aligned}
\end{equation*}
from which we deduce that $J_\alpha = J_\beta$ over $U_\alpha \cap U_\beta$. The local operators $J_\alpha$ can thus be patched up to form a globally defined charge conjugation operator $J_{M+}$ such that $J_{M+} = J_\alpha$ over $U_\alpha$.

\qed
\end{demo}

Note that the same technique can be used to prove the existence of local Robinson and anti-Robinson products on spin manifolds.

If $J_{M+}$ is a local charge conjugation operator, then all charge conjugation operators are necessarily of the form: $\rho J_{M+}$, with $\rho \in \mathcal{C}^\infty(M, U(1))$ a local phase. This is a consequence of the uniqueness, up to a phase, of charge conjugation operators for Clifford algebras.

Thanks to the existence of a chirality operator $\chi_M$, we can also define a globally defined graded charge conjugation operator:
\begin{equation}
J_{M-} = \rho \chi_M J_{M+},
\end{equation}
with $\rho$ a local phase. Hence the following corollary:
\begin{cor}
Let $M$ be a Spin manifold. Then there exists a local graded charge conjugation operator $J_{M-}$.
\end{cor}
Again, if $J_{M-}$ is a local graded charge conjugation operator, then all graded charge conjugation operators are necessarily of the form: $\rho J_{M-}$, with $\rho$ a local phase.

\section{Clifford connection and Dirac Operator}

We now construct the main objective of this chapter: the Dirac operator. To construct a Dirac operator, we will first need a Clifford connection:
\begin{defn}
On a manifold $M$ which admits a spinor bundle $S_M$, a Clifford connection $\nabla^S: \Gamma(M,S_M) \rightarrow \Gamma(M,S_M)$ is a connection:
\begin{equation}
\nabla_X^S(f \psi) = X(f) \psi + f \nabla_X^S \psi
\end{equation}
that lifts the Levi-Civita connection $\nabla$ to the spinor bundle $S$:
\begin{equation}
\nabla_X^S(\gamma(Y) \psi) = \gamma(\nabla_X Y) \psi + \gamma(Y) \nabla_X^S \psi
\end{equation}
for any $f \in \mathcal{C}^\infty (M, \mathbb{C})$, $X,Y \in \Gamma(M,TM)$ and $\psi \in \Gamma(M,S_M)$.

\end{defn}

Clifford connections have the following useful properties:
\begin{prop} \label{CliffordConnectionOmega}
Let $\nabla^S$ be a Clifford connection. Then $\nabla^{\prime S}$ is a Clifford connection if and only if there exists a complex-valued 1-form $\omega$ such that $\nabla^{\prime S}_X - \nabla^S_X = \omega(X)$.
\end{prop}

\begin{demo}

Proving the converse statement is straightforward. Let us prove the direct statement. Let us assume that $\nabla^{\prime S}$ is a Clifford connection. From the definition above, we have:
\begin{equation*}
\begin{aligned}
(\nabla^S_X - \nabla^{\prime S}_X) (f\psi) &= f \nabla_X^S \psi - f \nabla^{\prime S}_X \psi \\
&= f (\nabla^S_X - \nabla^{\prime S}_X) \psi.
\end{aligned}
\end{equation*}
This implies that $\nabla^S_X - \nabla^{\prime S}_X$ commutes with scalar functions. We deduce that it is a local endomorphism of the spinor bundle. We also have:
\begin{equation*}
\begin{aligned}
(\nabla^S_X - \nabla^{\prime S}_X) (\gamma(Y)\psi) &= \gamma(Y) \nabla_X^S \psi - \gamma(Y) \nabla^{\prime S}_X \psi \\
&= \gamma(Y) (\nabla^S_X - \nabla^{\prime S}_X) \psi,
\end{aligned}
\end{equation*}
which means that $(\nabla^S_X - \nabla^{\prime S}_X)$ commutes with all generators of $\Gamma(\mathbb{C}\mathrm{l}(T^\ast M))$. Since even Clifford algebras are simple, we deduce that $(\nabla^S_X - \nabla^{\prime S}_X)$ is a scalar function. Finally, since it is linear in $X$, there must exist a 1-form such that $\nabla^{\prime S}_X - \nabla^S_X = \omega(X)$.

\qed
\end{demo}

Although the existence of a global Clifford connection is not guaranteed, it is well-known that one can always construct one locally (see theorem 9.8 in \cite{GBVF}):
\begin{prop}
On a local trivialization $S_M|_U \simeq U \times S_0$, with $U$ a neighborhood small enough so that tangent and spinor bundles are both trivial, a locally defined Clifford connection is given by $\nabla_X^{lS} = \partial_X + \Gamma(X)$, where $\Gamma(X) = -\frac{1}{4} e^a(\nabla_X e_b) \gamma(e^b) \gamma(e_a)$. Here $(e_a)_a$ is a local pseudo-orthonormal basis of the tangent space, with dual basis $(e^a)_a$, such that the Clifford bundle elements $\gamma(e_a)$ and $\gamma(e^a)$ are constant over $U$.
\end{prop}

From the two previous properties, we deduce that any Clifford connection $\nabla^S$ takes the local form: $\nabla_X^{S} = \partial_X + \Gamma(X) + \omega(X)$. This results in the following theorem:
\begin{thm}
If $M$ is a spin$^c$ manifold that admits a Clifford connection $\nabla^S$, then the latter commutes with the chirality operator:
\begin{equation}
\left[\nabla_X^S, \chi_M\right] = 0,
\end{equation}
for all $X \in \Gamma(M,TM)$.

\end{thm}

\begin{demo}
Locally, on the trivialization used above, the chirality operator takes the form $\chi_M = \pm i^{(p-q)/2} \gamma(e^1) \dots \gamma(e^d)$. Since the $\gamma(e^a)$ are locally constant, $\chi_M$ and $\partial_X$ commute. Moreover, since $\Gamma(X) + \omega(X)$ is an even element of the Clifford bundle, it commutes with $\chi_M$ too. Thus, $\nabla_X^S$ commutes with $\chi_M$.

\qed
\end{demo}

In the following, we will: (i) define the Dirac operator, (ii) impose conditions on this operator related to the Spin$^c$ and Spin structures, (iii) deduce conditions on the Clifford connection, and (iv) prove that there always exists a Clifford connection, and thus a Dirac operator, that satisfies these axioms. There are two conventions for the Dirac operator, related to the East Coast and West Coast conventions for the metric (see section \ref{Dimensions}). We will treat them separately:

\subsection{The Real Convention}

The Dirac operator is an operator on smooth spinors with compact support: $\slashed{D} : \Gamma_c (M,S_M) \rightarrow \Gamma_c (M,S_M)$, defined by:
\begin{equation}
\slashed{D} = \gamma(\omega^a) \nabla_{e_a}^S
\label{DiracReal}
\end{equation}
(with an implicit sum over the index $a$), where $e_a$ is any local basis of the tangent space, and $\omega^a$ its unique dual basis. This expression is independent of the choice of local basis. One can in fact choose a local coordinate basis, and write:
\begin{equation}
\slashed{D} = \gamma(dx^\mu) \nabla_\mu^S
\end{equation}
where $\nabla_\mu^S \equiv \nabla_{\partial_\mu}^S$. From the commutation of $\chi_M$ and $\nabla^S$, one deduces that:
\begin{equation}
\{\slashed{D}, \chi_M\} = 0.
\end{equation}
We want the Dirac operator to have the same properties it has in spectral triples in the Riemannian case: it must be self-adjoint and must commute with some charge conjugation operator.

First, we consider a general invertible anti-linear operator $J$ that acts \emph{locally} on smooth spinor sections, and require that it commutes with $\slashed{D}$. We have the following theorem:
\begin{thm} \label{RealStructureThm}
The Dirac operator and anti-linear operator $J$ commute if and only if $[J,\gamma(X)] = [J,\nabla_X^S] = 0$ for all $X \in \Gamma(M,TM)$.
\end{thm}

\begin{demo}
The fact that $[J,\gamma(X)] = [J,\nabla_X^S] = 0$ are sufficient conditions is straightforward to prove.

Let us prove that they are necessary. We assume that $[J, \slashed{D}]=0$. Let $f$ be a smooth \emph{real}-valued function. Then $[J,f]=0$. From this we deduce that $J$ commutes with $[\slashed{D}, f] = \gamma(df)$. We infer easily that $J$ commutes with any real vector field or differential form. Next we consider the differential operator $\nabla^{\prime S}= J^{-1} \nabla^S J$. This operator is a Clifford connection. Indeed, for any $X,Y \in \Gamma(M,TM)$, $\psi \in \Gamma_c(M,S_M)$:
\begin{equation*}
\begin{aligned}
\nabla_X^{\prime S}(\gamma(Y) \psi) &= J^{-1} \nabla_X^S J(\gamma(Y) \psi) \\
&= J^{-1} \nabla_X^S (\gamma(Y) J\psi) \\
&= J^{-1} (\gamma(\nabla_X Y) J\psi + \gamma(Y) \nabla_X^S (J\psi)) \\
\nabla_X^{\prime S} (\gamma(Y) \psi) &= \gamma(\nabla_X Y) \psi + \gamma(Y) \nabla_X^{\prime S} \psi.
\end{aligned}
\end{equation*}
There thus exists a complex-valued one-form $\omega$ such that: $\nabla_X^{\prime S} - \nabla_X^S = \omega(X)$. This can also be written: $\omega(X) = J^{-1} [\nabla_X^S, J]$. We have:
\begin{equation*}
\begin{aligned}
0 &= [J, \slashed{D}] \\
&= [J, \gamma(dx^\mu) \nabla_\mu^S] \\
&= \gamma(dx^\mu)[J, \nabla_\mu^S] \\
0 &= J \gamma(dx^\mu) \omega_\mu \\
\end{aligned}
\end{equation*}
This implies that $\omega=0$, and thus that $[\nabla_X^S, J]=0$.

\qed
\end{demo}

From the condition $[J,\gamma(X)]=0$, we know that $J$ must be a charge conjugation operator. Such an operator necessarily squares to a real nonvanishing scalar function (see theorem \ref{CCThm}): $J^2 = \lambda$. Since $J$ must commute with the Clifford connection, so does $\lambda$: $\partial_\mu \lambda = [\nabla_\mu^S, \lambda] = 0$. The function $\lambda$ is thus constant, and one can normalize the charge conjugation operator $J$ globally: $J \rightarrow J/\sqrt{|\lambda|}$, without losing the commutation with the Clifford connection. One concludes that:
\begin{itemize}
	\item We must choose $J$ to be a normalized charge conjugation operator $J = J_{M+}$.
	\item The Clifford connection must commute with charge conjugation on spinors: $[\nabla_X^S, J_{M+}]=0$.
\end{itemize}

Next, we consider a general local product (\emph{i.e.} hermitian form) on the spinor bundle:
\begin{equation*}
B : x \longrightarrow (B_x : S_x \times S_x \rightarrow \mathbb{C})
\end{equation*}
that we integrate to construct a product on compactly supported smooth spinors:
\begin{equation*}
\begin{aligned}
B_M(\cdot, \cdot) : \Gamma_c(M,S_M) \times \Gamma_c(M,S_M) &\longrightarrow \mathbb{C} \\
(\psi,\varphi) &\longmapsto B_M(\psi,\varphi) = \int_{x \in M} B_x (\psi,\varphi) \sqrt{|g|} d^d x.
\end{aligned}
\end{equation*}
We want the Dirac operator to be symmetric with respect to this product, hence the following theorem:
\begin{thm} \label{ProductThm}
The Dirac operator is symmetric for $B_M$ if and only if real vector fields are anti-self-adjoint: 
\begin{equation*}
B(\gamma(X) \psi, \varphi) + B(\psi, \gamma(X) \varphi) = 0
\end{equation*}
for all $X \in \Gamma(M,TM)$, and $\psi,\varphi \in \Gamma_c(M,S_M)$, and the Clifford connection $\nabla^S$ is hermitian (or metric) for $B$:
\begin{equation*}
X(B(\psi, \varphi)) = B(\nabla_X^S \psi, \varphi) + B(\psi, \nabla_X^S \varphi).
\end{equation*}
\end{thm}

\begin{demo}
Let us assume that the Dirac operator is symmetric. Let $f$ be a real-valued smooth function. Then $f$ is symmetric. This implies that $[\slashed{D}, f] = \gamma(df)$ is anti-symmetric. We easily conclude from this that all real vector fields and all real-valued differential forms are anti-symmetric. In particular, the $\gamma(dx^\mu)$ are anti-symmetric. Let $\psi, \varphi$ be smooth spinors with compact support. We have:
\begin{equation*}
\begin{aligned}
0 &= B_M(\psi, \slashed{D} \varphi) - B_M(\slashed{D} \psi, \varphi) \\
&= \int \sqrt{|g|} \left[B(\psi, \gamma(dx^\mu) \nabla_\mu^S \varphi) - B(\gamma(dx^\mu) \nabla_\mu^S \psi, \varphi) \right] \\
0 &= \int \sqrt{|g|} \left[B(\psi, \gamma(dx^\mu) \nabla_\mu^S \varphi) + B(\nabla_\mu^S \psi, \gamma(dx^\mu) \varphi) \right] \\
\end{aligned}
\end{equation*}
We also have that: 
\begin{equation*}
\begin{aligned}
\left[\nabla_\mu^S, \gamma(dx^\mu)\right] &= \gamma(\nabla_\mu(dx^\mu)) \\
&= -\Gamma^\mu_{\mu\alpha} \gamma(dx^\alpha)\\
&= -\gamma(dx^\alpha) \frac{\partial_\alpha \sqrt{|g|}}{\sqrt{|g|}}
\end{aligned}
\end{equation*}
from which we infer that $\gamma(dx^\mu) \nabla_\mu^S = \nabla_\mu^S \gamma(dx^\mu) + \gamma(dx^\mu) (\partial_\mu \sqrt{|g|})/ \sqrt{|g|}$. Substituting in the integral above gives us:
\begin{equation*}
\int \left[ \sqrt{|g|} \left(B(\nabla_\mu^S \psi, \gamma(dx^\mu) \varphi) + B(\psi, \nabla_\mu^S (\gamma(dx^\mu) \varphi)) \right) + (\partial_\mu \sqrt{|g|}) B(\psi, \gamma(dx^\mu) \varphi)  \right] = 0.
\end{equation*}
Finally, an integration by part yields:
\begin{equation*}
\int \sqrt{|g|} \left[ B(\nabla_\mu^S \psi, \gamma(dx^\mu) \varphi) + B(\psi, \nabla_\mu^S (\gamma(dx^\mu) \varphi)) - \partial_\mu B(\psi, \gamma(dx^\mu) \varphi)  \right] = 0
\end{equation*}
for all $\psi, \varphi \in \Gamma_c(M,S_M)$. Now, the expression between brackets can be proven to be $\mathcal{C}^\infty (M, \mathbb{C})$-linear in $\varphi$ (and anti-linear in $\psi$ as well). Indeed, $B(\nabla_\mu^S \psi, \gamma(dx^\mu) \varphi)$ is clearly linear in $\varphi$. let $f \in \mathcal{C}^\infty (M, \mathbb{C})$. We replace $\varphi$ by $f \varphi$ in the two remaining terms:
\begin{equation*}
\begin{aligned}
B(\psi, \nabla_\mu^S (\gamma(dx^\mu) f\varphi)) - \partial_\mu B(\psi, \gamma(dx^\mu) f\varphi) =& B(\psi, \nabla_\mu^S [f (\gamma(dx^\mu) \varphi)]) - \partial_\mu [f B(\psi, \gamma(dx^\mu) \varphi)] \\
=& B(\psi, (\partial_\mu f) \gamma(dx^\mu) \varphi) + f \nabla_\mu^S (\gamma(dx^\mu) \varphi)) \\ 
&- (\partial_\mu f) B(\psi, \gamma(dx^\mu) \varphi) - f \partial_\mu B(\psi, \gamma(dx^\mu) \varphi) \\
=& f [B(\psi, \nabla_\mu^S (\gamma(dx^\mu) \varphi)) - \partial_\mu B(\psi, \gamma(dx^\mu) \varphi)].
\end{aligned}
\end{equation*}
Thus, for all $f \in \mathcal{C}^\infty (M, \mathbb{C})$ and $\psi, \varphi \in \Gamma_c(M,S_M)$:
\begin{equation*}
\int \sqrt{|g|} f \left[ B(\nabla_\mu^S \psi, \gamma(dx^\mu) \varphi) + B(\psi, \nabla_\mu^S (\gamma(dx^\mu) \varphi)) - \partial_\mu B(\psi, \gamma(dx^\mu) \varphi)  \right] = 0
\end{equation*}
which implies that:
\begin{equation}
B(\nabla_\mu^S \psi, \gamma(dx^\mu) \varphi) + B(\psi, \nabla_\mu^S (\gamma(dx^\mu) \varphi)) - \partial_\mu B(\psi, \gamma(dx^\mu) \varphi)=0.
\label{equationonHdivergence}
\end{equation}
Now, let $U$ be a sufficiently small open subset of $M$ such that the tangent, cotangent and spinor bundles become trivial: $S_M|_U \simeq U \times S_0$. The Hermitian form $B$ makes real vectors skew-symmetric. It is thus a local anti-Robinson product. Combining this with our local trivialization of the spinor bundle, we conclude that there exists a function $\lambda \in \mathcal{C}^\infty (U, \mathbb{R}^\ast)$ such that $B_x(\cdot, \cdot) = \lambda(x) [\cdot, \cdot]_0$ (the trivialization diffeomorphisms are implicit). Here $[\cdot, \cdot]_0$ is a particular realization of the anti-Robinson product on $S_0$ that makes real vectors locally anti-self-adjoint. It is constant, in the sense that:
\begin{equation}
\partial_\mu [\psi, \varphi]_0 = [\partial_\mu\psi, \varphi]_0 + [\psi, \partial_\mu\varphi]_0
\label{constantinnerproduct}
\end{equation}
Moreover, the Clifford connection takes the local form: $\nabla_X^{S} = \partial_X + \Gamma(X) + A(X)$.
Let us now replace the Clifford connection and indefinite product in equation (\ref{equationonHdivergence}) by their local forms:
\begin{equation*}
\begin{aligned}
\lambda[(\partial_\mu + \Gamma_\mu + A_\mu) \psi, \gamma(dx^\mu) \varphi]_0 & + \lambda[\psi, (\partial_\mu + \Gamma_\mu + A_\mu) (\gamma(dx^\mu) \varphi)]_0 \\
& - \partial_\mu (\lambda[\psi, \gamma(dx^\mu) \varphi]_0)=0.
\end{aligned}
\end{equation*}
Using equation (\ref{constantinnerproduct}), we find:
\begin{equation*}
\lambda[(\Gamma_\mu + A_\mu) \psi, \gamma(dx^\mu) \varphi]_0  + \lambda[\psi, (\Gamma_\mu + A_\mu) (\gamma(dx^\mu) \varphi)]_0 - (\partial_\mu \lambda) [\psi, \gamma(dx^\mu) \varphi]_0 =0.
\end{equation*}
One can prove that the $\Gamma_\mu$ are locally anti-self-adjoint, from which one infers that:
\begin{equation*}
[\lambda (A_\mu+\overline{A_\mu})-\partial_\mu \lambda] [\psi, \gamma(dx^\mu) \varphi]_0=0.
\end{equation*}
Since this is true for all $\psi,\varphi$, we infer that $\lambda (A_\mu+\overline{A_\mu})-\partial_\mu \lambda=0$. Let us finally prove that the Clifford connection is hermitian. We have:
\begin{equation*}
\begin{aligned}
\partial_\mu B(\psi, \varphi) &= \partial_\mu (\lambda [\psi, \varphi]_0) \\
&= (\partial_\mu \lambda) [\psi, \varphi]_0 + \lambda \partial_\mu [\psi, \varphi]_0 \\
&= (\partial_\mu \lambda) [\psi, \varphi]_0 + \lambda [\partial_\mu\psi, \varphi]_0 + \lambda [\psi, \partial_\mu\varphi]_0
\end{aligned}
\end{equation*}
where we have used equation (\ref{constantinnerproduct}). Using the equation $\lambda (A_\mu+\overline{A_\mu})-\partial_\mu \lambda=0$, and the anti-self-adjointness of the $\Gamma_\mu$, we find:
\begin{equation*}
\begin{aligned}
\partial_\mu B(\psi, \varphi) &= \lambda [(\partial_\mu + \Gamma_\mu + A_\mu)\psi, \varphi]_0 + \lambda [\psi, (\partial_\mu + \Gamma_\mu + A_\mu)\varphi]_0 \\
&= B(\nabla_\mu^S \psi, \varphi) + B(\psi, \nabla_\mu^S \varphi).
\end{aligned}
\end{equation*}
The Clifford connection is thus hermitian.

The converse can be proven easily following the same steps. It is in fact a standard result of spin geometry. See for example \cite{LawsonMichelsohn}.

\qed
\end{demo}

This has two consequences:
\begin{itemize}
	\item We must choose $B$ to be a local anti-Robinson product $G$, and the corresponding global product must be a global anti-Robinson product $[\cdot,\cdot]_M$.
	\item The Clifford connection must be hermitian for the anti-Robinson product: $X(G(\psi, \phi)) = G(\nabla_X^S \psi, \phi) + G(\psi, \nabla_X^S \phi)$.
\end{itemize}

To ensure the existence of the anti-Robinson product and charge conjugation operator, we will restrict ourselves to STO Spin manifolds. For these manifolds, we will prove that a Clifford connections that commutes with $J_{M+}$ and is hermitian for $G$ always exists. For this, we will need the following lemma:
\begin{lem} \label{TrivLemma}
Let $(M,g)$ be an STO Spin manifold, and let $J_{M+}$ and $G$ be a charge conjugation operator and anti-Robinson product. Then there exists a family of local trivializations of the spinor bundle $S$ with $\mathrm{Spin}_0 (q,p)$-valued transition functions such that $J_{M+}$ and $G$ are locally constant for every trivialization.
\end{lem}

\begin{demo}

We will use in this proof the same notations as in the proof of theorem \ref{CCSpinManifoldThm}. Let $(U_\alpha, \Phi_\alpha)_{\alpha \in I}$ be a family of trivializations of the bundle $S_M$, with all $U_\alpha$ simply connected. Let $J_0$ and $[\cdot, \cdot]_0$ be a charge conjugation operator and anti-Robinson product on $S_0$ respectively. Let $\alpha \in I$. By uniqueness of the anti-Robinson product and charge conjugation, there exist smooth maps $\lambda_\alpha \in \mathcal{C}^\infty (U_\alpha, \mathbb{R}^\ast)$ and $\rho_\alpha \in \mathcal{C}^\infty (U_\alpha, U(1))$ such that for all $\psi, \varphi \in S_M$ and $x = \pi(\psi) = \pi(\varphi) \in U_\alpha$:
\begin{equation*}
\begin{aligned}
G_x(\psi, \varphi) &= \lambda_\alpha(x) [\Pi_\alpha(\psi), \Pi_\alpha(\varphi)]_0 \\
\Pi_\alpha(J_{M+} \psi) &= \rho_\alpha(x) J_0 (\Pi_\alpha(\psi)).
\end{aligned}
\end{equation*}
Let $b_\alpha = \lambda_\alpha \rho_\alpha^{-1} \in \mathcal{C}^\infty (U_\alpha, \mathbb{C}^\ast)$. Since $U_\alpha$ is simply connected, a smooth square root $b_\alpha^{1/2}$ of $b_\alpha$ can be defined over all of $U_\alpha$. We now define a new trivialization of $S$ denoted $(U_\alpha, \Phi'_\alpha)_\alpha$ and defined by:
\begin{equation*}
\Pi'_\alpha(s) = b_\alpha^{1/2}(x) \Pi_\alpha(s),
\end{equation*}
with $x = \pi(s)$. Let us compute the local forms of $G$ and $J_{M+}$ in this new trivialization. We have:
\begin{equation*}
\begin{aligned}
G_x(\psi, \varphi) &= \lambda_\alpha(x) [\Pi_\alpha(\psi), \Pi_\alpha(\varphi)]_0 \\
&= \lambda_\alpha(x) [b_\alpha^{-1/2}(x) \Pi'_\alpha(\psi), b_\alpha^{-1/2}(x) \Pi'_\alpha(\varphi)]_0 \\
&= \lambda_\alpha(x) |b_\alpha(x)|^{-1} [\Pi'_\alpha(\psi), \Pi'_\alpha(\varphi)]_0 \\
G_x(\psi, \varphi) &= [\Pi'_\alpha(\psi), \Pi'_\alpha(\varphi)]_0
\end{aligned}
\end{equation*}
which proves that $G$ is locally equal to $[\cdot, \cdot]_0$. Substituting $\Pi'_\alpha(s) = b_\alpha^{1/2}(x) \Pi_\alpha(s)$ in the equation for the local form of $J_{M+}$, we find:
\begin{equation*}
b_\alpha^{-1/2}(x) \Pi'_\alpha(J_{M+} \psi) = \rho_\alpha(x) J_0 (b_\alpha^{-1/2}(x) \Pi'_\alpha(\psi)),
\end{equation*}
which yields:
\begin{equation*}
\begin{aligned}
\Pi'_\alpha(J_{M+} \psi) &= b_\alpha^{1/2}(x) \rho_\alpha(x) J_0 (b_\alpha^{-1/2}(x) \Pi'_\alpha(\psi)) \\
&= b_\alpha^{1/2}(x) \overline{b_\alpha^{-1/2}(x)} \rho_\alpha(x) J_0 (\Pi'_\alpha(\psi)) \\
\Pi'_\alpha(J_{M+} \psi) &= \pm J_0 (\Pi'_\alpha(\psi)),
\end{aligned}
\end{equation*}
which proves that $J_{M+}$ is locally constant.

To conclude, we need to compute the transition functions of the new trivializations. But first, we prove that $b_\alpha = b_\beta$ over $U_\alpha \cap U_\beta$ for any $\alpha,\beta \in I$ such that $U_\alpha \cap U_\beta \neq \emptyset$. Indeed, let $x \in U_\alpha \cap U_\beta$. We have:
\begin{equation*}
\begin{aligned}
G_x(\psi, \varphi) &= \lambda_\alpha(x) [\Pi_\alpha(\psi), \Pi_\alpha(\varphi)]_0 \\
G_x(\psi, \varphi) &= \lambda_\beta(x) [\Pi_\beta(\psi), \Pi_\beta(\varphi)]_0.
\end{aligned}
\end{equation*}
Using $\Pi_\beta (s) = g_{\beta \alpha}(x) \Pi_\alpha (s)$, we rewrite the second equation as:
\begin{equation*}
G_x(\psi, \varphi) = \lambda_\beta(x) [g_{\beta \alpha}(x) \Pi_\alpha(\psi), g_{\beta \alpha}(x) \Pi_\alpha(\varphi)]_0,
\end{equation*}
The $[\cdot, \cdot]_0$-unitarity of $\mathrm{Spin}_0 (q,p)$ group yields:
\begin{equation*}
G_x(\psi, \varphi) = \lambda_\beta(x) [\Pi_\alpha(\psi), \Pi_\alpha(\varphi)]_0,
\end{equation*}
from which we conclude that $\lambda_\alpha = \lambda_\beta$ over $U_\alpha \cap U_\beta$. One can prove in a similar fashion, and using the commutation of $J_0$ with any element of the Spin group, that $\rho_\alpha = \rho_\beta$ over $U_\alpha \cap U_\beta$. This proves that $b_\alpha = b_\beta$ over $U_\alpha \cap U_\beta$. We can now compute the new transition functions. We have:
\begin{equation*}
\begin{aligned}
\Pi_\beta (s) = g_{\beta \alpha}(x) \Pi_\alpha (s) &\Rightarrow b_\beta^{-1/2}(x) \Pi'_\beta (s) = g_{\beta \alpha}(x)  b_\alpha^{-1/2}(x) \Pi'_\alpha (s) \\
&\Rightarrow \Pi'_\beta (s) = b_\beta^{1/2}(x)  g_{\beta \alpha}(x)  b_\alpha^{-1/2}(x) \Pi'_\alpha (s) \\
&\Rightarrow \Pi'_\beta (s) = g_{\beta \alpha}(x) b_\beta^{1/2}(x) b_\alpha^{-1/2}(x) \Pi'_\alpha (s) \\
&\Rightarrow \Pi'_\beta (s) = g_{\beta \alpha}(x) \Pi'_\alpha (s)
\end{aligned}
\end{equation*}
which proves that the transition functions are the same. In particular, they are still $\mathrm{Spin}_0 (q,p)$-valued, and still satisfy the cocycle conditions.

\qed
\end{demo}

With this lemma in hand, we can now prove the following theorem\footnote{This is a generalization of theorem 9.8 in \cite{GBVF}.}:

\begin{thm} \label{SpinConnectionThm}
Let $(M,g)$ be an STO Spin manifold, and let $J_{M+}$ and $G$ be a charge conjugation operator and anti-Robinson product. Then there exists a unique Clifford connection $\nabla^S$ that commutes with $J_{M+}$ and is hermitian for $G$. We call this connection the canonical Clifford connection.
\end{thm}

\begin{demo}
According to the previous lemma, there exists a family of trivializations $(U_\alpha, \Phi_\alpha)_\alpha$ of the spinor bundle such that $G$ and $J_{M+}$ are constant on every open subset $U_\alpha$. One can assume, without any loss of generality, that the open sets $U_\alpha$ are sufficiently small for the tangent bundle to be trivializable over them\footnote{One can use a refinement of the open cover $(U_\alpha)_\alpha$ for that purpose.}. Let $\alpha \in I$. We define the local Clifford connection $\nabla^{S\alpha}$ over $U_\alpha$ whose local form is $\nabla^{S\alpha}_X = \partial_X + \Gamma(X)$. Since $J_{M+}$ is constant over $U_\alpha$, it commutes with $\partial_X$. Moreover, $\Gamma(X)$ commutes with $J_{M+}$. This proves that $\nabla^{S\alpha}$ commutes with $J_{M+}$. Let us now prove that $\nabla^{S\alpha}$ is metric with respect to $G$. Since $G$ is locally constant, we have:
\begin{equation*}
X(G(\psi, \varphi)) = G(\partial_X \psi, \varphi) + G(\psi, \partial_X \varphi)
\end{equation*}
(the trivialization diffeomorphism $\Phi_\alpha$ is implicit here). Using the anti-self-adjointness of $\Gamma$, we find:
\begin{equation*}
\begin{aligned}
X(G(\psi, \varphi)) &= G((\partial_X + \Gamma(X)) \psi, \varphi) + G(\psi, (\partial_X + \Gamma(X)) \varphi) \\
X(G(\psi, \varphi)) &= G(\nabla^{S\alpha}_X \psi, \varphi) + G(\psi, \nabla^{S\alpha}_X \varphi)
\end{aligned}
\end{equation*}
This proves that $\nabla^{S\alpha}$ is hermitian for $G$.

We proved that a local Clifford connection $\nabla^{S\alpha}$ satisfying the requirements of the theorem over every $U_\alpha$ exists, and is unique. Let us now prove that all the local Clifford connections coincide on overlaps of the open sets of the covering. Let $\alpha, \beta \in I$ such that $U_\alpha \cap U_\beta \neq \emptyset$. We have two Clifford connections $\nabla^{S\alpha}, \nabla^{S\beta}$ over $U_\alpha \cap U_\beta$. There thus exists a complex-valued $1$-form $B$ such that $\nabla_X^{S\alpha} - \nabla_X^{S\beta} = B(X)$. From the metricity of both connections, one finds:
\begin{equation*}
G((\nabla^{S\alpha}_X - \nabla^{S\beta}_X) \psi, \varphi) + G(\psi, (\nabla^{S\alpha}_X - \nabla^{S\beta}_X) \varphi) = 0,
\end{equation*}
from which we infer that $B(X)$ is imaginary. The commutation of $J_{M+}$ with both Clifford connections yields:
\begin{equation*}
[\nabla^{S\alpha}_X - \nabla^{S\beta}_X,J_{M+}] = 0,
\end{equation*}
which implies that $B(X)$ is real. We thus conclude that $B=0$, and that $\nabla^{S\alpha} = \nabla^{S\beta}$ over $U_\alpha \cap U_\beta$. The local Clifford connections $\nabla^{S\alpha}$ can thus be patched up to form a globally defined Clifford connection $\nabla^S$ that satisfies the requirements of the theorem. The uniqueness of the Clifford connection can be proven using its metricity and commutation with the charge conjugation operator, as it was done above to prove the coincidence of $\nabla^{S\alpha}$ and $\nabla^{S\beta}$.

\qed
\end{demo}

With this choice of Clifford connection, the Dirac operator has the sought after properties:
\begin{equation}
\begin{aligned}
\left[ \slashed{D}, J_{M+} \right] &= 0 \\
[\psi, \slashed{D} \varphi]_M &= [\slashed{D} \psi, \varphi]_M.
\end{aligned}
\end{equation}

It is important to notice that the canonical Clifford connection depends on the choice of anti-Robinson product and charge conjugation operator. Let $J_{M+},G$ and $J'_{M+},G'$ be two such choices. Then there exist smooth functions $\lambda \in \mathcal{C}^\infty (M, \mathbb{R}^\ast)$ and $\rho \in \mathcal{C}^\infty (M, U(1))$ such that $G' = \lambda G$ and $J_{M+}' = \rho J_{M+}$. Let $b = \lambda \rho^{-1} \in \mathcal{C}^\infty (M, \mathbb{C}^\ast)$. The resulting Clifford connections $\nabla^S$ and $\nabla^{\prime S}$ can be proven to be related by:
\begin{equation}
\nabla^{\prime S} - \nabla^S = \frac{db}{2b}. 
\end{equation}
Indeed, let $\nabla^{\prime\prime S} = \nabla^S + \frac{db}{2b}$ be a Clifford connection (see proposition \ref{CliffordConnectionOmega}). We will prove that it is hermitian for $G'$ and commutes with $J'_{M+}$. By the uniqueness of the canonical Clifford connection, one infers that it is equal to $\nabla^{\prime S}$, hence the result. The proof follows:
\begin{equation*}
\begin{aligned}
G'(\nabla^{\prime\prime S}_X \psi, \phi) + G'(\nabla^{\prime\prime S}_X \phi) &= \lambda [G((\nabla^{S}_X+\frac{X(b)}{2b}) \psi, \phi) + G(\psi, (\nabla^{S}_X+\frac{X(b)}{2b}) \phi)] \\
&= \lambda [G(\nabla^{S}_X \psi, \phi) + G(\psi, \nabla^{S}_X \phi) + \mathrm{Re}(\frac{X(b)}{b}) G(\psi, \phi)] \\
&= \lambda [X(G(\psi, \phi)) + \frac{X(\lambda)}{\lambda} G(\psi, \phi)] \\
G'(\nabla^{\prime\prime S}_X \psi, \phi) + G'(\nabla^{\prime\prime S}_X \phi) &= X(G'(\psi, \phi)),
\end{aligned}
\end{equation*}
and:
\begin{equation*}
\begin{aligned}
\left[\nabla^{\prime\prime S}_X, J_{M+}'\right] &= [\nabla^S_X + \frac{X(b)}{2b}, \rho J_{M+}] \\
&= [\nabla^S_X, \rho] J_{M+} + \rho [i\mathrm{Im}(\frac{X(b)}{b})] J_{M+} \\
&= X(\rho) J_{M+} + \rho (-\frac{X(\rho)}{\rho}) J_{M+} \\
[\nabla^{\prime\prime S}_X, J_{M+}'] &= 0.
\end{aligned}
\end{equation*}

The corresponding Dirac operators are thus related by:
\begin{equation}
\slashed{D}' - \slashed{D} = \frac{\slashed{\partial}b}{2b},
\end{equation}
where $\slashed{\partial} = \gamma(dx^\mu) \partial_\mu$. The two Clifford connections are equal if and only if $b$ is a constant function. That is, if and only if $J_{M+}$ and $J'_{M+}$ differ by a global phase and $G$ and $G'$ differ by a global constant factor. We can deduce from this that the space of Clifford connections is isomorphic to the multiplicative group $\mathcal{C}^\infty (M, \mathbb{C}^\ast) / \mathbb{C}^\ast$. In this group, two functions are in the same equivalence class if and only they differ by a global constant factor in $\mathbb{C}^\ast$. Given that the affine space of general Clifford connections is isomorphic to the space of complex-valued differential forms (see again proposition \ref{CliffordConnectionOmega}), we conclude that not all Clifford connections are canonical Clifford connections. In fact canonical Clifford connections are in one to one correspondence with differential forms of the form $\frac{db}{2b}$, and these are closed forms.

\begin{figure}[!h]
	\centering
		\includegraphics[scale=0.4]{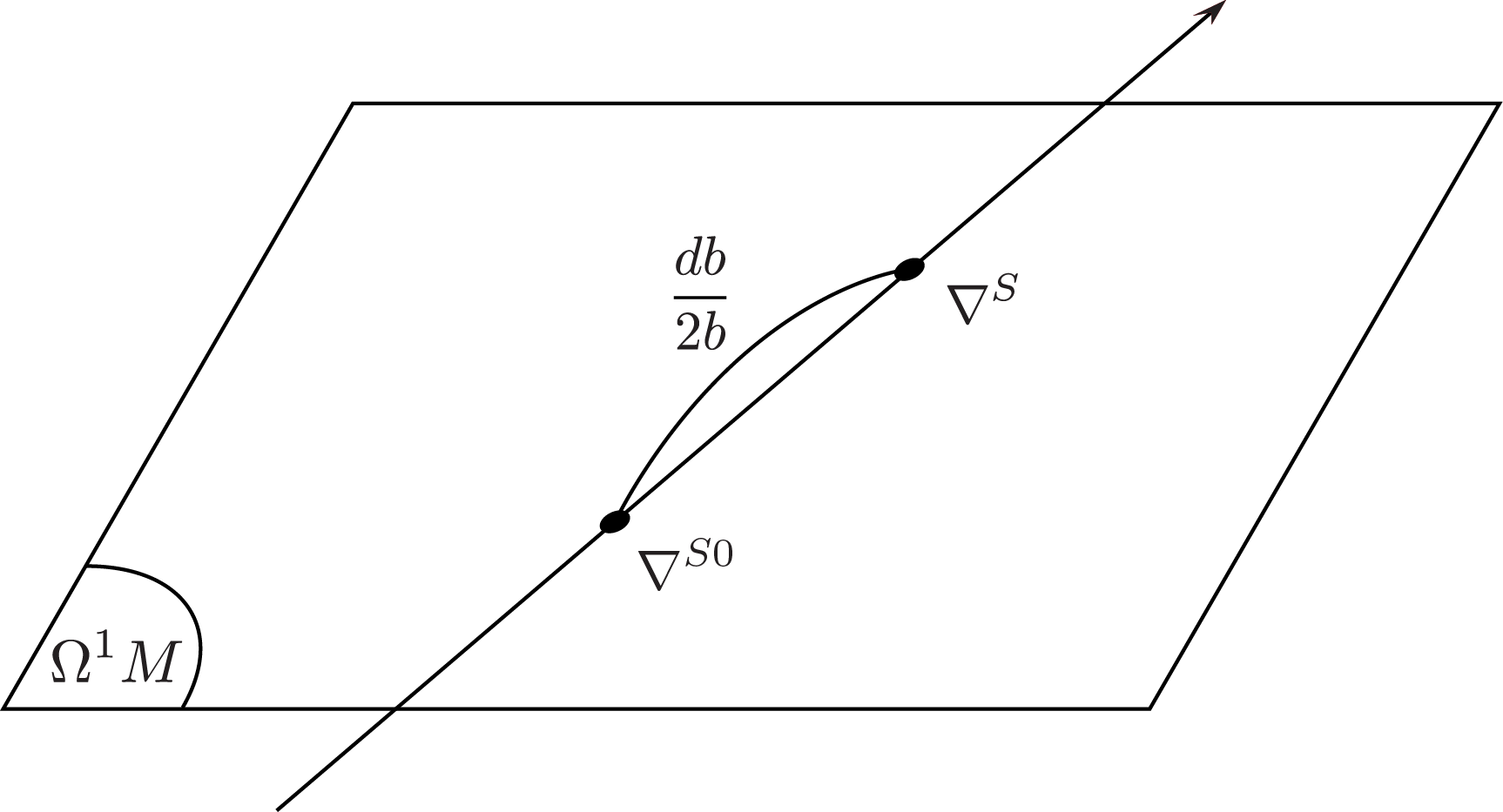}
	\label{fig:2D}
\end{figure}

In the context of Dirac field theory, the $A = db/2b$ term corresponds to a background gauge field. But its strength tensor is vanishing: $F = dA = 0$, which means that this background field should have no observable effect, except perhaps some contribution to topological effects.

\subsection{The Imaginary Convention}

One could also define the Dirac operator as:
\begin{equation}
\slashed{D} = i \gamma(\omega^a) \nabla_{e_a}^S = i \gamma(dx^\mu) \nabla_\mu^S.
\label{DiracImaginary}
\end{equation}

One can then repeat the same steps as above. The conclusions one arrives at are the following:
\begin{itemize}
	\item The manifold $M$ has to be an STO Spin manifold.
	\item One must choose the local Robinson product $H$ and local \emph{graded} charge conjugation operator $J_{M-}$.
	\item There exists a unique Clifford connection, called the canonical Clifford connection, that commutes with $J_{M-}$, and that is hermitian for $H$.
	\item With this Clifford connection, the Dirac operator satisfies:
		\begin{equation}
		\begin{aligned}
		\{\slashed{D}, \chi_M\} &= 0 \\
		\left[\slashed{D}, J_{M-} \right] &= 0 \\
		(\psi, \slashed{D} \varphi)_M &= (\slashed{D} \psi, \varphi)_M.
		\end{aligned}
		\end{equation}
\end{itemize}

We now have almost everything we need to define a commutative Semi-Riemannian spectral triple. We now only need a Krein space. From here on, $M$ will always be an STO Spin manifold.

\section{The Spinor Krein Space} \label{SectionSpinorKrein}

Our starting point to define a Krein space is the space of compactly supported smooth spinors $\Gamma_c (M,S_M)$. Indeed, all our operators act on it, and smooth functions can be represented on this space. And most importantly, it is equipped with an indefinite product relative to which all our operators have good properties (such as the Dirac operator being symmetric). It only remains to make it complete with respect to this indefinite product.

To achieve this, one must make several choices:
\begin{enumerate}
	\item The first step is to make a choice of convention of Dirac operator: with or without an $i$ factor.
	\item Then, one must choose a local anti-Robinson or Robinson product, and construct the corresponding global product.
	\item Next, one chooses a splitting of the tangent bundle $TM = E_+ \oplus E_-$. From theorems \ref{LocalRobinsonThm} and \ref{LocalAntiRobinsonThm} and remarks thereafter, we know that we can associate to each such splitting a fundamental symmetry $\eta_-$ or $\eta_+$, which we can use to construct an inner product $\langle \cdot, \cdot \rangle_{\eta_\pm}$.
	\item Finally, we construct our Krein space as the completion of $\Gamma_c (M,S_M)$ with respect to this inner product:
	\begin{equation}
		\mathcal{K}_M = \overline{\Gamma_c (M,S_M)}.
	\end{equation}
	This construction was first presented in \cite{Baum}.
\end{enumerate}
The resulting space, equipped with the global Robinson or anti-Robinson product, is a Krein space by construction, since it has a fundamental symmetry ($\eta_\pm$).

It is also not canonical, in the sense that the outcome generally depends on the choice of Robinson or anti-Robinson product, and the choice of splitting of the tangent bundle. We will study these two dependences separately. But first note that Robinson and anti-Robinson products are in one-to-one correspondence through an insertion of chirality:
\begin{equation*}
G(\cdot, \cdot) = H(\cdot, i^q \chi_M \cdot).
\end{equation*}
Note that $\chi_M$ satisfies: $\chi_M^\times (-1)^q = \chi_M$, with respect to the Robinson product. Notice also, from the local form of the fundamental symmetries $\eta_+$ used above, that: $\chi_M \eta_+ = (-1)^q \eta_+ \chi_M$. We deduce that: $\chi_M^{\dagger \eta_+} = \chi_M$. It is locally and globally Hilbert-self-adjoint. It is also involutive: $\chi_M^2 = 1$. We deduce that it is unitary, and thus bounded with bounded inverse. The different norms induced by $H$ and $G$ must therefore be equivalent. Thanks to this correspondence, we will restrict our study to Robinson products. 

We start with the choice of Robinson product:
\begin{thm}
Let $H$ and $H' = \lambda H$ be two Robinson products, with $\lambda \in \mathscr{C}^\infty (M, \mathbb{R}^\ast)$. For a given  splitting of the tangent bundle $TM = E_+ \oplus E_-$, the norms induced by $H$ and $H'$ are equivalent if and only if $\lambda$ is bounded with bounded inverse.
\end{thm}

\begin{demo}
Let $\langle \cdot, \cdot \rangle_{\eta_\pm}$ and $\langle \cdot, \cdot \rangle'_{\eta_\pm}$ be the two inner products constructed with $H$ and $H'$ respectively:
\begin{equation*}
\begin{aligned}
\langle \psi, \varphi \rangle_{\eta_\pm} &= \int_{x \in M} H_x (\psi,\eta_+ \varphi) \sqrt{|g|} d^d x \\
\langle \psi, \varphi \rangle'_{\eta_\pm} &= \int_{x \in M} |\lambda| H_x (\psi,\eta_+ \varphi) \sqrt{|g|} d^d x
\end{aligned}
\end{equation*}
(the absolute value comes from the reversal of sign of the fundamental symmetry for $H'$ when $\lambda$ is negative). If $\lambda$ is bounded with bounded inverse, \text{i.e.} there exist $a,b > 0$ such that $a \leq |\lambda| \leq b$, then it is easy to see that:
\begin{equation*}
a \langle \psi, \psi \rangle_{\eta_\pm} \leq \langle \psi, \psi \rangle'_{\eta_\pm} \leq b \langle \psi, \psi \rangle_{\eta_\pm},
\end{equation*}
which proves that the two norms are equivalent.

Conversely, let us assume that the equation above holds for all $\psi$. In particular, for $\psi \neq 0$:
\begin{equation*}
a \leq \frac{\int_{x \in M} |\lambda| H_x (\psi,\eta_+ \psi) \sqrt{|g|} d^d x}{\int_{x \in M} H_x (\psi,\eta_+ \psi) \sqrt{|g|} d^d x} \leq b.
\end{equation*}
Let $x_0 \in M$, and $U$ an open neighborhood of of $x_0$. We choose U sufficiently small so that the spinor bundle is trivial over it: $S|_U \cong U \times S_0$. Let $\psi_0 \in S_0$ be non-vanishing. Let $(f_k)_k$ be a sequence of smooth functions with support in $U$ that converges to the Dirac delta distribution $\delta_{x_0}$. We define the smooth spinor $\psi_k$ with compact support defined by:
\begin{equation*}
\psi_k(x) = \begin{cases}
f_k(x) \psi_0 \text{  if $x \in U$} \\
0 \text{  otherwise.}
\end{cases}
\end{equation*}
We then replace $\psi$ in the inequality above with $\psi_k$. Note that $H_{x_0} (\psi_0,\eta_+ \psi_0) > 0$. As a result, one can prove that the sequence inside the inequality converges to $|\lambda(x_0)|$, and we find that:
\begin{equation*}
a \leq |\lambda(x_0)| \leq b
\end{equation*}
for all $x_0 \in M$.

\qed
\end{demo}

Let us given an example where two Robinson products give unequivalent norms. We construct our example on the flat 2D spacetime $M = \mathbb{R}^{1,1}$. The corresponding Clifford algebra is $\mathrm{Cl}(1,1) = \mathrm{M}_2(\mathbb{R})$, and the corresponding spinor space is $S_0 = \mathbb{C}^2$. We choose a flat spinor bundle: $S_M = M \times S_0$, and a coordinate system $(t,x)$ such that the metric $g$ takes the form $g^{-1} = \mathrm{diag}(-1,1)$. We represent the Clifford bundle using the simple representation:
\begin{equation*}
\begin{aligned}
\gamma(dt) &= \begin{pmatrix} 0 & -1 \\ 1 & 0 \end{pmatrix} \\
\gamma(dx) &= \begin{pmatrix} 0 & 1 \\ 1 & 0 \end{pmatrix}.
\end{aligned}
\end{equation*}
The local Robinson product is necessarily of the form:
\begin{equation*}
H(\psi,\varphi) = \lambda \psi^\dagger \gamma(dx) \varphi,
\end{equation*}
with $\lambda \in \mathscr{C}^\infty (M, \mathbb{R}^\ast)$. Let $\psi_\tau \in \Gamma_c (M,S_M)$ be of the form:
\begin{equation*}
\psi_\tau(t,x) = f(t-\tau,x) \begin{pmatrix} 1 \\ 0 \end{pmatrix},
\end{equation*}
with $\tau \in \mathbb{R}$, and $f:\mathbb{R}^2 \rightarrow \mathbb{R}$ a smooth function with support inside the unit disk of $\mathbb{R}^2$, whose square integrates to 1:
\begin{equation*}
\int\int f(t,x)^2 dt dx = 1.
\end{equation*}

Let us choose two different Robinson products $H$ and $H'$ with the same fundamental symmetry $\eta_+ = \gamma(dx)$. For the first one we choose $\lambda(t,x) = 1$, and for the second one $\lambda'(t,x) = e^t$. For $H$ we find:
\begin{equation*}
\begin{aligned}
(\psi_\tau, \eta_+ \psi_\tau)_M &= \int\int dt dx \lambda(t,x) f(t-\tau,x)^2 \\
&= \int\int dt dx f(t-\tau,x)^2 \\
(\psi_\tau, \eta_+ \psi_\tau)_M &= 1.
\end{aligned}
\end{equation*}
We deduce that $\| \psi_\tau \|_{\eta_+} = 1$ for $H$, and does not depend on $\tau$. For $H'$, we find:
\begin{equation*}
\begin{aligned}
(\psi_\tau, \eta_+ \psi_\tau)'_M &= \int\int dt dx \lambda'(t,x) f(t-\tau,x)^2 \\
&= \int\int dt dx e^t f(t-\tau,x)^2 \\
&\geq \int\int dt dx e^{\tau-1} f(t-\tau,x)^2 \text{  (since $f(t-\tau,x)$ vanishes for $t-\tau \leq -1$)} \\
(\psi_\tau, \eta_+ \psi_\tau)'_M &\geq e^{\tau-1}.
\end{aligned}
\end{equation*}
We deduce that $\| \psi_\tau \|'_{\eta_+} \geq e^{(\tau-1)/2}$ for $H'$. It can be arbitrary large for arbitrarily large $\tau$. The norms $\| \cdot \|_{\eta_+}$ and $\| \cdot \|'_{\eta_+}$ are thus not equivalent. This is ultimately due to the fact that the function $\lambda$ is not bounded.

We now study the dependence with respect to the splitting of the tangent bundle. The following results are proven in \cite{BB2}. We only give the main results here. For a given splitting $TM = E_+ \oplus E_-$, we denote $R$ the corresponding fundamental symmetry on $TM$, and $g_R(\cdot,\cdot) = g(\cdot, R \cdot)$ the corresponding ("Wick-rotated") Riemannian metric. We choose a specific Robinson product $H$. Let $TM = E_+ \oplus E_-$ and $TM = E'_+ \oplus E'_-$ be two different splittings of the tangent bundle. We want to quantify the difference (or angle) between the two splittings. We first do this locally. Let $x \in M$. Let $\Lambda,\Lambda' \in \mathrm{SO}(T_x M,g_x)$ two special-orthogonal isomorphisms of the tangent fiber over $x$ that map $E_{x,\pm}$ to $E'_{x,\pm}$ isomorphically (see theorem \ref{FundamentalSymmetriesKreinUnitary}). Then one can prove (\cite{BB2}, Lemma 1) that they have the same norm with respect to $g_{R,x}$: $\|\Lambda\|_{g_{R,x}} = \|\Lambda' \|_{g_{R,x}}$. We thus use this to define the "angle" $\xi(x) = \|\Lambda\|_{g_{R,x}} = \|\Lambda' \|_{g_{R,x}}$ (also called Doppler shift factor in \cite{BB2}) between the two splittings. Around $x$, there exists a small neighborhood $U$ on which one can construct a smooth section $\Lambda \in \Gamma(U, \mathrm{SO}(TU,g|_U))$ with the properties above by trivializing the tangent bundle. Then $\xi$ is a smooth function over $U$. It is thus a smooth function over $x$, and thus over all $M$. We then have the following theorem:
\begin{thm} \textbf{\cite{BB2}, Theorem 2}
For a given Robinson product $H$, the norms on $\Gamma_c (M,S_M)$ induced by two splittings of the tangent bundle are equivalent if and only if their angle $\xi$ is bounded on $M$.
\end{thm}	
	
We now give an example where two different splittings give two unequivalent norms. Coming back to our previous example on 2D Minkowski, we choose the Robinson product with $\lambda  = 1$, and two fundamental symmetries. The first one is $\eta_+ = \gamma(dx)$. The second one is $\eta'_+ = \gamma(n)$, with $n$ the space-like differential form:
\begin{equation*}
n(t,x) = \cosh(t) dx + \sinh(t) dt.
\end{equation*}
The corresponding local inner product is:
\begin{equation*}
H(\psi,\varphi)_{\eta'_+} = \psi^\dagger \gamma(dx) \gamma(n) \varphi = \psi^\dagger \begin{pmatrix} e^t & 0 \\ 0 & e^{-t} \end{pmatrix} \varphi.
\end{equation*}
We already know that $\| \psi_\tau \|_{\eta_+} = 1$. We also have:
\begin{equation*}
\begin{aligned}
(\psi_\tau, \eta'_+ \psi_\tau)_M &= \int\int dt dx e^t f(t-\tau,x)^2 \\
&\geq \int\int dt dx e^{\tau-1} f(t-\tau,x)^2 \\
(\psi_\tau, \eta'_+ \psi_\tau)_M &\geq e^{\tau-1}.
\end{aligned}
\end{equation*}
We deduce that $\| \psi_\tau \|_{\eta'_+} \geq e^{(\tau-1)/2}$. Once again, it can be arbitrary large for arbitrarily large $\tau$, and the norms $\| \cdot \|_{\eta_+}$ and $\| \cdot \|_{\eta'_+}$ are not equivalent. One can prove that the angle between the splittings is $\xi(t,x) = e^{|t|}$, and this is clearly not bounded.

The key property of this example is that the manifold $M$ is non-compact. Indeed, the functions $\lambda, \lambda^{-1}$ and $\xi$ are smooth. They are thus automatically bounded if $M$ is compact, are can only be unbounded if $M$ is noncompact. We thus arrive at the following corollary:

\begin{cor} \label{Compact}
If $M$ is a \emph{compact} STO manifold, then the Krein space $\mathcal{K}_M = \overline{\Gamma_c (M,S_M)}$ does not depend on the choice of convention of the Dirac operator, the choice of Robinson or anti-Robinson product, or the choice of splitting of the tangent bundle. Moreover, all inner products obtained through the procedure above induce equivalent norms.
\end{cor}

Regardless of the lack of uniqueness of the Krein space in general, one can say a few things about the various operators constructed in the previous sections, and their action on the Krein space. We start with the Dirac operator. Its domain $\mathcal{D}(\slashed{D}) = \Gamma_c (M,S_M)$ is dense in $\mathcal{K}_M$ by construction. It is also symmetric with respect to the global Robinson or anti-Robinson product, by construction. Under certain assumptions, the Dirac operator is also Krein-self-adjoint:

\begin{thm} \label{DiracSelfAdjoint} \textbf{Theorem 3.19, \cite{Baum}}

If there exists a splitting of the tangent bundle $TM = E_- \oplus E_+$, of (globally defined) local fundamental symmetry $\eta$ on $TM$, such that $M$ is complete with respect to the "Wick-rotated" Riemannian metric $g(\cdot, \eta \cdot)$, then:
\begin{itemize}
	\item the Dirac operator is Krein-self-adjoint for the Krein space $(\mathcal{K}_M, [\cdot,\cdot]_M)$ constructed from this splitting, for the real convention
	\item the Dirac operator is Krein-self-adjoint for the Krein space $(\mathcal{K}_M, (\cdot,\cdot)_M)$ constructed from this splitting, for the imaginary convention.
\end{itemize}

\end{thm} 

See also \cite{Strohmaier} for a proof of this theorem.

As for the charge conjugation operators $J_{M,\pm}$ and chirality $\chi_M$, they are defined on $\Gamma_c (M,S_M)$, but can be extended into bounded operators on $\mathcal{K}_M$:
\begin{thm} \label{ExtensionJchi}
The charge conjugation operators $J_{M,\pm}$ and chirality $\chi_M$ can be extended into bounded operators on $\mathcal{K}_M$, that we denote the same way. These bounded operators have the same properties as their restrictions to the core domain $\Gamma_c (M,S_M)$:
\begin{equation}
\begin{aligned}
\chi_M^2 &= 1 \\
\chi_M^\times = \chi_M^+ &= (-1)^q \chi_M
\end{aligned}
\end{equation}
for chirality, and:
\begin{equation}
\begin{aligned}
J_{M \pm}^2 &= a(\pm (q-p)) \\
J_{M \pm}^\times &= a(\mp(p+q))J_{M \pm} \\
J_{M \pm}^+ &= a(\pm(p+q))J_{M \pm}
\end{aligned}
\end{equation}
for charge conjugation, and:
\begin{equation}
J_{M \pm} \chi_M = (-1)^{(p-q)/2} \chi_M J_{M\pm}.
\end{equation}

\end{thm}

\begin{demo}
We will prove the results above for the imaginary convention and Robinson product. They can be proven similarly for the anti-Robinson product. Let $\eta_+$ be the fundamental symmetry associated to the splitting of the tangent bundle used to construct the Krein space $\mathcal{K}_M$. We denote $\langle \cdot,\cdot \rangle = (\cdot, \eta_+ \cdot)_M$ the associated inner product, and the Hilbert-adjoints with respect to this product with a simple $\dagger$.

First, let us notice that all properties above are true for the restricted operators, according to the results of chapter \ref{ChapterClifford} (see table \ref{CCSummaryTable}). We saw in the proof of theorem \ref{Compact} that $\chi_M$ is bounded on $\Gamma_c (M,S_M)$. Indeed, from the local form of $\eta_+$:
\begin{equation*}
\eta_+ = \begin{cases}
\pm i^{(p-1)/2} \gamma(e_{q+1})...\gamma(e_d) &\text{for odd  } q,p \\
\pm i^{q/2} \gamma(e_1)...\gamma(e_q) &\text{for even  } q,p,
\end{cases}
\end{equation*}
we proved that $\chi_M \eta_+ = (-1)^q \eta_+ \chi_M$, and thus that $\chi_M^\dagger = \chi_M$. This implies that $\chi_M$ is unitary: $\chi_M^\dagger \chi_M= \chi_M \chi_M^\dagger =1$. It is thus bounded on $\Gamma_c (M,S_M)$. One can prove similarly that $J_{M+}$ and $J_{M-}$ are bounded on $\Gamma_c (M,S_M)$: mirroring the computation of $\kappa$ in section \ref{SectionCC}, one can prove that $J_{M+}$ and $J_{M-}$ are anti-unitary: $J_{M \pm}^\dagger J_{M \pm} = J_{M \pm} J_{M \pm}^\dagger = 1$. They are thus bounded on $\Gamma_c (M,S_M)$ too. The operators $\chi_M$ and $J_{M \pm}$ are thus densely-defined bounded operators, and it is well-known that such operators can be extended to bounded operators.

Now, let us prove that they have the same properties as their restrictions. We will prove this for chirality only. Extending the proof below to the charge conjugation operators is straightforward. Let $\psi \in \mathcal{K}_M$ be the limit of a Cauchy sequence $(\psi_n)_n$ of elements in $\Gamma_c (M,S_M)$. The vector $\chi_M \psi$ is then defined as the limit of the sequence $\chi_M \psi_n$. We have
\begin{equation*}
\chi_M^2 \psi = \lim_{n \rightarrow \infty} \chi_M^2 \psi_n = \lim_{n \rightarrow \infty} \psi_n = \psi,
\end{equation*}
which proves that the extended operator $\chi_M$ is involutive too. Let $\varphi \in \mathcal{K}_M$ be the limit of another Cauchy sequence $(\varphi_n)_n$ of elements in $\Gamma_c (M,S_M)$. Note that the inner product is continuous by construction, and that $\eta_+$ is bounded (as all fundamental symmetries are, see chapter \ref{ChapterKrein}). As a result, the indefinite product $(\cdot, \cdot)_M$ is continuous. We have:
\begin{equation*}
\begin{aligned}
(\psi, \chi_M \varphi)_M &= \lim_{n \rightarrow \infty} (\psi_n, \chi_M \varphi_n)_M \\
&= (-1)^q \lim_{n \rightarrow \infty} (\chi_M \psi_n, \varphi_n)_M \\
(\psi, \chi_M \varphi)_M &= (-1)^q (\chi_M \psi, \varphi)_M,
\end{aligned}
\end{equation*}
which proves that the extended operator satisfies: $\chi_M^\times = (-1)^q \chi_M$.

\qed
\end{demo}

\section{A Tentative Semi-Riemannian Spectral Triple} \label{SectionCommutativeTriple}

We have now constructed all the objects required for a spectral triple. Let us summarize these objects and their properties. We do this for both conventions for the Dirac operator:

\subsection{Real Convention}

The objects we have constructed so far are the following:
\begin{itemize}
	\item The spinor Krein space $(\mathcal{K}_M, [\cdot,\cdot]_M)$. We remind the reader that adjoints of operators with respect to $[\cdot,\cdot]_M$ are denoted with a $+$ superscript.
	\item The chirality operator $\chi_M$, a bounded involution: $\chi_M^2 = 1$. It is either self-adjoint or anti-self-adjoint:
	\begin{equation}
	\chi_M^+ = (-1)^q \chi_M.
	\end{equation}
	\item The charge conjugation operator $J_{M+}$. It is bounded, and has the following properties:
	\begin{equation}
	\begin{aligned}
	J_{M+}^2 &= a(q-p) \\
	J_{M+}^+ &= a(p+q)J_{M+} \\
	J_{M+} \chi_M &= (-1)^{(p-q)/2} \chi_M J_{M+}.
	\end{aligned}
	\end{equation}
	\item The Dirac operator $\slashed{D}$, of dense domain $\mathcal{D}(\slashed{D}) = \Gamma_c (M,S_M)$. It is symmetric, and self-adjoint under some conditions (see theorem \ref{DiracSelfAdjoint}). It has the following properties:
	\begin{equation}
	\begin{aligned}
	\{\chi_M, \slashed{D}\} &= 0 \\
	\left[ J_{M+}, \slashed{D} \right] &= 0.
	\end{aligned}
	\end{equation}

\end{itemize}
Finally, for this to be a commutative spectral triple, we need a suitable algebra of functions, that can be represented by bounded operators on the Krein space, and is possibly unital. This algebra $A_M$ will be a subalgebra of the algebra of smooth complex functions on $M$: $A_M \subset \mathcal{C}^\infty(M, \mathbb{C})$. Any such function can be represented on $\Gamma_c (M,S_M)$ by pointwise multiplication:
\begin{equation}
\pi_M(f) \psi = f\psi,
\end{equation}
for all $f \in A_M$ and $\psi \in \Gamma_c (M,S_M)$. We want this representation to extend into a representation on the entire Krein space. In other words, we want the representation to map $A_M$ to $B(\mathcal{K}_M)$. It is straightforward to prove that for any $f \in A_M$ we have: $\|\pi_M(f) \|_{\eta_+} = \sup_M |f|$, where $\eta_+$ is the fundamental symmetry used to construct the Krein space. The functions of $A_M$ must therefore be bounded. The simplest choices for $A_M$ are thus:
\begin{itemize}
	\item $A_M = \mathcal{C}_b^\infty(M, \mathbb{C})$, the algebra of smooth complex \emph{bounded} functions
	\item $A_M = \mathcal{C}_b^\infty(M, \mathbb{R})$, the algebra of smooth real bounded functions.
\end{itemize} 
Both algebras are unital algebras. The first one is a pre-C$^\ast$-algebra: it is a dense subalgebra of the algebra of bounded continuous functions on $M$. The first algebra is also a complexification of the second. In both cases, the representation $\pi_M$ is involutive:
\begin{equation*}
\pi_M(\overline{f}) = \pi_M(f)^+,
\end{equation*}
thanks to the local nature of the Robinson product\footnote{This is in fact a possible justification for the restriction to local products in theorem \ref{ProductThm}.}. In the next chapter, we will use all the objects above to give a possible axiomatization of Semi-Riemannian (or indefinite) noncommutative geometries. 

\subsection{The Imaginary Convention}

The corresponding objects are the following:
\begin{itemize}
	\item The spinor Krein space $(\mathcal{K}_M, (\cdot,\cdot)_M)$. We remind the reader that adjoints of operators with respect to $(\cdot,\cdot)_M$ are denoted with a $\times$ superscript.
	\item The chirality operator $\chi_M$, a bounded involution: $\chi_M^2 = 1$. It is either self-adjoint or anti-self-adjoint:
	\begin{equation}
	\chi_M^\times = (-1)^q \chi_M.
	\end{equation}
	\item The graded charge conjugation operator $J_{M-}$. It is bounded, and has the following properties:
	\begin{equation}
	\begin{aligned}
	J_{M-}^2 &= a(p-q) \\
	J_{M-}^\times &= a(p+q)J_{M-} \\
	J_{M-} \chi_M &= (-1)^{(p-q)/2} \chi_M J_{M-}.
	\end{aligned}
	\end{equation}
	\item The Dirac operator $\slashed{D}$, of dense domain $\mathcal{D}(\slashed{D}) = \Gamma_c (M,S_M)$. It is symmetric, and self-adjoint under some conditions. It has the following properties:
	\begin{equation}
	\begin{aligned}
	\{\chi_M, \slashed{D}\} &= 0 \\
	\left[ J_{M-}, \slashed{D} \right] &= 0.
	\end{aligned}
	\end{equation}
	\item Finally, we need an algebra: $A_M = \mathcal{C}_b^\infty(M, \mathbb{C})$ or $A_M = \mathcal{C}_b^\infty(M, \mathbb{R})$, with an involutive representation $\pi_M$ on the Krein space by pointwise multiplication $\pi_M(f) = f$.

\end{itemize}

\section{Tensor Products of Manifolds} \label{SectionCommutativeTensorProduct}

To conclude this chapter, we construct tensor products of spin geometries, using the rules of tensor products of Clifford algebras summarized in section \ref{SectionTensorSummary}. We will later use this to axiomatize the construction of tensor products of triples. Let $(M,g_M)$ and $(N,g_N)$ be two STO Spin manifolds of even dimensions and signatures $(p,q)$ and $(p',q')$ respectively. We construct the tensor manifold $M \times N$. Although the algebra $\mathcal{C}_b^\infty(M \times N, \mathbb{C})$ does not factorize to $\mathcal{C}_b^\infty(M, \mathbb{C}) \otimes_{\mathbb{C}} \mathcal{C}_b^\infty(N, \mathbb{C})$, the former is dense in the closure of the latter for the supremum norm. A similar result holds for the algebras of real smooth bounded functions. In both cases, $A_M \otimes A_N$ is \emph{almost-dense} in $A_{M \times N}$ in the sense that:
\begin{equation}
A_M \otimes A_N \subset A_{M \times N} \subset \overline{A_M \otimes A_N}.
\end{equation}

The total tangent bundle is the direct sum: $T(M\times N) = TM \oplus TN$. We equip it with the metric $g = g_M \oplus g_N$. The rules of tensor products for Clifford algebras tell us that the total Clifford bundle is a graded tensor product of bundles: $\mathrm{Cl}(T^\ast(M \times N)) \supset \mathrm{Cl}(T^\ast M) \hat{\otimes} \mathrm{Cl}(T^\ast N)$, with the latter almost-dense in the former, and with the embedding of the total cotangent space given by:
\begin{equation*}
\gamma(X \oplus Y) = \gamma(X) \hat{\otimes} 1 + 1 \hat{\otimes} \gamma(Y),
\end{equation*}
for any $X \in T^\ast M, Y \in T^\ast N$. The total Clifford bundle can be represented on the total spinor bundle:
\begin{equation*}
S_{M \times N} = S_M \hat{\otimes} S_N
\end{equation*}
(the bundles $S_M$ and $S_N$ are of course graded by their respective chirality operators). The total space of smooth spinors $\Gamma(M \times N, S_{M \times N})$ has as an almost-dense subset the tensor product $\Gamma(M,S_M) \hat{\otimes} \Gamma(N,S_N)$. The total chirality operator is given by:
\begin{equation}
\chi_{M \times N} = \pm \chi_M \hat{\otimes} \chi_N.
\end{equation}
If $H_M$ and $H_N$ are Robinson products for $M$ and $N$ respectively, then a local Robinson product for $M \times N$ is given by:
\begin{equation*}
H_{M \times N}(\psi_1 \hat{\otimes} \varphi_1, \psi_2 \hat{\otimes} \varphi_2) = H_M(\psi_1, \psi_2) H_N(\varphi_1, \beta \varphi_2),
\end{equation*}
where:
\begin{equation}
\beta = (i^{q'} \chi_N)^{q} = \begin{cases}
1 &\text{if $q$ is even} \\
\chi_N &\text{if $q$ is odd and $q'$ is even} \\
i \chi_N &\text{if $q$ and $q'$ are both odd.}
\end{cases}
\end{equation}
After integrating on $M\times N$, one finds the global Robinson product:
\begin{equation}
(\psi_1 \hat{\otimes} \varphi_1, \psi_2 \hat{\otimes} \varphi_2)_{M \times N} = (\psi_1, \psi_2)_M (\varphi_1, \beta \varphi_2)_N.
\end{equation}
The tensor product is similar for anti-Robinson products:
\begin{equation}
\begin{aligned}
G_{M \times N}(\psi_1 \hat{\otimes} \varphi_1, \psi_2 \hat{\otimes} \varphi_2) &= G_M(\psi_1, \psi_2) G_N(\varphi_1, \beta \varphi_2) \\
[\psi_1 \hat{\otimes} \varphi_1, \psi_2 \hat{\otimes} \varphi_2]_{M \times N} &= [\psi_1, \psi_2]_M [\varphi_1, \beta \varphi_2]_N.
\end{aligned}
\end{equation}
Finally, the total charge conjugation operators are given by:
\begin{equation}
J_{M \times N, \pm} = \chi_M^{(p'-q')/2} J_{M \pm} \hat{\otimes} \chi_N^{(p-q)/2} J_{N \pm}.
\end{equation}

Let us now construct the total Dirac operator. For this, we need the total canonical Clifford connection:
\begin{prop}
For the total Robinson or anti-Robinson product, and total charge conjugation operator constructed above, the canonical Clifford connection of $M \times N$ is given by:
\begin{equation*}
\nabla^S_{X \oplus Y} = \nabla^{S,M}_X \hat{\otimes} 1 + 1 \hat{\otimes} \nabla^{S,N}_Y,
\end{equation*}
where $\nabla^{S,M}$ and $\nabla^{S,N}$ are the canonical Clifford connections of $M$ and $N$ respectively.

\end{prop}

Let us notice that a Clifford connection always commutes with a chirality operator. It is thus an even operator: $|\nabla^{S,M}| = |\nabla^{S,N}| = 0$. The action of $\nabla^S$ is thus given by:
\begin{equation*}
\nabla^S_{X \oplus Y}(\psi \hat{\otimes} \varphi) = \nabla^{S,M}_X \psi \hat{\otimes} \varphi + \psi \hat{\otimes} \nabla^{S,N}_Y \varphi,
\end{equation*}
for $\psi \in \Gamma(M,S_M), \varphi \in \Gamma(N,S_N)$. This defines the action of $\nabla^S_{X \oplus Y}$ on $\Gamma(M,S_M) \hat{\otimes} \Gamma(N,S_N)$. It then extends uniquely to $\Gamma(M \times N, S_{M \times N})$ by the almost-density of the former space in the latter.

\begin{demo}

We only give an outline of the proof, without the gory details. We denote $\nabla^\otimes$ the right-hand side of the above equation: $\nabla^\otimes_{X \oplus Y} = \nabla^{S,M}_X \hat{\otimes} 1 + 1 \hat{\otimes} \nabla^{S,N}_Y$.

First, let $f \in \mathcal{C}^\infty(M, \mathbb{C}), g \in \mathcal{C}^\infty(N, \mathbb{C})$. It is easy to prove that:
\begin{equation*}
\nabla^\otimes_{X \oplus Y}(f \otimes g)(\psi \hat{\otimes} \varphi) = (X \oplus Y)(f \otimes g) (\psi \hat{\otimes} \varphi) + (f \otimes g) \nabla^\otimes_{X \oplus Y}(\psi \hat{\otimes} \varphi).
\end{equation*}
For $X' \in \Gamma(M,TM), Y' \in \Gamma(N,TN)$ it is also straightforward to prove that:
\begin{equation*}
\nabla^\otimes_{X \oplus Y}\gamma(X' \oplus Y')(\psi \hat{\otimes} \varphi) = \gamma(\nabla_{X \oplus Y}(X' \oplus Y')) (\psi \hat{\otimes} \varphi) + \gamma(X' \oplus Y') \nabla^\otimes_{X \oplus Y}(\psi \hat{\otimes} \varphi),
\end{equation*}
using the tensor rule for the Levi-Civita connection: $\nabla_{X \oplus Y}(X' \oplus Y') = \nabla_X (X') \oplus \nabla_Y (Y')$. Now, we know that $\mathcal{C}^\infty(M, \mathbb{C}) \otimes \mathcal{C}^\infty(N, \mathbb{C})$ is almost-dense in $\mathcal{C}^\infty(M \times N, \mathbb{C})$, and that $\Gamma(M,S_M) \hat{\otimes} \Gamma(N,S_N)$ is almost-dense in $\Gamma(M \times N, S_{M \times N})$. Note that the space $\Gamma(M \times N, T(M\times N))$ has as an almost-dense subspace the tensor product: $\mathcal{C}^\infty(M, \mathbb{R}) \otimes \Gamma(N,TN) \oplus \Gamma(M,TM) \otimes \mathcal{C}^\infty(N, \mathbb{R})$. We can thus extend the two identities above:
\begin{equation*}
\begin{aligned}
\nabla_X^S(f \psi) &= X(f) \psi + f \nabla_X^S \psi \\
\nabla_X^S(\gamma(Y) \psi) &= \gamma(\nabla_X Y) \psi + \gamma(Y) \nabla_X^S \psi,
\end{aligned}
\end{equation*}
for any $f \in \mathcal{C}^\infty(M \times N, \mathbb{C}), X,Y \in \Gamma(M \times N, T(M\times N))$, and $\psi \in \Gamma(M \times N, S_{M \times N})$. The differential operator $\nabla^\otimes$ is thus a Clifford connection. It remains to prove that it is canonical. It is straightforward to prove that $\nabla^\otimes$ commutes with $J_{M \times N, \pm}$, from the commutation of the Clifford connections of $M$ and $N$ with their respective chirality and charge conjugation operators. Proving that $\nabla^\otimes$ is hermitian for $H_{M \times N}$ or $G_{M \times N}$ is also straightforward, if one notices that $\nabla^{S,N}$ necessarily commutes with the $\beta$ operator.

\qed
\end{demo}

We can now construct the total Dirac operator. Let $(e_a)_a$ be a local basis of $TM$, with dual basis $(e^a)_a$, and $(f_b)_b$ be a local basis of $TN$ with dual basis $(f^b)_b$. Then $(u_i)_i = (e_a \oplus 0)_a \cup (0 \oplus f_b)_b$ is a local basis of $T(M \times N)$, with dual basis $(\omega^i)_i = (e^a \oplus 0)_a \cup (0 \oplus f^b)_b$. We thus have:
\begin{equation*}
\begin{aligned}
\gamma(\omega^i) \nabla^S_{u_i} =& \gamma(e^a \oplus 0) \nabla^S_{e_a \oplus 0} + \gamma(0 \oplus f^b) \nabla^S_{f_b \oplus 0} \\
=& (\gamma(e^a) \hat{\otimes} 1)(\nabla^{S,M}_{e_a} \hat{\otimes} 1) + (1 \hat{\otimes} \gamma(f^b))(1 \hat{\otimes} \nabla^{S,N}_{f_b}) \\
=& \gamma(e^a)\nabla^{S,M}_{e_a} \hat{\otimes} 1 + 1 \hat{\otimes} \gamma(f^b)\nabla^{S,N}_{f_b}.
\end{aligned}
\end{equation*}
We deduce that:
\begin{equation}
\slashed{D} = \slashed{D}_M \hat{\otimes} 1 + 1 \hat{\otimes} \slashed{D}_N
\end{equation}
for both conventions. Here $\slashed{D}_M$ and $\slashed{D}_N$ are the Dirac operators for $M$ and $N$ respectively, and $\slashed{D}$ is the Dirac operator for $M \times N$.

Finally, we construct the total Krein space. We do it for the imaginary convention, and the process is identical for the real convention. Let $TM = E_- \oplus E_+$ (resp. $TN = E'_- \oplus E'_+$) be a splitting of the tangent bundle of $M$ (resp. $N$), and $\eta_{M+}$ (resp. $\eta_{N+}$) be the corresponding fundamental symmetry. Let $\mathcal{K}_M$ (resp. $\mathcal{K}_N$) be the resulting Krein space. Then $T(M \times N) = (E_+ \oplus E'_+) \oplus (E_- \oplus E'_-)$ is a splitting of the total tangent bundle into positive and negative subspaces. Using the local forms of the fundamental symmetries, one can prove that the corresponding fundamental symmetry is:
\begin{equation*}
\eta_+ = i^{qq'} \chi_M^q \eta_{M+} \hat{\otimes} \chi_N^{q'} \eta_{N+},
\end{equation*}
and that the corresponding inner product is given by:
\begin{equation*}
(\cdot, \eta_+ \cdot)_{M \times N} = (\cdot, \eta_{M+} \cdot)_M \otimes (\cdot, \eta_{N+} \cdot)_N.
\end{equation*}
As a result, the completion of $\Gamma(M \times N, S_{M \times N})$, which has as an almost-dense subspace the tensor product $\Gamma(M,S_M) \hat{\otimes} \Gamma(N,S_N)$, is:
\begin{equation}
\mathcal{K}_{M \times N} = \overline{\mathcal{K}_M \hat{\otimes} \mathcal{K}_N}.
\end{equation}
When $M$ and $N$ are compact, the splittings used do not matter, and the identity above always holds by the uniqueness of the respective spinor Krein spaces.

\chapter{Indefinite Spectral Triples}

In this chapter, we use the results of the previous chapter to try to define a Semi-Riemannian generalization of spectral triples. This is also inspired from existing work on Semi-Riemannian NCG: see \cite{VdD, Barrett, PS, Strohmaier, VPR, VS2} for example. We will start with a definition of what we call an Indefinite Spectral Triple (IST), with a list of possible axioms. We will then relate these ISTs to Semi-Riemannian geometries to give them a "metric interpretation". This was reported in \cite{3B-IST}. Finally, we will define the tensor product of ISTs.

\section{Definition of Indefinite Spectral Triples}

We start this chapter with a \emph{tentative} definition of what an indefinite spectral triple should be:

\begin{defn} \label{ISTDef}
The family of objects $(A,\mathcal{K},D,\chi,J)$ is called an indefinite spectral triple if they have the following properties:
\begin{enumerate}
	\item $\mathcal{K}$ is a $\mathbb{Z}_2$-graded Krein space, with indefinite product $(\cdot,\cdot)$. The adjoint with respect to this product is denoted with a $\times$ superscript.
	\item $\chi$ is the grading operator of the above Krein space. It is either self-adjoint or anti-self-adjoint:
	\begin{equation}
	\chi^\times = (-1)^\sigma \chi,
	\end{equation}
	with $\sigma \in \{0,1\}$. We assume $\chi$ to be bounded. Equivalently, we assume that the homogeneous subspaces of $\mathcal{K}$ are closed.
	\item $A$ is a *-algebra with an involutive even representation $\pi$ on $\mathcal{K}$:
	\begin{equation}
	\pi: A \longrightarrow B(\mathcal{K})
	\end{equation}
	such that $\pi(a^\ast) = \pi(a)^\times$ and $[\chi,\pi(a)] = 0$, for any $a \in A$.
	\item $D$ is an odd, symmetric operator on $\mathcal{K}$ of dense domain $\mathcal{D}(D)$, called a generalized Dirac operator.
	\item Finally, the generalized charge conjugation operator $J$ is a homogeneous bounded anti-linear operator on $\mathcal{K}$ that commutes with the Dirac operator: $[J, D] = 0$. It squares to $\pm 1$ and is either self-adjoint or anti-self-adjoint:
	\begin{equation}
	\begin{aligned}
	J^2 &= \epsilon \\
	J^\times &= \kappa J,
	\end{aligned}
	\end{equation}
	with $\epsilon, \kappa \in \{0,1\}$. The grading $|J|$ of $J$ is then used to define the two additional signs:
	\begin{equation}
	\begin{aligned}
	\epsilon'' &= (-1)^{|J|} \\
	\kappa'' &= (-1)^{|J|+\sigma} (=(-1)^\sigma \epsilon'').
	\end{aligned}
	\end{equation}
	We thus have: $J\chi = \epsilon'' \chi J$.
	
\end{enumerate}

\end{defn}

The indefinite product on $\mathcal{K}$ is said to be homogeneous and even (resp. odd) when $\chi$ is self-adjoint (resp. anti-self-adjoint): its grading is $\sigma$. This is because the homogeneous subspaces of $\mathcal{K}$ are mutually orthogonal if $\chi$ is self-adjoint, and self-orthogonal if it is anti-self-adjoint. In other words, for $\psi,\varphi \in \mathcal{K}$ homogeneous vectors, the product $(\psi, \varphi)$ is nonvanishing only if $|\varphi| + |\psi| \equiv \sigma \modulo 2$. Let us see why:
\begin{equation*}
\begin{aligned}
(\psi,\varphi) &= (-1)^{|\psi|+|\varphi|} (\chi \psi, \chi \varphi) \\
&= (-1)^{|\psi|+|\varphi|} (\psi, \chi^\times \chi \varphi) \\
(\psi,\varphi) &= (-1)^{|\psi|+|\varphi|+\sigma} (\psi,\varphi)
\end{aligned}
\end{equation*}
hence the result.
Note that the homogeneity of the indefinite product ensures that an operator and its adjoint always have the same grading: $|T| = |T^\times|$.

One may want to supplement the IST with a \emph{privileged} fundamental symmetry of $\mathcal{K}$ that behaves "nicely". For example, one may seek a fundamental symmetry that either commutes or anti-commutes with $\chi$ - making it a homogeneous operator - and $J$. There would thus exist two signs $\alpha,\beta = \pm 1$ such that:
\begin{equation*}
\begin{aligned}
\chi \eta &= \alpha \eta \chi \\
J \eta &= \beta \eta J.
\end{aligned}
\end{equation*}
Let us determine those two signs. From the general relation $T^{\dagger \eta} = \eta T^\times \eta$, we deduce that:
\begin{equation*}
\begin{aligned}
\chi^{\dagger \eta} &= (-1)^\sigma \alpha \chi \\
J^{\dagger \eta} &= \kappa \beta J,
\end{aligned}
\end{equation*}
from which we deduce that:
\begin{equation*}
\begin{aligned}
\chi^{\dagger \eta} \chi &= (-1)^\sigma \alpha \\
J^{\dagger \eta}J &= \epsilon \kappa \beta.
\end{aligned}
\end{equation*}
The operators $\chi^{\dagger \eta} \chi$ and $J^{\dagger \eta}J$ are necessarily positive, and thus equal to $1$. We deduce that $\alpha = (-1)^\sigma = \epsilon'' \kappa''$, and $\beta = \epsilon \kappa$. This leads us the following definition:
\begin{defn} \label{PrivilegedEtaDef}
Let $(A,\mathcal{K},D,\chi,J)$ be an IST. A privileged fundamental symmetry $\eta$ is a homogeneous fundamental symmetry of the Krein space that either commutes or anti-commutes with $J$. It necessarily satisfies:
\begin{equation}
\begin{aligned}
\chi \eta &= \epsilon'' \kappa'' \eta \chi \\
J \eta &= \epsilon \kappa \eta J.
\end{aligned}
\end{equation}

\end{defn}
Note that the existence of a privileged fundamental symmetry necessarily implies that $\chi$ and $J$ are bounded operators, since it makes them $\eta$-unitary and $\eta$-anti-unitary operators respectively.

The correspondence with spin geometries is established by the following proposition (see section \ref{SectionCommutativeTriple}):
\begin{prop}
Let $(M,g)$ be an STO Spin manifold of even dimension. Then:
\begin{itemize}
	\item $(A_M, \mathcal{K}_M, \slashed{D}, \chi_M, J_{M+})$ is an IST, with $\mathcal{K}_M$ equipped with the global anti-Robinson product and $\slashed{D}$ defined according to the real convention
	\item $(A_M, \mathcal{K}_M, \slashed{D}, \chi_M, J_{M-})$ is an IST, with $\mathcal{K}_M$ equipped with the global Robinson product and $\slashed{D}$ defined according to the imaginary convention.
\end{itemize}
Such a triple is called a manifold IST.

\end{prop}

Note that a manifold IST always has a privileged fundamental symmetry: namely the fundamental symmetry $\eta_\pm$ used to construct the Krein space, and built from a splitting of the tangent bundle. Note also that a non-compact manifold admits more than one IST, since in that case the Krein space is not unique (see section \ref{SectionSpinorKrein}).

A few remarks are in order about definitions \ref{ISTDef} and \ref{PrivilegedEtaDef}. It is important to note that this list of axioms is not definitive, as there are properties of spin geometries that we have not axiomatized here. Some of these correspond to axioms of usual (\emph{i.e.} definite) spectral triples. For example, $A$ is not necessarily a C$^\ast$-algebra or a pre-C$^\ast$-algebra. In fact, $A$ could be a real algebra. Other axioms of (definite) spectral triples have also been abandoned, such as the requirement that $D$ is self-adjoint, and not simply symmetric. Another requirement we have dropped is that the commutator of $D$ with any element of $\pi(A)$ be a bounded operator: for manifold ISTs, this would require working with the algebra of smooth functions with compact support, which we think is too restrictive in the Semi-Riemannian case. Indeed, for a manifold IST, said commutator would be a differential form embedded in the Clifford bundle: $\gamma(df)$, whose operator norm depends on the "angle" with the fundamental symmetry $\eta_{M,\pm}$. Such a form could thus be unbounded, even if $f$ is a bounded function with bounded derivatives (see \ref{SectionSpinorKrein}), unless its support is required to be compact.

Finally, one could drop the requirement that the representation $\pi$ is involutive, in order to construct some particular ISTs of physical interest (see \cite{Besnard}).

In any case, these axioms do generalize the axioms of spectral triples, in the sense that \emph{a spectral triple is necessarily an IST} where $(\cdot,\cdot)$ is an inner product. From the axioms of ISTs, one can prove that:
\begin{equation*}
\begin{aligned}
\chi^\times \chi &= (-1)^\sigma = \epsilon'' \kappa'' \\
J^\times J &= \epsilon \kappa.
\end{aligned}
\end{equation*}
For $(\cdot,\cdot)$ positive definite, both $\chi^\times \chi$ and $J^\times J$ must be positive definite operators, and both must be equal to the identity as a consequence. This implies that $\kappa = \epsilon$ and $\kappa'' = \epsilon''$. The four independent signs $\epsilon, \epsilon'', \kappa, \kappa''$ thus reduce to the two independent signs $\epsilon$ and $\epsilon''$. This is consistent with what we know about spectral triples.

The notion of spectral triple equivalence can be generalized to a notion of IST equivalence:
\begin{defn}
Let $(A,\mathcal{K},D,\chi,J)$ and $(A',\mathcal{K}',D',\chi',J')$ be two ISTs. They are said to be isomorphic, or equivalent, up to orientation if there exists a *-algebra isomorphism $\alpha$:
\begin{equation}
\begin{aligned}
\alpha: A &\longrightarrow A' \\
\alpha(a^\ast) &= \alpha(a)^\ast, \; \forall a \in A \\
\alpha(ab) &= \alpha(a)\alpha(b), \; \forall a,b \in A
\end{aligned}
\end{equation}
and an isomorphism $U$ of vector spaces that preserves the Krein products up to a sign $\lambda$:
\begin{equation}
\begin{aligned}
U: \mathcal{K} &\longrightarrow \mathcal{K}' \\
(U \cdot, U \cdot) &= \lambda (\cdot,\cdot)',
\end{aligned}
\end{equation}
such that:
\begin{equation}
\begin{aligned}
U \pi(a) U^{-1} &= \pi'(\alpha(a)) \\
U \chi U^{-1} &= \delta \chi' \\
U D U^{-1} &= D' \\
U J U^{-1} &= J',
\end{aligned}
\end{equation}
where $\delta$ is a sign. The couple $(U,\alpha)$ is said to be an IST isomorphism.

If $\lambda=\delta=1$, then the two ISTs are simply said to be isomorphic, or equivalent.

\end{defn}

It is easy to see that IST equivalence and equivalence up to orientation are indeed equivalence relations (\textit{i.e.} symmetric, reflexive, and transitive). The reason for allowing the sign change for the Krein product is that indefinite products, such as the Robinson product, are typically defined up to a real non-vanishing factor. Similarly, grading is defined up to a sign. As a consequence, two ISTs that only differ by a \emph{positive} real factor $\lambda$ for the Krein product are equivalent, and two ISTs that only differ by a real factor $\lambda$ for the Krein product are equivalent up to orientation (with $U$ being a scalar operator equal to a square root of $\lambda$, and $\alpha$ being the identity).

The difference between equivalence and equivalence up to orientation manifests itself best for manifold ISTs, where the sign of Robinson or anti-Robinson product, and the sign of the chirality operator define a space- and time-orientation for the manifold. Two equivalent manifold ISTs will thus have the same space- and time-orientation, while two equivalent ISTs up to orientation may have a different choice of space- or time-orientation.

For definite spectral triples, it is easy to see that two equivalent triples as ISTs, are also unitarily equivalent as definite triples.

\section{KO, Metric, Space and Time Dimensions} \label{Dimensions}

To a (definite) spectral triple, one can associate a KO-dimension from the two signs $\epsilon$ and $\epsilon''$. For the canonical triple of a Riemannian manifold, this coincides with its dimension modulo 8. But for ISTs, there are four independent signs. We will associate to these four signs \emph{two} dimensions: a KO-dimension and a metric dimension. We will see below that for the manifold triple of a Semi-Riemannian manifold, these will coincide with the signature and dimension of the manifold respectively, modulo 8. Let us first define these two dimensions:
\begin{defn}
Let $(A,\mathcal{K},D,\chi,J)$ be an IST with signs $\epsilon, \epsilon'', \kappa, \kappa''$. The KO-dimension $n$ and metric dimension $m$ of the IST are the unique numbers in $2\mathbb{Z}/8\mathbb{Z} \cong \{0,2,4,6\}$ such that\footnote{We remind the reader that the $a$ function is defined in section \ref{SectionCC} by $a(n) = (-1)^{n(n+2)/8}$.}:
\begin{equation}
\begin{aligned}
\epsilon &= a(n) \\
\epsilon'' &= (-1)^{n/2} \\
\kappa &= a(m) \\
\kappa'' &= (-1)^{m/2}.
\end{aligned}
\label{KOMetricDef}
\end{equation}
The correspondence between signs and dimensions is also illustrated in table \ref{KOMetric}.

\end{defn}

The relations above are well-defined because both $a$ and $n\mapsto (-1)^{n/2}$ are periodic functions of period $8$. From table \ref{KOMetric}, it is easy to see that the maps $n \mapsto (\epsilon, \epsilon'')$ and $m \mapsto (\kappa, \kappa'')$ are bijective maps. Hence the existence and uniqueness of the KO and metric dimensions of any IST. Note that two equivalent ISTs necessarily have the same metric and KO-dimensions.

The KO-dimension $n$ is defined by the signs $\epsilon, \epsilon''$ the same way KO-dimension is defined for a (definite) spectral triple, hence the name KO-dimension. For an IST that is also a spectral triple, we know that $(\kappa, \kappa'') = (\epsilon, \epsilon'')$. The KO and metric dimensions are thus equal: $m=n$, and they coincide with the usual definition of the KO-dimension of a triple.

\begin{table}[!h]
	\centering
	\setlength{\tabcolsep}{2em}
	\begin{tabular}{c|cccc}
	$n$ (resp. $m$) & 0 & 2 & 4 & 6 \\
	\hline
	$\epsilon$ (resp. $\kappa$) & 1 & -1 & -1 & 1 \\
	$\epsilon''$ (resp. $\kappa''$) & 1 & -1 & 1 & -1
	\end{tabular}
	\caption{The signs $\epsilon, \epsilon'', \kappa, \kappa''$ are given as a function of $n,m$.}
	\label{KOMetric}
\end{table}

By comparing with the results of section \ref{SectionCommutativeTriple}, one arrives at the following result:
\begin{prop}
Let $(M,g)$ be an STO Spin manifold of even dimension and signature $(q,p)$. Then any corresponding manifold IST has metric dimension $m \equiv p+q \modulo 8$ and KO-dimension:
\begin{equation*}
n \equiv \begin{cases}
q-p \modulo 8 \text{  for the real convention} \\
p-q \modulo 8 \text{  for the imaginary convention.}
\end{cases}
\end{equation*}

\end{prop}

The relation $m \equiv p+q \modulo 8$ between $m$ and the dimension $d=p+q$ of the manifold $M$ is what suggests calling $m$ the metric dimension.

\begin{demo}
From the results of section \ref{SectionCommutativeTriple}, we know that:
\begin{equation*}
\begin{aligned}
\sigma &= q \modulo 2 \\
\epsilon'' &= (-1)^{(p-q)/2} \\
\kappa &= a(p+q)
\end{aligned}
\end{equation*}
for both conventions. We deduce that $\kappa'' = (-1)^\sigma \epsilon'' = (-1)^{(p+q)/2}$, and thus that $m \equiv p+q \modulo 8$. We also know that:
\begin{equation*}
\epsilon = \begin{cases}
a(q-p) \text{  for the real convention} \\
a(p-q) \text{  for the imaginary convention,}
\end{cases}
\end{equation*}
from which one deduces $n \equiv q-p \modulo 8$ for the real convention, and $n \equiv p-q \modulo 8$ for the imaginary convention.

\qed
\end{demo}

\subsection{Space and Time Dimensions}

Let $(M,g)$ be an STO Spin manifold of signature $(q,p)$. We know that the metric dimension of any corresponding manifold IST is $m \equiv p+q \modulo 8$ and its KO-dimension is $n \equiv \pm(q-p) \modulo 8$, depending on the convention. One is tempted to invert and generalize these relations, in order to associate a total signature $(q,p)$ to any IST, starting from its KO and metric dimensions. One might also be tempted to interpret $q$ and $p$ as space and time dimensions. It turns out that such an interpretation simplifies the extraction of $(q,p)$ from $m$ and $n$. We will thus start with that.

To find space and time, one starts with what one knows best: Lorentzian manifolds. There are two conventions for Lorentzian manifolds: the so-called West Coast and East Coast conventions. For each convention, we will consider a Lorentzian manifold $M$ of even dimension $d$ that admits an IST, and relate its signature $(q,p)$ to the number $t$ of time dimensions and the number $s$ of space dimensions. By definition, $(t,s) = (1,d-1)$ for a Lorentzian manifold.

\begin{itemize}
	\item \textbf{East Coast convention:} A West Coast metric has a signature of the form $(- + \cdots +)$. We thus have: $(q,p) = (1,d-1)$. We deduce that for such a metric: $(t,s) = (q,p)$.
	\item \textbf{West Coast convention:} A West Coast metric has a signature of the form $(+ - \cdots -)$. We thus have: $(q,p) = (d-1,1)$. We deduce that for such a metric: $(t,s) = (p,q)$.
\end{itemize}

As it turns out, these two conventions are in one-to-one correspondence with the real and imaginary conventions of the Dirac operator. This can be seen by solving the corresponding Dirac equation for a \emph{flat} Lorentzian manifold $M$:

\begin{itemize}
	\item \textbf{Real convention:} In this case, the IST can be chosen so that the Dirac takes the form: $\slashed{D} = \gamma(dx^\mu) \partial_\mu$. The Dirac equation for a free fermion of mass $m$ is then: 
\begin{equation*}
(\gamma(dx^\mu) \partial_\mu + m)\psi = 0
\end{equation*}
Substituting a plane wave solution $\psi(x) = e^{i p_\mu x^\mu} u$ then gives the equation:
\begin{equation*}
(i \gamma(p) + m)u=0,
\end{equation*}
which gives the dispersion relation:
\begin{equation*}
p^2 = -m^2 = \vec{p}^2 - E^2.
\end{equation*}
We deduce that the corresponding metric has the signature $(- + \cdots +)$. Thus, the real convention necessarily implies using the East Coast convention. From now on, \emph{we will equate real convention and East Coast convention}, and use either name indiscriminately.

	\item \textbf{Imaginary convention:} The Dirac operator can be taken of the form: $\slashed{D} = i\gamma(dx^\mu) \partial_\mu$. The additional $i$ factor changes the dispersion relation to:
\begin{equation*}
p^2 = m^2 = E^2 - \vec{p}^2.
\end{equation*}
We deduce that the corresponding metric has the signature $(+ - \cdots -)$. The imaginary convention necessarily implies using the West Coast convention. From now on, \emph{we will equate imaginary convention and West Coast convention}, and use either name indiscriminately.

\end{itemize}

Let us now generalize the relations we found above to general Semi-Riemannian manifolds and their manifold ISTs, and relate KO and metric dimensions to space and time dimensions:

\begin{itemize}
	\item East Coast convention: For this convention, we know that $n \equiv q-p \modulo 8$ and $m \equiv p+q \modulo 8$. We also established that $(t,s) = (q,p)$. We deduce that $n \equiv t-s \modulo 8$ and $m \equiv t+s \modulo 8$.
	\item West Coast convention: For this convention, we know that $n \equiv p-q \modulo 8$ and $m \equiv p+q \modulo 8$. We also established that $(t,s) = (p,q)$. We deduce that $n \equiv t-s \modulo 8$ and $m \equiv t+s \modulo 8$.
\end{itemize}

We thus have for both conventions:
\begin{equation}
\begin{aligned}
n &\equiv t-s \modulo 8 \\
m &\equiv t+s \modulo 8.
\end{aligned}
\label{SpaceTimeKOMetric}
\end{equation}
For a manifold IST, the relation between space and time dimensions on one hand, and KO and metric dimensions on the other, \emph{does not depend on the convention for the Dirac operator}. This makes the following definition meaningful:
\begin{defn}
Let $(A,\mathcal{K},D,\chi,J)$ be an IST of KO-dimension $n$ and metric dimension $m$. A pair $(t,s) \in \mathbb{N}^2$ of space and time dimensions for the IST is a solution to the equations \eqref{SpaceTimeKOMetric}.
\end{defn}
Note that for a manifold IST, a possible solution is:
\begin{equation*}
(t,s) = \begin{cases}
(q,p) \text{  for the East Coast convention} \\
(p,q) \text{  for the West Coast convention.}
\end{cases}
\end{equation*}

For given $m$ and $n$, let us find the general solution to the equations \eqref{SpaceTimeKOMetric}. By taking the sum and difference of the two equations, one finds:
\begin{equation*}
\begin{aligned}
m+n &\equiv 2t \modulo 8 \\
m-n &\equiv 2s \modulo 8.
\end{aligned}
\end{equation*}
Since each side of each equation is even, one can divide by 2:
\begin{equation*}
\begin{aligned}
t &\equiv \frac{m+n}{2} \modulo 4 \\
s &\equiv \frac{m-n}{2} \modulo 4.
\end{aligned}
\end{equation*}
There thus exist two integers $j,k$ such that:
\begin{equation}
\begin{aligned}
t &= \frac{m+n}{2} + 4j \\
s &= \frac{m-n}{2} + 4k.
\end{aligned}
\label{Solutions}
\end{equation}
summing both equation gives us:
\begin{equation*}
t+s \equiv m + 4(j+k) \modulo 8.
\end{equation*}
We know that $t+s \equiv m \modulo 8$. We deduce that $4(j+k) \equiv 0 \modulo 8$. This is true if and only if $j + k \equiv 0 \modulo 2$. Thus all solutions to \eqref{SpaceTimeKOMetric} are of the form \eqref{Solutions}, with $j,k$ of the same parity. Proving the converse is immediate. For the reader's convenience, the solutions are presented for all values of $n$ and $m$ in table \ref{TableSpaceTime}. Note that $t$ and $s$ necessarily have the same parity.

\begin{table}[!h]
	\centering
	\setlength{\tabcolsep}{1.5em}
	\begin{tabular}{|c|c|c|c|c|}
	\hline
	 & n=0 & 2 & 4 & 6 \\
	\hline m=0 & (0,0) (4,4) & (1,7) (5,3) & (2,6) (6,2) & (3,5) (7,1) \\
	\hline 2 & (1,1) (5,5) & (2,0) (6,4) & (3,7) (7,3) & (4,6) (0,2) \\
	\hline 4 & (2,2) (6,6) & (3,1) (7,5) & (4,0) (0,4) & (5,7) (1,3) \\
	\hline 6 & (3,3) (7,7) & (4,2) (0,6) & (5,1) (1,5) & (6,0) (2,4) \\
	\hline
	\end{tabular}
	\caption{Smallest values of $(t,s)$ that solve equations \eqref{SpaceTimeKOMetric}, and correspond to $j,k=0$ or $1$ in \eqref{Solutions}. The general solution is of the form $(t+8a,s+8b)$.}
	\label{TableSpaceTime}
\end{table}

\subsection{The Cardinal Conventions}

When we defined $t$ and $s$ for an IST, we did not make use of the Dirac operator: we only used the Krein space, its indefinite product, $\chi$ and $J$. To these objects, one can always associate KO and metric dimensions, as well as space and time dimensions, according to equations \eqref{KOMetricDef},\eqref{SpaceTimeKOMetric}, and the defining equations of the signs $\epsilon, \epsilon'', \kappa, \kappa''$ in definition \ref{ISTDef}. The physical interpretation of these dimensions is then established by computing them for the manifold ISTs of some Lorentzian manifolds, for the East Coast and West Coast metrics.

Given the triple of objects $(\mathcal{K}, \chi, J)$, if a privileged fundamental symmetry $\eta$ exists, one says that $(\mathcal{K}, \eta, \chi, J)$ is a CPT co-representation\footnote{The term co-representation is used when dealing with a representation that involves anti-linear operators.} (see \cite{3B-IST} for more). This is because for manifold triples, the operators $\eta, \chi, J$ generate the same group as the lift of the C,P,T transformations to the spinor bundle, represented here by the spinor Krein space.

Without the Dirac operator to constrain the charge conjugation and indefinite product, there are two more conventions that we have not yet considered in section \ref{SectionCommutativeTriple}, that do not correspond to manifold ISTs but correspond to CPT co-representations (and can still be used to give a physical interpretation to some ISTs):

\begin{itemize}
	\item \textbf{South Coast convention:} In this convention, one uses the global Robinson product $(\cdot,\cdot) = (\cdot,\cdot)_M$ as an indefinite product, and the ungraded charge conjugation operator $J = J_{M+}$ as a generalized charge conjugation. The chirality is of course $\chi = \chi_M$. The Dirac operator constructed in the previous chapter will either be antisymmetric (real convention) or anticommute with $J$ (imaginary convention). Nonetheless, $J$ and $\chi$ satisfy (see theorem \ref{ExtensionJchi}):
\begin{equation*}
\begin{aligned}
\chi_M^\times &= (-1)^q \chi_M \\
J_{M+} \chi_M &= (-1)^{(p-q)/2} \chi_M J_{M+} \\
J_{M+}^2 &= a(q-p) \\
J_{M+}^\times &= a(-(p+q))J_{M+}.
\end{aligned}
\end{equation*}
We deduce that:
\begin{equation*}
\begin{aligned}
\epsilon &= a(q-p) \\
\epsilon'' &= (-1)^{(p-q)/2} \\
\kappa &= a(-(p+q))\\
\kappa'' &= (-1)^{(p+q)/2}.
\end{aligned}
\end{equation*}
This gives the KO dimension $n \equiv q-p \modulo 8$ and metric dimension $m \equiv -(q+p) \modulo 8$. Comparison with equations \eqref{SpaceTimeKOMetric} shows that possible pairs of space and time dimensions are given by $(t,s) \equiv (-p, -q) \modulo 8$. This is obviously unphysical for a manifold IST.

	\item \textbf{North Coast convention:} This convention is opposite to the South Coast convention. Here, one uses the global anti-Robinson product $(\cdot,\cdot) = [\cdot,\cdot]_M$ as an indefinite product, and the graded charge conjugation operator $J = J_{M-}$ as a generalized charge conjugation. The chirality is of course $\chi = \chi_M$. The Dirac operator constructed in the previous chapter will either be antisymmetric (imaginary convention) or anticommute with $J$ (real convention). Similarly to what was done above, one finds the KO dimension $n = p-q \modulo 8$ and the metric dimension $m \equiv -(q+p) \modulo 8$. Possible pairs of space and time dimensions are given by $(t,s) \equiv (-q, -p) \modulo 8$. This is also unphysical for a manifold IST.

\end{itemize}

All four cardinal conventions are summarized in table \ref{CardinalSummary}.

\begin{table}[!h]
\centering
\setlength{\tabcolsep}{1.5em}
\begin{tabular}{|l||c|c|c|c|}
\hline
Convention & $m$ & $n$ & $(t,s)$ \\
\hline
\hline East-coast  & $p+q$  & $q-p$ & $(q,p)$ \\
\hline West-coast  & $p+q$  & $p-q$ & $(p,q)$ \\
\hline South-coast  & $-p-q$  & $q-p$ & $(-p, -q)$ \\
\hline North-coast  & $-p-q$  & $p-q$ & $(-q, -p)$ \\
\hline
\end{tabular}
\caption{Metric, KO, and allowed space and time dimensions for all four conventions. All dimensions are given modulo 8.}
\label{CardinalSummary}
\end{table}

\section{Tensor Products of Triples} \label{SectionISTTensorProduct}

We conclude this chapter with a recipe for constructing tensor products of ISTs, that generalizes the existing recipe for (definite) spectral triples. This is inspired from the results of section \ref{SectionCommutativeTensorProduct}, and from the work \cite{VdD}.

\begin{thm} \label{ISTTensorProduct}
Let $(A_1,\mathcal{K}_1,D_1,\chi_1,J_1)$ and $(A_2,\mathcal{K}_2,D_2,\chi_2,J_2)$ be two ISTs, such that the algebras $A_1$ and $A_2$ are either both real or both complex. Then the following family $(A,\mathcal{K},D,\chi,J)$ of objects is an IST:
\begin{itemize}
	\item $\mathcal{K} = \overline{\mathcal{K}_1 \otimes \mathcal{K}_2}$ is the topological tensor product of the Krein spaces $\mathcal{K}_1$ and $\mathcal{K}_2$. We equip it with the indefinite product:
	\begin{equation}
	(\varphi_1 \hat{\otimes} \varphi_2, \psi_1 \hat{\otimes} \psi_2) = (\varphi_1, \psi_1)_1 (\varphi_2, \beta \psi_2)_2
	\label{tensorproductinner}
	\end{equation}
	where:
	\begin{equation}
	\beta = (i^{\sigma_2} \chi_2)^{\sigma_1} = \begin{cases}
	1 &\text{if $(\cdot, \cdot)_1$ is even} \\
	\chi_2 &\text{if $(\cdot, \cdot)_1$ is odd and $(\cdot, \cdot)_2$ is even} \\
	i \chi_2 &\text{if $(\cdot, \cdot)_1$ and $(\cdot, \cdot)_2$ are both odd}
	\end{cases}
	\label{tensorproductbeta}
	\end{equation}
	
	\item The grading of $\mathcal{K}$ is the bounded extension of the operator:
	\begin{equation}
	\chi = \chi_1 \hat{\otimes} \chi_2
	\end{equation}
	
	\item The Dirac operator is given by:
	\begin{equation}
	D = D_1 \hat{\otimes} 1 + 1 \hat{\otimes} D_2
	\end{equation}
	with domain $\mathcal{D}(D) = \mathcal{D}(D_1) \hat{\otimes} \mathcal{D}(D_2)$.
	
	\item The charge conjugation is given by
	\begin{equation}
	J = \chi_1^{|J_2|} J_1 \hat{\otimes} \chi_2^{|J_1|} J_2 = J_1 \chi_1^{|J_2|} \hat{\otimes} J_2 \chi_2^{|J_1|}
	\end{equation}
	
	\item Finally, the algebra $A$ is any *-algebra such that $A_1 \otimes A_2$ is almost-dense in it, that is:
	\begin{equation}
	A_1 \otimes A_2 \subset A \subset \overline{A_1 \otimes A_2},
	\end{equation}
	with involution defined by:
	\begin{equation*}
	(a \otimes b)^\ast = a^\ast \otimes b^\ast,
	\end{equation*}
	and a representation given by:
	\begin{equation}
	\pi(a \otimes b) = \pi_1(a) \hat{\otimes} \pi_2(b).
	\end{equation}
	The topology on $A_1 \otimes A_2$ is induced by $\pi$ and the topology on $\mathcal{K}$.
	
\end{itemize}

\end{thm}

Note that this tensor product is not unique, since the algebra $A$ is not unique. Once again, we denote the graded tensor product with the symbol $\hat{\otimes}$. One ought not to confuse this with the topological tensor product, sometimes denoted similarly in the literature. We remind the reader that the rules of the graded tensor product are as follows:
\begin{equation}
\begin{aligned}
(T_1 \hat{\otimes} T_2) (S_1 \hat{\otimes} S_2) &= (-1)^{|T_2||S_1|}(T_1 S_1 \hat{\otimes} T_2 S_2) \\
(T_1 \hat{\otimes} T_2) (\psi_1 \hat{\otimes} \psi_2) &= (-1)^{|T_2||\psi_1|}(T_1 \psi_1 \hat{\otimes} T_2 \psi_2)
\end{aligned}
\label{gradedproduct}
\end{equation}
for any operators $S_1,T_1$ and $S_2,T_2$ acting on $\mathcal{K}_1$ and $\mathcal{K}_2$ respectively, and vectors $\psi_1$ and $\psi_2$ in those spaces. The grading of the operator $T_1 \hat{\otimes} T_2$ is simply the sum of the gradings of $T_1$ and $T_2$. An important property of the graded tensor product is that:
\begin{equation}
(T_1 \hat{\otimes} T_2)^{-1} = (-1)^{|T_1||T_2|}(T_1^{-1} \hat{\otimes} T_2^{-1})
\label{tensorinverse}
\end{equation}
for $T_1$ and $T_2$ invertible. This can be checked with a straightforward computation.

The two different forms of $J$ given above can be proven to be equal using the identities:
\begin{equation*}
\begin{aligned}
J_1 \chi_1^{|J_2|} =& (-1)^{|J_1||J_2|} \chi_1^{|J_2|} J_1 \\
J_2 \chi_2^{|J_1|} =& (-1)^{|J_2||J_1|} \chi_2^{|J_1|} J_2,
\end{aligned}
\end{equation*}
from which one infers that:
\begin{equation*}
\chi_1^{|J_2|} J_1 \hat{\otimes} \chi_2^{|J_1|} J_2 = (-1)^{2|J_1||J_2|} J_1 \chi_1^{|J_2|} \hat{\otimes} J_2 \chi_2^{|J_1|} = J_1 \chi_1^{|J_2|} \hat{\otimes} J_2 \chi_2^{|J_1|}
\end{equation*}

In order to prove our theorem, we will need two simple but useful lemmas. The first one gives a property of the indefinite product constructed above:

\begin{lem} \label{tensoradjoint}
Let $T_1, T_2$ be two homogeneous linear operators on $\mathcal{K}_1, \mathcal{K}_2$ respectively. Then:
\begin{equation}
(T_1 \hat{\otimes} T_2)^\times = (-1)^{|T_1||T_2|} T_1^\times \hat{\otimes} T_2^\times
\label{tensorproductadjoint}
\end{equation}
For homogeneous \emph{anti}-linear operators $K_1, K_2$, the rule is:
\begin{equation}
(K_1 \hat{\otimes} K_2)^\times = (-1)^{\sigma_1 \sigma_2 + |K_1||K_2|} K_1^\times \hat{\otimes} K_2^\times
\label{tensorproductadjointantilinear}
\end{equation}
\end{lem}

\begin{demo}
Let us compute the adjoint of $T_1 \hat{\otimes} T_2$:
\begin{equation*}
\begin{aligned}
((T_1 \hat{\otimes} T_2)(\varphi_1 \hat{\otimes} \varphi_2), \psi_1 \hat{\otimes} \psi_2) &= (-1)^{|T_2||\varphi_1|} (T_1 \varphi_1 \hat{\otimes} T_2 \varphi_2, \psi_1 \hat{\otimes} \psi_2) \\
&= (-1)^{|T_2||\varphi_1|} (T_1 \varphi_1, \psi_1)_1 (T_2 \varphi_2, \beta \psi_2)_2 \\
&= (-1)^{|T_2||\varphi_1|} (\varphi_1, T_1^\times \psi_1)_1 (\varphi_2, T_2^\times \beta \psi_2)_2 \\
\end{aligned}
\end{equation*}
From the definition \eqref{tensorproductbeta} of $\beta$, one can can see that $T_2 \beta = (-1)^{\sigma_1 |T_2|} \beta T_2$. The same holds for $T_2^\times$, thanks to the self-adjointness of $\beta$. We now have:
\begin{equation*}
\begin{aligned}
((T_1 \hat{\otimes} T_2)(\varphi_1 \hat{\otimes} \varphi_2), \psi_1 \hat{\otimes} \psi_2) &= (-1)^{|T_2|(|\varphi_1|+\sigma_1)} (\varphi_1, T_1^\times \psi_1)_1 (\varphi_2, \beta T_2^\times \psi_2)_2 \\
&= (-1)^{|T_2|(|\varphi_1|+|\psi_1|+\sigma_1)} ((\varphi_1 \hat{\otimes} \varphi_2), (T_1 \hat{\otimes} T_2)(\psi_1 \hat{\otimes} \psi_2))
\end{aligned}
\end{equation*}
The product $(T_1 \varphi_1, \psi_1)_1$ is non vanishing only if $|T_1|+|\varphi_1| \cong |\psi_1|+\sigma_1$. This implies that $(-1)^{|T_2|(|\varphi_1|+|\psi_1|+\sigma_1)} = (-1)^{|T_1||T_2|}$, and the lemma follows.

The proof is similar for antilinear operators, with the exception that the formula $T_2 \beta = (-1)^{\sigma_1 |T_2|} \beta T_2$ is replaced by $K_2 \beta = (-1)^{\sigma_1 (\sigma_2 + |K_2|)} \beta K_2$, due to the $i^{\sigma_1 \sigma_2}$ factor in $\beta$.

\qed
\end{demo}

The second lemma gives a property of the charge conjugation operator:

\begin{lem} \label{TensorCCAdjointThm}
Let $T_1, T_2$ be two linear operators on $\mathcal{K}_1, \mathcal{K}_2$ respectively. Then:
\begin{equation}
J(T_1 \hat{\otimes} T_2)J^{-1} = J_1 T_1 J_1^{-1} \hat{\otimes} J_2 T_2 J_2^{-1}.
\label{TensorCCAdjoint}
\end{equation}

\end{lem}

\begin{demo}
We assume, without any loss in generality, that $T_1$ and $T_2$ are homogeneous. First, the inverse of $J$ is given by equation \eqref{tensorinverse} as:
\begin{equation*}
J^{-1} = (-1)^{|J_1||J_2|}(\chi_1^{|J_2|} J_1^{-1} \hat{\otimes} \chi_2^{|J_1|} J_2^{-1}).
\end{equation*}
Now, we have:
\begin{equation*}
\begin{aligned}
J(T_1 \hat{\otimes} T_2)J^{-1} =& (-1)^{|J_1||J_2|}(J_1 \chi_1^{|J_2|} \hat{\otimes} J_2 \chi_2^{|J_1|})(T_1 \hat{\otimes} T_2)(\chi_1^{|J_2|} J_1^{-1} \hat{\otimes} \chi_2^{|J_1|} J_2^{-1}) \\
=& (-1)^{|J_1||J_2|+|J_1||T_2|} (J_1 \chi_1^{|J_2|} \hat{\otimes} J_2 \chi_2^{|J_1|})(T_1 \chi_1^{|J_2|} J_1^{-1} \hat{\otimes} T_2 \chi_2^{|J_1|} J_2^{-1}) \\
=& (-1)^{|J_1||T_2|+|J_2||T_1|+2|J_1||J_2|} (J_1 \chi_1^{|J_2|} T_1 \chi_1^{|J_2|} J_1^{-1} \hat{\otimes} J_2 \chi_2^{|J_1|} T_2 \chi_2^{|J_1|} J_2^{-1}) \\
=& (-1)^{2|J_1||T_2|+2|J_2||T_1|} (J_1 \chi_1^{2|J_2|} T_1 J_1^{-1} \hat{\otimes} J_2 \chi_2^{2|J_1|} T_2 J_2^{-1}) \\
J(T_1 \hat{\otimes} T_2)J^{-1} =& (J_1 T_1 J_1^{-1} \hat{\otimes} J_2 T_2 J_2^{-1}).
\end{aligned}
\end{equation*}

\qed
\end{demo}

We can now carry on with the proof of the theorem:

\begin{demo}
First, let us observe that $\mathcal{K} = \overline{\mathcal{K}_1 \otimes \mathcal{K}_2}$ is a Krein space when equipped with the direct tensor indefinite product:
\begin{equation*}
(\varphi_1 \hat{\otimes} \varphi_2, \psi_1 \hat{\otimes} \psi_2)' = (\varphi_1, \psi_1)_1 (\varphi_2, \psi_2)_2.
\end{equation*}
This product and the one constructed in equation \eqref{tensorproductinner} differ by an insertion of $\beta$. Since $\chi_2$ is bounded, so is $\beta$, and the two indefinite products necessarily define the same topology on $\mathcal{K} = \overline{\mathcal{K}_1 \otimes \mathcal{K}_2}$. This means that $(\mathcal{K}, (\cdot,\cdot))$ is a Krein space. 

The operator $\chi = \chi_1 \hat{\otimes} \chi_2$ is necessarily bounded for $(\cdot,\cdot)'$, and thus for $(\cdot,\cdot)$, since $\chi_1$ and $\chi_2$ are bounded. It is also easy to see that it is involutive:
\begin{equation*}
\chi^2 = \chi_1^2 \hat{\otimes} \chi_2^2 = 1.
\end{equation*}
It thus defines a grading on $\mathcal{K}$. Finally, using lemma \ref{tensoradjoint}, we have:
\begin{equation*}
\begin{aligned}
\chi^\times =& \chi_1^\times \hat{\otimes} \chi_2^\times \\
=& (-1)^{\sigma_1 + \sigma_2} \chi_1 \hat{\otimes} \chi_2 \\
\chi^\times =& (-1)^{\sigma_1 + \sigma_2} \chi,
\end{aligned}
\end{equation*}
from which we deduce that $\sigma \equiv \sigma_1 + \sigma_2 \modulo 2$.

We now turn our attention to the algebra. Since $\pi_1$ and $\pi_2$ are bounded even representations, so is $\pi = \pi_1 \hat{\otimes} \pi_2$. To prove that it is involutive, note that $\pi_1$ and $\pi_2$ are even. As a result, we have:
\begin{equation*}
\begin{aligned}
\pi((a \otimes b)^\ast) =& \pi(a^\ast \otimes b^\ast) \\
=& \pi_1(a^\ast) \hat{\otimes} \pi_2(b^\ast) \\
=& \pi_1(a)^\times \hat{\otimes} \pi_2(b)^\times \\
=& (\pi_1(a) \hat{\otimes} \pi_2(b))^\times \\
\pi((a \otimes b)^\ast) =& \pi(a \otimes b)^\times.
\end{aligned}
\end{equation*}

We know that $\mathcal{D}(D_1)$ and $\mathcal{D}(D_2)$ are dense in $\mathcal{K}_1$ and $\mathcal{K}_2$ respectively, and that the topology of $\mathcal{K}$ is the product of the topologies of $\mathcal{K}_1$ and $\mathcal{K}_2$. As a result, the domain $\mathcal{D}(D) = \mathcal{D}(D_1) \hat{\otimes} \mathcal{D}(D_2)$ of $D$ is dense. $D$ is clearly odd. Let us prove that it is symmetric:
\begin{equation*}
\begin{aligned}
D^\times &= (D_1 \hat{\otimes} 1 + 1 \hat{\otimes} D_2)^\times \\
&= (-1)^{|D_1||1|} D_1^\times \hat{\otimes} 1 + (-1)^{|D_2||1|} 1 \hat{\otimes} D_2^\times \\
&= D_1 \hat{\otimes} 1 + 1 \hat{\otimes} D_2 \\
D^\times &= D
\end{aligned}
\end{equation*}

To conclude this proof, we look at the charge conjugation operator $J$. It is clearly anti-linear and bounded. It is also homogeneous, of grading $|J| = |J_1|+|J_2|$. We deduce that:
\begin{equation*}
\epsilon'' = \epsilon''_1 \epsilon''_2.
\end{equation*}
Let us prove that $J$ squares to $\pm 1$:
\begin{equation*}
\begin{aligned}
J^2 &= (\chi_1^{|J_2|} J_1 \hat{\otimes} \chi_2^{|J_1|} J_2)^2 \\
&= (-1)^{|J_1||J_2|} (\chi_1^{|J_2|} J_1)^2 \hat{\otimes} (\chi_2^{|J_1|} J_2)^2 \\
&= (-1)^{|J_1||J_2|} \chi_1^{2|J_2|} J_1^2 \hat{\otimes} \chi_2^{2|J_1|} J_2^2 \\
J^2 &= (-1)^{|J_1||J_2|} \epsilon_1 \epsilon_2
\end{aligned}
\end{equation*}
Using the identity:
\begin{equation}
(-1)^{ab} = \frac{1}{2}[1+(-1)^a+(-1)^b-(-1)^{a+b}]
\label{SignProduct}
\end{equation}
we write this as:
\begin{equation*}
\epsilon = \frac{1}{2}\epsilon_1 \epsilon_2 (1 + \epsilon''_1 + \epsilon''_2 - \epsilon''_1 \epsilon''_2)
\end{equation*}
We also prove that $D$ and $J$ commute:
\begin{equation*}
\begin{aligned}
JDJ^{-1} =& J(D_1 \hat{\otimes} 1 + 1 \hat{\otimes} D_2)J^{-1} \\
=& J_1 D_1 J_1^{-1} \hat{\otimes} 1 + 1 \hat{\otimes} J_2 D_2 J_2^{-1} \\
=& D_1 \hat{\otimes} 1 + 1 \hat{\otimes} D_2 \\
JDJ^{-1} =& D.
\end{aligned}
\end{equation*}
Finally, we prove that $J$ is self-adjoint or anti-self-adjoint, using lemma \ref{tensoradjoint}:
\begin{equation*}
\begin{aligned}
J^\times &= (\chi_1^{|J_2|} J_1 \hat{\otimes} \chi_2^{|J_1|} J_2)^\times \\
&= (-1)^{\sigma_1 \sigma_2 + |J_1||J_2|} (J_1^\times (\chi_1^\times)^{|J_2|} \hat{\otimes} J_2^\times (\chi_2^\times)^{|J_1|}) \\
&= (-1)^{\sigma_1 \sigma_2 + |J_1||J_2|+\sigma_1 |J_2| +\sigma_2 |J_1|}(J_1^\times \chi_1^{|J_2|} \hat{\otimes} J_2^\times \chi_2^{|J_1|}) \\
&= (-1)^{\sigma_1 \sigma_2 + |J_1||J_2|+\sigma_1 |J_2| +\sigma_2 |J_1|} \kappa_1 \kappa_2 (J_1 \chi_1^{|J_2|} \hat{\otimes} J_2 \chi_2^{|J_1|}) \\
&= (-1)^{\sigma_1 \sigma_2 + |J_1||J_2|+\sigma_1 |J_2| +\sigma_2 |J_1|} \kappa_1 \kappa_2 (\chi_1^{|J_2|} J_1 \hat{\otimes} \chi_2^{|J_1|} J_2) \\
J^\times  &= (-1)^{(\sigma_1 + |J_1|) (\sigma_2 + |J_2|)} \kappa_1 \kappa_2 J = \pm J.
\end{aligned}
\end{equation*}
Using once again the identity \ref{SignProduct}, we rewrite this as:
\begin{equation*}
\kappa = \frac{1}{2}\kappa_1 \kappa_2 (1 + \kappa''_1 + \kappa''_2 - \kappa''_1 \kappa''_2) 
\end{equation*}

\qed
\end{demo}

A tensor product rule exists for privileged fundamental symmetries:

\begin{prop} \label{TensorEtaProp}
Let $(A_1,\mathcal{K}_1,D_1,\chi_1,J_1)$ and $(A_2,\mathcal{K}_2,D_2,\chi_2,J_2)$ be two ISTs, such that the algebras $A_1$ and $A_2$ are either both real or both complex. Let $(A,\mathcal{K},D,\chi,J)$ be their tensor product as in theorem \ref{ISTTensorProduct}. If $\eta_1$ and $\eta_2$ are privileged fundamental symmetries of the two ISTs, then the following operator:
\begin{equation}
\eta = i^{\sigma_1 \sigma_2} \chi_1^{\sigma_2} \eta_1 \hat{\otimes} \chi_2^{\sigma_1} \eta_2 = i^{\sigma_1 \sigma_2} \eta_1 \chi_1^{\sigma_2} \hat{\otimes} \eta_2 \chi_2^{\sigma_1}
\label{TensorEta}
\end{equation}
is a privileged fundamental symmetry of the total IST, and its $\eta$-adjoint is given by:
\begin{equation}
(T_1 \hat{\otimes} T_2)^{\dagger\eta} = (-1)^{|T_1||T_2|} T_1^{\dagger\eta_1} \hat{\otimes} T_2^{\dagger\eta_2}
\label{Etatensorproductadjoint}
\end{equation}
for $T_1, T_2$ two homogeneous linear operators on $\mathcal{K}_1, \mathcal{K}_2$ respectively, and:
\begin{equation}
(K_1 \hat{\otimes} K_2)^{\dagger\eta} = (-1)^{|K_1||K_2|} K_1^{\dagger\eta_1} \hat{\otimes} K_2^{\dagger\eta_2}
\label{Etatensorproductadjointantilinear}
\end{equation}
for $K_1, K_2$ two homogeneous anti-linear operators.

\end{prop}

That the two forms above for $\eta$ are equal can be proven using the fact that the gradings of $\eta_1$ and $\eta_2$ are $\sigma_1$ and $\sigma_2$ respectively.

\begin{demo}
Let $\eta_1$ and $\eta_2$ be privileged fundamental symmetries of the two factor ISTs. That $\eta = i^{\sigma_1 \sigma_2} \chi_1^{\sigma_2} \eta_1 \hat{\otimes} \chi_2^{\sigma_1} \eta_2$ commutes or anticommutes with $J$ and $\chi$ is easy to see. What needs to be proven is that it is a fundamental symmetry. First, note that it is bounded, since $\eta_1$ and $\eta_2$ are bounded. Next, using that the gradings of $\eta_1$ and $\eta_2$ are $\sigma_1$ and $\sigma_2$, we prove that $\eta$ is involutive:
\begin{equation*}
\begin{aligned}
\eta^2 =& (-1)^{\sigma_1 \sigma_2} (\chi_1^{\sigma_2} \eta_1 \hat{\otimes} \chi_2^{\sigma_1} \eta_2)^2 \\
=& (-1)^{2\sigma_1 \sigma_2} (\chi_1^{\sigma_2} \eta_1)^2 \hat{\otimes} (\chi_2^{\sigma_1} \eta_2)^2 \\
=& (-1)^{4\sigma_1 \sigma_2} (\chi_1^{2\sigma_2} \eta_1^2) \hat{\otimes} (\chi_2^{2\sigma_1} \eta_2^2) \\
\eta^2 =& 1.
\end{aligned}
\end{equation*}
The operator $\eta$ thus splits the Krein space into two closed eigenspaces of eigenvalues $\pm 1$. It remains to prove that these eigenspaces are definite. Equivalently, we need to prove that $\langle \cdot, \cdot \rangle_\eta = (\cdot, \eta \cdot)$ is positive definite. We have:
\begin{equation*}
\begin{aligned}
\langle \varphi_1 \hat{\otimes} \varphi_2, \psi_1 \hat{\otimes} \psi_2 \rangle_\eta =& (\varphi_1 \hat{\otimes} \varphi_2, \eta(\psi_1 \hat{\otimes} \psi_2)) \\
=& i^{\sigma_1 \sigma_2} (\varphi_1 \hat{\otimes} \varphi_2, (\chi_1^{\sigma_2} \eta_1 \hat{\otimes} \chi_2^{\sigma_1} \eta_2)(\psi_1 \hat{\otimes} \psi_2)) \\
=& i^{\sigma_1 \sigma_2} (-1)^{|\psi_1|\sigma_2} (\varphi_1 \hat{\otimes} \varphi_2, \chi_1^{\sigma_2} \eta_1 \psi_1 \hat{\otimes} \chi_2^{\sigma_1} \eta_2 \psi_2) \\
\langle \varphi_1 \hat{\otimes} \varphi_2, \psi_1 \hat{\otimes} \psi_2 \rangle_\eta =& i^{\sigma_1 \sigma_2} (-1)^{|\psi_1|\sigma_2} (\varphi_1, \chi_1^{\sigma_2} \eta_1 \psi_1)_1 (\varphi_2, \beta \chi_2^{\sigma_1} \eta_2 \psi_2)_2.
\end{aligned}
\end{equation*}
We now substitute with $\beta = i^{\sigma_1 \sigma_2} \chi_2^{\sigma_1}$:
\begin{equation*}
\begin{aligned}
\langle \varphi_1 \hat{\otimes} \varphi_2, \psi_1 \hat{\otimes} \psi_2 \rangle_\eta =& (-1)^{|\psi_1|\sigma_2 + \sigma_1 \sigma_2} (\varphi_1, \chi_1^{\sigma_2} \eta_1 \psi_1)_1 (\varphi_2, \eta_2 \psi_2)_2 \\
=& (-1)^{|\psi_1|\sigma_2} (\varphi_1, \eta_1 \chi_1^{\sigma_2} \psi_1)_1 (\varphi_2, \eta_2 \psi_2)_2 \\
\langle \varphi_1 \hat{\otimes} \varphi_2, \psi_1 \hat{\otimes} \psi_2 \rangle_\eta =& (\varphi_1, \eta_1 \psi_1)_1 (\varphi_2, \eta_2 \psi_2)_2,
\end{aligned}
\end{equation*}
which proves that $\langle \cdot, \cdot \rangle_\eta = (\cdot, \eta \cdot)$ is positive definite. This also proves that $\langle \cdot, \cdot \rangle_\eta$ induces the appropriate topology on $\mathcal{K}$.

Now, let $T_1$ and $T_2$ be two homogeneous linear operators on $\mathcal{K}_1$ and $\mathcal{K}_2$. We have:
\begin{equation*}
\begin{aligned}
(T_1 \hat{\otimes} T_2)^{\dagger\eta} =& \eta (T_1 \hat{\otimes} T_2)^\times \eta \\
=& (-1)^{|T_1||T_2|+\sigma_1 \sigma_2} (\eta_1 \chi_1^{\sigma_2} \hat{\otimes} \eta_2 \chi_2^{\sigma_1}) (T_1^\times \hat{\otimes} T_2^\times) (\chi_1^{\sigma_2} \eta_1 \hat{\otimes} \chi_2^{\sigma_1} \eta_2) \\
=& (-1)^{|T_1||T_2|+\sigma_1 \sigma_2+\sigma_1 |T_2|} (\eta_1 \chi_1^{\sigma_2} \hat{\otimes} \eta_2 \chi_2^{\sigma_1}) (T_1^\times \chi_1^{\sigma_2} \eta_1 \hat{\otimes} T_2^\times \chi_2^{\sigma_1} \eta_2) \\
=& (-1)^{|T_1||T_2|+2\sigma_1 \sigma_2+\sigma_1 |T_2|+\sigma_2|T_1|} (\eta_1 \chi_1^{\sigma_2} T_1^\times \chi_1^{\sigma_2} \eta_1 \hat{\otimes} \eta_2 \chi_2^{\sigma_1} T_2^\times \chi_2^{\sigma_1} \eta_2) \\
=& (-1)^{|T_1||T_2|+2\sigma_1 |T_2|+2\sigma_2|T_1|} (\eta_1 \chi_1^{2\sigma_2} T_1^\times \eta_1 \hat{\otimes} \eta_2 \chi_2^{2\sigma_1} T_2^\times \eta_2) \\
=& (-1)^{|T_1||T_2|} \eta_1 T_1^\times \eta_1 \hat{\otimes} \eta_2 T_2^\times \eta_2 \\
(T_1 \hat{\otimes} T_2)^{\dagger\eta} =& (-1)^{|T_1||T_2|} T_1^{\dagger\eta_1} \hat{\otimes} T_2^{\dagger\eta_2}.
\end{aligned}
\end{equation*}

The proof for anti-linear operators is similar, with an additional $(-1)^{\sigma_1 \sigma_2}$ sign that comes from the $i^{\sigma_1 \sigma_2}$ phase of $\eta$. This sign cancels an identical sign that comes from equation \eqref{tensorproductadjointantilinear}.

\qed
\end{demo}

This tensor product of ISTs has a few interesting properties:

\begin{thm}
The tensor product of ISTs defined in theorem \ref{ISTTensorProduct} is additive for the metric and KO-dimensions, as well as the grading of the indefinite product.
\end{thm}

\begin{demo}
Let $(A_1,\mathcal{K}_1,D_1,\chi_1,J_1)$ and $(A_2,\mathcal{K}_2,D_2,\chi_2,J_2)$ be two ISTs, such that the algebras $A_1$ and $A_2$ are either both real or both complex, and of metric and KO-dimensions $(m_1,n_1)$ and $(m_2,n_2)$ respectively. From the proof of theorem \ref{ISTTensorProduct}, we know that for a product IST $(A,\mathcal{K},D,\chi,J)$ we have:
\begin{equation*}
\begin{aligned}
\epsilon &= \frac{1}{2}\epsilon_1 \epsilon_2 (1 + \epsilon''_1 + \epsilon''_2 - \epsilon''_1 \epsilon''_2) \\
\epsilon'' &= \epsilon''_1 \epsilon''_2.
\end{aligned}
\end{equation*}
From the definition of the metric and KO-dimensions, we have:
\begin{equation*}
\epsilon'' = \epsilon''_1 \epsilon''_2 = (-1)^{(n_1+n_2)/2}.
\end{equation*}
For $\epsilon$, we have:
\begin{equation*}
\epsilon = \frac{1}{2} a(n_1) a(n_2) (1 + (-1)^{n_1 /2} + (-1)^{n_2 /2} - (-1)^{(n_1+n_2)/2}).
\end{equation*}
From equation \eqref{SignProduct}, we infer that:
\begin{equation*}
\epsilon = a(n_1) a(n_2) (-1)^{n_1 n_2 /4}.
\end{equation*}
It is easy to prove that:
\begin{equation*}
\frac{n_1(n_1+2)}{8}+\frac{n_2(n_2+2)}{8}+\frac{n_1 n_2}{4} = \frac{(n_1+n_2)(n_1+n_2+2)}{8},
\end{equation*}
which, with the definition of $a$, implies that:
\begin{equation*}
\epsilon = a(n_1+n_2).
\end{equation*}
From this and the identity: $\epsilon'' = (-1)^{(n_1+n_2)/2}$, we deduce that the KO-dimension of the product IST is $n \equiv n_1 + n_2 \modulo 8$.

From the proof of theorem \ref{ISTTensorProduct}, we also know that:
\begin{equation*}
\begin{aligned}
\kappa &= \frac{1}{2}\kappa_1 \kappa_2 (1 + \kappa''_1 + \kappa''_2 - \kappa''_1 \kappa''_2) \\
\sigma &\equiv \sigma_1 + \sigma_2 \modulo 2.
\end{aligned}
\end{equation*}
From the second identity and $\epsilon'' = \epsilon''_1 \epsilon''_2$, we deduce that $\kappa'' = \kappa''_1 \kappa''_2$. A computation similar to the one performed above for the KO-dimension then proves that the metric dimension of the product IST is $m \equiv m_1 + m_2 \modulo 8$

\qed
\end{demo}

Although the tensor product is not symmetric, it is symmetric \emph{up to equivalence}:

\begin{thm}
Let $(A,\mathcal{K},D,\chi,J)$ be the tensor product of two ISTs $(A_1,\mathcal{K}_1,D_1,\chi_1,J_1)$ and $(A_2,\mathcal{K}_2,D_2,\chi_2,J_2)$ \emph{in that order}, and $(A',\mathcal{K}',D',\chi',J')$ the tensor product \emph{in the reverse order}. Let $\alpha$ be the unique extension of the *-algebra isomorphism:
\begin{equation}
\begin{aligned}
\alpha: A_1 \otimes A_2 &\longrightarrow A_2 \otimes A_1 \\
a \otimes b &\longmapsto b \otimes a
\end{aligned}
\end{equation}
with respect to the closures $\overline{A_1 \otimes A_2} \cong \overline{A_2 \otimes A_1}$. If $\alpha$ is a *-isomorphism from $A$ to $A'$, then the two tensor product ISTs are equivalent.

\end{thm}

Note that the condition on $A$ and $A'$ is automatically satisfied if we simply take:
\begin{equation*}
\begin{aligned}
A &= A_1 \otimes A_2 \\
A' &= A_2 \otimes A_1.
\end{aligned}
\end{equation*}

\begin{demo}
Let $U$ be the following isomorphism of vector spaces:
\begin{equation*}
\begin{aligned}
U: \mathcal{K}_1 \hat{\otimes} \mathcal{K}_2 &\longrightarrow \mathcal{K}_2 \hat{\otimes} \mathcal{K}_1 \\
\psi_1 \hat{\otimes} \psi_2 &\longmapsto (-1)^{|\psi_1||\psi_2|} \psi_2 \hat{\otimes} \psi_1.
\end{aligned}
\end{equation*}
Let us prove that it preserves the indefinite product up to a sign. Let $\psi_1,\varphi_1 \in \mathcal{K}_1$ and $\psi_2,\varphi_2 \in \mathcal{K}_2$ be homogeneous vectors. Recall that the tensor indefinite products are given by (see \eqref{tensorproductinner}):
\begin{equation*}
\begin{aligned}
(\varphi_1 \hat{\otimes} \varphi_2, \psi_1 \hat{\otimes} \psi_2) &= i^{\sigma_1 \sigma_2} (-1)^{\sigma_1 |\psi_2|} (\varphi_1, \psi_1)_1 (\varphi_2, \psi_2)_2 \\
(\varphi_2 \hat{\otimes} \varphi_1, \psi_2 \hat{\otimes} \psi_1)' &= i^{\sigma_1 \sigma_2} (-1)^{\sigma_2 |\psi_1|} (\varphi_2, \psi_2)_2 (\varphi_1, \psi_1)_1.
\end{aligned}
\end{equation*}
We thus have:
\begin{equation*}
\begin{aligned}
(U(\varphi_1 \hat{\otimes} \varphi_2), U(\psi_1 \hat{\otimes} \psi_2))' =& (-1)^{|\psi_1||\psi_2|+|\varphi_1||\varphi_2|} (\varphi_2 \hat{\otimes} \varphi_1, \psi_2 \hat{\otimes} \psi_1)' \\
=& i^{\sigma_1 \sigma_2} (-1)^{|\psi_1||\psi_2|+|\varphi_1||\varphi_2|+\sigma_2 |\psi_1|} (\varphi_2, \psi_2)_2 (\varphi_1, \psi_1)_1.
\end{aligned}
\end{equation*}
This product is nonvanishing only if $|\psi_1| + |\varphi_1| \equiv \sigma_1 \modulo 2$ and $|\psi_2| + |\varphi_2| \equiv \sigma_2 \modulo 2$. We have:
\begin{equation*}
\begin{aligned}
|\psi_1||\psi_2|+|\varphi_1||\varphi_2|+\sigma_2 |\psi_1| \equiv & |\psi_1|(|\psi_2|+\sigma_2) + |\varphi_1||\varphi_2| \modulo 2 \\
\equiv & |\psi_1||\varphi_2| + |\varphi_1||\varphi_2| \modulo 2 \\
\equiv & (|\psi_1|+|\varphi_1|)|\varphi_2| \modulo 2 \\
\equiv & \sigma_1 |\varphi_2| \modulo 2 \\
\equiv & \sigma_1 (\sigma_2 + |\psi_2|) \modulo 2 \\
|\psi_1||\psi_2|+|\varphi_1||\varphi_2|+\sigma_2 |\psi_1| \equiv & \sigma_1 \sigma_2 + \sigma_2 |\psi_2| \modulo 2.
\end{aligned}
\end{equation*}
We infer that:
\begin{equation*}
\begin{aligned}
(U(\varphi_1 \hat{\otimes} \varphi_2), U(\psi_1 \hat{\otimes} \psi_2))' =& i^{\sigma_1 \sigma_2} (-1)^{\sigma_1 \sigma_2 + \sigma_2 |\psi_2|} (\varphi_2, \psi_2)_2 (\varphi_1, \psi_1)_1 \\
(U(\varphi_1 \hat{\otimes} \varphi_2), U(\psi_1 \hat{\otimes} \psi_2))' =& (-1)^{\sigma_1 \sigma_2} (\varphi_1 \hat{\otimes} \varphi_2, \psi_1 \hat{\otimes} \psi_2).
\end{aligned}
\end{equation*}
Since $U$ maps one indefinite product to the other, it can be extended to a bounded operator from $\mathcal{K}$ to $\mathcal{K}'$.

Let us compute the adjoint action of $U$ on operators. Let $T_1$ and $T_2$ be two homogeneous linear operators on $\mathcal{K}_1$ and $\mathcal{K}_2$. We have:
\begin{equation*}
\begin{aligned}
U(T_1 \hat{\otimes} T_2)(\psi_1 \hat{\otimes} \psi_2) =& (-1)^{|T_2||\psi_1|} U(T_1 \psi_1 \hat{\otimes} T_2 \psi_2) \\
=& (-1)^{|T_2||\psi_1|+(|T_1| + |\psi_1|)(|T_2| + |\psi_2|)} (T_2 \psi_2 \hat{\otimes} T_1 \psi_1) \\
=& (-1)^{|T_1||\psi_2|+|T_1||T_2|+|\psi_1||\psi_2|} (T_2 \psi_2 \hat{\otimes} T_1 \psi_1) \\
=& (-1)^{|T_1||T_2|+|\psi_1||\psi_2|} (T_2 \hat{\otimes} T_1)(\psi_2 \hat{\otimes} \psi_1) \\
U(T_1 \hat{\otimes} T_2)(\psi_1 \hat{\otimes} \psi_2) =& (-1)^{|T_1||T_2|}(T_2 \hat{\otimes} T_1) U (\psi_1 \hat{\otimes} \psi_2),
\end{aligned}
\end{equation*}
from which we deduce that $U(T_1 \hat{\otimes} T_2) = (-1)^{|T_1||T_2|}(T_2 \hat{\otimes} T_1) U$ and thus that:
\begin{equation}
U(T_1 \hat{\otimes} T_2) U^{-1} = (-1)^{|T_1||T_2|}(T_2 \hat{\otimes} T_1).
\end{equation}
Using this identity, it is straightforward to prove that:
\begin{equation*}
\begin{aligned}
U \pi(\cdot) U^{-1} &= \pi' \circ \alpha(\cdot) \\
U \chi U^{-1} &= \chi' \\
U D U^{-1} &= D' \\
U J U^{-1} &= (-1)^{|J_1||J_2|}J'.
\end{aligned}
\end{equation*}

\qed
\end{demo}

\subsection{Non-Graded Representation of The Tensor Product}

Similarly to what was done for the tensor product of Clifford algebras in section \ref{SectionTensorSummary}, we rewrite the tensor product of ISTs in a non-graded form. That is, a representation of graded tensor products of operators using \emph{non}-graded tensor products of operators. This takes the form of an algebra isomorphism $\Pi : \mathrm{End}(\mathcal{K}_1) \hat{\otimes} \mathrm{End}(\mathcal{K}_2) \longrightarrow \mathrm{End}(\mathcal{K}_1 \hat{\otimes} \mathcal{K}_2)$ that satisfies:
\begin{equation*}
\begin{aligned}
\Pi(T_1 \hat{\otimes} T_2) \Pi(S_1 \hat{\otimes} S_2)&= (-1)^{|T_2||S_1|}\Pi(T_1 S_1 \hat{\otimes} T_2 S_2) \\
\Pi(T_1 \hat{\otimes} T_2) (\psi_1 \otimes \psi_2) &= (-1)^{|T_2||\psi_1|} (T_1 \psi_1 \otimes T_2 \psi_2)
\end{aligned}
\end{equation*}
A possible solution is the following:
\begin{equation}
\Pi(T_1 \hat{\otimes} T_2) = T_1 \chi_1^{|T_2|} \otimes T_2
\label{nongradedrep}
\end{equation}
This means that $T_1 \hat{\otimes} T_2$ and $T_1 \chi_1^{|T_2|} \otimes T_2$ act the same way on the vector space $\mathcal{K}_1 \hat{\otimes} \mathcal{K}_2$. The inverse isomorphism is easily found to be:
\begin{equation*}
\Pi^{-1}(T_1 \otimes T_2) = T_1 \chi_1^{|T_2|} \hat{\otimes} T_2
\end{equation*}

The tensor indefinite product defined in \eqref{tensorproductinner} can be rewritten as the basic product of two products. We thus write:
\begin{equation}
(\varphi_1 \otimes \varphi_2, \psi_1 \otimes \psi_2) = (\varphi_1 \hat{\otimes} \varphi_2, \psi_1 \hat{\otimes} \psi_2) = (\varphi_1, \psi_1)_1 (\varphi_2, \psi_2)_{2 \beta}
\label{nongradedinner}
\end{equation}
where $(\cdot, \cdot)_{2 \beta} = (\cdot, \beta \cdot)_2$ is an ``effective'' product on the second vector space. The adjoint of an operator $T_2$ with respect to this altered product will be denoted $T_2^{\times\beta} = (-1)^{\sigma_1 |T_2|}T_2^\times$. In particular, the Dirac operator $D_2$ is either self-adjoint or anti-self-adjoint with respect to the effective indefinite product, depending on whether $\chi_1$ is self-adjoint or anti-self-adjoint: $D_2^{\times\beta} = (-1)^{\sigma_1} D_2^\times$. This is compensated by the appearance of $\chi_1$ in the non-graded representation of $1 \hat{\otimes} D_2$. It is indeed given by $\Pi(1 \hat{\otimes} D_2) = \chi_1 \otimes D_2$, which is always self-adjoint, as expected.

The total spectral triple defined in theorem \ref{ISTTensorProduct} has the following non-graded representation:
\begin{equation}
\begin{aligned}
A_1 \otimes A_2 &\subset A \subset \overline{A_1 \otimes A_2} \\
\mathcal{K} &= \mathcal{K}_1 \otimes \mathcal{K}_2 \\
\Pi(D) &= D_1 \otimes 1 + \chi_1 \otimes D_2 \\
\Pi \circ \pi &= \pi_1 \otimes \pi_2 \\
\Pi(J) &= J_1 \otimes J_2 \chi_2^{|J_1|} \\
(\cdot,\cdot) &= (\cdot,\cdot)_1 \otimes (\cdot,\cdot)_{2 \beta}
\end{aligned}
\label{tensorproductnongraded}
\end{equation}
where $(\cdot, \cdot)_{2 \beta} = (\cdot, \beta \cdot)_2$. For privileged fundamental symmetries, this gives (see proposition \ref{TensorEtaProp}):
\begin{equation*}
\Pi(\eta) = i^{\sigma_1 \sigma_2} \eta_1 \otimes \eta_2 \chi_2^{\sigma_1} = \eta_1 \otimes \beta^{-1} \eta_2,
\end{equation*}
and $\beta^{-1} \eta_2$ is easily seen to be a fundamental symmetry for the altered product $(\cdot,\cdot)_{2 \beta}$.

This definition of the tensor product thus explains the ones that can be found in the literature. For example, in \cite{Vanhecke}, the tensor product for usual spectral triples is built using the following recipe\footnote{We restrict ourselves to the even case, as it is the only case we are dealing with here}:
\begin{equation*}
\begin{aligned}
A &= A_1 \otimes A_2 \\
\mathcal{H} &= \mathcal{H}_1 \otimes \mathcal{H}_2 \\
D &= D_1 \otimes 1 + \chi_1 \otimes D_2 \\
\pi &= \pi_1 \otimes \pi_2 \\
J &= J_1 \otimes \chi_2^{|J_1|} J_2 \\
\langle \cdot,\cdot \rangle &= \langle \cdot,\cdot \rangle_1 \otimes \langle \cdot,\cdot \rangle_2
\end{aligned}
\end{equation*}
This coincides\footnote{Note that there is an additional sign in $J$ that comes from interchanging $\chi_2$ and $J_2$. This sign does not matter, since the charge conjugation operator of a triple is defined up to a phase only.} with \eqref{tensorproductnongraded}, as $\beta=1$ whenever the products $(\cdot,\cdot)_1$ and $(\cdot,\cdot)_2$ are positive definite. See also \cite{DD}.

\chapter{A Case Against The Spectral Action}

In this chapter, we take a quick look at the spectral action, and give a brief explanation of why it might not work in the Lorentzian case. To simplify the discussion, we ignore all gauge degrees of freedom, and use a manifold IST $(A_M, \mathcal{K}_M, \slashed{D}, \chi_M, J_{M+})$ for a compact manifold $M$ of signature $(t,s)$ and dimension $d$, defined with the East Coast convention. The spectral action of this IST is the following functional of the Dirac operator:
\begin{equation}
S[\slashed{D}] = \mathrm{Tr}_{\mathcal{K}_M} f \left(\frac{\slashed{D}^2}{\Lambda^2} \right),
\end{equation}
where $f$ is a cut-off like function that decreases sufficiently fast at infinity (such as a Gaussian), and $\Lambda$ is a cut-off parameter. In the Riemannian case ($t=0$), the functional $S$ admits an asymptotic expansion for large values of $\Lambda$, whose first two terms are:
\begin{equation*}
S[\slashed{D}] = \int_M (f_0 \Lambda^d + f_2 \Lambda^{d-2} R) \sqrt{|g|} d^d x + O(\Lambda^{d-2}),
\end{equation*}
where $R$ is the scalar curvature, and $f_0, f_2$ are real numbers that depend of the function $f$. The spectral action is thus an interesting way to obtain the Einstein-Hilbert action with cosmological constant. It is also known to generate the action of a gauge theory for the appropriate spectral triple.

The key issue here is the convergence of the spectral action. One can already see in the Riemannian case that the action might diverge: the cosmological constant term is proportional to the volume $V(M) = \int_M \sqrt{|g|} d^d x$ of the manifold. The convergence of the spectral action thus requires the manifold to be of finite volume (or even compact, as was assumed above)\footnote{Another less physical and less covariant, but quite useful solution is to insert a cut-off function $h$ on the manifold $M$ in the spectral action: $S' = \mathrm{Tr} (h \cdot f(\slashed{D}^2 / \Lambda^2)) $, see \cite{Vass}}. But we will see that in the general Pseudo-Riemannian case, a finite volume is not sufficient for the convergence of the spectral action. For this, we turn our attention to momentum space for a semi-quantitative argument, and then study a specific example for a more quantitative treatment.

\section{Position Space vs. Momentum Space}

To simplify the discussion, we will focus on the second order component of the squared Dirac operator:
\begin{equation*}
\slashed{D}^2 = g^{\mu \nu} \partial_\mu \partial_\nu + \text{1\textsuperscript{st} order terms}
\end{equation*}
on some given trivialization of the spinor bundle. This component does not depend on spinorial degrees of freedom. We will thus simply be working with scalar functions. The second simplification we will make is to assume that the manifold is flat, but not necessarily of the form $\mathbb{R}^{t,s}$, as it could have a different topology. We will, in fact, consider such a manifold in the next section. The flatness of the manifold means that the spectral action will only give a cosmological term, should the action be well-defined.

We thus consider the Hilbert space $\mathcal{H} = \overline{\mathcal{C}_c(M,\mathbb{C})}$, obtained by completing compactly supported functions with respect to the inner product:
\begin{equation*}
<u,v> = \int_{M} \overline{u} v.
\end{equation*}
On this space, we define the symmetric operator:
\begin{equation}
\Delta = - g^{\mu \nu} \partial_\mu \partial_\nu
\end{equation}
here called generalized Laplacian, where $g^{\mu \nu}$ is a flat metric of signature $(t,s)$:
\begin{equation}
g^{\mu \nu} = \begin{pmatrix} -I_t & 0 \\ 0 & I_s \end{pmatrix}.
\end{equation}
Note that the purpose of the minus sign in the definition of $\Delta$ is to make it a positive operator when $g$ is Riemannian ($t=0$).

The simplified spectral action we consider is:
\begin{equation}
S[\Delta] = \mathrm{Tr}_{\mathcal{H}} f \left(-\frac{\Delta}{\Lambda^2} \right).
\end{equation}
To give an estimate of this trace, we write it as a sum over the (real) eigenvalues of the Laplacian:
\begin{equation*}
S[\Delta] = \sum_{\lambda \in \mathrm{Sp}(\Delta)} n(\lambda) f \left(-\frac{\lambda}{\Lambda^2} \right),
\end{equation*}
where $n(\lambda)$ is the multiplicity of the eigenvalue $\lambda$. For most flat manifolds, one can construct "eigenfunctions" of the form $u(x) = C e^{ikx}$, with eigenvalues $\lambda = k^2$. The vector $k$ is usually constrained by the topology of the manifold $M$. We denote $K \subset \mathbb{R}^{t,s}$ the momentum space. That is, the set of all allowed vectors $k \in \mathbb{R}^{t,s}$. We thus have:
\begin{equation}
S[\Delta] = \sum_{k \in K} f \left(-\frac{k^2}{\Lambda^2} \right)
\end{equation}
(each vector $k$ determines at most one linearly independent eigenfunction). When the distance between points of $K$ is small enough (or vanishing) it can be useful to rewrite this sum as an integral: 
\begin{equation}
S[\Delta] \approx \int_\mathbb{R} \rho(m) f \left(\frac{m}{\Lambda^2} \right) dm.
\end{equation}
Here $\rho(m)$ is the density of the quantity $m = -k^2$ on the real line. That is, for a $\delta m$ of an appropriate intermediary scale:
\begin{equation}
\rho(m) \delta  m \approx \# \{ k \in K \text{ such that } m \leq -k^2 < m + \delta m \}. 
\end{equation}

If $f$ is an integrable cut-off function, then it is enough for the convergence of $S$ that $\rho$ be finite at all points. This is typically not the case when the momentum space $K$ is continuous. such as for the manifold $\mathbb{R}^{t,s}$, for which $K =\mathbb{R}^{t,s}$. The convergence of $S$ thus requires the discreteness of the momentum space. This is usually the result of boundary conditions on the eigenfunctions, resulting from the compactness of the position space $M$. This is coherent with the requirement that the volume of $M$ be finite.

\begin{figure}[!h]

\centering
\subfloat[Riemannian or anti-Riemannian momentum space]{\label{MomentumSpaceRiemannian}{\includegraphics[width=0.45\textwidth]{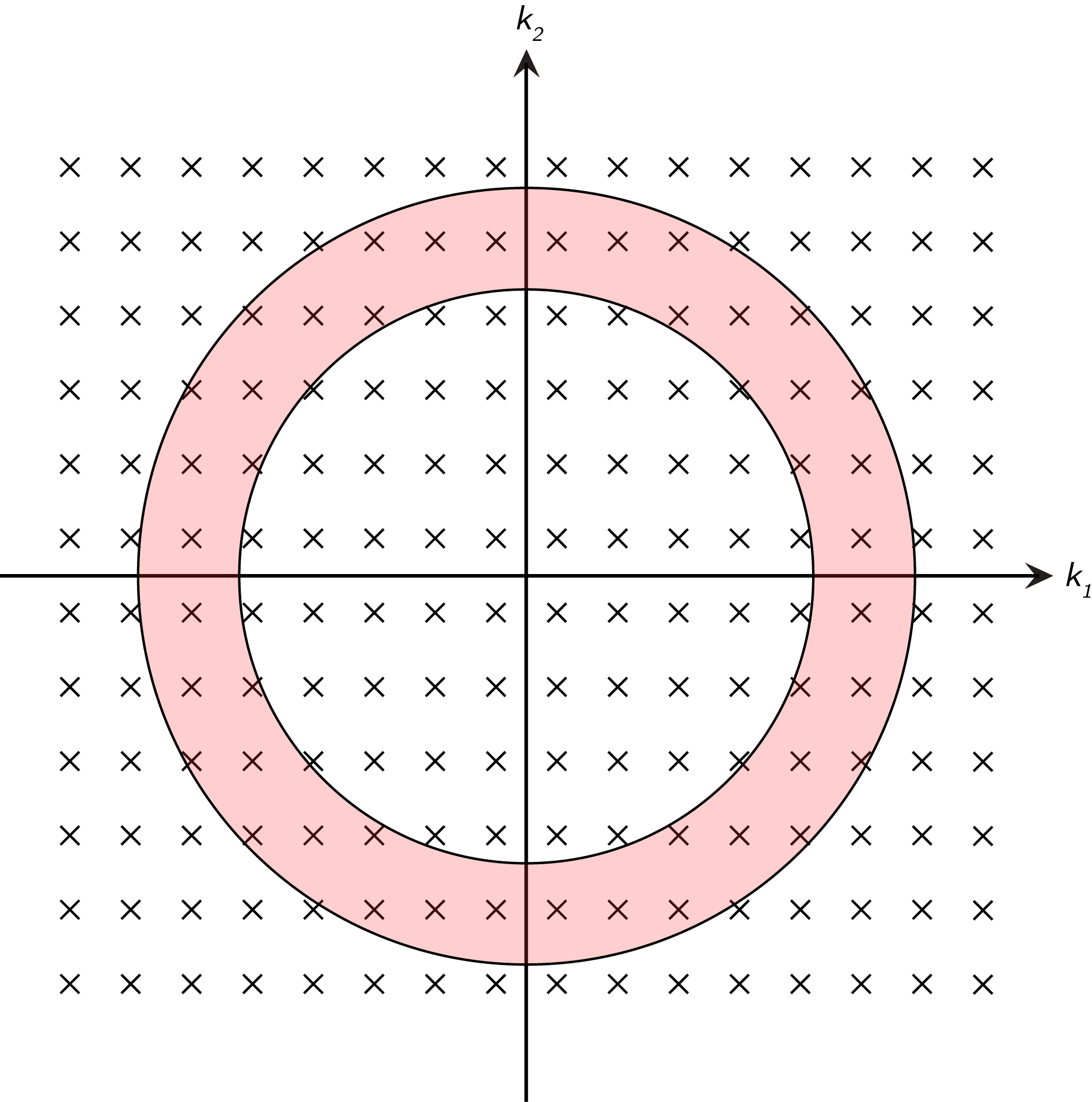}}} \hfill
\subfloat[Lorentzian momentum space]{\label{MomentumSpaceLorentzian}{\includegraphics[width=0.45\textwidth]{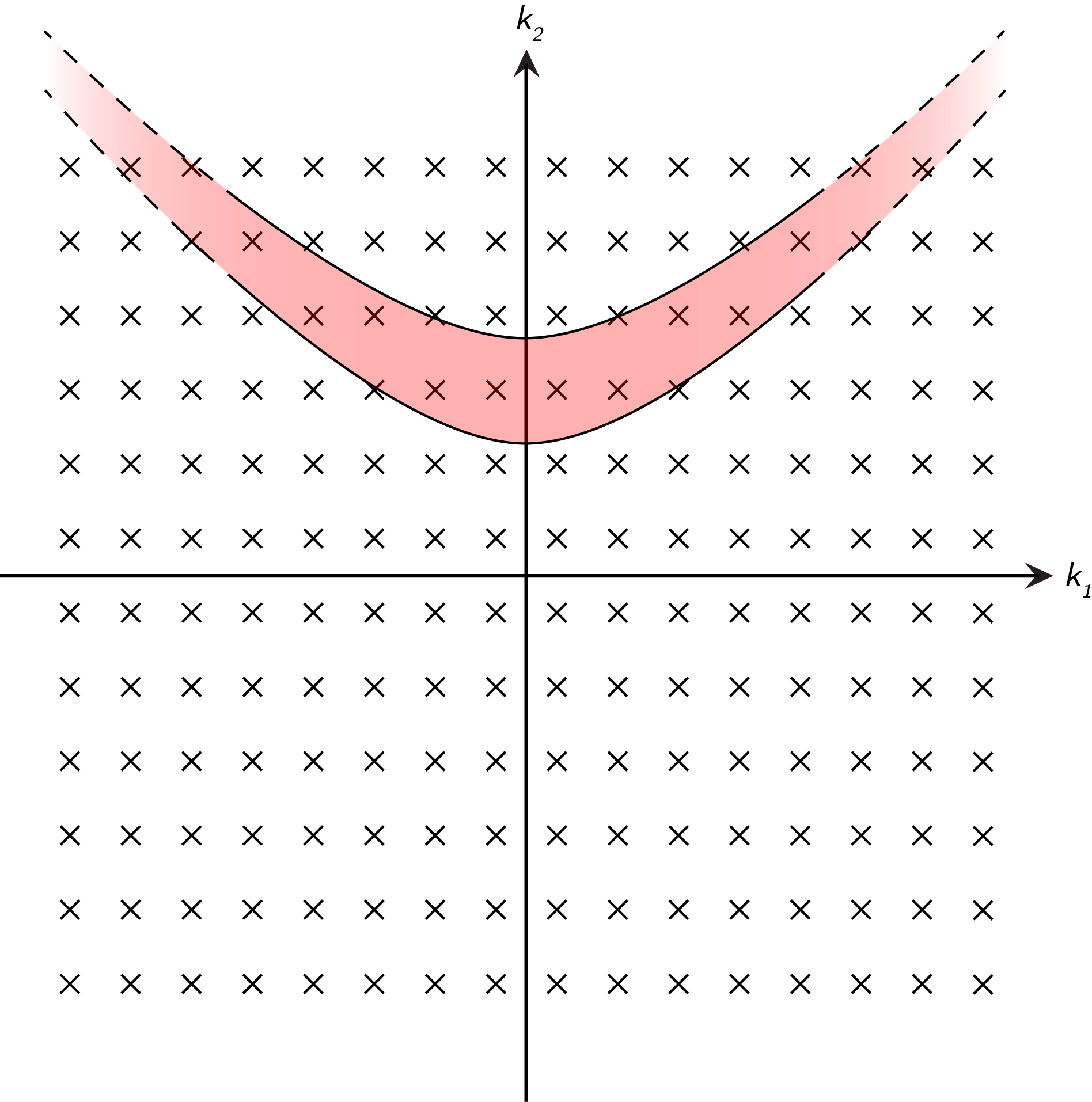}}}

\caption{Momentum space for a 2-dimensional manifold with (\ref{MomentumSpaceRiemannian}) Riemannian or anti-Riemannian signature and (\ref{MomentumSpaceLorentzian}) Lorentzian signature respectively. The crosses are the points of the momentum space $K$ inside the vector space $\mathbb{R}^{t,s}$. The curves are defined by their equations $k^2 = -m$ and $k^2 = -m+\delta m$. The red region contains all points of $K$ such that $m \leq -k^2 < m + \delta m$. The number of those points is thus approximately equal to $\rho(m) \delta m$.}
\label{MomentumSpace}
\end{figure}

But this is still not enough. Indeed, $\rho$ can still be infinite even if $K$ is discrete. To understand why, let us take a look at figure \ref{MomentumSpace} that depicts momentum space for a 2-dimensional manifold.

Let us assume that $K$ is homogeneously distributed in $\mathbb{R}^{t,s}$ (this is the case for a flat torus, see next section). We see that $\rho(m) \delta m$ is approximately proportional to the area between the two curves of equations $k^2 = -m$ and $k^2 = -m+\delta m$. It is easy to prove that the distance between the two neighboring curves is proportional to $\delta m$ at first order. The density $\rho(m)$ is thus the ratio of the area between the curves to their distance. It is thus proportional to the length of the curves. We thus conclude that $\rho(m)$ is finite for the Riemannian and anti-Riemannian signatures, and infinite for the Lorentzian one! To solve this issue, the simplest solution is prevent $K$ from being homogeneously distributed. At worst, the density of points of $K$ must decrease sufficiently fast at infinity. At best, $K$ is finite, and thus bounded. The last option is the most natural, and easiest to realize. To impose a cut-off on momentum space, it is enough to assume that the manifold is actually a continuous approximation of a discrete space.

To summarize, we need momentum space to be discrete and bounded respectively. The simplest way to achieve that is to require that position space be itself bounded and discrete respectively. In the next section, we explore the consequences of these requirements on the spectral action through a simple example.

\section{A Simple Example: Flat, Discrete Tori}

The simplest finite, discrete manifold we could use is the discrete torus. Since a torus is nothing but the product of circles, we start with a discrete circle. We start with a circle of size (\emph{i.e.} perimeter) $L$. We divide it in $N$ intervals, each of length $a = \frac{L}{N}$. The result is the set of points $\mathbb{Z}/N\mathbb{Z}$. The space of scalar functions on this discrete circle is isomorphic to the vector space $\mathbb{C}^N$. We equip this circle with a discrete second derivative $D^2$, whose action on the space of scalar functions is given by the matrix:
\begin{equation}
D^2 = \frac{C-2}{a^2}
\label{D2Def}
\end{equation}
where $C$ is the adjacency matrix of the circle. That is, the matrix that takes the value 1 between neighboring points, and 0 elsewhere:
\begin{equation}
C_{ij} = \begin{cases}
1 &\text{ if } i \equiv j \pm 1 \modulo N \\
0 &\text{ otherwise.}
\end{cases}
\end{equation}

We now construct the $d$-dimensional flat torus $M$ we seek. It is the product of $d$ discrete circles: $M = (\mathbb{Z}/N\mathbb{Z})^d$. The space of scalar functions on $M$ is isomorphic to $(\mathbb{C}^N)^{\otimes d}$. On this space, we construct a discrete, self-adjoint, flat Laplacian of signature $(t,s)$:
\begin{equation}
\begin{aligned}
\Delta_N =& \left( D^2 \otimes 1^{\otimes (d-1)} + \dots + 1^{\otimes (t-1)} \otimes D^2 \otimes 1^{\otimes s} \right) \\ & \;\; - \left( 1^{\otimes t} \otimes D^2 \otimes 1^{\otimes (s-1)} + \dots + 1^{\otimes (d-1)} \otimes D^2 \right) .
\end{aligned}
\end{equation}
It will also be convenient to define the corresponding Euclidean Laplacian:
\begin{equation}
\Delta_{NE} = -\left( D^2 \otimes 1^{\otimes (d-1)} + \dots + 1^{\otimes (d-1)} \otimes D^2 \right),
\end{equation}
which would correspond to the continuous Laplacian:
\begin{equation*}
\Delta_E = - \delta^{\mu \nu} \partial_\mu \partial_\nu
\end{equation*}
We will see that the spectral action of $\Delta_N$ can expressed using that of $\Delta_{NE}$, which as we saw can be made convergent very easily in the limit of vanishing $a$.

As a first step towards the computation of the spectral action of $\Delta_N$, we compute the quantity $\mathrm{Tr}(e^{\theta \Delta_N})$, with $\theta \in \mathbb{C}$. Since $\Delta_N$ is the sum of terms that all commute with each other, we can factorize the exponential:
\begin{equation*}
\begin{aligned}
e^{\theta \Delta_N} =& e^{\theta [\left( D^2 \otimes 1^{\otimes (d-1)} + \dots + 1^{\otimes (t-1)} \otimes D^2 \otimes 1^{\otimes s} \right) - \left( 1^{\otimes t} \otimes D^2 \otimes 1^{\otimes (s-1)} + \dots + 1^{\otimes (d-1)} \otimes D^2 \right)]} \\
=& e^{\theta (D^2 \otimes 1^{\otimes (d-1)})} \otimes \dots \otimes e^{-\theta (1^{\otimes (d-1)} \otimes D^2)}.
\end{aligned}
\end{equation*}
Through a simple series expansion of each exponential, one can prove that:
\begin{equation*}
e^{\theta (D^2 \otimes 1^{\otimes (d-1)})} = e^{\theta D^2} \otimes 1^{\otimes (d-1)},
\end{equation*}
and similarly for the other factors. As a result, we have:
\begin{equation*}
e^{\theta \Delta_N} = (e^{\theta D^2})^{\otimes t} \otimes (e^{-\theta D^2})^{\otimes s}.
\end{equation*}
The trace of this exponential is thus equal to:
\begin{equation}
\mathrm{Tr}(e^{\theta \Delta_N}) = \mathrm{Tr}(e^{\theta D^2})^t \mathrm{Tr}(e^{-\theta D^2})^s.
\end{equation}
one finds, through the substitution $(t,s) = (0,d)$, a similar result for $\Delta_{NE}$:
\begin{equation}
\mathrm{Tr}(e^{\theta \Delta_{NE}}) = \mathrm{Tr}(e^{-\theta D^2})^d.
\end{equation}
The ratio of the two is thus equal to:
\begin{equation*}
\frac{\mathrm{Tr}(e^{\theta \Delta_N})}{\mathrm{Tr}(e^{\theta \Delta_{NE}})} = \left( \frac{\mathrm{Tr}(e^{\theta D^2})}{\mathrm{Tr}(e^{-\theta D^2})} \right)^t.
\end{equation*}
We will now see that this ratio has a simple expression. Indeed, using equation \eqref{D2Def}, we see that:
\begin{equation*}
\mathrm{Tr}(e^{\theta D^2}) = e^{-2\theta / a^2} \mathrm{Tr}(e^{\theta C / a^2}).
\end{equation*}
Substituting above gives:
\begin{equation*}
\frac{\mathrm{Tr}(e^{\theta \Delta_N})}{\mathrm{Tr}(e^{\theta \Delta_{NE}})} = e^{-4t\theta / a^2} \left( \frac{\mathrm{Tr}(e^{\theta C / a^2})}{\mathrm{Tr}(e^{-\theta C / a^2})} \right)^s.
\end{equation*}
Notice that traces of odd powers of $C$ always vanish\footnote{This is easily proved using the fact $C$ is the adjacency matrix of the discrete circle: it shifts an "observer" from one point of the circle to one of its neighbors. The matrix $C^k$ thus takes one from one point to points that are $k$ jumps away from the starting point. The diagonal elements of this matrix are thus nonzero only if one can return to the starting point after $k$ jumps. One can easily convince oneself that the number $k$ of jumps must be even.}: $\mathrm{Tr}(C^{2k+1}) = 0$. A simple series expansion of $e^{\theta C / a^2}$ then proves that it must be even in $\theta$, and so must its trace be. We thus find that:
\begin{equation}
\frac{\mathrm{Tr}(e^{\theta \Delta_N})}{\mathrm{Tr}(e^{\theta \Delta_{NE}})} = e^{-4t\theta / a^2}.
\end{equation}
We thus have the relation:
\begin{equation}
\mathrm{Tr}(e^{\theta \Delta_N}) = e^{-4t\theta / a^2} \mathrm{Tr}(e^{\theta \Delta_{NE}}) = \mathrm{Tr}(e^{\theta (\Delta_{NE} - 4t/a^2)})
\end{equation}
For $\theta$ real and negative, the trace $\mathrm{Tr}(e^{\theta \Delta_{NE}})$ is a heat kernel \cite{Vass}, and is known to converge to:
\begin{equation*}
\lim_{N \rightarrow \infty} \mathrm{Tr}(e^{\theta \Delta_{NE}}) \approx \left( \frac{L}{\sqrt{4 \pi |\theta|}} \right)^d.
\end{equation*}
The trace $\mathrm{Tr}(e^{\theta \Delta_N})$ thus diverges as $e^{4 t |\theta| / a^2}$ as $a$ goes to zero.

Let us now evaluate a general spectral action defined with a cut-off function $f$:
\begin{equation}
S[\Delta_N] = \mathrm{Tr} f \left(-\frac{\Delta_N}{\Lambda^2} \right).
\end{equation}
Since $f$ is a cut-off function, it is natural to assume that it is square-integrable, in which case, it admits an integral Fourier representation:
\begin{equation*}
f(u) = \int_{\mathbb{R}} h(k) e^{iku} dk,
\end{equation*}
with $h$ the Fourier transform of $f$. We substitute in the spectral action to find:
\begin{equation*}
\begin{aligned}
S[\Delta_N] =& \mathrm{Tr} f \left(-\frac{\Delta_N}{\Lambda^2} \right) \\
=& \int_{\mathbb{R}} h(k) \mathrm{Tr}(e^{-i \frac{k}{\Lambda^2}\Delta_N}) dk \\
=& \int_{\mathbb{R}} h(k) \mathrm{Tr}(e^{-i \frac{k}{\Lambda^2}(\Delta_{NE} - 4t/a^2)}) dk \\
S[\Delta_N] =& \mathrm{Tr} f \left(-\frac{\Delta_{NE} - \frac{4t}{a^2}}{\Lambda^2} \right) = S[\Delta_{NE} - 4t/a^2].
\end{aligned}
\end{equation*}
We now use the approximation of the spectral action as an integral given in the previous section:
\begin{equation*}
S[\Delta_{NE}] \approx \int_\mathbb{R} \rho_{NE}(m) f \left(\frac{m}{\Lambda^2} \right) dm,
\end{equation*}
where $\rho_{NE}$ is the density of the spectrum of $\Delta_{NE}$. Substituting for $f$ with the function $u \mapsto f(u + 4t/(a\Lambda)^2)$ gives:
\begin{equation*}
S[\Delta_N] = S[\Delta_{NE} - 4t/a^2] \approx \int_\mathbb{R} \rho_{NE}(m) f \left(\frac{m + \frac{4t}{a^2}}{\Lambda^2} \right) dm.
\end{equation*}
A change of integration variable then gives us:
\begin{equation*}
S[\Delta_N] \approx \int_\mathbb{R} \Lambda^2 \rho_{NE} \left(\Lambda^2 u - \frac{4t}{a^2} \right) f(u) du.
\end{equation*}
In the limit of vanishing $a$, the spectrum density $\rho_{NE}$ converges towards $\rho_E$, the spectrum density of the continuous Euclidean Laplacian. From the estimations of the previous section, we know that $\rho_E(m)$ must vanish for positive $m$. For negative $m$, it must be proportional to the volume of the sphere of radius $|m|^{1/2}$, times the density of the points of momentum space:
\begin{equation*}
\rho_E(m) \approx A_d L^d |m|^{(d-1)/2},
\end{equation*}
where $A_d$ is a constant of order unity that depends only on $d$. We thus have that:
\begin{equation*}
S[\Delta_N] \approx A_d L^d \Lambda^2 \int_\infty^{4t/(a\Lambda)^2} \left| \Lambda^2 u - \frac{4t}{a^2} \right|^{(d-1)/2} f(u) du.
\end{equation*}
If we assume that $f$ is indeed a cut-off function, then the main contribution to the integral will come from small values of $u$. We deduce that:
\begin{equation*}
S[\Delta_N] \approx A_d \Lambda^2 (4t)^{(d-1)/2} \frac{L^d}{a^{d-1}} \left( \int_\mathbb{R} f \right). 
\end{equation*}
This is of course still divergent, even if less so. It is interesting that the speed of divergence is the same for all nonzero values of $t$:
\begin{equation*}
S[\Delta_N] \propto \frac{1}{a^{d-1}}.
\end{equation*}

We now see after regularization that the divergence of the spectral action is in a divergent factor, and not a divergent summand. Moreover, this factor is proportional to the volume of the \emph{flat} manifold, and thus corresponds to a cosmological constant-like contribution, which is an important part of the expansion of the spectral action. It is also a contribution that highly depends on the metric. From all of this, we are tempted to deduce (or rather, conjecture) that a redefinition of the spectral action by simply subtracting its value for a given background geometry might not actually eliminate its divergence. 

Even if one assumes the cut-off scale $a$ to be physically real, its extremely small value might give an unexpectedly large or small value to the different parameters of the action. For all these reasons, we will be looking at a different approach in the next chapters.

\chapter{Noncommutative Gauge Theories}

Before the spectral action, there already was a noncommutative gauge theory: the Connes-Lott model \cite{CL}. This model relies on noncommutative differential forms to generalize the usual bosonic action of non-abelian gauge theories. We will thus start this chapter with a presentation of noncommutative differential forms for ISTs. They generalize in a straightforward manner from definite triples to ISTs, and the discussion will thus be succint (see also \cite{Strohmaier}). Next, we present noncommutative gauge theory; once again, the generalization to ISTs is straightforward. The fermionic action we use has already appeared in the litterature \cite{VdD}. Finally, we construct a bosonic action for this noncommutative gauge theory. A few subtleties relative to the indefinite signature of the Krein space will appear. Luckily, they can be taken care of, following the prescriptions of Elsner (see \cite{Elsner, ENU}). We thus call this model the Connes-Lott-Elsner (CLE) model.

\section{Noncommutative Differential Forms}

Let $(A,\mathcal{K},D,\chi,J)$ be an IST. The construction of the associated algebra of differential forms, or Differential Graded Algebra (DGA), goes through two steps:

\subsection{Step 1: Universal Differential Forms}

Let $A$ be an algebra. The universal DGA $\Omega_U (A) = \bigoplus_{n \in \mathbb{N}} \Omega_U^n (A)$ associated to $A$ is constructed as follows \cite{ConnesRed, Masson}: its order zero component $\Omega_U^0 (A)$ is simply the algebra $A$ itself:
\begin{equation}
\Omega_U^0 (A) = A.
\end{equation}
Its order 1 component is generated by the elements of $A$ and their differentials:
\begin{equation}
\Omega_U^1 (A) = \mathrm{Span} \left\{ \sum_i a_i d_U b_i | a_i, b_i \in A \right\},
\end{equation}
where the sum is finite. All differentials $d_U a$ are assumed to be linearly independent, except differentials of elements proportional to unity, should the algebra be unital. The universal differential is assumed to obey the Leibniz rule:
\begin{equation}
d_U (ab) = (d_U a) b + a (d_U b),
\end{equation}
allowing us to define a right action of $A$ on $\Omega_U^1 (A)$ (the left action is obviously defined):
\begin{equation}
(d_U a) b = d_U (ab) - a (d_U b).
\end{equation}
We similarly define the $n$-th order component of $\Omega_U (A)$ as generated by $n \geq 2$ differentials:
\begin{equation}
\Omega_U^n (A) = \mathrm{Span} \left( \sum_i a_i d_U b_i^1 \dots d_U b_i^n | a_i, b_i^1, \dots, b_i^n \in A \right).
\end{equation}
One can also define it as the following tensor product:
\begin{equation*}
\Omega_U^n (A) = \Omega_U^1 (A)^{\otimes_A n},
\end{equation*}
using the Leibniz rule. Using this same rule, we can define the product $\Omega_U^p (A) \times \Omega_U^q (A) \rightarrow \Omega_U^{p+q} (A)$ of universal differential forms as follows:
\begin{equation}
\begin{aligned}
(a_0 d_U a_1 \dots d_U a_p)&(b_0 d_U b_1 \dots d_U b_p) = (-1)^p a_0 b_0 d_U a_1 \dots d_U a_p d_U b_1 \dots d_U b_p \\ &+ \sum_{i=1}^p (-1)^{p-i} a_0 d_U a_1 \dots d_U a_{i-1} d_U(a_i b_0) d_U a_{i+2} \dots d_U a_p d_U b_1 \dots d_U b_p.
\end{aligned}
\end{equation}
We can now define the universal differential at all orders:
\begin{equation}
\begin{aligned}
d_U : \Omega_U^p (A) &\longrightarrow \Omega_U^{p+1} (A) \\
a_0 d_U a_1 \dots d_U a_p &\longmapsto d_U a_0 d_U a_1 \dots d_U a_p.
\end{aligned}
\end{equation}
One can prove that $d_U$ is indeed a differential. That is, it is nilpotent: $d_U^2 = 0$, and satisfies the graded Leibniz rule:
\begin{equation*}
d_U (\omega \theta) = (d_U \omega) \theta +(-1)^p \omega (d_U \theta)
\end{equation*}
for $\omega \in \Omega_U^p (A)$ and $\theta \in \Omega_U^q (A)$. As a result, the graded algebra $(\Omega_U (A), d_U)$ is a DGA. Finally, if $A$ is involutive, then $\Omega_U (A)$ can be equipped with the involution:
\begin{equation}
\begin{aligned}
(d_U a)^\ast &= - d_U (a^\ast) \\
(a_0 d_U a_1 \dots d_U a_p)^\ast &= (d_U a_p)^\ast \dots (d_U a_1)^\ast a_0^\ast = (-1)^p d_U (a_p^\ast) \dots d_U (a_1^\ast) a_0^\ast.
\end{aligned}
\end{equation}

The importance of universal DGAs lies in the following theorem (see \cite{Masson}):
\begin{thm}
Let $(\Omega, d)$ be a DGA generated by the element of $\Omega^0$ and their differentials $da \in \Omega^1$ (for any $a \in \Omega^0)$. Then there exists a differential graded ideal $J$ of $\Omega_U (\Omega^0)$ such that $(\Omega, d)$ is the quotient of $\Omega_U (\Omega^0)$ by $J$:
\begin{equation*}
\Omega \cong \Omega_U (\Omega^0) / J.
\end{equation*}
\end{thm}
We remind the reader that a differential graded two-sided ideal of a DGA is a two sided ideal that is stable by differentiation. The differential can then be passed on to the quotient, making it a DGA itself. Note also another important property of universal DGAs: their cohomology is trivial. That is, every closed universal form is exact.

\subsection{Step 2: The "Junk"}

We now construct the DGA of noncommutative forms associated to the IST $(A,\mathcal{K},D,\chi,J)$ as a quotient of $\Omega_U (A)$ by a differential graded two-sided ideal $J$ subsequently called the "junk". We will construct this junk using the Dirac operator in way that enables us to recover ordinary differential forms when applied to a manifold IST. This construction generalizes the one given in \cite{CL, ConnesRed}.

We start with the universal DGA $\Omega_U (A)$. We define the following map:
\begin{equation*}
\begin{aligned}
\pi_D: &\; \Omega_U^n (A) &&\longrightarrow \mathrm{End}(\mathcal{K}) \\
&a_0 d_U a_1 \dots d_U a_n &&\longmapsto \pi(a_0) [D, \pi(a_1)] \dots [D, \pi(a_n)].
\end{aligned}
\end{equation*}
Using the fact that $[D, \cdot]$ is a derivation, one can prove that $\pi_D$ is a (non-graded) representation of $\Omega_U (A)$ on $\mathrm{End}(\mathcal{K})$. When $D$ is self-adjoint, one can prove that the representation $\pi_D$ is involutive, since we have:
\begin{equation*}
\begin{aligned}
\pi_D ((d_U a)^\ast) &= -\pi_D( d_U (a^\ast)) \\
&= -[D, \pi(a^\ast)] \\
&= -[D, \pi(a)^\times ] \\
&= [D, \pi(a)]^\times \\
\pi_D ((d_U a)^\ast) &= \pi_D( d_U a)^\times.
\end{aligned}
\end{equation*}

We now construct the "junk" ideal:
\begin{prop}
The kernel $J_0 = \mathrm{ker} \pi_D$ of $\pi_D$ is a graded two-sided ideal of $\Omega_U (A)$, and $J = J_0 +  d_U J_0$ is a graded differential two-sided ideal. Both ideals are stable by involution. The ideal $J$ is called the ideal of junk forms, or junk.
\end{prop}

\begin{demo}
First, notice that $J_0$ is a graded two-sided ideal, since it is the kernel of an algebra homomorphism. Next, we prove that $J$ is a two-sided ideal as well. Let $\omega = \omega_1 + d_U \omega_2 \in J^p$, with $\omega_1 \in \Omega_U^p (A)$ and $\omega_2 \in \Omega_U^{p-1} (A)$ such that $\pi(\omega_1) = \pi(\omega_2) = 0$. Let $\omega' \in \Omega_U (A)$. We have:
\begin{equation*}
\begin{aligned}
\omega \omega' &= \omega_1 \omega' + (d_U \omega_2) \omega' \\
&= \omega_1 \omega' + d_U (\omega_2 \omega') - (-1)^p \omega_2 (d_U \omega')
\end{aligned}
\end{equation*} 
which is an element of $J$, since the first and third term are in $J_0$, while the second term is an element of $d_U J_0$. The ideal $J$ is thus a left-sided ideal. One can similarly prove that it is a right-sided ideal, and thus a two-sided ideal. Finally, it is a differential ideal, since it is stable by differentiation: $d_U J = d_U J_0 \subset J$.

The fact that $J_0$ is stable by involution is simply a consequence of the facts that $\pi_D$ is involutive. So is $J$ since $(d_U J_0)^\ast = d_U (J_0^\ast) = d_U J_0$.

\qed
\end{demo}

We now construct the DGA of noncommutative forms:

\begin{defn}
The differential graded algebra of noncommutative forms associated to the IST $(A,\mathcal{K},D,\chi,J)$ is the quotient:
\begin{equation}
\Omega_D (A) \equiv \Omega_U (A) / J.
\end{equation}
It is equipped with the differential:
\begin{equation*}
d[\omega] = [d_U \omega],
\end{equation*}
and the involution:
\begin{equation*}
[\omega]^\ast = [\omega^\ast].
\end{equation*}

\end{defn}

If the representation $\pi$ is faithful (\emph{i.e.} injective), one can write:
\begin{equation}
\Omega_D^n (A) = \Omega_U^n (A) / J^n \cong \pi_D(\Omega_U^n (A)) / \pi_D(d_U J_0^{n-1}) .
\end{equation}
For physical applications, we are interested in 0-forms, 1-forms, and 2-forms. So let us take a look at these. Because we assumed the representation to be faithful, we have:
\begin{equation}
\Omega_D^0 (A) = A \cong \pi(A).
\end{equation}
We also have $J_0^0 = \{0\}$, so:
\begin{equation}
\Omega_D^1 (A) \cong \pi_D(\Omega_U^1 (A)).
\end{equation}
There are no simplifications for 2-forms, and we simply have:
\begin{equation}
\Omega_D^2 (A) \cong \pi_D(\Omega_U^2 (A)) / \pi_D(d_U J_0^1) .
\end{equation}
We thus need to compute $\pi_D(d_U J_0^1)$ to determine the space of 2-forms. Note that $\pi_D(d_U J_0^1)$ is spanned by all elements of the form:
\begin{equation*}
\sum_i [D,\pi(a_i)] [D, \pi(b_i)]
\end{equation*}
such that $\sum_i \pi(a_i) [D, \pi(b_i)] = 0$, and for these elements, we have:
\begin{equation}
\sum_i [D,\pi(a_i)] [D, \pi(b_i)] = - \sum_i \pi(a_i) [D^2, \pi(b_i)].
\end{equation}
As a consequence of the fact that $J$ is a two sided ideal, the space $\pi_D(d_U J_0^1)$ is a bimodule over $\pi(A)$.

\subsection{An Important Example: The Manifold IST}

As a check of the significance of this construction, one can apply it to the "canonical" case: a manifold IST $(A_M, \mathcal{K}_M, \slashed{D}, \chi_M, J_{M -})$ of a pseudo-Riemannian manifold $(M,g)$, with the West Coast convention, for which the representation $\pi_M$ is faithful. The result and its proof are a straightforward generalization of the Riemannian case \cite{Madore}. For simplicity, we choose $A_M$ to be the algebra of smooth real bounded function $A_M = \mathcal{C}_b^\infty(M, \mathbb{R})$. If $A_M$ is the algebra of complex functions, then everything below must be complexified. We denote the corresponding DGA $\Omega_\slashed{D} (M)$, and its junk $J_M$. In that case, the order zero component of the DGA is:
\begin{equation}
\Omega_\slashed{D}^0 (M) = A_M = \mathcal{C}_b^\infty(M, \mathbb{R}).
\end{equation}
The first order component is:
\begin{equation}
\Omega_\slashed{D}^1 (M) \cong \pi_\slashed{D}(\Omega_U^1 (A_M)) = \mathrm{Span} \left\{ \sum_i (f_i \partial_\mu g_i) i\gamma^\mu | f_i, g_i \in \mathcal{C}_b^\infty(M, \mathbb{R}) \right\}.
\end{equation}
It is almost-dense in, but not necessarily isomorphic to the space of one-forms in the Clifford bundle: $i\gamma(\Gamma(M, T^\ast M))$. If $M$ is a compact manifold covered by a finite number of charts, then one can prove the isomorphism. To determine two-forms, we need to determine $\pi_\slashed{D} (d_U J_{M,0}^1)$. It is spanned by elements of the form:
\begin{equation*}
j = \sum_i [\slashed{D},\pi(f_i)] [\slashed{D}, \pi(g_i)] = -\sum_i \gamma^\mu \gamma^\nu (\partial_\mu f_i) (\partial_\nu g_i),
\end{equation*}
with $f_i, g_i \in \mathcal{C}_b^\infty(M, \mathbb{R})$ such that $\sum_i f_i \partial_\mu g_i = 0$. We thus have:
\begin{equation*}
\begin{aligned}
j &= -\sum_i \gamma^\mu \gamma^\nu (\partial_\mu f_i) (\partial_\nu g_i) \\
&= \sum_i \gamma^\mu \gamma^\nu f_i \partial_\mu \partial_\nu g_i \\
j &= \sum_i g^{\mu\nu} f_i \partial_\mu \partial_\nu g_i.
\end{aligned}
\end{equation*}
The space $\pi_\slashed{D} (d_U J_{M,0}^1)$ thus contains smooth real functions. It is not hard to convince one's self that $\pi_\slashed{D} (d_U J_{M,0}^1)$ is almost dense in the space of real smooth \emph{unbounded} functions, and that it is isomorphic for a compact manifold covered by a finite number of charts. One can then prove that noncommutative two-forms are spanned by elements of the form:
\begin{equation}
\Omega_\slashed{D}^2 (M) \cong \pi_\slashed{D}(\Omega_U^1 (A_M)) = \mathrm{Span} \left( \sum_i [\gamma^\mu, \gamma^\nu] (\partial_\mu f_i) (\partial_\nu g_i) | f_i, g_i \in \mathcal{C}_b^\infty(M, \mathbb{R}) \right),
\end{equation}
and that $\Omega_\slashed{D}^2 (M)$ is almost-dense in the space of two-forms in the Clifford bundle.

\section{Gauge Theory}

We now turn to the construction of noncommutative gauge theories. This chapter will be concerned with the choice of IST necessary to recover gauge theory, and the form this gauge theory takes. The fermionic action (including the coupling of fermions and bosons) will be given in this section, while the construction of the purely bosonic part of the action will be postponed to the next section.

\subsection{Almost-Commutative ISTs}

In noncommutative gauge theory of the Connes-Lott or spectral model type, the necessary IST will be the tensor product of two ISTs:
\begin{enumerate}
	\item The first IST is the manifold IST of a 4D Lorentzian Spin space-time. We will take the IST to satisfy the West Coast convention: $(A_M, \mathcal{K}_M, \slashed{D}, \chi_M, J_{M-})$. This determines the signature of the manifold to be $(+---)$. That is, $(q,p)=(3,1)$. We thus have $n_M = 6$ and $m_M=4$, and $\sigma_M  = 1$. The algebra $A_M$ is taken to be real or complex depending on the second IST. The purpose of this IST is to represent the background geometry on which the gauge theory lives. In the spectral model, it can also be used to recover General Relativity.
	
	Note that the choice of convention of the manifold IST will not matter: the metric and KO-dimensions of the IST only depend on the number of space dimensions. Both East and West Coast conventions give $n_M = 6$ and $m_M=4$, and this is what will matter the most when determining the finite IST.
	
	\item The second IST is a "finite" IST $(A_F, \mathcal{K}_F, D_F, \chi_F, J_F)$, meaning that its Krein space $\mathcal{K}_F$ is finite-dimensional. Unlike the algebra of a manifold IST, the finite algebra $A_F$ is allowed (and even required) to be noncommutative. Its purpose is to represent the internal, or point-wise, structure of a gauge theory. We will discuss the construction of this finite IST in the remainder of this section. 
\end{enumerate}

The total triple we will work with $(A, \mathcal{K}, D, \chi, J)$ is the result of a tensor product (see section \ref{SectionISTTensorProduct}) of the ISTs above in that order. The result, in non-graded form, is the following IST:
\begin{equation}
\begin{aligned}
A &= A_M \otimes A_F \cong \mathcal{C}_b^\infty(M, A_F) \\
\mathcal{K} &= \mathcal{K}_M \otimes \mathcal{K}_F \\
D &= \slashed{D} \otimes 1 + \chi_M \otimes D_F \\
\pi &= \pi_M \otimes \pi_F \\
J &= J_{M-} \otimes J_F \chi_F \\
(\cdot,\cdot) &= (\cdot,\cdot)_M \otimes (\cdot,\cdot)_{F \beta} = (\cdot,\cdot)_M \otimes (\cdot,\beta \cdot)_F \\
\beta &= i^{\sigma_F} \chi_F.
\end{aligned}
\label{AlmostCommutativeIST}
\end{equation}

Before we study the finite IST closely, let us give a few useful results about the DGA of noncommutative forms for the total triple. To simplify without losing too much in generality, we will assume that the manifold can be covered by a finite number of charts (see previous section). One can prove the following results, which are a straightforward generalization of the results that hold for (definite) spectral triples. We find that (see \cite{Martin}):
\begin{equation}
\begin{aligned}
\Omega_D^0 (A) &= A = \mathcal{C}_b^\infty(M, A_F) \\
\Omega_D^1 (A) &= \Omega_\slashed{D}^1 (M) \otimes \pi_F(A_F) \oplus A_M \chi_M \otimes \Omega_{D_F}^1 (A_F).
\end{aligned}
\end{equation}
Thus a general 1-form $\omega \in \Omega_D^1 (A)$ is given by a $\pi_F(A_F)$-valued differential form $B$, and a $\Omega_{D_F}^1 (A_F)$-valued scalar field $H$, and we write:
\begin{equation}
\omega = i\gamma^\mu \otimes B_\mu + \chi_M \otimes H
\end{equation}
(note that this notation is slightly abusive, as the manifold part of $H$ and $B_\mu$ should be on the other side of the tensor product). As for junk two-forms, we have:
\begin{equation}
\pi_D (d_U J_0^1) = \pi_M^2(d_U J_{M,0}^1) \otimes \pi_F(A_F) + A_M \otimes \pi_{D_F}(d_U J_{F,0}^1),
\end{equation}
from which we deduce that:
\begin{equation}
\pi_D (d_U J_0^1) = A_M \otimes ( \pi_F(A_F) + \pi_{D_F}(d_U J_{F,0}^1) ).
\end{equation}

\subsection{The Finite IST and Gauge Theory}

Let us now focus on the finite triple. The finite space $\mathcal{K}_F$ represents the space of fermion species, while the algebra $A_F$ replaces the gauge group, which can be recovered as the unitary group of the algebra (its automorphism group, to be precise). The resulting gauge group is typically a product of unitary and/or orthogonal groups. For a factor such as $SU(2)$ or $SU(3)$, as in the Standard Model, the fermions are necessarily in the fundamental representation, and each fermion is in the representation of one group, while being a singlet for the others. This is an issue for theories such as the Standard Model, where fermions can be in the representation of multiple components of the gauge group. One standard solution is to make the fermion species in $\mathcal{K}_F$ (anti-linearly) dependent: this is the so-called fermion doubling, that will enable us to put a fermion species in the representation of two components of a gauge group\footnote{In the Standard Model, the left-handed quarks are in the representation of all three subgroups. This requires an additional "trick": the unimodularity condition, which is introduced for an entirely different reason.}.

Concretely, the space $\mathcal{K}_F$ is built as the sum of two copies of the space of fermion species. Because these spaces are antilinearly related, they will be on conjugate representations of the gauge group. For this reason, they are abusively called the particle and antiparticle subspaces. For chiral theories (such as the Standard Model), each subspace must be split into left and right particles (resp. antiparticles), resulting into a second fermion doubling. So we have:
\begin{equation}
\mathcal{K}_F = \mathcal{K}_R \oplus \mathcal{K}_L \oplus \mathcal{K}_{\overline{R}} \oplus \mathcal{K}_{\overline{L}}.
\end{equation}
According to the arguments above, the gauge group, and thus the algebra, should act separately on each subspace. For any $a \in A_F$, we can write:
\begin{equation}
\pi_F(a) = \begin{pmatrix} a_R &&& \\ & a_L && \\ && a_{\overline{R}} & \\ &&& a_{\overline{L}} \end{pmatrix}.
\end{equation}
To implement a Weyl condition on fermions and get a chiral theory, we need to define a chirality operator on $\mathcal{K}_F$, and we will ask from it to coincide with the grading $\chi_F$:
\begin{equation}
\chi_F = \begin{pmatrix} 1 &&& \\ & -1 && \\ && -1 & \\ &&& 1 \end{pmatrix}.
\end{equation}
The choice of chirality for antiparticles is related to how charge conjugation inverts the chirality of spinors in Lorentzian 4D signature. Finally, to implement the fermion doubling procedure, we need an antilinear operator that swaps particles and antiparticles, and we will chose it as a real or quaternionic structure $J_F$:
\begin{equation}
J_F = \begin{pmatrix} && \epsilon_F & 0 \\ && 0 & \epsilon_F \\ 1 & 0 && \\ 0 & 1 && \end{pmatrix} \circ CC,
\end{equation}
where $CC$ represents complex conjugation in a basis that we will specify below. A quick computation shows that\footnote{This is related to the fact that $\epsilon''_M = -1$, that is that charge conjugation must invert chirality.} $\epsilon''_F=-1$, which means that the KO dimension $n_F$ of the finite triple is either 2 if $\epsilon_F=-1$ or 6 if $\epsilon_F=+1$. Note that in order to write these matrix representations, we have to choose a basis of the finite Krein space. The basis is usually chosen so that the representation of the algebra $A_F$ is the simplest, \textit{e.g.} an element of a matrix algebra would be represented by multiple copies of itself. Using this (almost) canonical choice of basis, one can define an involution on the algebra $A_F$ in the following way:
\begin{equation}
\pi_F(a^\ast) = \pi_F(a)^\dagger.
\label{AFinvolution}
\end{equation}

This is not always possible, and is thus a constraint on the IST\footnote{Note that if $A_F$ is a $C^\ast$-algebra, such a basis always exists for a given involution.}. An additional justification for such a constraint is that the unitary group of the algebra is compact. As a result, the gauge group is compact as well. Let us say something about the physical meaning of this representation. If $\{p\}$ is the set of fermions of our theory (\textit{e.g.} $\nu, e, $ etc.), then a basis of $\mathcal{K}_R$ is given by a family of vectors that we will denote $(p_R)$. The same goes for the other three subspaces. We choose these four bases to all have the same order. This convention will be useful when choosing a basis for the Standard Model. Let us give an example of the use of this basis. The action of $J_F$ in this basis is given by:
\begin{equation}
\begin{aligned}
J_F p_R &= p_{\overline{R}} \\
J_F p_L &= p_{\overline{L}} \\
J_F p_{\overline{R}} &= \epsilon_F p_R \\
J_F p_{\overline{L}} &= \epsilon_F p_L \\
\end{aligned}
\end{equation}
For ease of notation, we will sometimes treat the four subspaces as if they were identical (and not just isomorphic), using the ``canonical'' isomorphisms that map all four vectors $p_R, p_L, p_{\overline{R}}, p_{\overline{L}}$ to the same vector $p$. We will denote $\mathcal{K}_0$ the vector space spanned by the basis $(p)$, and $i_R$ the map from $\mathcal{K}_R$ on $\mathcal{K}_0$. Let us see how this vector space can be used to simplify notations. Let $u = u_R \oplus u_L \oplus u_{{\overline{R}}} \oplus u_{{\overline{L}}}$, and $A$ be an operator on $\mathcal{K}_F$ represented by the matrix:
\begin{equation*}
A = \begin{pmatrix} 0&1&& \\ 0&0&& \\ &&0& \\ &&&0 \end{pmatrix}.
\end{equation*}
Then by the natural identification of the four subspaces and their vectors, one can write that $A u = u_L \oplus 0 \oplus 0 \oplus 0$, instead of writing $A u = i_R^{-1} i_L u_L \oplus 0 \oplus 0 \oplus 0$. Note that this is nothing more than \textit{splitting vectors in four blocks (\textit{i.e.} subvectors) and moving those blocks around}. Similarly, the action of $J_F$ is given by:
\begin{equation*}
J_F u = \epsilon_F CC u_{{\overline{R}}} \oplus \epsilon_F CC u_{{\overline{L}}} \oplus CC u_R \oplus CC u_L,
\end{equation*}
where $CC$ is now complex conjugation on $\mathcal{K}_0$ in the real basis $(p)$.

What remains to be discussed is the indefinite product $(\cdot,\cdot)_F$ on $\mathcal{K}_F$. As for any hermitian form, there exists a self-adjoint matrix $\eta_F$ such that for any $u,v \in \mathcal{K}_F$:
\begin{equation}
(u,v)_F = u^\dagger \eta_F v.
\label{etaF}
\end{equation}
This notation is suggestive: $\eta_F$ will turn out to be a fundamental symmetry, but for the moment it is just a self-adjoint matrix: $\eta_F^\dagger = \eta_F$. The total product on the total IST is:
\begin{equation*}
(\varphi \otimes u, \psi \otimes v) = (\varphi, \psi) u^\dagger \eta_F \beta v.
\end{equation*}
In QFT, the different fermions species, when put into the same space, are considered to be the orthonormal basis vectors of a Hilbert space. The fermionic part of the action is thus usually written using the "trivial" product:
\begin{equation}
(\varphi \otimes u, \psi \otimes v)_0 = (\varphi, \psi)_M u^\dagger v
\end{equation}
which amounts to constructing a spinor product for each fermion species, and then summing over all species. We can thus write:
\begin{equation}
(\Phi , \Psi) = (\Phi, (1 \otimes \varpi) \Psi)_0,
\end{equation}
where $\varpi = \eta_F \beta$ is an effective signature matrix for the finite part. The $(\cdot,\cdot)_0$ will be useful when comparing the action of our theory to the action of a desired gauge theory. One can also define the product $(\cdot, \cdot)_0$ on $\mathcal{K}_M \otimes \mathcal{K}_0$, using the same definition as above. One can then write, for example, that:
\begin{equation*}
(\Phi, (1 \otimes A) \Psi)_0 = (\Phi_R, \Psi_L)_0.
\end{equation*}
This stems from the fact that: $u^\dagger A v = u_R^\dagger v_L$, where $u_R$ and $v_L$ are now vectors in $\mathcal{K}_0$.

There are still two signs to discuss: $\kappa_F$ and $\kappa''_F$ (or equivalently, $\sigma_F$). We must have the relations:
\begin{equation*}
\begin{aligned}
\chi_F^\times &= (-1)^{\sigma_F} \chi_F \\
J_F^\times &= \kappa_F J_F,
\end{aligned}
\end{equation*}
but we also have:
\begin{equation*}
\begin{aligned}
\chi_F^\dagger &= \chi_F \\
J_F^\dagger &= \epsilon_F J_F.
\end{aligned}
\end{equation*}
From this and equation \eqref{etaF}, we deduce that:
\begin{equation*}
\begin{aligned}
\chi_F \eta_F &= (-1)^{\sigma_F} \eta_F \chi_F \\
J_F \eta_F &= (\epsilon_F \kappa_F) \eta_F J_F.
\end{aligned}
\end{equation*}
Finally from equation \eqref{AFinvolution} and the fact that $\pi$ is involutive, we deduce that $\eta_F$ must commute with the representation of the algebra:
\begin{equation*}
\left[\eta_F, \pi_F(A_F) \right] = 0.
\end{equation*}
To summarize, $\eta_F$ has to satisfy the following requirements:
\begin{equation*}
\begin{aligned}
\eta_F^\dagger &= \eta_F \\
\chi_F \eta_F &= (-1)^{\sigma_F} \eta_F \chi_F \\
J_F \eta_F &= (\epsilon_F \kappa_F) \eta_F J_F \\
\left[\eta_F, \pi_F(A_F) \right] &= 0.
\end{aligned}
\end{equation*}

We now anticipate a bit on the next subsection, and use a small part of the fermionic action, and discuss its relation with fermion quadrupling. We will see that their interactions determine $\eta_F$ uniquely, as well as the dimensions $n_F$ and $m_F$. If one ignores bosonic fields and mass terms, then the fermionic action contains only kinetic terms\footnote{The total fermionic action will differ only by a linear addition of the boson fields to the Dirac kinetic term, as in usual gauge theory. The resulting operator, also called the fluctuated Dirac operator, has the same properties as the basic Dirac $D$. Thus, all the results of this section will apply to the total Fermionic action as well.}. 

Let $\Psi \in \mathcal{K}$ be the (classical) fermionic field. Due to the fermion quadrupling, the components of the field are not independent. We will assume $\Psi$ to be of the form \cite{Elsner, ENU, 3B-IST}:
\begin{equation}
\Psi = (1+J)\Psi_0
\label{PsiDef}
\end{equation}
where $\Psi_0 \in \mathcal{K}_M \otimes (\mathcal{K}_R \oplus \mathcal{K}_L)$ satisfies $\chi \Psi_0 = \Psi_0$. In other words, $\Psi_0$ is a collection of right spinors in $\mathcal{K}_R$ and left spinors in $\mathcal{K}_L$. This reduced spinor represents the particle fields. The total spinor $\Psi$ contains then these spinors and their charge conjugates: this is the anti-linear dependency we previously discussed.

The kinetic part of the fermionic action is:
\begin{equation}
S_\mathrm{Kin} = \frac{1}{2}(\Psi, (\slashed{D} \otimes 1) \Psi).
\end{equation}
with $\Psi \in \mathcal{K}$ the (classical) fermionic field. We substitute for $\Psi$ with the form above:
\begin{equation*}
\begin{aligned}
S_\mathrm{Kin} =& \frac{1}{2}(\Psi, (\slashed{D} \otimes 1) \Psi) \\
=& \frac{1}{2}((1+J)\Psi_0, (\slashed{D} \otimes 1) (1+J)\Psi_0) \\
S_\mathrm{Kin} =& \frac{1}{2}(\Psi_0, (\slashed{D} \otimes 1) \Psi_0) + \frac{1}{2}(J\Psi_0, (\slashed{D} \otimes 1) J\Psi_0) \\ &+ \frac{1}{2}(J\Psi_0, (\slashed{D} \otimes 1) \Psi_0) + \frac{1}{2}(\Psi_0, (\slashed{D} \otimes 1) J\Psi_0).
\end{aligned}
\end{equation*}
We will now use the properties of $J$ in order to simplify the dependency of $S_\mathrm{Kin}$ with respect to the matter field $\Psi_0$, in particular the defining relations $J^2 = \epsilon$ and $J^\times = \kappa J$.

But a subtlety needs to be addressed first. Indeed the operator $J$ is antilinear, which means that its adjoint is defined the following way:
\begin{equation*}
(\Psi, J \Phi) = (\Phi, J^\times \Psi).
\end{equation*}
Notice how the order of the two fields $\Psi$ and $\Phi$ is reversed. In QFT, fermionic fields are usually anticommuting, which mean that the relation above would actually be:
\begin{equation*}
(\Psi, J \Phi) = {\color{red}-}(\Phi, J^\times \Psi).
\end{equation*}
In NCG, it is not clear whether the spin $1/2$ field $\Psi$ should be anticommuting. The question remains open as of yet, and one needs to consider both possibilities. For this reason, we introduce the parameter $s = \pm 1$ that takes the value $s=1$ if the field $\Psi$ is a commuting element of $\mathcal{K}$, and $s=-1$ if it is anticommuting. We now have:
\begin{equation*}
(\Psi, J \Phi) = s(\Phi, J^\times \Psi) = \kappa s (\Phi, J \Psi).
\end{equation*}
Note that the two choices $s = \pm 1$ can have distinct physical consequences. It is known in QFT that on 4D manifolds, any Majorana mass term would be trivial if fermions were commuting fields (see for example \cite{Schwartz}, on p.179). Indeed, a Majorana mass term couples a particle to its conjugate, and is thus of the form:
\begin{equation*}
S_\mathrm{Maj} = a (\psi, J_{M-} \psi)_M,
\end{equation*}
with $\psi \in \mathcal{K}_M$. On a 4D manifold, we have $m_M = 4$, from which we deduce that $J_{M-}^\times = -J_{M-}$. We thus have:
\begin{equation*}
S_\mathrm{Maj} = -s a (\psi, J_{M-} \psi)_M = -s S_\mathrm{Maj}.
\end{equation*}
The Majorana mass term vanishes for $s=+1$. Whether or not this holds for our model depends, once again, on when the quantization of the action occurs.

Let us now go back to the kinetic terms. We have:
\begin{equation*}
\begin{aligned}
S_\mathrm{Kin} =& \frac{1}{2}(\Psi_0, (\slashed{D} \otimes 1) \Psi_0) + \frac{1}{2}(J\Psi_0, (\slashed{D} \otimes 1) J\Psi_0) + \frac{1}{2}(J\Psi_0, (\slashed{D} \otimes 1) \Psi_0) \\ &+ \frac{1}{2}(\Psi_0, (\slashed{D} \otimes 1) J\Psi_0) \\
=& \frac{1}{2}(\Psi_0, (\slashed{D} \otimes 1) \Psi_0) + \frac{1}{2}(J\Psi_0, J(\slashed{D} \otimes 1) \Psi_0) + \frac{1}{2}(J\Psi_0, (\slashed{D} \otimes 1) \Psi_0) \\ &+ \frac{1}{2}(\Psi_0, J(\slashed{D} \otimes 1) \Psi_0) \\
=& \frac{1}{2}(\Psi_0, (\slashed{D} \otimes 1) \Psi_0) + \frac{1}{2}\kappa s ((\slashed{D} \otimes 1) \Psi_0, J^2 \Psi_0) + \frac{1}{2}(J\Psi_0, (\slashed{D} \otimes 1) \Psi_0) \\ &+ \frac{1}{2}\kappa s ((\slashed{D} \otimes 1) \Psi_0, J \Psi_0) \\
S_\mathrm{Kin} =& \frac{1}{2}(1+\kappa \epsilon s)(\Psi_0, (\slashed{D} \otimes 1) \Psi_0) + \frac{1}{2}(J\Psi_0, (\slashed{D} \otimes 1) \Psi_0) + \frac{1}{2}\kappa s ((\slashed{D} \otimes 1) \Psi_0, J \Psi_0).
\end{aligned}
\end{equation*}
We require from this action to contain kinetic terms for fermions (\text{i.e.} for $\Psi_0$), and no kinetic terms that couple fermions to their charge conjugates ($\Psi_0$ and $J\Psi_0$). Let us start with the first requirement. This implies that the first term in $S_\mathrm{Kin}$ does not vanish, from which we deduce that $\kappa \epsilon s = 1$. Moreover, we know that $\chi \Psi_0 = \Psi_0$. This means that:
\begin{equation*}
\begin{aligned}
(\Psi_0, (\slashed{D} \otimes 1) \Psi_0) &= (\chi \Psi_0, (\slashed{D} \otimes 1) \chi \Psi_0) \\
&= -(\Psi_0, \chi^\times \chi (\slashed{D} \otimes 1) \Psi_0) \\
&= -(-1)^\sigma (\Psi_0, (\slashed{D} \otimes 1) \Psi_0)\\
(\Psi_0, (\slashed{D} \otimes 1) \Psi_0) &= -(-1)^\sigma (\Psi_0, (\slashed{D} \otimes 1) \Psi_0).
\end{aligned}
\end{equation*}
The kinetic term is nontrivial if and only if $\sigma=1$, and we conclude that $\kappa''=-\epsilon''$. We also deduce that $\sigma_F \cong \sigma - \sigma_M \cong 0$. We now have:
\begin{equation*}
S_\mathrm{Kin} = (\Psi_0, (\slashed{D} \otimes 1) \Psi_0) + \frac{1}{2}(J\Psi_0, (\slashed{D} \otimes 1) \Psi_0) + \frac{1}{2} \epsilon ((\slashed{D} \otimes 1) \Psi_0, J \Psi_0).
\end{equation*}
We will take care of the second and third terms later. To summarize, the kinetic terms are non-trivial if and only if:
\begin{equation}
\begin{aligned}
\kappa &= s \epsilon \\
\kappa'' &= -\epsilon''
\end{aligned}
\end{equation}
There are two possible cases:
\begin{enumerate}
	\item \underline{$s=+1$ :} we have $\kappa = \epsilon$ and $\kappa'' = -\epsilon''$. Using table \ref{KOMetric} one can prove that this is equivalent to $n+m = 6 \modulo 8$ for the total triple. From the fact $n_M + m_M = 2 \modulo 8$, we infer that $n_F + m_F = 4 \modulo 8$. Since $n_F=2$ or $6$, there are only two possible cases: $(n_F, m_F) = (2,2)$ or $(6,6)$, and in both cases, we have:
\begin{equation*}
\begin{aligned}
\kappa_F &= \epsilon_F \\
\kappa''_F &= \epsilon''_F
\end{aligned}
\end{equation*}
	\item \underline{$s=-1$ :} similarly, we find that: $m+n = 2 \modulo 8$. We infer that $n_F + m_F = 0 \modulo 8$, and the two possible cases are $(n_F, m_F) = (2,6)$ or $(6,2)$, and in both cases we have:
\begin{equation*}
\begin{aligned}
\kappa_F &= -\epsilon_F \\
\kappa''_F &= \epsilon''_F
\end{aligned}
\end{equation*}
\end{enumerate}
Both cases can thus be summarized in the following way:
\begin{equation}
\begin{aligned}
\kappa_F &= s\epsilon_F \\
\kappa''_F &= \epsilon''_F
\end{aligned}
\end{equation}
with the second identity being equivalent to $\sigma_F = 0$. We can now determine $\eta_F$. The requirements on $\eta_F$ now take the form:
\begin{equation}
\begin{aligned}
\eta_F^\dagger &= \eta_F \\
\chi_F \eta_F &= \eta_F \chi_F \\
J_F \eta_F &= s \eta_F J_F \\
\left[\eta_F, \pi_F(A_F) \right] &= 0.
\end{aligned}
\end{equation}
From the first three requirements, we find:
\begin{equation*}
\eta_F = \begin{pmatrix}
R & 0 & 0 & T^\dagger \\
0 & L & \epsilon_F s\overline{T} & 0 \\
0 & \epsilon_F s T^T & s\overline{R} & 0 \\
T & 0 & 0 & s\overline{L}
\end{pmatrix},
\end{equation*}
with $R$ and $L$ self-adjoint matrices. In the last section of this paper, we will consider specific choices of finite triples for which we are able to solve the fourth requirement on $\eta_F$. For all considered triples there, the matrix $T$ will have to vanish. Indeed, the $\overline{R}-L$ block of the equation $\left[\eta_F, \pi_F(a) \right] = 0$ gives $T^T a_L - a_{\overline{R}} T^T = 0$. And for all considered triples, it will be possible to choose en element $a$ of the algebra such that $a_L=1$ and $a_{\overline{R}}=0$. The constraint on $T$ then becomes $T^T = 0$. The self-adjoint $R$ and $L$ will also be constrained by the algebra. However, from the assumed block diagonal form of the representation of the algebra, we can see that real scalar matrices are allowed solutions for $R$ and $L$. There are thus solutions to these requirements, but $\eta_F$ is still underdetermined. The effective signature matrix $\varpi$ is now:
\begin{equation*}
\begin{aligned}
\varpi &= \beta \eta_F \\
&= i^{\sigma_F} \chi_F \eta_F \\
&= \chi_F \eta_F \\
\varpi &= \begin{pmatrix} R &&& \\ & -L && \\ && -s\overline{R} & \\ &&& s\overline{L} \end{pmatrix}.
\end{aligned}
\end{equation*}

The last step is to compute the kinetic fermionic action, and deduce $R$ and $L$ from it. The charge conjugation operator is $J = J_{M-} \otimes J_F \chi_F$, and we deduce from this that $\Psi$ takes the form:
\begin{equation*}
\Psi = (1+J)\Psi_0 = \begin{pmatrix}
\Psi_R \\
\Psi_L \\
\Psi_{{\overline{R}}} \\
\Psi_{\overline{L}}
\end{pmatrix}
\end{equation*}
where:
\begin{equation*}
\begin{aligned}
\Psi_{\overline{R}} &= (J_{M-} \otimes CC) \Psi_R \\
\Psi_{\overline{L}} &= -(J_{M-} \otimes CC) \Psi_L \\
\end{aligned}
\end{equation*}
We can now compute the kinetic terms:
\begin{equation*}
\begin{aligned}
S_\mathrm{Kin} =&  \frac{1}{2}(\Psi, (\slashed{D} \otimes 1) \Psi) \\
=& \frac{1}{2}(\Psi, (\slashed{D} \otimes \varpi) \Psi)_0 \\
=& \frac{1}{2}(\Psi_R, (\slashed{D} \otimes R) \Psi_R)_0 - \frac{1}{2}(\Psi_L, (\slashed{D} \otimes L) \Psi_L)_0 \\ & - \frac{s}{2}(\Psi_{\overline{R}}, (\slashed{D} \otimes \overline{R}) \Psi_{\overline{R}})_0 + \frac{s}{2}(\Psi_{\overline{L}}, (\slashed{D} \otimes \overline{L}) \Psi_{\overline{L}})_0 \\
S_\mathrm{Kin} =& \frac{1}{2}(\Psi_R, (\slashed{D} \otimes R) \Psi_R)_0 - \frac{1}{2}(\Psi_L, (\slashed{D} \otimes L) \Psi_L)_0 \\ & - \frac{s}{2}(J_{M-} \Psi_R, (\slashed{D} \otimes \overline{R}) J_{M-} \Psi_R)_0 + \frac{s}{2}(J_{M-} \Psi_L, (\slashed{D} \otimes \overline{L}) J_{M-} \Psi_L)_0.
\end{aligned}
\end{equation*}
For the manifold triple, we have $J_{M-}^\times J_{M-} = -1$, and using this we deduce that:
\begin{equation*}
S_\mathrm{Kin} = (\Psi_R, (\slashed{D} \otimes R) \Psi_R)_0 - (\Psi_L, (\slashed{D} \otimes L) \Psi_L)_0.
\end{equation*}
We know that the action should be:
\begin{equation*}
S_\mathrm{Kin} = (\Psi_R, (\slashed{D} \otimes 1) \Psi_R)_0 + (\Psi_L, (\slashed{D} \otimes 1) \Psi_L)_0.
\end{equation*}
We infer from this that $R=1$ and $L=-1$, and that:
\begin{equation}
\eta_F = \begin{pmatrix} 1 &&& \\ & -1 && \\ && s & \\ &&& -s \end{pmatrix}.
\end{equation}
and
\begin{equation}
\varpi = \begin{pmatrix} 1 &&& \\ & 1 && \\ && -s & \\ &&& -s \end{pmatrix}.
\end{equation}
It is clear that $\eta_F$ is a fundamental symmetry. Note that for $s=-1$ we have $\eta_F = \chi_F$ and $\varpi = 1$, making the finite part effectively Euclidean.

To conclude this section, we apply the axioms of ISTs to the finite Dirac $D_F$ to constrain it; we must have:
\begin{equation*}
\begin{aligned}
D_F^\times &= D_F \\
\left[J_F, D_F\right] &= 0 \\
\{ \chi_F, D_F \} &= 0.
\end{aligned}
\end{equation*}
The allowed finite Dirac operators are of the form:
\begin{equation}
D_F = \begin{pmatrix}
0 & -Y^\dagger & \epsilon_F \overline{M} & 0 \\
Y & 0 & 0 & \epsilon_F \overline{Z} \\
M & 0 & 0 & -Y^T \\
0 & Z & \overline{Y} & 0
\end{pmatrix},
\end{equation}
where $M$ and $Z$ satisfy:
\begin{equation}
\begin{aligned}
M^T &= s \epsilon_F M \\
Z^T &= s \epsilon_F Z.
\end{aligned}
\end{equation}

\subsection{The Fermionic Action and Gauge Transformations}

We now use fermion quadrupling to construct gauge transformations, and derive the full fermionic action from it. The justification for the correct of gauge transformations in NCG and the resulting conditions on the IST are a classic result of definite noncommutative geometry, and generalize to indefinite ISTs straightforwardly. We summarize the most important points here. Let $\Psi \in \mathcal{K}$ be the (classical) fermionic field. The action of a gauge transformation on this "multi-spinor" should be of the form:
\begin{equation}
\Psi \longmapsto U \Psi,
\end{equation}
where $U$ is a (Krein-) unitary operator on $\mathcal{K}$. We pointed out above that the algebra $A$ replaces the gauge group, and that the gauge group should be recovered from the unitary group $U(A) = \mathcal{C}^\infty(M, U(A_F))$ of the algebra. We thus want to be able to associate to each $u \in U(A)$ a Krein unitary operator $U$ whose adjoint action on $U(A)$ (and thus $A$) is the same as that of $u$. Another condition one might want to require is that $U$ commute with $J$ and $\chi$, in order to preserve the special form \eqref{PsiDef} of the field $\Psi$ of fermions\footnote{Another good reason for this requirement is that $U$ is then an \emph{inner} IST isomorphism, that acts on the algebra like an inner automorphism, and preserves everything but the algebra and the Dirac operator.}. This is always possible if the IST satisfies the so-called order zero condition.

We thus now introduce the order zero condition, and assume from now on that our gauge IST satisfies this requirement. Let $T$ be an operator on $\mathcal{K}$. We associate to it an "opposite" operator:
\begin{equation}
T^\circ = J T^\times J^{-1}.
\end{equation}
The operation:
\begin{equation}
T \longmapsto T^\circ = J T^\times J^{-1}
\end{equation}
is a \emph{linear} antiautomorphism of operators. The \emph{order zero condition} is the requirement that:
\begin{equation}
\left[ \pi(a), \pi(b)^\circ \right] = 0,
\label{OrderZero}
\end{equation}
for all $a,b \in A$. We will determine later when it is satisfied for almost-commutative ISTs. When this condition is satisfied, we can construct gauge transformations according to the following rule:
\begin{equation}
u\in U(A) \longmapsto U = \pi(u) \pi(u^\ast)^\circ = \pi(u) J \pi(u) J^{-1}.
\end{equation}
The interpretation of this action of $u$ on $\Psi$ is that it acts linearly, but through both the direct and conjugate representation, in another manifestation of the fermion quadrupling discussed above.

We now turn our attention to the gauge fields. In analogy with usual gauge theory, the bosonic fields take the form of a differential 1-form $\omega \in \Omega_D^1 (A)$. The reality of the gauge fields is here replaced with the self-adjointness of the 1-form: $\omega^\ast = \omega^\times = \omega$. The bosonic 1-form takes the form:
\begin{equation*}
\omega = i\gamma^\mu \otimes B_\mu + \chi_M \otimes H.
\end{equation*}
It is self-adjoint if and only if:
\begin{equation*}
\begin{aligned}
B_\mu^\dagger &= -\varpi B_\mu \varpi \\
H^\dagger &= -\varpi H \varpi.
\end{aligned}
\end{equation*}
We know that $B_\mu$ is $A_F$-valued, and that $\pi_F(A_F)$ commutes with $\omega$. We deduce that:
\begin{equation}
B_\mu^\dagger = -B_\mu.
\end{equation}
The differential form $i B_\mu$ takes value in the self-adjoint part of the algebra $A_F$. This means that $i B_\mu$ is a gauge field for the unitary group of $A_F$. In order to have the unimodular unitary group as a gauge group, as required for the Standard Model, $i B_\mu$ needs to be "traceless", hence the so-called (and unfortunately \emph{ad-hoc}) unimodularity condition (see \cite{VS} for a discussion):
\begin{equation}
\mathrm{tr}_{\mathcal{K}_F}(B_\mu) = 0.
\end{equation}
The equivalent for gauge transformations is the restriction to transformations $u$ that satisfy:
\begin{equation}
\mathrm{det}_{\mathcal{K}_F}(\pi_F (u)) = 1.
\end{equation}
The scalar field $H$ on the other hand will turn out to be a Higgs field.

We can now postulate a fermionic action (see \cite{VdD} among others):
\begin{equation*}
S_f = \frac{1}{2}(\Psi, (D+\Omega) \Psi),
\end{equation*}
where $\Omega$ is a linear function of $\omega$. We know that $\Psi$ transforms as $\Psi \mapsto U \Psi$. Let $\Omega'$ be the transform of $\Omega$ by a gauge transformation. We want to determine the $\Omega'$ that leaves the action (and thus the fermionic action) gauge invariant. The fermionic action transforms to:
\begin{equation*}
\begin{aligned}
S'_f &= \frac{1}{2}(U\Psi, (D+\Omega') U\Psi) \\
&= \frac{1}{2}(\Psi, [U^{-1} (D+\Omega') U] \Psi) \\
S'_f &= \frac{1}{2}(\Psi, [D + (U^{-1}DU-D) + U^{-1}\Omega' U] \Psi) \\
\end{aligned}
\end{equation*}
We infer that:
\begin{equation*}
\Omega =U^{-1}DU-D + U^{-1}\Omega' U ,
\end{equation*}
and thus that:
\begin{equation*}
\Omega' = (UDU^{-1}-D) + U\Omega U^{-1} = U\Omega U^{-1} + [U,D]U^{-1}.
\end{equation*}
The gauge transformation rule for $\Omega$ is thus:
\begin{equation}
\Omega \longmapsto  U\Omega U^{-1} + [U,D]U^{-1}.
\end{equation}

To simplify further this expression, we need to introduce the so-called \emph{first order condition}:
\begin{equation}
\left[[D, \pi(a)], \pi(b)^\circ \right] = 0,
\label{OrderOne}
\end{equation}
for all $a,b \in A$. We now assume this condition holds for our gauge IST, and will determine later when it is satisfied for almost-commutative ISTs. Its purpose will become clear after we perform the simplification of the gauge transformation rule of $\Omega'$. We have:
\begin{equation*}
\begin{aligned}
{[}U,D{]}U^{-1} &= [\pi(u) J \pi(u) J^{-1},D] (\pi(u) J \pi(u) J^{-1})^{-1} \\
&= ([\pi(u),D] J \pi(u) J^{-1} + \pi(u) J [\pi(u),D] J^{-1}) (J \pi(u^\ast) J^{-1} \pi(u^\ast)) \\
&= [\pi(u),D] \pi(u^\ast) + \pi(u) J([\pi(u),D] \pi(u^\ast))J^{-1} \pi(u^\ast) \\
&= [\pi(u),D] \pi(u^\ast) + \pi(u)\pi(u^\ast) J([\pi(u),D] \pi(u^\ast))J^{-1} \text{ (by \eqref{OrderZero} and \eqref{OrderOne})} \\
{[}U,D{]}U^{-1} &= [\pi(u),D] \pi(u^\ast) + J([\pi(u),D] \pi(u^\ast))J^{-1}.
\end{aligned}
\end{equation*}
The first-order condition thus separates the affine part $[U,D]U^{-1}$ of the gauge transformation in two pieces that correspond to the linear and "opposite" representations of the gauge group that are merged in $U = \pi(u) J \pi(u) J^{-1}$, consistently with what one would expect from fermion quadrupling. Note that there are models where this condition is not used (see for example the Noncommutative Pati-Salam Model \cite{CCvS}). Let us remark that: $[\pi(u),D] \pi(u^\ast)$ is a 1-form, so this could be, in principle, part of the gauge transformation of $\omega$. So let us assume that $\omega$ transforms as in usual gauge theory:
\begin{equation}
\omega \longrightarrow  \pi(u) \omega \pi(u)^{-1} + [\pi(u),D] \pi(u^\ast) \equiv u\omega u^{-1} - d(u) u^{-1}.
\end{equation}
Then $\Omega$ has all the desired properties if it is of the form:
\begin{equation}
\Omega = \omega + J \omega J^{-1}.
\end{equation}
Indeed, let $\omega' = \pi(u) \omega \pi(u)^{-1} + [\pi(u),D] \pi(u^\ast)$ be the transform of $\omega$. Then the desired transform of $\Omega$ is:
\begin{equation*}
\begin{aligned}
\Omega' =& U\Omega U^{-1} + [U,D]U^{-1} \\
=& (\pi(u) J \pi(u) J^{-1}) (\omega + J \omega J^{-1})(J \pi(u^\ast) J^{-1} \pi(u^\ast)) + [\pi(u),D] \pi(u^\ast) \\ &+ J([\pi(u),D] \pi(u^\ast))J^{-1} \\
=& \pi(u) \omega \pi(u^\ast) + J \pi(u) \omega \pi(u^\ast) J^{-1} + [\pi(u),D] \pi(u^\ast) + J([\pi(u),D] \pi(u^\ast))J^{-1} \\
\Omega' =& \omega' + J \omega' J^{-1},
\end{aligned}
\end{equation*}
as desired. We conclude that the fermionic action is given by:
\begin{equation}
S_f = \frac{1}{2}(\Psi, (D+\omega + J\omega J^{-1}) \Psi).
\end{equation}

Let us expand this fermionic action in terms of $B_\mu$ and $H$. We substitute $\omega$ and $D$ with their explicit forms. First, he have:
\begin{equation*}
\begin{aligned}
J \omega J^{-1} &= (J_{M-} \otimes J_F \chi_F) (i\gamma^\mu \otimes B_\mu + \chi_M \otimes H) (J_{M-} \otimes J_F \chi_F)^{-1} \\
&= J_{M-} (i\gamma^\mu) J_{M-}^{-1} \otimes (J_F \chi_F) B_\mu (J_F \chi_F)^{-1} + J_{M-} \chi_M J_{M-}^{-1} \otimes (J_F \chi_F) H (J_F \chi_F)^{-1} \\
J \omega J^{-1} &= i\gamma^\mu \otimes J_F B_\mu J_F^{-1} + \chi_M \otimes J_F H J_F^{-1},
\end{aligned}
\end{equation*}
where we used what we know about the commutative triple, and the fact that $B_\mu$ is even while $H$ is odd. We can substitute in the fermionic action, and the result is that the action splits in three terms:
\begin{equation}
S_f = S_\mathrm{Kin} + S_g + S_h.
\end{equation}
The first term is the kinetic term (without mass terms):
\begin{equation*}
S_\mathrm{Kin} = \frac{1}{2}(\Psi, (\slashed{D} \otimes 1) \Psi),
\end{equation*}
and we have already computed it:
\begin{equation*}
S_\mathrm{Kin} = (\Psi_R, (\slashed{D} \otimes 1) \Psi_R)_0 + (\Psi_L, (\slashed{D} \otimes 1) \Psi_L)_0.
\end{equation*}
The second term is a coupling term to the gauge fields:
\begin{equation}
S_g = \frac{1}{2}(\Psi, i\gamma^\mu \otimes (B_\mu + J_F B_\mu J_F^{-1}) \Psi),
\label{FermionGaugeCoupling}
\end{equation}
and the third term is a coupling term to the Higgs (plus the finite Dirac contribution):
\begin{equation}
S_h = \frac{1}{2}(\Psi, \chi_M \otimes (D_F + H + J_F H J_F^{-1}) \Psi).
\label{FermionHiggsCoupling}
\end{equation}

We conclude this section with a discussion of the order zero and one conditions for almost commutative triples. By linearity, the order zero condition holds for the total IST if and only if:
\begin{equation*}
\left[ \pi(f \otimes a), \pi(b \otimes g)^\circ \right] = 0,
\end{equation*}
for all $f,g \in A_M$ and $a,b \in A_F$. This is equivalent to:
\begin{equation*}
\left[ \pi_M(f) \otimes \pi_F(a), (\pi_M(g) \otimes \pi_F(b))^\circ \right] = 0.
\end{equation*}
Using the results of theorems \ref{tensoradjoint} and \ref{TensorCCAdjointThm}, one can prove that $(\pi_M(g) \otimes \pi_F(b))^\circ = \pi_M(g)^\circ \otimes \pi_F(b)^\circ$. For the manifold IST, we have: $\pi_M(g)^\circ = \pi_M(g)$ (and the order zero condition holds automatically). The order zero condition for the total IST now takes the form:
\begin{equation*}
\pi_M(fg) \otimes [\pi_F(a), \pi_F(b)^\circ ] = 0,
\end{equation*}
thus the necessary and sufficient condition:
\begin{equation*}
[\pi_F(a), \pi_F(b)^\circ ] = 0
\end{equation*}
that the order zero condition hold for the finite IST. One can similarly prove that a necessary and sufficient condition for the order one condition to hold for the total IST is that it holds for the finite IST:
\begin{equation}
\left[[D_F, \pi_F (a)], \pi_F (b)^\circ \right] = 0,
\end{equation}
for all $a,b \in A_F$

\section{The Bosonic Action}

We now complete our noncommutative gauge theory by constructing the purely bosonic part $S_b$ of the action. This will be a functional of the gauge 1-form $\omega$, and will contain its kinetic and self-coupling terms.

\subsection{Gauge Fields and Curvature}

The key idea is to formulate a Yang-Mills type Lagrangian, as is done in the Connes-Lott model, but in a Lorentzian setting, as in the work of Elsner \& \emph{al.}. We introduce the curvature form $\rho$ that generalizes gauge curvature:
\begin{equation}
\rho = d\omega + \omega^2 \in \Omega_D^2 (A).
\end{equation}
Let us see how this transforms under a gauge transformation:
\begin{equation*}
\begin{aligned}
\rho' =& d\omega' + \omega'^2 \\
=& d(u\omega u^{-1} - d(u) u^{-1}) + (u\omega u^{-1} - d(u) u^{-1})^2 \\
=& du \omega u^{-1} + u d\omega u^{-1} - u \omega d(u^{-1}) + du d(u^{-1}) \\ &+ u \omega^2 u^{-1} - du \omega u^{-1} - u \omega u^{-1} (du) u^{-1} + (du) u^{-1} (du) u^{-1}.
\end{aligned}
\end{equation*}
Using the identity $d(u^{-1}) = - u^{-1} (du) u^{-1}$, this simplifies to:
\begin{equation*}
\begin{aligned}
\rho' =& du \omega u^{-1} + u d\omega u^{-1} + u \omega u^{-1} (du) u^{-1} - (du) u^{-1} (du) u^{-1} \\ &+ u \omega^2 u^{-1} - du \omega u^{-1} - u \omega u^{-1} (du) u^{-1} + (du) u^{-1} (du) u^{-1} \\
=& u d\omega u^{-1} + u \omega^2 u^{-1} \\
\rho' =& u \rho u^{-1}.
\end{aligned}
\end{equation*}

Let us compute this curvature form. One finds (see \cite{CL, Martin}) that\footnote{Once again, the notation is abusive, in that the dependence on the manifold point $x$ is sometimes on the second part of the tensor product when it should be in the first part.}:
\begin{equation}
\rho = -\frac{1}{4} [\gamma^\mu, \gamma^\nu] \otimes F_{\mu\nu} - i\chi_M \gamma^\mu \otimes D_\mu H + 1 \otimes (H^2 + d_{F,U} H) + j,
\end{equation}
where:
\begin{equation*}
\begin{aligned}
F_{\mu \nu} &= \partial_\mu B_\nu - \partial_\nu B_\mu +[B_\mu, B_\nu] \\
D_\mu H &= \partial_\mu H + [B_\mu, D_F+H], \\
j &\in \pi_D(d_U J_0^1)
\end{aligned}
\end{equation*}
implying that $H$ is in an adjoint representation of the gauge group. Let us prove it. Let us write:
\begin{equation*}
\omega = \sum_i \pi(a_i) [D, \pi(b_i)]
\end{equation*}
with $a_i,b_i \in A$. We find:
\begin{equation*}
\omega = i\gamma^\mu \otimes \pi_F(\sum_i a_i \partial_\mu b_i) + \chi_M \otimes \sum_i \pi_F(a_i) [D_F, \pi_F(b_i)], 
\end{equation*}
from which we deduce that:
\begin{equation*}
\begin{aligned}
B_\mu &= \pi_F(\sum_i a_i \partial_\mu b_i) \\
H &= \sum_i \pi_F(a_i) [D_F, \pi_F(b_i)].
\end{aligned}
\end{equation*}
Now, we have:
\begin{equation*}
\begin{aligned}
d \omega =& \sum_i [D, \pi(a_i)] [D, \pi(b_i)] \\
=& \sum_i \left(i\gamma^\mu \otimes \pi_F(\partial_\mu a_i) + \chi_M \otimes [D_F, \pi_F(a_i)]\right) \left(i\gamma^\nu \otimes \pi_F(\partial_\nu b_i) + \chi_M \otimes [D_F, \pi_F(b_i)]\right) \\
d \omega =& -\gamma^\mu \gamma^\nu \otimes \sum_i \pi_F(\partial_\mu a_i \partial_\nu b_i) + 1 \otimes \sum_i [D_F, \pi_F(a_i)] [D_F, \pi_F(b_i)] \\ &+ i\gamma^\mu \chi_M \otimes \sum_i \left( \pi_F(\partial_\mu a_i)[D_F, \pi_F(b_i)] - [D_F, \pi_F(a_i)] \pi_F(\partial_\mu b_i) \right)
\end{aligned}
\end{equation*}
There are three piece here. The first piece is:
\begin{equation*}
-\gamma^\mu \gamma^\nu \otimes \sum_i \pi_F(\partial_\mu a_i \partial_\nu b_i).
\end{equation*}
Using the junk, one can replace $\gamma^\mu \gamma^\nu$ with the commutator $[\gamma^\mu, \gamma^\nu]/2$, and one finds:
\begin{equation*}
\begin{aligned}
-\gamma^\mu \gamma^\nu \otimes \sum_i \pi_F(\partial_\mu a_i \partial_\nu b_i) &\equiv - \frac{[\gamma^\mu, \gamma^\nu]}{2} \otimes \sum_i \pi_F(\partial_\mu a_i \partial_\nu b_i) + j_1\\
&= -\frac{[\gamma^\mu, \gamma^\nu]}{4} \otimes \sum_i \pi_F(\partial_\mu a_i \partial_\nu b_i - \partial_\nu a_i \partial_\mu b_i) + j_1\\
&= -\frac{[\gamma^\mu, \gamma^\nu]}{4} \otimes (\partial_\mu B_\nu - \partial_\nu B_\mu) + j_1.
\end{aligned}
\end{equation*}
The second piece:
\begin{equation*}
\sum_i [D_F, \pi_F(a_i)] [D_F, \pi_F(b_i)],
\end{equation*}
is simply $d_U H + j_2$. Finally, the third piece is given by the sum:
\begin{equation*}
\begin{aligned}
\sum_i & \left( \pi_F(\partial_\mu a_i)[D_F, \pi_F(b_i)] - [D_F, \pi_F(a_i)] \pi_F(\partial_\mu b_i) \right) \\
&= \sum_i (\pi_F(\partial_\mu a_i)[D_F, \pi_F(b_i)] - [D_F, \pi_F(a_i)\pi_F(\partial_\mu b_i)] + \pi_F(a_i)[D_F, \pi_F(\partial_\mu b_i)]) \\
&= \partial_\mu(\sum_i \pi_F(a_i) [D_F, \pi_F(b_i)]) - [D_F,  \sum_i \pi_F(a_i \partial_\mu b_i)] \\
&= \partial_\mu H - [D_F, B_\mu].
\end{aligned}
\end{equation*}
We thus have, up to the junk:
\begin{equation*}
d \omega = -\frac{[\gamma^\mu, \gamma^\nu]}{4} \otimes (\partial_\mu B_\nu - \partial_\nu B_\mu) + i\gamma^\mu \chi_M \otimes (\partial_\mu H + [B_\mu, D_F]) + 1 \otimes d_U H + j_3.
\end{equation*}
For $\omega^2$, one finds:
\begin{equation*}
\begin{aligned}
\omega^2 =& (i\gamma^\mu \otimes B_\mu + \chi_M \otimes H)^2 \\
=& -\gamma^\mu \gamma^\nu \otimes B_\mu B_\nu + i\gamma^\mu \chi_M \otimes [B_\mu, H] + 1 \otimes H^2.
\end{aligned}
\end{equation*}
We use the junk:
\begin{equation*}
\begin{aligned}
\omega^2 =& -\frac{[\gamma^\mu, \gamma^\nu]}{2} \otimes B_\mu B_\nu + i\gamma^\mu \chi_M \otimes [B_\mu, H] + 1 \otimes H^2 + j_4 \\
=& -\frac{[\gamma^\mu, \gamma^\nu]}{4} \otimes [B_\mu, B_\nu] + i\gamma^\mu \chi_M \otimes [B_\mu, H] + 1 \otimes H^2 + j_4.
\end{aligned}
\end{equation*}
The final result is:
\begin{equation*}
\begin{aligned}
\rho =& -\frac{1}{4} [\gamma^\mu, \gamma^\nu] \otimes (\partial_\mu B_\nu - \partial_\nu B_\mu + [B_\mu, B_\nu]) - i\chi_M \gamma^\mu \otimes (\partial_\mu H + [B_\mu, H+D_F]) \\ &+ 1 \otimes (H^2 + d_{F,U} H) + j,
\end{aligned}
\end{equation*}
as expected.

Remember now that $\pi_D(d_U J_0^1) = A_M \otimes ( \pi_F(A_F) + \pi_{D_F}(d_U J_{F,0}^1) )$, so the quotient is a local quotient, and it only affects scalar functions. The first two terms of $\rho$ will thus not be affected, and only $H^2 + d_{F,U} H$ must be quotiented locally by the vector space:
\begin{equation}
Q_F = \pi_F(A_F) + \pi_F^2(d_U J_{F,0}^1),
\end{equation}
allowing us to write (abusively):
\begin{equation}
\rho = -\frac{1}{4} [\gamma^\mu, \gamma^\nu] \otimes F_{\mu\nu} - i\chi_M \gamma^\mu \otimes D_\mu H + 1 \otimes (H^2 + d_{F,U} H + q_F),
\end{equation}
with $q_F \in \pi_D(d_U J_0^1) = A_M \otimes Q_F$. Note that $Q_F$ is a bimodule over $\pi_F(A_F)$.

\subsection{The Connes-Lott-Elsner Model}

We know construct an action that is both gauge invariant and quadratic in $\rho$. This action must also take the form of an integral of a Lagrangian density. The simplest way to achieve this is to take the action proportional to a trace over the Krein space $\mathcal{K}$. One can thus write:
\begin{equation*}
S_b \equiv -\mathrm{Tr}(\rho^2) \equiv -\int_M  \mathrm{tr}_x (\rho(x)^2) \sqrt{|g|} d^d x,
\end{equation*}
where the first trace is over $\mathcal{K}$, and the second trace is a local trace over the spinor fiber $S_x$ and $\mathcal{K}_F$. There is of course an obvious problem: $\rho(x)$ does not act unambiguously on $\mathcal{K}_x \cong S_x \otimes \mathcal{K}_F$. It is indeed an operator on $\mathcal{K}_x$ up to a quotient by $1 \otimes Q_F$, making the expression $\mathrm{tr} (\rho(x)^2)$ meaningless. We need to choose a representative of the equivalence class of $H^2 + d_{F,U} H$ to put in the above trace. One way to choose it is to pick the unique representative element of the equivalence class that is orthogonal to the space $Q_F$. This is nothing but the projection of $H^2 + d_{F,U} H$ on the orthogonal of $Q_F$.

We still have to specify the product with respect to which we project! The obvious choice is the indefinite product:
\begin{equation}
\begin{aligned}
\mathrm{End}(\mathcal{K}_F) \times \mathrm{End}(\mathcal{K}_F) &\longrightarrow \mathbb{C} \\
S \times T &\longmapsto \mathrm{tr}_x (\varpi S^\dagger \varpi T).
\end{aligned}
\end{equation}
Notice that we use $\varpi S^\dagger \varpi$ instead of $S^\times = \eta_F S^\dagger \eta_F$ because what we ought to project is the entire curvature form $\rho$. But as we explained above, only the scalar term is projected. This hermitian form guarantees gauge invariance, but it also suffers from two drawbacks:
\begin{itemize}
	\item The projection is well-defined if and only if the form is non-degenerate on the space $Q_F$. A sufficient condition for this is that the form be definite on $Q_F$. Fortunately, this will hold for the IST of the Standard Model (see next chapter).
	\item If the algebra $A_F$ is real, then so is the space $Q_F$. The projection is then only well-defined for a real bilinear form, and not a hermitian form. The obvious solution is to extract a bilinear from the hermitian form by taking its real part \cite{Elsner}. We will thus use the following product when the algebra is real:
\begin{equation}
\begin{aligned}
\mathrm{End}(\mathcal{K}_F) \times \mathrm{End}(\mathcal{K}_F) &\longrightarrow \mathbb{C} \\
S \times T &\longmapsto \mathrm{Re}(\mathrm{tr}_x (\varpi S^\dagger \varpi T)).
\end{aligned}
\end{equation}
\end{itemize}

Once a local projection of $P(H^2 + d_{F,U} H)$ has been chosen, we have a representative $P\rho$ of the curvature form:
\begin{equation}
P\rho = -\frac{1}{4} [\gamma^\mu, \gamma^\nu] \otimes F_{\mu\nu} - i\chi_M \gamma^\mu \otimes D_\mu H + 1 \otimes P(H^2 + d_{F,U} H),
\end{equation}
which can then be used to express the bosonic action:
\begin{equation*}
S_b = -\mathrm{Tr}((P\rho)^2) \equiv -\int_M \mathrm{tr}_x ((P\rho(x))^2) \sqrt{|g|} d^d x,
\end{equation*}
For this action to be well-defined, we require $(P\rho)^2$ to be a trace-class operator. The bosonic Lagrangian density is given by:
\begin{equation*}
\mathscr{L}_b = -\mathrm{tr}_x ((P\rho(x))^2).
\end{equation*}

The bosonic action defined above is not the most general one. It is in fact more constrained than one would want it to be, as the couplings of the gauge group will not be independent (see next chapter). A more general action would be of the form:
\begin{equation*}
S_b = -\mathrm{Tr}(Z(P\rho)^2) \equiv -\int_M \mathrm{tr}_x (Z(x) (P\rho(x))^2) \sqrt{|g|} d^d x,
\end{equation*}
with $Z$ an arbitrary \emph{local} operator. Since $P\rho(x)$ is self-adjoint, a sufficient condition for the reality of the action is $Z^\times = Z$. Local Poincaré covariance requires $Z$ to be a constant scalar, that is an operator of the form $Z = 1 \otimes z = 1 \otimes \mathrm{End}(\mathcal{K}_F)$. We also need to ensure local gauge invariance. We know that $\rho$ transforms like $\rho \rightarrow u\rho u^\ast$. Since we are projecting with respect to an $A$-bimodule, the projection $P\rho$ transforms the same way:
\begin{equation*}
P\rho \rightarrow \pi(u) P\rho \pi(u)^{-1}
\end{equation*}
The bosonic Lagrangian density transforms like:
\begin{equation*}
\mathscr{L}_b = -\mathrm{tr}_x (Z(P\rho(x))^2) \rightarrow -\mathrm{tr}_x (Z \pi(u) (P\rho(x))^2 \pi(u)^{-1}) = -\mathrm{tr}_x (\pi(u)^{-1} Z \pi(u) (P\rho(x))^2).
\end{equation*}
The simplest way to ensure gauge invariance is to require that:
\begin{equation*}
z = \pi_F(u)^{-1} z \pi_F(u)
\end{equation*}
for all unitaries $u$ of $A_F$. The infinitesimal version of that requirement is that $z$ must commute with all anti-self-adjoint elements of $\pi_F(A_F)$. For complex algebras $A_F$, a simple multiplication by $i$ implies that $z$ must commute with all self-adjoint, and thus \emph{all} elements of $\pi_F(A_F)$. Things are not so simple for real algebras, but in the case of the Standard Model studied in the next chapter, we will prove that $z$ must commute with all elements of $\pi_F(A_F)$.

This time, the convergence of the action requires $Z(P\rho)^2$ to be a trace-class operator. A sufficient condition for this is that $(P\rho)^2$ be a trace-class operator once again. If $Z$ is invertible, then its inverse is bounded, and $(P\rho)^2$ \emph{must} be a trace-class operator.

Let us expand the bosonic Lagrangian density:
\begin{equation*}
\begin{aligned}
\mathscr{L}_b =& -\mathrm{tr}_x (Z(P\rho(x))^2) \\
&= -\mathrm{tr}_x \left( (1 \otimes z) \left( -\frac{1}{4} [\gamma^\mu, \gamma^\nu] \otimes F_{\mu\nu} - i\chi_M \gamma^\mu \otimes D_\mu H + 1 \otimes P(H^2 + d_{F,U} H) \right)^2 \right).
\end{aligned}
\end{equation*}
There are a number of Clifford traces that need to be evaluated. First, let us note that the following traces are vanishing:
\begin{equation*}
\mathrm{tr}([\gamma^\mu, \gamma^\nu] \chi_M \gamma^\lambda) = \mathrm{tr}([\gamma^\mu, \gamma^\nu]) = \mathrm{tr}(\chi_M \gamma^\lambda) = 0,
\end{equation*}
since the unit is the only element of a Clifford algebra with a nonvanishing trace. This means that the three terms in the curvature are orthogonal:
\begin{equation*}
\begin{aligned}
\mathscr{L}_b =& -\frac{1}{16}\mathrm{tr}([\gamma^\mu, \gamma^\nu][\gamma^\lambda, \gamma^\rho])\mathrm{tr}(F_{\mu\nu}F_{\lambda\rho}) + \mathrm{tr}(\chi_M \gamma^\mu \chi_M \gamma^\nu) \mathrm{tr}(D_\mu H D_\nu H) \\ &- 4 \mathrm{tr}(P(H^2 + d_{F,U} H)^2).
\end{aligned}
\end{equation*}

Let us compute the Clifford traces. The first trace is $\mathrm{tr}([\gamma^\mu, \gamma^\nu][\gamma^\lambda, \gamma^\rho])$. Note that this 4 times the antisymmetrization of $\mathrm{tr}(\gamma^\mu \gamma^\nu \gamma^\lambda \gamma^\rho)$ with respect to $\mu,\nu$, and $\lambda,\rho$ as well. We have:
\begin{equation*}
\mathrm{tr}(\gamma^\mu \gamma^\nu \gamma^\lambda \gamma^\rho) = 4(g^{\mu\nu}g^{\lambda\rho} - g^{\mu\lambda} g^{\nu\rho} + g^{\mu\rho} g^{\nu\lambda}).
\end{equation*}
After antisymmetrizing with respect to $\mu,\nu$, one finds:
\begin{equation*}
\mathrm{tr}([\gamma^\mu, \gamma^\nu] [\gamma^\lambda \gamma^\rho]) = 8(- g^{\mu\lambda} g^{\nu\rho} + g^{\mu\rho} g^{\nu\lambda}).
\end{equation*}
Since this is already antisymmetric for $\lambda,\rho$, we find:
\begin{equation*}
\mathrm{tr}([\gamma^\mu, \gamma^\nu][\gamma^\lambda, \gamma^\rho]) = 16(- g^{\mu\lambda} g^{\nu\rho} + g^{\mu\rho} g^{\nu\lambda}).
\end{equation*}
The second trace is $\mathrm{tr}(\chi_M \gamma^\mu \chi_M \gamma^\nu)$. We have:
\begin{equation*}
\begin{aligned}
\mathrm{tr}(\chi_M \gamma^\mu \chi_M \gamma^\nu) &= -\mathrm{tr}(\chi_M^2 \gamma^\mu \gamma^\nu) \\
&= -\mathrm{tr}(\gamma^\mu \gamma^\nu) \\
\mathrm{tr}(\chi_M \gamma^\mu \chi_M \gamma^\nu) &= -4 g^{\mu\nu}.
\end{aligned}
\end{equation*}
The Lagrangian density is now:
\begin{equation*}
\mathscr{L}_b = (g^{\mu\lambda} g^{\nu\rho} - g^{\mu\rho} g^{\nu\lambda})\mathrm{tr}(F_{\mu\nu}F_{\lambda\rho}) -4 g^{\mu\nu} \mathrm{tr}(D_\mu H D_\nu H) - 4 \mathrm{tr}(P(H^2 + d_{F,U} H)^2).
\end{equation*}
Using the antisymmetry of the field strength tensor, we find:
\begin{equation}
\mathscr{L}_b = 2\mathrm{tr}(z F_{\mu\nu}F^{\mu\nu}) -4 \mathrm{tr}(z D_\mu H D^\mu H) - 4 \mathrm{tr}(z P(H^2 + d_{F,U} H)^2).
\label{BosonicLagrangianDensity}
\end{equation}

\subsection{Summary of Noncommutative Gauge Theory}

We now summarize our noncommutative gauge theory for the reader's convenience. The IST of our gauge theory is the tensor product of two ISTs. The first IST is a manifold IST $(A_M, \mathcal{K}_M, \slashed{D}, \chi_M, J_{M-})$ with the West Coast convention, for a 4D Lorenztian Spin manifold. We have for this IST: $n_M = 6$ and $m_M = 4$.

The second IST is a finite-dimensional IST $(A_F, \mathcal{K}_F, D_F, \chi_F, J_F)$ of metric and KO-dimension $n_F,m_F = 2$ or $6$. Its components are the following:
\begin{itemize}
	\item a finite Krein space that is the sum of four isomorphic subspaces:
	\begin{equation*}
\mathcal{K}_F = \mathcal{K}_R \oplus \mathcal{K}_L \oplus \mathcal{K}_{\overline{R}} \oplus \mathcal{K}_{\overline{L}}
\end{equation*}
equipped with the indefinite product:
\begin{equation*}
(u,v)_F = u^\dagger \eta_F v,
\end{equation*}
where:
\begin{equation*}
\eta_F = \begin{pmatrix} 1 &&& \\ & -1 && \\ && s & \\ &&& -s \end{pmatrix};
\end{equation*}
The definition of $s = \pm 1$ is recalled below
	\item a finite algebra $A_F$ with an involutive block-diagonal representation:
	\begin{equation*}
\pi_F(a) = \begin{pmatrix} a_R &&& \\ & a_L && \\ && a_{\overline{R}} & \\ &&& a_{\overline{L}} \end{pmatrix}
\end{equation*}
that satisfies the zero order condition:
\begin{equation*}
\left[\pi_F (a), \pi_F (b)^\circ \right] = 0
\end{equation*}
	\item a finite Dirac operator of the form:
	\begin{equation*}
D_F = \begin{pmatrix}
0 & -Y^\dagger & \epsilon_F \overline{M} & 0 \\
Y & 0 & 0 & \epsilon_F \overline{Z} \\
M & 0 & 0 & -Y^T \\
0 & Z & \overline{Y} & 0
\end{pmatrix},
\end{equation*}
where $M$ and $Z$ satisfy:
\begin{equation*}
\begin{aligned}
M^T &= s \epsilon_F M \\
Z^T &= s \epsilon_F Z,
\end{aligned}
\end{equation*}
	that satisfies the order one condition:
	\begin{equation*}
\left[[D_F,\pi_F (a)], \pi_F (b)^\circ \right] = 0
\end{equation*}
	\item a grading operator:
	\begin{equation*}
\chi_F = \begin{pmatrix} 1 &&& \\ & -1 && \\ && -1 & \\ &&& 1 \end{pmatrix}
\end{equation*}
	\item a charge conjugation operator:
	\begin{equation*}
J_F = \begin{pmatrix} && \epsilon_F & 0 \\ && 0 & \epsilon_F \\ 1 & 0 && \\ 0 & 1 && \end{pmatrix} \circ CC.
\end{equation*}
\end{itemize}

The total IST is:
\begin{equation*}
\begin{aligned}
A &= A_M \otimes A_F \cong \mathcal{C}_b^\infty(M, A_F) \\
\mathcal{K} &= \mathcal{K}_M \otimes \mathcal{K}_F \\
D &= \slashed{D} \otimes 1 + \chi_M \otimes D_F \\
\pi &= \pi_M \otimes \pi_F \\
J &= J_{M-} \otimes J_F \chi_F \\
(\cdot , \cdot) &= (\cdot, (1 \otimes \varpi) \cdot)_0
\end{aligned}
\end{equation*}
where:
\begin{equation*}
\begin{aligned}
(\varphi \otimes u, \psi \otimes v)_0 = (\varphi, \psi)_M u^\dagger v \\
\varpi = \begin{pmatrix} 1 &&& \\ & 1 && \\ && -s & \\ &&& -s \end{pmatrix}.
\end{aligned}
\end{equation*}

The field content of the theory is as follows:
\begin{itemize}
	\item the fermionic field is a multi-spinor $\Psi \in \mathcal{K}$ of either commuting or anti-commuting nature, depending on the value of the parameter $s = \pm 1$. It is of the particular form:
	\begin{equation*}
\Psi = (1+J)\Psi_0 = \begin{pmatrix}
\Psi_R \\
\Psi_L \\
\Psi_{{\overline{R}}} \\
\Psi_{\overline{L}}
\end{pmatrix}
\end{equation*}
where:
\begin{equation*}
\begin{aligned}
\Psi_{\overline{R}} &= (J_{M-} \otimes CC) \Psi_R \\
\Psi_{\overline{L}} &= -(J_{M-} \otimes CC) \Psi_L \\
\end{aligned}
\end{equation*}
	\item the bosonic field $\omega$ is of the form:
	\begin{equation*}
\omega = i\gamma^\mu \otimes B_\mu + \chi_M \otimes H.
\end{equation*}
where $B$ is a $\pi_F(A_F)$-valued differential form, and $H$ a $\Omega_{D_F}^1 (A_F)$-valued scalar field such that:
\begin{equation*}
\begin{aligned}
B_\mu^\dagger &= -B_\mu \\
\mathrm{tr}_{\mathcal{K}_F}(B_\mu) &= 0 \\
H^\dagger &= -\varpi H \varpi.
\end{aligned}
\end{equation*}
\end{itemize}

The action is the sum of two contributions:
\begin{itemize}
\item a fermionic action given by:
	\begin{equation*}
\begin{aligned}
S_f =& \frac{1}{2}(\Psi, (\slashed{D} \otimes 1 + i\gamma^\mu \otimes (B_\mu + J_F B_\mu J_F^{-1}) + \chi_M \otimes (D_F + H + J_F H J_F^{-1})) \Psi)
\end{aligned}
\end{equation*}

	\item a bosonic action, of Lagrangian density:
	\begin{equation*}
\mathscr{L}_b = 2\mathrm{tr}(z F_{\mu\nu}F^{\mu\nu}) -4 \mathrm{tr}(z D_\mu H D^\mu H) - 4 \mathrm{tr}(z P(H^2 + d_{F,U} H)^2).
\end{equation*}
Here $z$ is a self-adjoint operator on $\mathcal{K}_F$ that commutes with anti-self-adjoint elements of $\pi_F (A_F)$, and:
\begin{equation*}
\begin{aligned}
F_{\mu \nu} &= \partial_\mu B_\nu - \partial_\nu B_\mu +[B_\mu, B_\nu] \\
D_\mu H &= \partial_\mu H + [B_\mu, D_F+H].
\end{aligned}
\end{equation*}
The operator $P(H^2 + d_{F,U} H)^2$ is the projection of $\pi_{D_F}(H^2 + d_{F,U} H)$ orthogonally to the space $Q_F = \pi_F(A_F) + \pi_F^2(d_U J_{F,0}^1)$ with respect to the product $S \times T \mapsto \mathrm{Re}(\mathrm{tr}_x (\varpi S^\dagger \varpi T))$ or$S \times T \mapsto \mathrm{tr}_x (\varpi S^\dagger \varpi T)$, depending on whether $A_F$ is real or complex respectively.
\end{itemize}

\chapter{The Noncommutative Standard Model}

We now apply the results of the previous chapter to a specific IST that will enable us to recover the Standard Model of particle physics. The IST is described in the first section. In the second section, we compute the corresponding fields. In the third section, we compute the corresponding action and compare to that of the Standard Model.

\section{The Standard Model IST} \label{SectionSMIST}

To recover the Standard Model, we will use the usual finite algebra and its usual representation. The algebra is thus $A_F = \mathbb{C} \oplus \mathbb{H} \oplus \mathrm{M}_3 (\mathbb{C})$. We will parametrize quaternions using complex numbers. Each quaternion $q$ will be seen as a $2 \times 2$ matrix parametrized by two complex numbers $\alpha$ and $\beta$:
\begin{equation*}
q=\alpha+\beta j =\begin{pmatrix}
\alpha & \beta \\
-\overline{\beta} & \overline{\alpha}	
\end{pmatrix}
\in \mathrm{M}_2(\mathbb{C}).
\end{equation*}

To represent the algebra, we use the following space:
\begin{equation}
\mathcal{K}_0 = \mathcal{K}_R = \mathcal{K}_L = \mathcal{K}_{\overline{R}} = \mathcal{K}_{\overline{L}} = \left(\mathbb{C}_l^2 \oplus \mathbb{C}_q^2 \otimes \mathbb{C}_c^3 \right) \otimes \mathbb{C}_g^N,
\end{equation}
where:
\begin{itemize}
	\item $\mathbb{C}_l^2$ is a lepton doublet of orthonormal basis $(\nu,e)$
	\item $\mathbb{C}_q^2$ is a quark doublet of orthonormal basis $(u,d)$
	\item $\mathbb{C}_c^3$ is the space of colors, of orthonormal basis $(r,g,b)$
	\item $\mathbb{C}_g^N$ is the space of generations (we leave the number $N$ of generations unspecified for the moment), of orthonormal basis $(f_i)_{i=1,\dots,N}$.
\end{itemize}
Let $T$ be a $2\times 2$ complex matrix. We associate to it the operator:
\begin{equation}
\tilde{T} = (T \oplus T \otimes I_3) \otimes I_N
\end{equation}
which acts on the doublets of any of the four spaces above. We also define the following useful matrix:
\begin{equation*}
Q_\lambda=\begin{pmatrix}
\lambda & 0 \\
0 & \overline{\lambda}	
\end{pmatrix}
\in \mathrm{M}_2(\mathbb{C})
\end{equation*} 
for any $\lambda \in \mathbb{C}$. Note that this nothing but $\lambda$ viewed as a quaternion.

We can define the representation of $A_F$. Let $a = (\lambda, q, m) \in A_F$. We have:
\begin{equation}
\begin{aligned}
a_R &= \tilde{Q}_\lambda = (Q_\lambda \oplus Q_\lambda \otimes I_3) \otimes I_N \\
a_L &= \tilde{q} = (q \oplus q \otimes I_3) \otimes I_N \\
a_{\overline{R}} &= a_{\overline{L}} = (\lambda I_2 \oplus I_2 \otimes m) \otimes I_N.
\end{aligned}
\end{equation}
One can prove that this representation satisfies the order zero condition.

Finally, we choose our finite Dirac to be of the form:
\begin{equation}
\begin{aligned}
Y &= (E_{\nu \nu} \otimes Y_\nu + E_{ee} \otimes Y_e) \oplus (E_{uu} \otimes I_3 \otimes Y_u + E_{dd} \otimes I_3 \otimes Y_d) \\
M &= E_{\nu \nu} \otimes Y_R \\
Z &= 0,
\end{aligned}
\end{equation}
where $E_{\nu \nu}$ is the projector on the basis vector $\nu$, etc. One can prove that this Dirac satisfies the order one condition\footnote{Note that it is not the unique Dirac operator that satisfies the requirements. This particular form of $D_F$ is chosen empirically, on the basis that $D_F$ can be interpreted as a mass matrix of the standard model. But one can constrain $D_F$ further theoretically, with the semi-empirical massless photon condition, or the second order condition (see \cite{3B-DGA, Boyle}).}. For any $a \in A_F$, we have the following useful identities:
\begin{equation}
\begin{aligned}
\left[Y, a_R \right] &= [Y, a_{\overline{R}}] = [Y, a_{\overline{L}}] = 0 \\
M a_R &= a_{\overline{R}} M \\
M a_{\overline{R}} &= a_R M, 
\end{aligned}
\end{equation}
with the same identities holding for the transpose, complex conjugate, and adjoint of the matrices $Y$ and $M$ respectively. The matrix $Y_R$ must satisfy:
\begin{equation}
Y_R^T = s \epsilon_F Y_R. 
\end{equation}
The reader who is familiar with the noncommutative Standard Model knows that $Y_R$ is related to the Majorana mass matrix of right neutrinos. For $s \epsilon_F =-1$ and $N = 3$ generations, $Y_R$ is antisymmetric, and thus not invertible. This means that at least one of the right neutrinos does not get a Majorana mass term (more on this below), making it harder for the see-saw mechanism to operate. Note that this argument holds even if the fermions are quantized after the removal of the doubling. We are thus inclined to assume that $s \epsilon_F = 1$.

We also define for future use the self-adjoint matrices: $m_\nu = Y_\nu Y_\nu^\dagger$, etc. of squared masses.

\subsection{Finite Differential Forms}

We now study the space of noncommutative differential forms of the finite IST. Let us compute the space of finite 1-forms $\Omega_{D_F}^1(A_F)$. We start with an exact 1-form. Let $a = (\lambda,q,m) \in A_F$. Thanks to the identities above, we have:
\begin{equation}
d_{F,U}a = [D_F,\pi_F(a)] = \begin{pmatrix} 0 & Y^\dagger (\tilde{Q}_\lambda -\tilde{q}) && \\ (\tilde{Q}_\lambda -\tilde{q}) Y & 0 && \\ && 0 & \\ &&& 0 \end{pmatrix}.
\end{equation}
Notice that this 1-form commutes with $\omega$. Since $\pi_F(A_F)$ does as well, we conclude that finite differential forms commute with $\omega$. From this we infer that the hermitian and bilinear forms defined earlier are positive definite on all differential forms. Let $a_i = (\lambda_i, q_i, m_i), b_i = (\mu_i, p_i, n_i) \in A_F$. A general 1-form $\varphi$ is given by the linear combination:
\begin{equation*}
\varphi = \sum_i \pi_F(a_i) [D_F, \pi_F(b_i)] = \begin{pmatrix} 0 & Y^\dagger \sum_i \tilde{Q}_{\lambda_i} (\tilde{Q}_{\mu_i} -\tilde{p}_i) && \\ \sum_i \tilde{q}_i (\tilde{Q}_{\mu_i} -\tilde{p}_i) Y & 0 && \\ && 0 & \\ &&& 0 \end{pmatrix}.
\end{equation*}
Notice that $q_1 = \sum_i Q_{\lambda_i} (Q_{\mu_i} -p_i)$ and $q_2 = \sum_i q_i (Q_{\mu_i} - p_i)$ are both quaternions. Since $Q_{\lambda_i}$ and $q_i$ are independent, one can see that $q_1$ are $q_2$ are arbitrary, independent quaternions. The space of 1-forms is thus isomorphic to $\Omega_{D_F}^1(A_F) \cong \mathbb{H} \oplus \mathbb{H}$, and a general 1-form is of the form:
\begin{equation*}
\varphi = \begin{pmatrix} 0 & Y^\dagger \tilde{q}_1 && \\  \tilde{q}_2 Y & 0 && \\ && 0 & \\ &&& 0 \end{pmatrix}.
\end{equation*}

Let us now compute the elements of the space $\pi_F^2(d_U J_{F,0}^1)$. Let $a_i = (\lambda_i, q_i, m_i), b_i = (\mu_i, p_i, n_i) \in A_F$ such that $\varphi = \sum_i \pi_F(a_i) [D_F, \pi_F(b_i)] = 0$. This is equivalent to:
\begin{equation}
\begin{aligned}
q_1 &= \sum_i Q_{\lambda_i} (Q_{\mu_i} -p_i) = 0 \\
q_2 &= \sum_i q_i (Q_{\mu_i} - p_i) = 0.
\label{junkcon1}
\end{aligned}
\end{equation}
The most general element of $\pi_F^2(d_U J_{F,0}^1)$ is of the form:
\begin{equation*}
\begin{aligned}
d_{F,U}\varphi =& \sum_i [D_F,\pi_F(a_i)] [D_F, \pi_F(b_i)] \\
=& \begin{pmatrix} Y^\dagger \sum_i (\tilde{Q}_{\lambda_i}-\tilde{q}_i) (\tilde{Q}_{\mu_i} -\tilde{p}_i) Y &&& \\ & \sum_i (\tilde{Q}_{\lambda_i} - \tilde{q}_i) Y Y^\dagger (\tilde{Q}_{\mu_i} -\tilde{p}_i) && \\ && 0 & \\ &&& 0 \end{pmatrix}.
\end{aligned}
\end{equation*}
Note that:
\begin{equation*}
\sum_i (\tilde{Q}_{\lambda_i}-\tilde{q}_i) (\tilde{Q}_{\mu_i} -\tilde{p}_i) = \tilde{q}_1 - \tilde{q}_2 = 0,
\end{equation*}
and we are left with:
\begin{equation}
\begin{aligned}
d_{F,U}\varphi =& \sum_i [D_F,\pi_F(a_i)] [D_F, \pi_F(b_i)] \\
=& \begin{pmatrix} 0 &&& \\ & \sum_i (\tilde{q}_i - \tilde{Q}_{\lambda_i}) Y Y^\dagger (\tilde{p}_i - \tilde{Q}_{\mu_i}) && \\ && 0 & \\ &&& 0 \end{pmatrix} \in \pi_F^2(d_U J_{F,0}^1).
\end{aligned}
\label{junk1}
\end{equation}
Note also that in $d_{F,U}\varphi$ only the linear combinations $\tilde{q}_i - \tilde{Q}_{\lambda_i}$ and $\tilde{p}_i - \tilde{Q}_{\mu_i}$ occur, and these are all quaternions. One can thus assume without any loss in generality that $\lambda_i = \mu_i = 0$, effectively absorbing $\tilde{Q}_{\lambda_i}$ in $\tilde{q}_i$ and $\tilde{Q}_{\mu_i}$ in $\tilde{p}_i$. This leaves us with the condition:
\begin{equation}
\sum_i q_i p_i = 0
\label{junkcon2}
\end{equation}
and the form:
\begin{equation}
d_{F,U}\varphi = \begin{pmatrix} 0 &&& \\ & \sum_i \tilde{q}_i Y Y^\dagger \tilde{p}_i && \\ && 0 & \\ &&& 0 \end{pmatrix} \in \pi_F^2(d_U J_{F,0}^1).
\label{junk2}
\end{equation}
Indeed, any element of the form \eqref{junk2} with \eqref{junkcon2} is of the form \eqref{junk1} with \eqref{junkcon1}, with $\lambda_i = \mu_i = 0$. Conversely, by redefining $q_i$ to become $q_i - Q_{\lambda_i}$ and $p_i$ to become $p_i - Q_{\mu_i}$, an element of the form \eqref{junk1} with \eqref{junkcon1} becomes an element of the form \eqref{junk2} with \eqref{junkcon2}. Let us write:
\begin{equation*}
q_i=\begin{pmatrix} \alpha_i & \beta_i \\ -\overline{\beta}_i & \overline{\alpha}_i	\end{pmatrix},
\end{equation*}
and:
\begin{equation*}
p_i=\begin{pmatrix} \gamma_i & \delta_i \\ -\overline{\delta}_i & \overline{\gamma}_i	\end{pmatrix}.
\end{equation*}
We have:
\begin{equation*}
\sum_i q_i p_i = \sum_i \begin{pmatrix} \alpha_i \gamma_i - \beta_i \overline{\delta}_i & \alpha_i \delta_i + \beta_i \overline{\gamma_i} \\ -\overline{\beta}_i \gamma_i - \overline{\alpha}_i \overline{\delta}_i & -\overline{\beta}_i \delta_i + \overline{\alpha}_i \overline{\gamma}_i	\end{pmatrix} = 0,
\end{equation*}
which is equivalent to:
\begin{equation*}
\begin{aligned}
z_1 = \sum_i \alpha_i \gamma_i = \sum_i \beta_i \overline{\delta}_i \\
z_2 = \sum_i \alpha_i \delta_i = - \sum_i \beta_i \overline{\gamma_i}.
\end{aligned}
\end{equation*}
It is straightforward to prove that $z_1$ and $z_2$ are arbitrary and independent. We will see that these two complex number are our only degrees of freedom. Let us now simplify $d\varphi$ using these constraints. Its only nonvanishing block is the $LL$ block that we denote $(d_{F,U}\varphi)_{LL}$:
\begin{equation*}
\begin{aligned}
(d_{F,U}\varphi)_{LL} =& \sum_i \tilde{q}_i Y Y^\dagger \tilde{p}_i \\
=& \sum_i [(q_i \oplus q_i \otimes I_3) \otimes I_N] \\ &\times [(E_{\nu \nu} \otimes m_\nu + E_{ee} \otimes m_e) \oplus (E_{uu} \otimes I_3 \otimes m_u + E_{dd} \otimes I_3 \otimes m_d)]\\ &\times [(p_i \oplus p_i \otimes I_3) \otimes I_N] \\
=& \sum_i [q_i \otimes I_N][E_{\nu \nu} \otimes m_\nu + E_{ee} \otimes m_e][p_i \otimes I_N] \\
 & \oplus \sum_i [q_i \otimes I_3 \otimes I_N][E_{uu} \otimes I_3 \otimes m_u + E_{dd} \otimes I_3 \otimes m_d][p_i \otimes I_3 \otimes I_N] \\
(d_{F,U}\varphi)_{LL} =& [(\sum_i q_i E_{\nu \nu} p_i) \otimes m_\nu + (\sum_i q_i E_{ee} p_i) \otimes m_e] \\ & \oplus [(\sum_i q_i E_{uu} p_i) \otimes I_3 \otimes m_u + (\sum_i q_i E_{dd} p_i) \otimes I_3 \otimes m_d]. 
\end{aligned}
\end{equation*}
We have:
\begin{equation*}
\begin{aligned}
\sum_i q_i E_{\nu \nu} p_i &= \sum_i \begin{pmatrix} \alpha_i & \beta_i \\ -\overline{\beta}_i & \overline{\alpha}_i	\end{pmatrix} \begin{pmatrix} 1&0 \\ 0&0 \end{pmatrix} \begin{pmatrix} \gamma_i & \delta_i \\ -\overline{\delta}_i & \overline{\gamma}_i	\end{pmatrix} \\ 
&= \sum_i \begin{pmatrix} \alpha_i \gamma_i & \alpha_i \delta_i \\ -\overline{\beta}_i \gamma_i & -\overline{\beta}_i \delta_i	\end{pmatrix}.
\end{aligned}
\end{equation*}
Thanks to the conditions above, we have:
\begin{equation*}
\sum_i q_i E_{\nu \nu} p_i = \begin{pmatrix} z_1 & z_2 \\ \overline{z}_2 & -\overline{z}_1 \end{pmatrix}.
\end{equation*}
Let $q = (-iz_1) +(-iz_2) j$ be the quaternion parametrized by $-iz_1$ and $-iz_2$. One can then prove that:
\begin{equation*}
\sum_i q_i E_{\nu \nu} p_i = iq.
\end{equation*}
One can similarly prove that:
\begin{equation*}
\begin{aligned}
\sum_i q_i E_{ee} p_i &= -iq \\
\sum_i q_i E_{uu} p_i &= iq \\
\sum_i q_i E_{dd} p_i &= -iq.
\end{aligned}
\end{equation*}
We thus have:
\begin{equation}
(d_{F,U}\varphi)_{LL} = iq \otimes (m_\nu - m_e) \oplus iq \otimes I_3 \otimes (m_u - m_d) \equiv j_q,
\end{equation}
and we infer from this that $\pi_F^2(d_U J_{F,0}^1) \cong \mathbb{H}$. From this one can check explicitly that $\pi_F^2(d_U J_{F,0}^1)$ is a bimodule over $\pi_F(A_F)$. Note that the symbol $j_q$, unlike $\tilde{q}$, does not define a representation of quaternions.

\section{The Gauge and Higgs Fields}

In this section, we compute the gauge and Higgs fields $B_\mu$ and $H$, and compute the projection of the curvature of $H$. We will use the results of this section to compute the action of the theory in the next section.

\subsection{Gauge Group and Gauge Fields}

Before we describe the gauge fields, let us describe the gauge group itself. We start with the unitary group of the algebra. Let $u = (\lambda, q, m) \in A_F$ such that $uu^\ast = 1$. Then $\lambda$ is a phase, while $q \in \mathrm{SU}(2)$ and $m \in \mathrm{U}(3)$. We have:
\begin{equation*}
\pi_F(u) = \begin{pmatrix} \tilde{Q}_\lambda &&& \\ & \tilde{q} && \\ && (\lambda I_2 \oplus I_2 \otimes m) \otimes I_N & \\ &&& (\lambda I_2 \oplus I_2 \otimes m) \otimes I_N \end{pmatrix}.
\end{equation*}
To take into account fermion doubling, we can associate to it the unitary $U = \pi_F(u) J_F \pi_F(u) J_F^{-1}$. We have:
\begin{equation*}
J_F \pi_F(u) J_F^{-1} =  \begin{pmatrix} (\overline{\lambda} I_2 \oplus I_2 \otimes \overline{m}) \otimes I_N &&& \\ & (\overline{\lambda} I_2 \oplus I_2 \otimes \overline{m}) \otimes I_N && \\ && \tilde{Q}_{\overline{\lambda}} & \\ &&& \tilde{\overline{q}} \end{pmatrix}.
\end{equation*}
We thus have:
\begin{equation*}
U = \begin{pmatrix} \overline{\lambda} Q_\lambda \oplus Q_\lambda \otimes \overline{m} &&& \\ & \overline{\lambda} q \oplus q \otimes \overline{m} && \\ && \lambda Q_{\overline{\lambda}} I_2 \oplus Q_{\overline{\lambda}} \otimes m & \\ &&& \lambda \overline{q} \oplus \overline{q} \otimes m \end{pmatrix} \otimes I_N .
\end{equation*}
Note that:
\begin{equation*}
\overline{\lambda} Q_\lambda=\begin{pmatrix}
1 & 0 \\
0 & \overline{\lambda}^2	
\end{pmatrix}.
\end{equation*} 
The determinant of $m$ is so far arbitrary. One can write $m = \alpha g$, with $\alpha$ a phase, and $g \in \mathrm{SU(3)}$. We have $\mathrm{det}(\pi_F(u)) = (\lambda \alpha^3)^{4N}$. By imposing unimodularity: $\mathrm{det}(\pi_F(u)) = 1$, we find: $\alpha = e^{2ik\pi/4N} \lambda^{-1/3}$. The $e^{2ik\pi/4N}$ factor is a root of unity, and contributes to the gauge group by a constant discrete factor. But a "physical" gauge transformation is a transformation that varies from point to point on the manifold: this is how the Lie algebra - where the gauge fields live - emerges. This constant factor can be absorbed by a global gauge transformation without affecting the physics of the gauge or fermionic fields. We can thus assume, without loss in generality, that $\alpha = \lambda^{-1/3}$. Substituting in $U$, one finds:
\begin{equation*}
\setlength\arraycolsep{-2pt}
\renewcommand{\arraystretch}{1.5}
U = \begin{pmatrix} \overline{\lambda} Q_\lambda \oplus \lambda^{1/3}Q_\lambda \otimes \overline{g} &&& \\ & \overline{\lambda} q \oplus \lambda^{1/3}q \otimes \overline{g} && \\ && \lambda Q_{\overline{\lambda}} I_2 \oplus \lambda^{-1/3}Q_{\overline{\lambda}} \otimes g & \\ &&& \lambda \overline{q} \oplus \lambda^{-1/3}\overline{q} \otimes g \end{pmatrix} \otimes I_N.
\end{equation*}

Let us now turn to the gauge field $B_\mu$. We know that $B_\mu(x) \in \pi_F(A_F)$. We will thus write $B_\mu(x) = \pi_F(B_\mu^Y(x), B_\mu^W(x), B_\mu^{C^\prime}(x))$. Since $B_\mu(x)$ is anti-selfadjoint, we deduce that $B_\mu^Y$ is imaginary, $B_\mu^W \in \mathrm{su}(2)$ (anti-self adjoint quaternions are always traceless), and $B_\mu^C \in \mathrm{u}(3)$. We have:
\begin{equation*}
B_\mu = \begin{pmatrix} \tilde{Q}_{B_\mu^Y} &&& \\ & \widetilde{B_\mu^W} && \\ && (B_\mu^Y I_2 \oplus I_2 \otimes B_\mu^{C^\prime}) \otimes I_N & \\ &&& (B_\mu^Y I_2 \oplus I_2 \otimes B_\mu^{C^\prime}) \otimes I_N \end{pmatrix}.
\end{equation*}
Let us now impose unimodularity: $B_\mu$ needs to be traceless. This will fix the trace of $B_\mu^{C^\prime}$. Let us write $B_\mu^{C^\prime} = B_\mu^C + f_\mu I_3$ where $B_\mu^C \in \mathrm{su}(3)$ is traceless, and $f_\mu$ is a vector field. We have:
\begin{equation*}
\mathrm{tr}_F(B_\mu) = 4N(B_\mu^Y + \mathrm{tr}_F(B_\mu^{C^\prime})) = 4N(B_\mu^Y + 3 f_\mu) = 0
\end{equation*}
and we infer that $f_\mu = -\frac{1}{3} B_\mu^Y$. We thus have:
\begin{equation*}
\setlength\arraycolsep{-2pt}
\renewcommand{\arraystretch}{1.5}
B_\mu = \begin{pmatrix} \tilde{Q}_{B_\mu^Y} &&& \\ & \widetilde{B_\mu^W} && \\ && (B_\mu^Y I_2 \oplus I_2 \otimes (B_\mu^C -\frac{1}{3} B_\mu^Y I_3)) \otimes I_N & \\ &&& (B_\mu^Y I_2 \oplus I_2 \otimes (B_\mu^C -\frac{1}{3} B_\mu^Y I_3)) \otimes I_N \end{pmatrix}.
\end{equation*}
Now, the actual gauge fields are contained in the field $A_\mu = iB_\mu$. We thus define the fields according to:
\begin{equation*}
A_\mu^j = iB_\mu^j
\end{equation*}
where $i = Y,W,C$. We thus have:
\begin{equation*}
\setlength\arraycolsep{-2pt}
\renewcommand{\arraystretch}{1.5}
A_\mu = \begin{pmatrix} A_\mu^Y \tilde{\sigma}_z &&& \\ & \widetilde{A_\mu^W} && \\ && (A_\mu^Y I_2 \oplus I_2 \otimes (A_\mu^C -\frac{1}{3} A_\mu^Y I_3)) \otimes I_N & \\ &&& (A_\mu^Y I_2 \oplus I_2 \otimes (A_\mu^C -\frac{1}{3} A_\mu^Y I_3)) \otimes I_N \end{pmatrix},
\end{equation*}
where we used the identity $Q_i = i\sigma_z$, and the fact that $A_\mu^Y$ is now real. Let us compute the field strength tensor $F_{\mu\nu}$. We have:
\begin{equation*}
\begin{aligned}
F_{\mu \nu} &= \partial_\mu B_\nu - \partial_\nu B_\mu +[B_\mu, B_\nu] \\
&= \partial_\mu (-iA_\nu) - \partial_\nu(-iA_\mu) +[(-iA_\mu), (-iA_\nu)] \\
&= -i \left( \partial_\mu A_\nu - \partial_\nu A_\mu - i[A_\mu, A_\nu] \right).
\end{aligned}
\end{equation*}
Let $F^i_{\mu\nu} = \partial_\mu A^i_\nu - \partial_\nu A^i_\mu - i[A^i_\mu, A^i_\nu]$ be the strength of the field $A^i_\mu$ (with $i = Y,W,C$). Through a straightforward computation one finds:
\begin{equation}
\setlength\arraycolsep{-6pt}
\renewcommand{\arraystretch}{2}
F_{\mu \nu} = -i \begin{pmatrix} F^Y_{\mu\nu} \tilde{\sigma}_z &&& \\ & \widetilde{F^W_{\mu\nu}} && \\ && (F^Y_{\mu\nu} I_2 \oplus I_2 \otimes (F^C_{\mu\nu} -\frac{1}{3} F^Y_{\mu\nu} I_3)) \otimes I_N & \\ &&& (F^Y_{\mu\nu} I_2 \oplus I_2 \otimes (F^C_{\mu\nu} -\frac{1}{3} F^Y_{\mu\nu} I_3)) \otimes I_N \end{pmatrix},
\label{GaugeCurvature}
\end{equation}

\subsection{The Higgs Field Curvature and its Projection}

Let us determine the scalar field $H$, its gauge covariant derivative $D_\mu H$ and the projection of $d_{F,U}H +H^2$. The operator $H$ is a 1-form, and is thus of the form:
\begin{equation*}
H = \begin{pmatrix} 0 & Y^\dagger \tilde{q}_1 && \\  \tilde{q}_2 Y & 0 && \\ && 0 & \\ &&& 0 \end{pmatrix},
\end{equation*}
with $q_1, q_2$ arbitrary quaternions. The constraint $H^\dagger = -\omega H \omega = -H$ tells us then that $q_2 = - q_1^\dagger$. We conclude that:
\begin{equation}
H = \begin{pmatrix} 0 & -Y^\dagger \tilde{q}_H^\dagger && \\  \tilde{q}_H Y & 0 && \\ && 0 & \\ &&& 0 \end{pmatrix},
\end{equation}
with $q_H = \alpha_H+\beta_H j$ a quaternion.

A quick computation shows that the covariant derivative $D_\mu H$ of $H$ is given by:
\begin{equation}
D_\mu H = \partial_\mu H + [B_\mu, D_F+H] = \begin{pmatrix} 0 & -Y^\dagger \widetilde{D_\mu q_\Phi}^\dagger && \\  \widetilde{D_\mu q_\Phi} Y & 0 && \\ && 0 & \\ &&& 0 \end{pmatrix},
\label{HiggsDerivative}
\end{equation}
where $q_\Phi = 1 + q_H$ is the quaternion that describes the Higgs field, and:
\begin{equation}
D_\mu q_\Phi = \partial_\mu q_\Phi - iA_\mu^W q_\Phi + A_\mu^Y q_\Phi Q_i
\end{equation}
is its covariant derivative.

Let us compute $d_{F,U}H$. One can prove that:
\begin{equation*}
H = \pi_F(1,0,0) [D_F, \pi_F(0, q_H^\dagger, 0)] + \pi_F(0,1,0) [D_F, \pi_F(0,-q_H,0)].
\end{equation*}
We deduce that:
\begin{equation*}
\begin{aligned}
d_{F,U}H &= [D_F,\pi_F(1,0,0)] [D_F, \pi_F(0, q_H^\dagger, 0)] + [D_F,\pi_F(0,1,0)] [D_F, \pi_F(0,-q_H,0)] \\
&= -2 \mathrm{Re}(q_H) \begin{pmatrix} Y^\dagger Y &&& \\ & Y Y^\dagger && \\ && 0 & \\ &&& 0 \end{pmatrix},
\end{aligned}
\end{equation*}
where $\mathrm{Re}(q_H) = \mathrm{Re}(\alpha_H)$. We also have:
\begin{equation*}
H^2 = \begin{pmatrix} -|q_H|^2 Y^\dagger Y &&& \\ & -\tilde{q}_H Y Y^\dagger \tilde{q}_H^\dagger && \\ && 0 & \\ &&& 0 \end{pmatrix},
\end{equation*}
where $|q_H|^2 = |\alpha_H|^2 + |\beta_H|^2$. To prove the formula above we used the identity $q_H^\dagger q_H = q_H q_H^\dagger = |q_H|^2 I_2$. We will use it again to simplify the result further. Indeed writing this as $q_H \times q_H^\dagger - |q_H|^2 \times I_2 = 0$, and comparing with equations \eqref{junkcon2} and \eqref{junk2}, we see that the following operator has to be in the space $\pi_F^2(d_U J_{F,0}^1)$:
\begin{equation*}
\begin{pmatrix} 0 &&& \\ & \tilde{q}_H Y Y^\dagger \tilde{q}_H^\dagger - |q_H|^2 Y Y^\dagger && \\ && 0 & \\ &&& 0 \end{pmatrix},
\end{equation*}
allowing us to write $H^2$ as:
\begin{equation*}
H^2 = -|q_H|^2 \begin{pmatrix} Y^\dagger Y &&& \\ & Y Y^\dagger && \\ && 0 & \\ &&& 0 \end{pmatrix} + j,
\end{equation*}
with $ j \in \pi_F^2(d_U J_{F,0}^1)$. We thus have:
\begin{equation*}
d_{F,U}H + H^2 = -(|q_H|^2+2 \mathrm{Re}(q_H)) \begin{pmatrix}  Y^\dagger Y &&& \\ & YY^\dagger && \\ && 0 & \\ &&& 0 \end{pmatrix} + j.
\end{equation*}

Let us now compute the projection $P(d_{F,U}H +H^2)$. Generally, the projection $P\varphi$ of a 2-form $\varphi$ is the unique operator $P\varphi = \varphi + \pi_F(a) + j$, with $a \in A_F$ and $j \in \pi_F^2(d_U J_{F,0}^1)$, such that:
\begin{equation}
\mathrm{Re \; tr}(\omega (\pi_F(b)+k)^\dagger \omega (\varphi+\pi_F(a)+j))=\mathrm{Re \; tr}((\pi_F(b)+k)^\dagger (\varphi+\pi_F(a)+j))=0
\end{equation}
for all $b \in A_F$ and $k \in \pi_F^2(d_U J_{F,0}^1)$ ($\omega$ canceled out because it commutes with all differential forms). This operation is linear. We will thus decompose $d_{F,U}H +H^2$ in three parts, and project each part separately. We write:
\begin{equation}
d_{F,U}H +H^2 = -(|q_H|^2+2 \mathrm{Re}(q_H)) (\varphi_1 + \varphi_2)+j.
\end{equation}
where:
\begin{equation*}
\begin{aligned}
\varphi_1 &= \begin{pmatrix} Y^\dagger Y &&& \\ & 0 && \\ && 0 & \\ &&& 0 \end{pmatrix} \\
\varphi_2 &= \begin{pmatrix} 0 &&& \\ & YY^\dagger && \\ && 0 & \\ &&& 0 \end{pmatrix}.
\end{aligned}
\end{equation*}
Note that the projection of $j$ is easy to determine: since it is in $Q_F$, its projection is clearly vanishing. Let us compute the projection of each of the remaining 2-forms:
\begin{itemize}

	\item \underline{$\varphi_1$}: This is the simplest one. Indeed, its only nonvanishing block is the $RR$ block. It is thus orthogonal to $\pi_F^2(d_U J_{F,0}^1)$, as well as the $\mathbb{H}$ and $\mathrm{M}_3 (\mathbb{C})$ parts of $A_F$. We thus only have to project it orthogonally to $\mathbb{C} \subset A_F$. We are thus looking for $\lambda \in \mathbb{C}$ such that:
\begin{equation*}
\begin{aligned}
0 =&\mathrm{Re \; tr}(\pi_F(\mu,0,0)^\dagger (\varphi_1+\pi_F(\lambda,0,0))) \\
=& \mathrm{Re \; tr} \left(\tilde{Q}_\mu^\dagger (Y^\dagger Y + \tilde{Q}_\lambda) \right) + 2 \mathrm{Re \; tr}(\overline{\mu} \lambda I_2 \otimes I_N)
\end{aligned}
\end{equation*}
for all $\mu \in \mathbb{C}$.
We have:
\begin{equation*}
\begin{aligned}
Y^\dagger Y + \tilde{Q}_\lambda =& ((E_{\nu \nu} \otimes Y_\nu^\dagger Y_\nu + E_{ee} \otimes Y_e^\dagger Y_e) \oplus (E_{uu} \otimes I_3 \otimes Y_u^\dagger Y_u + E_{dd} \otimes I_3 \otimes Y_d^\dagger Y_d)) \\ &+ (Q_\lambda \oplus Q_\lambda \otimes I_3) \otimes I_N \\
=& \left(E_{\nu \nu} \otimes (Y_\nu^\dagger Y_\nu + \lambda) + E_{ee} \otimes (Y_e^\dagger Y_e + \overline{\lambda}) \right) \\ &\oplus \left(E_{uu} \otimes I_3 \otimes (Y_u^\dagger Y_u + \lambda) + E_{dd} \otimes I_3 \otimes (Y_d^\dagger Y_d + \overline{\lambda}) \right).
\end{aligned}
\end{equation*}
Multiplication by $\tilde{Q}_\mu^\dagger$ then yields the operator:
\begin{equation*}
\begin{aligned}
\tilde{Q}_\mu^\dagger(Y^\dagger Y + \tilde{Q}_\lambda) =& \left(E_{\nu \nu} \otimes \overline{\mu}(Y_\nu^\dagger Y_\nu + \lambda) + E_{ee} \otimes \mu (Y_e^\dagger Y_e + \overline{\lambda}) \right) \\ &\oplus \left(E_{uu} \otimes I_3 \otimes \overline{\mu}(Y_u^\dagger Y_u + \lambda) + E_{dd} \otimes I_3 \otimes \mu (Y_d^\dagger Y_d + \overline{\lambda}) \right)
\end{aligned}
\end{equation*}
whose trace is:
\begin{equation*}
\mathrm{Re \; tr} \left(\tilde{Q}_\mu^\dagger (Y^\dagger Y + \tilde{Q}_\lambda) \right) = 8N \mathrm{Re}(\overline{\mu} \lambda) + \mathrm{Re}(\mu) \mathrm{tr}(Y_\nu^\dagger Y_\nu + Y_e^\dagger Y_e + 3 Y_u^\dagger Y_u + 3 Y_d^\dagger Y_d).
\end{equation*}
We thus have the equation:
\begin{equation*}
12N \mathrm{Re}(\overline{\mu} \lambda) + \mathrm{Re}(\mu) \mathrm{tr}(Y_\nu^\dagger Y_\nu + Y_e^\dagger Y_e + 3 Y_u^\dagger Y_u + 3 Y_d^\dagger Y_d)=0.
\end{equation*}
Substituting $\mu = i$ gives us the equation $\mathrm{Re}(i\lambda)=0$, implying that $\lambda$ is real. Substituting $\mu = 1$ gives us then the value of $\lambda$:
\begin{equation*}
\lambda = -\frac{1}{12N} C_1,
\end{equation*}
where:
\begin{equation}
C_1 = \mathrm{tr}(Y_\nu^\dagger Y_\nu + Y_e^\dagger Y_e + 3 Y_u^\dagger Y_u + 3 Y_d^\dagger Y_d) = \mathrm{tr}(m_\nu + m_e + 3 m_u + 3 m_d).
\end{equation}
We thus have:
\begin{equation}
P\varphi_1 = \begin{pmatrix} Y^\dagger Y - \frac{1}{12N} C_1 I_{8N} &&& \\ & 0 && \\ && - \frac{1}{12N} C_1 I_2 \otimes I_N \oplus 0 & \\ &&& - \frac{1}{12N} C_1 I_2 \otimes I_N \oplus 0 \end{pmatrix}.
\end{equation}
Note that this is a traceless matrix.

	\item \underline{$\varphi_2$}: Its only nonvanishing block is the $LL$ block. It is thus already orthogonal to the $\mathbb{C}$ and $\mathrm{M}_3 (\mathbb{C})$ parts of $A_F$. We thus have to project it orthogonally to the $\mathbb{H}$ part of $A_F$ and $\pi_F^2(d_U J_{F,0}^1)$. We are thus looking for $q, q' \in \mathbb{H}$ such that:
\begin{equation*}
\mathrm{Re \; tr} \left( (\tilde{p} + j_{p'})^\dagger (YY^\dagger + \tilde{q} + j_{q'}) \right) = 0
\end{equation*}
for all $p,p' \in \mathbb{H}$. This can be split in two equations:
\begin{equation*}
\mathrm{Re \; tr} \left( \tilde{p}^\dagger (YY^\dagger + \tilde{q} + j_{q'}) \right) = \mathrm{Re \; tr} \left( j_{p'}^\dagger (YY^\dagger + \tilde{q} + j_{q'}) \right) = 0
\end{equation*}
Let us pick a parametrization for our quaternions. We parametrize $q$ with $\alpha,\beta$, $p$ with $\gamma,\delta$, etc. 

The first equation can be written:
\begin{equation*}
\mathrm{Re \; tr} \left( \tilde{p}^\dagger YY^\dagger + \widetilde{p^\dagger q} + j_{p^\dagger q'}) \right) = 0
\end{equation*}
using the properties of the symbols $\tilde{q}$ and $j_q$. It is straightforward to prove that generally:
\begin{equation*}
\begin{aligned}
\mathrm{Re \; tr} \tilde{q} &= 8N \mathrm{Re}(q) \\
\mathrm{Re \; tr} j_{q} &= 0,
\end{aligned}
\end{equation*}
which means that:
\begin{equation*}
\begin{aligned}
\mathrm{Re \; tr} \left(\widetilde{p^\dagger q} + j_{p^\dagger q'}) \right) &= 8N \mathrm{Re}(p^\dagger q) \\
&= 8N \mathrm{Re}(\alpha \overline{\gamma} + \overline{\beta} \delta).
\end{aligned}
\end{equation*}
We also have:
\begin{equation*}
YY^\dagger = (E_{\nu \nu} \otimes m_\nu + E_{ee} \otimes m_e) \oplus (E_{uu} \otimes I_3 \otimes m_u + E_{dd} \otimes I_3 \otimes m_d)
\end{equation*}
and this implies that:
\begin{equation*}
\begin{aligned}
\mathrm{Re \; tr} (\tilde{p}^\dagger YY^\dagger) =& \mathrm{Re \; tr} [ (p^\dagger E_{\nu \nu} \otimes m_\nu + p^\dagger E_{ee} \otimes m_e) \\ &\oplus (p^\dagger E_{uu} \otimes I_3 \otimes m_u + p^\dagger E_{dd} \otimes I_3 \otimes m_d)] \\
=& \mathrm{Re \; tr}(p^\dagger E_{\nu \nu}) \mathrm{tr}(m_\nu) + \mathrm{Re \; tr}(p^\dagger E_{ee}) \mathrm{tr}(m_e) + 3\mathrm{Re \; tr}(p^\dagger E_{uu}) \mathrm{tr}(m_u) \\ &+ 3\mathrm{Re \; tr}(p^\dagger E_{dd}) \mathrm{tr}(m_d) \\
=& \mathrm{tr}(m_\nu+m_e+3m_u+3m_d) \mathrm{Re}(\gamma) \\
\mathrm{Re \; tr} (\tilde{p}^\dagger YY^\dagger) =& C_1 \mathrm{Re}(\gamma).
\end{aligned}
\end{equation*}
The first equation takes the form:
\begin{equation*}
8N \mathrm{Re}(\alpha \overline{\gamma} + \overline{\beta} \delta) + C_1 \mathrm{Re}(\gamma) = 0.
\end{equation*}
First, by choosing $\gamma = 0$, we see that $\beta$ must vanish. Next, choosing $\gamma=i$ tells us that $\alpha$ is real. Finally, choosing $\gamma=1$ gives us $\alpha$:
\begin{equation*}
\alpha = -\frac{1}{8N} C_1.
\end{equation*}
We deduce that:
\begin{equation*}
q = -\frac{1}{8N} C_1 I_2.
\end{equation*}

The second equation is:
\begin{equation*}
\mathrm{Re \; tr} \left( j_{p'}^\dagger (YY^\dagger + \tilde{q} + j_{q'}) \right) = \mathrm{Re \; tr} \left( j_{p'}^\dagger YY^\dagger + j_{p'}^\dagger \tilde{q} + j_{p'}^\dagger j_{q'} \right) = 0.
\end{equation*}
Note that:
\begin{equation*}
\mathrm{Re \; tr} \left( j_{p'}^\dagger \tilde{q} \right) = -\mathrm{Re \; tr} \left( j_{p'^\dagger q} \right)=0.
\end{equation*}
The equation is now:
\begin{equation*}
\mathrm{Re \; tr} \left( j_{p'}^\dagger YY^\dagger + j_{p'}^\dagger j_{q'} \right) = 0.
\end{equation*}
Similarly to what we have computed above, we have:
\begin{equation*}
\begin{aligned}
\mathrm{Re \; tr} \left( j_{p'}^\dagger YY^\dagger \right) =& -\mathrm{Re \; tr} [ (ip'^\dagger E_{\nu \nu} \otimes (m_\nu-m_e)m_\nu + ip'^\dagger E_{ee} \otimes (m_\nu-m_e)m_e) \\ &\oplus (ip'^\dagger E_{uu} \otimes I_3 \otimes (m_u-m_d)m_u + ip'^\dagger E_{dd} \otimes I_3 \otimes (m_u-m_d)m_d)] \\
=& -\mathrm{Re \; tr} (ip'^\dagger E_{\nu \nu}) \mathrm{tr}[(m_\nu-m_e)m_\nu] - \mathrm{Re \; tr}(ip'^\dagger E_{ee}) \mathrm{tr}[(m_\nu-m_e)m_e] \\ &- 3\mathrm{Re \; tr}(ip'^\dagger E_{uu}) \mathrm{tr}[(m_u-m_d)m_u] - 3\mathrm{Re \; tr}(ip'^\dagger E_{dd}) \mathrm{tr}[(m_u-m_d)m_d)] \\
\mathrm{Re \; tr} \left( j_{p'}^\dagger YY^\dagger \right) =& -\mathrm{Im}(\gamma') \mathrm{tr}[(m_\nu-m_e)^2 + 3 (m_u-m_d)^2].
\end{aligned}
\end{equation*}
We also have:
\begin{equation*}
\begin{aligned}
\mathrm{Re \; tr} \left(j_{p'}^\dagger j_{q'} \right) =& \mathrm{Re \; tr} \left( p'^\dagger q' \otimes (m_\nu-m_e)^2 \oplus p'^\dagger q' \otimes I_3 \otimes (m_u-m_d)^2 \right) \\
=& \mathrm{tr}[(m_\nu-m_e)^2 + 3 (m_u-m_d)^2] \mathrm{Re \; tr}(p'^\dagger q') \\
\mathrm{Re \; tr} \left(j_{p'}^\dagger j_{q'} \right) =& 2 \mathrm{tr}[(m_\nu-m_e)^2 + 3 (m_u-m_d)^2] \mathrm{Re}(\alpha' \overline{\gamma}' + \overline{\beta}' \delta').
\end{aligned}
\end{equation*}
The $\mathrm{tr}[(m_\nu-m_e)^2 + 3 (m_u-m_d)^2]$ cancels out and the equation is finally:
\begin{equation*}
-\mathrm{Im}(\gamma') + 2 \mathrm{Re}(\alpha' \overline{\gamma}' + \overline{\beta}' \delta') = 0.
\end{equation*}
Substituting $\gamma'=0$, we deduce that $\beta'=0$. Substituting $\gamma'=1$ tells us that $\alpha'$ is imaginary. Finally, substituting $\gamma'=i$, we find $\alpha'$:
\begin{equation*}
\alpha' = \frac{i}{2}.
\end{equation*}
We thus have:
\begin{equation}
q' = \frac{Q_i}{2}.
\end{equation}

The projection of $\varphi_2$ is:
\begin{equation*}
P\varphi_2 = \begin{pmatrix} 0 &&& \\ & YY^\dagger - \frac{C_1}{8N} I_{8N} + \frac{j_{Q_i}}{2} && \\ && 0 & \\ &&& 0 \end{pmatrix}.
\end{equation*}
As above, this is a traceless matrix. Note that from:
\begin{equation*}
j_{Q_i} = (-E_{\nu\nu} + E_{ee}) \otimes (m_\nu - m_e) \oplus (-E_{uu} + E_{dd}) \otimes I_3 \otimes (m_u - m_d),
\end{equation*}
we can deduce that:
\begin{equation*}
YY^\dagger+\frac{j_{Q_i}}{2} = I_2 \otimes \frac{m_\nu + m_e}{2} \oplus I_2 \otimes I_3 \otimes \frac{m_u + m_d}{2}
\end{equation*}

\end{itemize}

Gathering the previous results, we find:
\begin{equation}
\setlength\arraycolsep{-2pt}
\renewcommand{\arraystretch}{1.5}
\begin{aligned}
P(d_{F,U}H +H^2) =& -(|q_H|^2+2 \mathrm{Re}(q_H)) \\
&\times \begin{pmatrix} Y^\dagger Y - \frac{C_1}{12N}I_{8N} &&& \\ & YY^\dagger + \frac{j_{Q_i}}{2} - \frac{C_1}{8N} I_{8N} && \\ && - \frac{1}{12N} C_1 I_{2N} \oplus 0 & \\ &&& - \frac{1}{12N} C_1 I_{2N} \oplus 0 \end{pmatrix}.
\end{aligned}
\label{HiggsProjection}
\end{equation}

\section{The Bosonic and Fermionic Actions}

We finally have at our disposal all the necessary tools to compute the action of our noncommutative gauge theory. We will compute the bosonic and fermionic actions separately, then compare them to the action of the Standard Model.

\subsection{The Fermionic Action}

We have already determined the kinetic part of the fermionic action, we now need to compute the couplings to gauge fields \eqref{FermionGaugeCoupling} and Higgs fields \eqref{FermionHiggsCoupling}. Before we compute these two terms, let us make a preliminary calculation. Let $T \in \mathrm{End}(\mathcal{K}_F)$ of the form:
\begin{equation*} 
T = \begin{pmatrix} A & B \\ C & D \end{pmatrix}
\end{equation*}
where the blocks act on the subspaces $\mathcal{K}_R \oplus \mathcal{K}_L$ and $\mathcal{K}_{\overline{R}} \oplus \mathcal{K}_{\overline{L}}$. Then:
\begin{equation*}
\begin{aligned}
J_F T J_F^{-1} &= \begin{pmatrix} 0 & \epsilon_F \\ 1 & 0 \end{pmatrix} CC \begin{pmatrix} A & B \\ C & D \end{pmatrix} CC \begin{pmatrix} 0 & 1 \\ \epsilon_F & 0 \end{pmatrix} \\
&= \begin{pmatrix} 0 & \epsilon_F \\ 1 & 0 \end{pmatrix} \begin{pmatrix} \overline{A} & \overline{B} \\ \overline{C} & \overline{D} \end{pmatrix} \begin{pmatrix} 0 & 1 \\ \epsilon_F & 0 \end{pmatrix} \\
J_F T J_F^{-1} &= \begin{pmatrix} \overline{D} & \epsilon_F \overline{C} \\ \epsilon_F \overline{B} & \overline{A} \end{pmatrix}.
\end{aligned}
\end{equation*}

Let us compute the gauge term. We first express it using the field $A_\mu = iB_\mu$:
\begin{equation}
S_g = \frac{1}{2}(\Psi, \gamma^\mu \otimes (A_\mu - J_F A_\mu J_F^{-1}) \Psi).
\end{equation}
We write first $A_\mu$ in a block diagonal form:
\begin{equation*}
A_\mu = \begin{pmatrix} A_\mu^R &&& \\ & A_\mu^L && \\ && A_\mu^{\overline{R}} & \\ &&& A_\mu^{\overline{L}} \end{pmatrix}.
\end{equation*}
We then have:
\begin{equation*}
A_\mu - J_F A_\mu J_F^{-1} = \begin{pmatrix} A_\mu^R - \overline{A_\mu^{\overline{R}}} &&& \\ & A_\mu^L - \overline{A_\mu^{\overline{L}}} && \\ && A_\mu^{\overline{R}} - \overline{A_\mu^R} & \\ &&& A_\mu^{\overline{L}} - \overline{A_\mu^L} \end{pmatrix}.
\end{equation*}
We have:
\begin{equation*}
\begin{aligned}
S_g =& \frac{1}{2}(\Psi, \gamma^\mu \otimes (A_\mu - J_F A_\mu J_F^{-1}) \Psi) \\
=& \frac{1}{2}(\Psi, \gamma^\mu \otimes \varpi (A_\mu - J_F A_\mu J_F^{-1}) \Psi)_0 \\
=& \frac{1}{2} (\Psi_R, \gamma^\mu \otimes (A_\mu^R - \overline{A_\mu^{\overline{R}}}) \Psi_R)_0 + \frac{1}{2} (\Psi_L, \gamma^\mu \otimes (A_\mu^L - \overline{A_\mu^{\overline{L}}}) \Psi_L)_0 \\ & -\frac{s}{2} (\Psi_{\overline{R}}, \gamma^\mu \otimes (A_\mu^{\overline{R}} - \overline{A_\mu^R}) \Psi_{\overline{R}})_0 -\frac{s}{2} (\Psi_{\overline{L}}, \gamma^\mu \otimes (A_\mu^{\overline{L}} - \overline{A_\mu^L}) \Psi_{\overline{L}})_0.
\end{aligned}
\end{equation*}
We will simplify this as we have done for the kinetic terms earlier. We have (abusively):
\begin{equation*}
\begin{aligned}
(\Psi_{\overline{R}}, \gamma^\mu \otimes (A_\mu^{\overline{R}} - \overline{A_\mu^R}) \Psi_{\overline{R}})_0 &= ((J_{M-} \otimes CC) \Psi_R, \gamma^\mu \otimes (A_\mu^{\overline{R}} - \overline{A_\mu^R}) (J_{M-} \otimes CC) \Psi_R)_0 \\
&= -((J_{M-} \otimes CC) \Psi_R, (J_{M-} \otimes CC) \gamma^\mu \otimes (\overline{A_\mu^{\overline{R}}} - A_\mu^R) \Psi_R)_0
\end{aligned}
\end{equation*}
where we used the anticommutation of $J_{M-}$ and $\gamma^\mu$. Using $J_{M-}^\times J_{M-} = -1$ we have:
\begin{equation*}
(\Psi_{\overline{R}}, \gamma^\mu \otimes (A_\mu^{\overline{R}} - \overline{A_\mu^R}) \Psi_{\overline{R}})_0 = s(\gamma^\mu \otimes (\overline{A_\mu^{\overline{R}}} - A_\mu^R) \Psi_R, \Psi_R)_0.
\end{equation*}
The self-adjointness of $A_\mu$ and $\gamma_\mu$ then implies that:
\begin{equation*}
\begin{aligned}
(\Psi_{\overline{R}}, \gamma^\mu \otimes (A_\mu^{\overline{R}} - \overline{A_\mu^R}) \Psi_{\overline{R}})_0 &= s(\Psi_R, \gamma^\mu \otimes (\overline{A_\mu^{\overline{R}}} - A_\mu^R) \Psi_R)_0 \\
(\Psi_{\overline{R}}, \gamma^\mu \otimes (A_\mu^{\overline{R}} - \overline{A_\mu^R}) \Psi_{\overline{R}})_0 &= -s(\Psi_R, \gamma^\mu \otimes (A_\mu^R - \overline{A_\mu^{\overline{R}}}) \Psi_R)_0 \\
\end{aligned}
\end{equation*}
Similarly, we have:
\begin{equation*}
(\Psi_{\overline{L}}, \gamma^\mu \otimes (A_\mu^{\overline{L}} - \overline{A_\mu^L}) \Psi_{\overline{L}})_0 = -s(\Psi_L, \gamma^\mu \otimes (A_\mu^L - \overline{A_\mu^{\overline{L}}}) \Psi_L)_0.
\end{equation*}
We thus have:
\begin{equation}
S_g = (\Psi_R, \gamma^\mu \otimes (A_\mu^R - \overline{A_\mu^{\overline{R}}}) \Psi_R)_0 + (\Psi_L, \gamma^\mu \otimes (A_\mu^L - \overline{A_\mu^{\overline{L}}}) \Psi_L)_0.
\end{equation}
We have:
\begin{equation*}
\begin{aligned}
A_\mu^R - \overline{A_\mu^{\overline{R}}} &= A_\mu^Y \tilde{\sigma}_z - \overline{(A_\mu^Y I_2 \oplus I_2 \otimes (A_\mu^C -\frac{1}{3} A_\mu^Y I_3)) \otimes I_N} \\
&= A_\mu^Y \tilde{\sigma}_z - (A_\mu^Y I_2 \oplus I_2 \otimes (\overline{A_\mu^C} -\frac{1}{3} A_\mu^Y I_3)) \otimes I_N \\
&= A_\mu^Y (\sigma_z \otimes I_N \oplus \sigma_z \otimes I_3 \otimes I_N) - (A_\mu^Y I_2 \oplus I_2 \otimes (\overline{A_\mu^C} -\frac{1}{3} A_\mu^Y I_3)) \otimes I_N \\
&= A_\mu^Y [(\sigma_z - I_2) \otimes I_N \oplus (\sigma_z + \frac{1}{3} I_2) \otimes I_3 \otimes I_N] - I_2 \otimes \overline{A_\mu^C} \otimes I_N.
\end{aligned}
\end{equation*}
Let us denote:
\begin{equation}
T_Y^R = (\sigma_z - I_2) \oplus (\sigma_z + \frac{1}{3} I_2) \otimes I_3
\end{equation}
the generator of hypercharge on right particles. Note that:
\begin{equation}
T_Y^R = \begin{pmatrix} 0 & \\ & -2 \end{pmatrix} \oplus \begin{pmatrix} \frac{4}{3} & \\ & -\frac{2}{3} \end{pmatrix} \otimes I_3,
\end{equation}
We now have:
\begin{equation}
A_\mu^R - \overline{A_\mu^{\overline{R}}} = (A_\mu^Y T_Y^R - I_2 \otimes \overline{A_\mu^C}) \otimes I_N.
\end{equation}
Similarly, we have:
\begin{equation}
A_\mu^L - \overline{A_\mu^{\overline{L}}} = (A_\mu^Y T_Y^L + (A_\mu^W \oplus A_\mu^W \otimes I_3) - I_2 \otimes \overline{A_\mu^C}) \otimes I_N,
\end{equation}
where:
\begin{equation}
T_Y^L = -I_2 \oplus \frac{1}{3} I_2 \otimes I_3
\end{equation}
generates hypercharge on left fermions. We will prove later that $T_Y^R$ and $T_Y^L$ do indeed generate hypercharge. We will substitute the previous expressions in $S_g$ and expand later, when comparing to the Standard Model.

Let us now turn to the Higgs coupling term:
\begin{equation}
S_h = \frac{1}{2}(\Psi, \chi_M \otimes (D_F + H + J_F H J_F^{-1}) \Psi).
\end{equation}
We have:
\begin{equation*}
\begin{aligned}
D_F + H + J_F H J_F^{-1} =& \begin{pmatrix} 0 & -Y^\dagger & \epsilon_F \overline{M} & 0 \\ Y & 0 & 0 & 0 \\ M & 0 & 0 & -Y^T \\ 0 & 0 & \overline{Y} & 0 \end{pmatrix} + \begin{pmatrix} 0 & -Y^\dagger \tilde{q}_H^\dagger & 0 & 0 \\  \tilde{q}_H Y & 0 & 0 & 0 \\ 0 & 0 & 0 & 0 \\ 0 & 0 & 0 & 0 \end{pmatrix} \\ &+ \begin{pmatrix} 0 & 0 & 0 & 0 \\ 0 & 0 & 0 & 0 \\ 0 & 0 & 0 & -\tilde{q}_H^T Y^T \\ 0 & 0 & \tilde{\overline{q}}_H \overline{Y} & 0 \end{pmatrix} \\
=& \begin{pmatrix} 0 & -Y^\dagger \widetilde{(I_2+q_H)}^\dagger & \epsilon_F \overline{M} & 0 \\ \widetilde{(I_2+q_H)} Y & 0 & 0 & 0 \\ M & 0 & 0 & -Y^T\widetilde{(I_2+q_H)} ^T \\ 0 & 0 & \widetilde{\overline{(I_2+q_H)}}\overline{Y} & 0 \end{pmatrix}.
\end{aligned}
\end{equation*}
This suggests that we define the quaternion:
\begin{equation}
q_\Phi = 1 + q_H
\end{equation}
that should later correspond to the Higgs field. We thus have:
\begin{equation*}
D_F + H + J_F H J_F^{-1} = \begin{pmatrix} 0 & -Y^\dagger \tilde{q}_\Phi^\dagger & \epsilon_F \overline{M} & 0 \\ \tilde{q}_\Phi Y & 0 & 0 & 0 \\ M & 0 & 0 & -Y^T \tilde{q}_\Phi^T \\ 0 & 0 & \tilde{\overline{q}}_\Phi \overline{Y} & 0 \end{pmatrix}.
\end{equation*}
Let us now substitute in $S_h$:
\begin{equation*}
\begin{aligned}
S_h =& \frac{1}{2}(\Psi, \chi_M \otimes (D_F + H + J_F H J_F^{-1}) \Psi) \\
=& \frac{1}{2}(\Psi, \chi_M \otimes \varpi(D_F + H + J_F H J_F^{-1}) \Psi)_0.
\end{aligned}
\end{equation*}
We have established earlier that $n_M = 6$ and $n_F = 2$ or $6$. This implies that $n = 0$ or $4$, and hence that $J$ and $\chi$ commute. Since $\Psi_0$ is an eigenvector of $\chi$ with eigenvalue $1$, then so is $\Psi = (1+J) \Psi_0$. This means that $(\chi_M \otimes 1)\Psi = (1 \otimes \chi_F) \Psi$. We now have:
\begin{equation*}
\begin{aligned}
S_h =& \frac{1}{2}(\Psi, \chi_M \otimes \varpi(D_F + H + J_F H J_F^{-1}) \Psi)_0 \\
=& \frac{1}{2}(\Psi, (1 \otimes \varpi(D_F + H + J_F H J_F^{-1}))(\chi_M \otimes 1) \Psi)_0 \\
=& \frac{1}{2}(\Psi, (1 \otimes \varpi(D_F + H + J_F H J_F^{-1}))(1 \otimes \chi_F) \Psi)_0 \\
=& \frac{1}{2}(\Psi, 1 \otimes (-\eta_F(D_F + H + J_F H J_F^{-1})) \Psi)_0,
\end{aligned}
\end{equation*}
where:
\begin{equation}
-\eta_F = \begin{pmatrix} -1 &&& \\ & 1 && \\ && -s & \\ &&& s \end{pmatrix}.
\end{equation}
We thus have:
\begin{equation*}
-\eta_F (D_F + H + J_F H J_F^{-1}) = \begin{pmatrix} 0 & Y^\dagger \tilde{q}_\Phi^\dagger & -\epsilon_F \overline{M} & 0 \\ \tilde{q}_\Phi Y & 0 & 0 & 0 \\ -sM & 0 & 0 & sY^T \tilde{q}_\Phi^T \\ 0 & 0 & s\tilde{\overline{q}}_\Phi \overline{Y} & 0 \end{pmatrix}.
\end{equation*}
A straightforward substitution then gives us:
\begin{equation*}
\begin{aligned}
S_h =& \frac{1}{2}(\Psi_L, \tilde{q}_\Phi Y \Psi_R)_0 + \frac{1}{2}(\Psi_R, Y^\dagger \tilde{q}_\Phi^\dagger \Psi_L)_0 \\ &+ \frac{s}{2}(\Psi_{\overline{L}}, \tilde{\overline{q}}_\Phi \overline{Y} \Psi_{\overline{R}})_0 + \frac{s}{2}(\Psi_{\overline{R}}, Y^T \tilde{q}_\Phi^T \Psi_{\overline{L}})_0 \\ &- \frac{s}{2} (\Psi_{\overline{R}}, M \Psi_R)_0 - \frac{\epsilon_F}{2} (\Psi_R, \overline{M} \Psi_{\overline{R}})_0.
\end{aligned}
\end{equation*}

Similarly to what was proven for gauge terms, we have:
\begin{equation*}
\begin{aligned}
(\Psi_{\overline{L}}, \tilde{\overline{q}}_\Phi \overline{Y} \Psi_{\overline{R}})_0 &= -((J_{M-} \otimes CC) \Psi_L, \tilde{\overline{q}}_\Phi \overline{Y} (J_{M-} \otimes CC) \Psi_R)_0 \\
&= -((J_{M-} \otimes CC) \Psi_L, (J_{M-} \otimes CC) \tilde{q}_\Phi Y \Psi_R)_0 \\
&= s(\tilde{q}_\Phi Y \Psi_R, \Psi_L)_0 \\
(\Psi_{\overline{L}}, \tilde{\overline{q}}_\Phi \overline{Y} \Psi_{\overline{R}})_0 &= s(\Psi_R, Y^\dagger \tilde{q}_\Phi^\dagger \Psi_L)_0.
\end{aligned}
\end{equation*}
One can similarly prove that:
\begin{equation*}
(\Psi_{\overline{R}}, Y^T \tilde{q}_\Phi^T \Psi_{\overline{L}})_0 = s(\Psi_L, \tilde{q}_\Phi Y \Psi_R)_0.
\end{equation*}
Finally, we have:
\begin{equation*}
\begin{aligned}
(\Psi_R, \overline{M} \Psi_{\overline{R}})_0 &= (\Psi_R, \overline{M} (J_{M-} \otimes CC) \Psi_R)_0 \\
&= (\Psi_R, (J_{M-} \otimes CC) M \Psi_R)_0 \\
(\Psi_R, \overline{M} \Psi_{\overline{R}})_0 &= -s(M \Psi_R, (J_{M-} \otimes CC) \Psi_R)_0.
\end{aligned}
\end{equation*}
To simplify, we will from now on use the notation $J_{M-} \equiv J_{M-} \otimes CC$. The Higgs term is:
\begin{equation*}
\begin{aligned}
S_h =& (\Psi_L, \tilde{q}_\Phi Y \Psi_R)_0 + (\Psi_R, Y^\dagger \tilde{q}_\Phi^\dagger \Psi_L)_0 \\ &- \frac{s}{2} (J_{M-} \Psi_R, M \Psi_R)_0 + \frac{s\epsilon_F}{2} (M\Psi_R, J_{M-} \Psi_R)_0.
\end{aligned}
\end{equation*}
We know that the action must be real. The first two terms are the complex conjugate of each other. Therefore, the last two terms have to be as well. This is obviously true if $\epsilon_F= -1$. If $\epsilon_F = +1$, the Majorana mass term $(J_{M-} \Psi_R, M \Psi_R)_0$ must vanish. Let us check that it is indeed the case. Using $M^T = s\epsilon_F M$, we have:
\begin{equation*}
\begin{aligned}
\epsilon_F (M\Psi_R, J_{M-} \Psi_R)_0 &= s(M^T \Psi_R, J_{M-} \Psi_R)_0 \\
&= -(\Psi_R, J_{M-} M^T \Psi_R)_0 \\
&= -(\Psi_R, M^\dagger J_{M-} \Psi_R)_0 \\
&= -(M\Psi_R, J_{M-} \Psi_R)_0.
\end{aligned}
\end{equation*}
This proves that $(M\Psi_R, J_{M-} \Psi_R)_0 = 0$ for $\epsilon_F = +1$.

To have a nonvanishing Majorana mass term, we need to have $\epsilon_F = -1$. As we saw earlier, a proper seesaw mechanism requires that $s\epsilon_F=1$, and thus that $s=-1$. From previous results, we see that $(n_F,m_F) = (2,6)$. This means that $(n,m) = (0,2)$, and that $J$ and $\chi$ commute, and that $J$ squares to 1. One can then prove that the special form \eqref{PsiDef} of $\Psi$ is equivalent to the so-called Majorana-Weyl conditions \cite{Barrett}:
\begin{equation}
\begin{aligned}
J \Psi =& \Psi \\
\chi \Psi =& \Psi.
\end{aligned}
\end{equation}
We also see that \emph{$\Psi$ needs to be an anticommuting variable}. Let us now summarize the Fermionic action we just computed:
\begin{equation*}
\begin{aligned}
S_f =& (\Psi_R, (\slashed{D} \otimes 1) \Psi_R)_0 + (\Psi_L, (\slashed{D} \otimes 1) \Psi_L)_0 + (\Psi_R, \gamma^\mu \otimes [(A_\mu^Y T_Y^R - I_2 \otimes \overline{A_\mu^C}) \otimes I_N] \Psi_R)_0 \\ &+ (\Psi_L, \gamma^\mu \otimes [(A_\mu^Y T_Y^L + (A_\mu^W \oplus A_\mu^W \otimes I_3) - I_2 \otimes \overline{A_\mu^C}) \otimes I_N] \Psi_L)_0 + (\Psi_L, \tilde{q}_\Phi Y \Psi_R)_0 \\ &+ (\Psi_R, Y^\dagger \tilde{q}_\Phi^\dagger \Psi_L)_0 + \frac{1}{2} (J_{M-} \Psi_R, M \Psi_R)_0 + \frac{1}{2} (M\Psi_R, J_{M-} \Psi_R)_0.
\end{aligned}
\end{equation*}

We will now extract a fermionic Lagrangian density out of this action. For this purpose, we write the product $(\cdot, \cdot)_0$ as an integral:
\begin{equation*}
(\varphi \otimes u, \psi \otimes v)_0 = (\varphi, \psi)_M u^\dagger v = \int_{x \in M} H_x (\varphi,\psi) u^\dagger v \sqrt{|g|} d^d x.
\end{equation*}
One can thus write:
\begin{equation}
(\Phi, \Psi)_0 =\int_{x \in M} \overline{\Phi}(x) \Psi(x) \sqrt{|g|} d^d x,
\end{equation}
where:
\begin{equation}
\begin{aligned}
S_x \otimes \mathcal{K}_0 &\longrightarrow S_x^\ast \otimes \mathcal{K}_0^\ast \\
\varphi(x) \otimes u &\longmapsto \overline{\varphi \otimes u}(x) = H_x (\varphi, \cdot) \otimes u^\dagger.
\end{aligned}
\end{equation}
is a map that associates to each multi-spinor a "multi-cospinor". In the physics literature, the co-spinor $H_x (\varphi, \cdot)$ with respect to the Robinson product is usually notated $\overline{\varphi}(x)$. The operation above generalizes this duality to multi-spinors that contain multiple particle species. We can now write the fermionic action above as the integral of the following Lagrangian density:
\begin{equation}
\begin{aligned}
\mathscr{L}_f =& \overline{\Psi_R}(\slashed{D} \otimes 1)\Psi_R + \overline{\Psi_L}(\slashed{D} \otimes 1)\Psi_L + \overline{\Psi_R} \left[\gamma^\mu \otimes (A_\mu^Y T_Y^R - I_2 \otimes \overline{A_\mu^C}) \otimes I_N \right] \Psi_R \\ &+ \overline{\Psi_L} \left[\gamma^\mu \otimes (A_\mu^Y T_Y^L + (A_\mu^W \oplus A_\mu^W \otimes I_3) - I_2 \otimes \overline{A_\mu^C}) \otimes I_N \right] \Psi_L + \overline{\Psi_L} \tilde{q}_\Phi Y \Psi_R \\ &+ \overline{\Psi_R} Y^\dagger \tilde{q}_\Phi^\dagger \Psi_L + \frac{1}{2} \overline{J_{M-} \Psi_R} M \Psi_R + \frac{1}{2} \overline{M \Psi_R} J_{M-} \Psi_R.
\end{aligned}
\end{equation}

\subsection{The Bosonic Action}

Let us compute the bosonic action. In order to do that, we first constrain the $z$ matrix. Let us first recall that the bosonic Lagrangian density is given by:
\begin{equation*}
\mathscr{L}_b = 2\mathrm{tr}(z F_{\mu\nu}F^{\mu\nu}) -4 \mathrm{tr}(z D_\mu H D^\mu H) - 4 \mathrm{tr}(z P(H^2 + d_{F,U} H)^2).
\end{equation*}
We saw in equations \eqref{GaugeCurvature} and \eqref{HiggsProjection} that the operators $F_{\mu\nu}$ (and thus $F_{\mu\nu}F^{\mu\nu}$) and $P(H^2 + d_{F,U} H)^2$ are block diagonal. From \eqref{HiggsDerivative}, one can prove that $D_\mu H D^\mu H$ is block diagonal as well. Therefore, through the properties of the Hilbert-Schimdt product of matrices, only the diagonal blocks of $z$ contribute to the Lagrangian. We will from now on assume that $z$ is of the form:
\begin{equation}
z = \begin{pmatrix} z_R &&& \\ & z_L && \\ && z_{\overline{R}} & \\ &&& z_{\overline{L}} \end{pmatrix}.
\end{equation}
We see that it commutes with $\chi_F, \eta_F$ and $\varpi$. The requirement $z^\times = z$ is thus equivalent to $z^\dagger = z$. We saw in the previous chapter that $z$ must commute with all anti-self-adjoint elements of $\pi_F(A_F)$. One such element is $I = \pi_F (i, Q_i, iI_3)$, which by multiplication turns any anti-self-adjoint element into a self-adjoint element of $\pi_F(A_F)$, with which $z$ must commute. We deduce that $z$ must commute with all elements of $\pi_F(A_F)$. The commutation of $z$ with the algebra is equivalent to the set of equations:
\begin{equation}
\forall a \in A_F:
\begin{aligned}
{[}a_R,z_R] &= 0 \\
[a_L,z_L] &= 0 \\
[a_{\overline{R}},z_{\overline{R}}] &= 0 \\
[a_{\overline{L}},z_{\overline{L}}] &= 0,
\end{aligned}
\end{equation}
while the self-adjointness condition $z^\dagger = z$ is equivalent to the self-adjointness of the blocks.

To solve these equations and determine $z$, It is better to use a slightly different representation for the remainder of this subsection. Instead of writing:
\begin{equation*}
\mathcal{K}_0 = \left(\mathbb{C}_l^2 \oplus \mathbb{C}_q^2 \otimes \mathbb{C}_c^3 \right) \otimes \mathbb{C}_g^N,
\end{equation*}
we will put the spaces $\mathbb{C}_l^2$ and $\mathbb{C}_q^2$ as a common isospin factor $\mathbb{C}_W^2$, where the neutrino and up quark are mapped to the same basis vector $u$, and the electron and down quark are mapped to the other basis vector $d$:
\begin{equation}
\mathcal{K}_0 = \mathbb{C}_W^2 \otimes \left(\mathbb{C} \oplus \mathbb{C}_c^3 \right) \otimes \mathbb{C}_g^N.
\end{equation}
This is nothing but the Pati-Salam representation, where the leptons are unified with quarks as an additional white quark. For $a = (\lambda, q, m) \in A_F$, we have:
\begin{equation}
\begin{aligned}
a_R &= \tilde{Q}_\lambda \\
a_L &= \tilde{q} \\
a_{\overline{R}} &= a_{\overline{L}} = I_2 \otimes (\lambda \oplus m) \otimes I_N.
\end{aligned}
\end{equation}
where this time: $\tilde{T} = T \otimes (1 \oplus I_3) \otimes I_N$.

By linearity, the commutation equations of $z$ must hold for the complexification of the algebra. In particular, on can replace $Q_\lambda$ with an arbitrary $2\times 2$ complex diagonal matrix, and $q$ with an arbitrary $2 \times 2$ complex matrix. The first equation we have to solve is thus:
\begin{equation*}
\left[ \begin{pmatrix} \lambda & \\ & \mu \end{pmatrix} \otimes (1 \oplus I_3) \otimes I_N,  z_R \right] = 0,
\end{equation*}
with $\lambda, \mu$ arbitrary complex numbers. The block $z_R$ thus has to be diagonal with respect to the first factor of the tensor product, and we write this as:
\begin{equation}
z_R = E_{uu} \otimes A + E_{dd} \otimes B,
\end{equation}
with $A$, $B$ self-adjoint operators on $\left(\mathbb{C} \oplus \mathbb{C}_c^3 \right) \otimes \mathbb{C}_g^N$. The second equation is:
\begin{equation*}
\left[ T \otimes (1 \oplus I_3) \otimes I_N,  z_L \right] = 0,
\end{equation*}
with $T$ an arbitrary $2 \times 2$ complex matrix. We infer that $z_L$ has to be scalar for the first tensor factor:
\begin{equation}
z_L = I_2 \otimes C,
\end{equation}
with $C$ a self-adjoint operator on $\left(\mathbb{C} \oplus \mathbb{C}_c^3 \right) \otimes \mathbb{C}_g^N$. Finally, the last two equations are identical. For $z_{\overline{R}}$ it is:
\begin{equation*}
\left[ I_2 \otimes (\lambda \oplus m) \otimes I_N,  z_{\overline{R}} \right] = 0.
\end{equation*}
Since $\lambda$ and $m$ are independent, $z_{\overline{R}}$ has be to block diagonal for the second tensor factor $\left(\mathbb{C} \oplus \mathbb{C}_c^3 \right)$. In other words, it is of the form:
\begin{equation*}
z_{\overline{R}} = E_{uu} \otimes (\alpha_{uu} \otimes D_{1uu} \oplus \beta_{uu} \otimes D'_{1uu}) + \dots
\end{equation*}
with similar terms for $E_{ud}, E_{du}, E_{dd}$. Here $\beta_{uu}, \dots$ are operators on $\mathbb{C}_c^3$ and $D_{1uu}, D'_{1uu}, \dots$ are operators on $\mathbb{C}_g^N$. Since $\beta_{uu}, \dots$ have to commute with $m$, they must be scalar, and we conclude that:
\begin{equation}
z_{\overline{R}} = E_{uu} \otimes (1 \otimes D_{1uu} \oplus I_3 \otimes D'_{1uu}) + \dots,
\end{equation}
(we absorbed the $\alpha_{uu}, \beta_{uu}, \dots$ in the $D_{1uu}, D'_{1uu}, \dots$). The selfadjointness of $z_{\overline{R}}$ is equivalent to:
\begin{equation}
\begin{aligned}
D_{1uu}^\dagger &= D_{1uu}, &D_{1uu}^{\prime\dagger} &= D'_{1uu} \\
D_{1dd}^\dagger &= D_{1dd}, &D_{1dd}^{\prime\dagger} &= D'_{1dd} \\
D_{1ud}^\dagger &= D_{1du}, &D_{1ud}^{\prime\dagger} &= D'_{1du}.
\end{aligned}
\end{equation}
The block $z_{\overline{L}}$ has a similar form:
\begin{equation}
z_{\overline{L}} = E_{uu} \otimes (1 \otimes D_{2uu} \oplus I_3 \otimes D'_{2uu}) + \dots.
\end{equation}
We define for future use the operators $D_{uu} = D_{1uu}+D_{2uu}, \dots$, such that:
\begin{equation}
z_{\overline{R}} + z_{\overline{L}} = E_{uu} \otimes (1 \otimes D_{uu} \oplus I_3 \otimes D'_{uu}) + \dots.
\end{equation}

Let us now compute the Lagrangian. The first term we have to compute is the gauge kinetic term $\mathrm{tr}(z F_{\mu\nu}F^{\mu\nu})$. Recall that:
\begin{equation*}
\setlength\arraycolsep{-6pt}
\renewcommand{\arraystretch}{2}
F_{\mu \nu} = -i \begin{pmatrix} F^Y_{\mu\nu} \tilde{\sigma}_z &&& \\ & \widetilde{F^W_{\mu\nu}} && \\ && I_2 \otimes (F^Y_{\mu\nu} \oplus  (F^C_{\mu\nu} -\frac{1}{3} F^Y_{\mu\nu} I_3)) \otimes I_N & \\ &&& I_2 \otimes (F^Y_{\mu\nu} \oplus  (F^C_{\mu\nu} -\frac{1}{3} F^Y_{\mu\nu} I_3)) \otimes I_N \end{pmatrix}.
\end{equation*}
Substituting we find:
\begin{equation*}
\begin{aligned}
\mathrm{tr}(z F_{\mu\nu}F^{\mu\nu}) =& -\mathrm{tr}(z_R \tilde{\sigma}_z^2) F^Y_{\mu\nu} F^{Y \mu\nu} - \mathrm{tr}(z_L \widetilde{F^W_{\mu\nu}F^{W \mu\nu}}) \\ &- \mathrm{tr}[(z_{\overline{R}} + z_{\overline{L}})(I_2 \otimes (F^Y_{\mu\nu} \oplus  (F^C_{\mu\nu} -\frac{1}{3} F^Y_{\mu\nu} I_3)) \otimes I_N)^2].
\end{aligned}
\end{equation*}
Since $\sigma_z^2 = 1$, the factor in the first gauge term is simply $\mathrm{tr}(z_R \tilde{\sigma}_z^2) = \mathrm{tr}(z_R) = \mathrm{tr}(A+B)$. To compute the two other terms, let us observe that the trace is multiplicative with respect to tensor products: $\mathrm{tr}(S \otimes T) = \mathrm{tr}(S) \mathrm{tr}(T)$. The second gauge term is:
\begin{equation*}
\mathrm{tr}(z_L \widetilde{F^W_{\mu\nu}F^{W \mu\nu}}) = \mathrm{tr}(F^W_{\mu\nu}F^{W \mu\nu} \otimes C) = \mathrm{tr}(C) \mathrm{tr}(F^W_{\mu\nu}F^{W \mu\nu}).
\end{equation*}
Similarly, the third gauge term is:
\begin{equation*}
\begin{aligned}
\mathrm{tr}[(z_{\overline{R}} + z_{\overline{L}}) & (I_2 \otimes (F^Y_{\mu\nu} \oplus  (F^C_{\mu\nu} -\frac{1}{3} F^Y_{\mu\nu} I_3)) \otimes I_N)^2 ] \\ =& \mathrm{tr} [(E_{uu} \otimes (1 \otimes D_{uu} \oplus I_3 \otimes D'_{uu}) + \dots) \\ &\times (I_2 \otimes (F^Y_{\mu\nu}F^{Y \mu\nu} \oplus  (F^C_{\mu\nu} -\frac{1}{3} F^Y_{\mu\nu} I_3)^2) \otimes I_N) ] \\
=& \mathrm{tr} [E_{uu} \otimes (F^Y_{\mu\nu}F^{Y \mu\nu} \otimes D_{uu} \oplus (F^C_{\mu\nu} -\frac{1}{3} F^Y_{\mu\nu} I_3)^2 \otimes D'_{uu}) + \dots] \\
=& \mathrm{tr}(E_{uu} \otimes D_{uu}+\dots) F^Y_{\mu\nu}F^{Y \mu\nu} + \mathrm{tr}(E_{uu} \otimes D'_{uu}+\dots) \mathrm{tr}((F^C_{\mu\nu} -\frac{1}{3} F^Y_{\mu\nu} I_3)^2) \\
=& \mathrm{tr}(D_{uu}+D_{dd}) F^Y_{\mu\nu}F^{Y \mu\nu} + \mathrm{tr}(D'_{uu}+D'_{dd}) \left[\frac{1}{3} F^Y_{\mu\nu} F^{Y \mu\nu} + \mathrm{tr}(F^C_{\mu\nu} F^{C \mu\nu}) \right],
\end{aligned}
\end{equation*}
where used the tracelessness of $F^C_{\mu\nu}$. As a result:
\begin{equation*}
\begin{aligned}
\mathrm{tr} & \left[(z_{\overline{R}} + z_{\overline{L}})(I_2 \otimes (F^Y_{\mu\nu} \oplus  (F^C_{\mu\nu} -\frac{1}{3} F^Y_{\mu\nu} I_3)) \otimes I_N)^2 \right] \\ 
&= \mathrm{tr} \left( D_{uu}+D_{dd}+\frac{D'_{uu}+D'_{dd}}{3} \right) F^Y_{\mu\nu}F^{Y \mu\nu} + \mathrm{tr}(D'_{uu}+D'_{dd}) \mathrm{tr}(F^C_{\mu\nu} F^{C \mu\nu}).
\end{aligned}
\end{equation*}
The total gauge term is thus:
\begin{equation}
\begin{aligned}
\mathrm{tr}(z F_{\mu\nu}F^{\mu\nu}) =& -\left[ \mathrm{tr}(A+B) + \mathrm{tr}\left( D_{uu}+D_{dd}+\frac{D'_{uu}+D'_{dd}}{3} \right) \right] F^Y_{\mu\nu}F^{Y \mu\nu} \\ &- \mathrm{tr}(C) \mathrm{tr}(F^W_{\mu\nu}F^{W \mu\nu}) - \mathrm{tr}(D'_{uu}+D'_{dd}) \mathrm{tr}(F^C_{\mu\nu} F^{C \mu\nu}).
\end{aligned}
\end{equation}
We compute next the Higgs kinetic term $\mathrm{tr}(z D_\mu H D^\mu H)$. We have:
\begin{equation*}
D_\mu H = \begin{pmatrix} 0 & -Y^\dagger \widetilde{D_\mu q_\Phi}^\dagger && \\  \widetilde{D_\mu q_\Phi} Y & 0 && \\ && 0 & \\ &&& 0 \end{pmatrix},
\end{equation*}
and thus that:
\begin{equation*}
D_\mu H D^\mu H = \begin{pmatrix} -Y^\dagger \widetilde{(D_\mu q_\Phi)^\dagger D^\mu q_\Phi} Y &&& \\  & -\widetilde{D_\mu q_\Phi} YY^\dagger \widetilde{D_\mu q_\Phi}^\dagger && \\ && 0 & \\ &&& 0 \end{pmatrix}.
\end{equation*}
We infer that:
\begin{equation*}
\mathrm{tr}(z D_\mu H D^\mu H) = -\mathrm{tr}(z_R Y^\dagger \widetilde{(D_\mu q_\Phi)^\dagger D^\mu q_\Phi} Y) - \mathrm{tr}(z_L \widetilde{D_\mu q_\Phi} YY^\dagger \widetilde{D_\mu q_\Phi}^\dagger).
\end{equation*}
Note that $z_L$ is a scalar for doublet degrees of freedom, and thus commutes with $\widetilde{D_\mu q_\Phi}$. The second term can thus be rewritten (using the cyclic property of the trace): $\mathrm{tr}(z_L YY^\dagger \widetilde{D_\mu q_\Phi}^\dagger \widetilde{D_\mu q_\Phi})$. The total kinetic term is now:
\begin{equation*}
\mathrm{tr}(z D_\mu H D^\mu H) = -\mathrm{tr}(z_R Y^\dagger \widetilde{(D_\mu q_\Phi)^\dagger D^\mu q_\Phi} Y) - \mathrm{tr}(z_L YY^\dagger \widetilde{(D_\mu q_\Phi)^\dagger D^\mu q_\Phi}).
\end{equation*}
The quaternion $q_\Phi$ can be parametrized by two complex numbers $\alpha,\beta$:
\begin{equation*}
q=\begin{pmatrix} \alpha & \beta \\ -\overline{\beta} & \overline{\alpha}	\end{pmatrix}.
\end{equation*}
From the covariant derivative $D_\mu q_\Phi$, we see that the columns of $q_\Phi$ transform as $\mathrm{SU}(2)$ doublets. We will thus use the doublet:
\begin{equation}
\Phi=\begin{pmatrix} \beta \\ \overline{\alpha} \end{pmatrix}
\end{equation}
as a Higgs field. Note that the second column of $q_\Phi Q_i$ is $-i\Phi$. We will take the second column of $D_\mu q_\Phi$ to be its covariant derivative, and it is given by:
\begin{equation}
D_\mu \Phi = (\partial_\mu - iA_\mu^W - iA_\mu^Y) \Phi. 
\end{equation}
Note that $(D^\mu q_\Phi)^\dagger D_\mu q_\Phi = (D_\mu \Phi)^\dagger (D^\mu \Phi) I_2$. Indeed, $D_\mu q_\Phi = \partial_\mu q_\Phi - iA_\mu^W q_\Phi + A_\mu^Y q_\Phi Q_i$ is a quaternion (since $Q_i$ and $iA_\mu^W$ are quaternions too). And like any quaternion, its norm is that of its first column or second column. For example, for $q_\Phi$ itself we have: $q_\Phi^\dagger q_\Phi = |\Phi|^2 I_2$. We thus have:
\begin{equation*}
\widetilde{(D_\mu q_\Phi)^\dagger D^\mu q_\Phi} = (D_\mu \Phi)^\dagger (D^\mu \Phi) [I_2 \otimes (1 \oplus I_2) \otimes I_N].
\end{equation*}
We conclude that:
\begin{equation}
\mathrm{tr}(z D_\mu H D^\mu H) = -\mathrm{tr}(z_R Y^\dagger Y + z_L YY^\dagger) (D_\mu \Phi)^\dagger D^\mu \Phi.
\end{equation}
The matrix $Y$ is given by:
\begin{equation*}
Y = E_{uu} \otimes (1 \otimes Y_\nu \oplus I_3 \otimes Y_u) + E_{dd} \otimes (1 \otimes Y_e \oplus I_3 \otimes Y_d).
\end{equation*}
A straightforward computation then shows that:
\begin{equation}
\begin{aligned}
\mathrm{tr}(z_R Y^\dagger Y + z_L YY^\dagger) =& \mathrm{tr}[ A (1 \otimes Y_\nu^\dagger Y_\nu \oplus I_3 \otimes Y_u^\dagger Y_u) + B (1 \otimes Y_e^\dagger Y_e \oplus I_3 \otimes Y_d^\dagger Y_d) \\ &+ C (1 \otimes (Y_\nu Y_\nu^\dagger + Y_e Y_e^\dagger) \oplus I_3 (Y_u Y_u^\dagger + Y_d Y_d^\dagger)) ].
\end{aligned}
\end{equation}
We finally compute the last term $\mathrm{tr}(z P(H^2 + d_{F,U} H)^2)$. We know that $P(H^2 + d_{F,U} H)$ is of the form:
\begin{equation*}
P(H^2 + d_{F,U} H) = (|q_H|^2 + 2\mathrm{Re}(q_H)) \times M_H
\end{equation*}
with $M_H$ a block diagonal operator on $\mathcal{K}_F$. As a result:
\begin{equation*}
\mathrm{tr}(z P(H^2 + d_{F,U} H)^2) = \mathrm{tr}(z M_H^2) (|q_H|^2 + 2\mathrm{Re}(q_H))^2
\end{equation*}
We also have:
\begin{equation*}
\begin{aligned}
|\Phi|^2 &= |q_\Phi|^2 \\
&= q_\Phi^\dagger q_\Phi \\
&= (1 + q_H^\dagger)(1 + q_H) \\
&= 1 + (q_H^\dagger + q_H) + q_H^\dagger q_H \\
|\Phi|^2 &= 1+ 2\mathrm{Re}(q_H) + |q_H|^2.
\end{aligned}
\end{equation*}
We thus have:
\begin{equation*}
\mathrm{tr}(z P(H^2 + d_{F,U} H)^2) = \mathrm{tr}(z M_H^2) (|\Phi|^2 - 1)^2
\end{equation*}

The complete Lagrangian density is thus of the form:
\begin{equation}
\mathscr{L}_b = - a F^Y_{\mu\nu}F^{Y\mu\nu} - b \mathrm{tr}(F^W_{\mu\nu}F^{W\mu\nu}) - c \mathrm{tr}(F^C_{\mu\nu}F^{C\mu\nu}) + d (D_\mu \Phi)^\dagger D^\mu \Phi - e (|\Phi|^2 - 1)^2,
\end{equation}
with:
\begin{equation}
\begin{aligned}
a &= 2\mathrm{tr}(A+B) + 2\mathrm{tr}(D_{uu}+D_{dd}+\frac{D'_{uu}+D'_{dd}}{3}) \\
b &= 2\mathrm{tr}(C) \\
c &= 2\mathrm{tr}(D'_{uu}+D'_{dd}) \\
d &= 4\mathrm{tr}(z_R Y^\dagger Y + z_L YY^\dagger) \\
e &= 4 \mathrm{tr}(z M_H^2).
\end{aligned}
\end{equation}
Barring any accident, these coefficients are all linearly independent in $z$. Before we prove that, let us determine their dimensions, and thus the dimension of $z$, in a unit system where $\hbar = c = 1$. For simplicity, we choose a coordinate system where all coordinate functions have dimensions of length. The metric will thus be dimensionless. The action must be dimensionless, which means that the bosonic Lagrangian density must have the dimension: $[\mathscr{L}_b] = L^{-4}$, and so does its second term $b \mathrm{tr}(F^W_{\mu\nu}F^{W\mu\nu})$. We know that $F^W_{\mu\nu} = \partial_\mu A^W_\nu - \partial_\nu A^W_\mu - i[A^W_\mu, A^W_\nu]$, and by comparing the first and last terms, we see that $[A^W_\mu] = L^{-1}$. We deduce (taking into account the dimensionlessness of the metric) that $[F^W_{\mu\nu}F^{W\mu\nu}] = L^{-4}$, which means that $b$ must be dimensionless. From its expression, we deduce that $\mu, \nu$ are dimensionless. We conclude that $z$ is dimensionless (we are thus allowed to choose $z=1$). We infer that $\alpha,\beta,\dots$ are dimensionless, and that $a,b,c$ are as well. We also infer that $[d] = M^2$ and $[e] = M^4$. Note that in this unit system $M = L^{-1}$.

To prove that the parameters of the Lagrangian are indeed all independent, let us pick a specific $z$ with just enough degrees of freedom. To reduce the degrees of freedom, we will try to pick a $z$ that satisfies a certain number of (strong) constraints that appeared in the literature. These constraints are the following:
\begin{equation}
\begin{aligned}
{[}\pi_F(A_F), z{]} &= 0 \\
[\chi_F, z] &= 0 \\
[J_F, z] &= 0 \\
[D_F, z] &= 0.
\end{aligned}
\end{equation}
Note that these constraints automatically imply that:
\begin{equation}
\begin{aligned}
{[}\pi(A), Z {]} &= 0 \\
[\chi, Z] &= 0 \\
[J, Z] &= 0 \\
[D, Z] &= 0.
\end{aligned}
\end{equation}
We will also see that the last constraint is not compatible with a simple solution. We have already imposed the first two constraints on $z$. The third constraint means that we must have:
\begin{equation}
\begin{aligned}
z_{\overline{R}} &= \overline{z_R} \\
z_{\overline{L}} &= \overline{z_L}.
\end{aligned}
\end{equation}
Recall now that:
\begin{equation}
\begin{aligned}
z_R &= E_{uu} \otimes A + E_{dd} \otimes B \\
z_L &= I_2 \otimes C \\
z_{\overline{R}} &= E_{uu} \otimes (1 \otimes D_{1uu} \oplus I_3 \otimes D'_{1uu}) + \dots \\
z_{\overline{L}} &= E_{uu} \otimes (1 \otimes D_{2uu} \oplus I_3 \otimes D'_{2uu}) + \dots
\end{aligned}
\end{equation}
By comparing $z_R$ and $z_{\overline{R}}$, we deduce that:
\begin{equation}
\begin{aligned}
A &= 1 \otimes \overline{D_{1uu}} \oplus I_3 \otimes \overline{D'_{1uu}} \\
B &= 1 \otimes \overline{D_{1dd}} \oplus I_3 \otimes \overline{D'_{1dd}} \\
D_{1ud} &= D'_{1ud} = D_{1du} = D'_{1du} = 0.
\end{aligned}
\end{equation}
Similarly, we find for $z_L$ and $z_{\overline{L}}$ that:
\begin{equation}
\begin{aligned}
C &= 1 \otimes \overline{D_{2uu}} \oplus I_3 \otimes \overline{D'_{2uu}} \\
D_{2uu} &= D_{2dd} \\
D'_{2uu} &= D'_{2dd} \\
D_{1ud} &= D'_{1ud} = D_{1du} = D'_{1du} = 0.
\end{aligned}
\end{equation}
We can now write that:
\begin{equation}
\begin{aligned}
z_R =& E_{uu} \otimes (1 \otimes \overline{D_{1uu}} \oplus I_3 \otimes \overline{D'_{1uu}}) \\ &+ E_{dd} \otimes (1 \otimes \overline{D_{1dd}} \oplus I_3 \otimes \overline{D'_{1dd}}) \\
z_L =& I_2 \otimes (1 \otimes \overline{D_{2uu}} \oplus I_3 \otimes \overline{D'_{2uu}}) \\
z_{\overline{R}} =& \overline{z_R} \\
z_{\overline{L}} =& \overline{z_L}.
\end{aligned}
\end{equation}
The remaining degrees of freedom are the self-adjoint matrices $D_{1uu}, D'_{1uu}, D_{1dd}, D'_{1dd}, D_{2uu}$, $D'_{2uu}$. Let us now impose the commutation of $z$ with $D_F$. We have:
\begin{equation}
{[}D_F, z{]} = \begin{pmatrix}
0 & -Y^\dagger z_L + Y^\dagger z_R & \epsilon_F (\overline{M} \overline{z_R} - z_R \overline{M}) & 0 \\
Y z_R - z_L Y & 0 & 0 & \epsilon_F \overline{Z} \\
M z_R - \overline{z_R} M & 0 & 0 & -Y^T \overline{z_L} + \overline{z_R} Y^T \\
0 & Z & \overline{Y} \overline{z_R} - \overline{z_L} \overline{Y} & 0
\end{pmatrix}.
\end{equation}
The commutator vanishes if and only if:
\begin{equation}
\begin{aligned}
Y z_R - z_L Y &= 0 \\
M z_R - \overline{z_R} M &= 0,
\end{aligned}
\end{equation}
with all other the equations being a consequence of these two through complex conjugation, transposition, and of course hermitian conjugation. We have:
\begin{equation}
\begin{aligned}
Y &= E_{uu} \otimes (1 \otimes Y_{\nu} \oplus I_3 \otimes Y_u) + E_{dd} \otimes (1 \otimes Y_e \oplus I_3 \otimes Y_d) \\
M &= E_{uu} \otimes (1 \oplus 0) \otimes Y_R,
\end{aligned}
\end{equation}
from which we compute that:
\begin{equation}
\begin{aligned}
Y z_R - z_L Y =& E_{uu} \otimes (1 \otimes (Y_{\nu} \overline{D_{1uu}} - \overline{D_{2uu}} Y_{\nu}) \oplus I_3 \otimes (Y_u \overline{D'_{1uu}} - \overline{D'_{2uu}} Y_u)) \\ &+ E_{dd} \otimes (1 \otimes (Y_e \overline{D_{1dd}} - \overline{D_{2uu}} Y_e) \oplus I_3 \otimes (Y_d \overline{D'_{1dd}} - \overline{D'_{2uu}} Y_d)) \\
M z_R - \overline{z_R} M =&  E_{uu} \otimes (1 \oplus 0) \otimes (Y_R \overline{D_{1uu}} - D_{1uu} Y_R).
\end{aligned}
\end{equation}
For simplicity, we want to assume that $z$ is generation-blind. That is, that all six matrices $D_{1uu}, \dots$ are scalar matrices. Since they all have to be self-adjoint, they must be real scalar matrices. We thus choose six real numbers $\alpha, \beta, \gamma, \delta, \mu, \nu \in \mathbb{R}$ such that:
\begin{equation}
\begin{aligned}
D_{1uu} &= \alpha I_N \\
D'_{1uu} &= \beta I_N \\
D_{1dd} &= \gamma I_N \\
D'_{1dd} &= \delta I_N \\
D_{2uu} &= \mu I_N \\
D'_{2uu} &= \nu I_N.
\end{aligned}
\end{equation}
The two equations above become:
\begin{equation}
\begin{aligned}
Y z_R - z_L Y =& E_{uu} \otimes (1 \otimes (\alpha-\mu) Y_{\nu} \oplus I_3 \otimes (\beta-\nu)Y_u) \\ &+ E_{dd} \otimes (1 \otimes (\gamma-\mu)Y_e \oplus I_3 \otimes (\delta-\nu)Y_d) \\
M z_R - \overline{z_R} M =&  0.
\end{aligned}
\end{equation}
The second equation is automatically satisfied, but the first one imposes that:
\begin{equation}
\begin{aligned}
\alpha &= \gamma = \mu \\
\beta &= \delta = \nu.
\end{aligned}
\end{equation}
We thus only have two degrees of freedom left, and the parameters $a,b,c,d,e$ cannot be independent, with the assumptions of generation-blindness and commutation of $D_F$ and $z$. We will thus drop the second assumption.

Our solution for $z$ is thus:
\begin{equation}
\begin{aligned}
z_R = z_{\overline{R}} =& E_{uu} \otimes (\alpha \oplus \beta I_3) \otimes I_N + E_{dd} \otimes (\gamma \oplus \delta I_3) \otimes I_N \\
z_L = z_{\overline{L}} =& I_2 \otimes (\mu \oplus \nu I_3) \otimes I_N.
\end{aligned}
\end{equation}
Let us now compute the parameters of the bosonic Lagrangian $a,b,c,d,e$. The matrices $A,B,C$ are now:
\begin{equation}
\begin{aligned}
A =& (\alpha \oplus \beta I_3) \otimes I_N \\
B =& (\gamma \oplus \delta I_3) \otimes I_N \\
C =& (\mu \oplus \nu I_3) \otimes I_N.
\end{aligned}
\end{equation}
and the reduced $D$ matrices are:
\begin{equation}
\begin{aligned}
D_{uu} &= (\alpha + \mu) \otimes I_N \\
D'_{uu} &= (\beta + \nu) \otimes I_N \\
D_{dd} &= (\gamma + \mu) \otimes I_N \\
D'_{dd} &= (\delta + \nu) \otimes I_N.
\end{aligned}
\end{equation}
The first three parameters can be computed straightforwardly:
\begin{equation}
\begin{aligned}
a &= \frac{4N}{3}(3 \alpha+ 5 \beta + 3 \gamma + 5 \delta + 3 \mu + \nu) \\
b &= 2N (\mu + 3 \nu) \\
c &= 2N (\beta + \delta + 2 \nu).
\end{aligned}
\end{equation}
The parameter $d$ is given by $d = 4\mathrm{tr}(z_R Y^\dagger Y + z_L YY^\dagger)$. Given the block diagonal forms of $Y, z_R, z_L$, we conclude that these 3 matrices commute with each other, and that:
\begin{equation}
\begin{aligned}
d &= 4\mathrm{tr}(z_R Y^\dagger Y + z_L YY^\dagger) \\
&= 4\mathrm{tr}(Y z_R Y^\dagger + z_L YY^\dagger) \\
&= 4\mathrm{tr}(z_R YY^\dagger + z_L YY^\dagger) \\
d &= 4\mathrm{tr}((z_R + z_L) YY^\dagger).
\end{aligned}
\end{equation}
Using:
\begin{equation}
\begin{aligned}
YY^\dagger &= E_{uu} \otimes (1 \otimes m_{\nu} \oplus I_3 \otimes m_u) + E_{dd} \otimes (1 \otimes m_e \oplus I_3 \otimes m_d) \\
z_R + z_L &= E_{uu} \otimes ((\alpha+\mu) \oplus (\beta+\nu) I_3) \otimes I_N + E_{dd} \otimes ((\gamma+\mu) \oplus (\delta+\nu) I_3) \otimes I_N,
\end{aligned}
\end{equation}
we find:
\begin{equation}
\begin{aligned}
d =& 4\mathrm{tr}((z_R + z_L) YY^\dagger) \\
=& 4(\alpha+\mu)\mathrm{tr}(m_{\nu}) + 12(\beta+\nu)\mathrm{tr}(m_u) + 4(\gamma+\mu)\mathrm{tr}(m_e) + 12(\delta+\nu)\mathrm{tr}(m_d) \\
d =& \mathrm{tr}(4 m_{\nu}) \alpha + \mathrm{tr}(12 m_u) \beta + \mathrm{tr}(4 m_e) \gamma + \mathrm{tr}(12 m_d) \delta + \mathrm{tr}(4 (m_{\nu} + m_e)) \mu \\ &+ \mathrm{tr}(12(m_u + m_d)) \nu.
\end{aligned}
\end{equation}
Finally, let us compute $e$. We know that $e = 4 \mathrm{tr}(z M_H^2)$, and:
\begin{equation}
\setlength\arraycolsep{-6pt}
\renewcommand{\arraystretch}{1.5}
M_H = \begin{pmatrix} Y^\dagger Y - \frac{C_1}{12N}I_{8N} &&& \\ & YY^\dagger + \frac{j_{Q_i}}{2} - \frac{C_1}{8N} I_{8N} && \\ && - \frac{1}{12N} C_1 I_{2} \otimes (1 \oplus 0) \otimes I_N & \\ &&& - \frac{1}{12N} C_1 I_{2} \otimes (1 \oplus 0) \otimes I_N \end{pmatrix}.
\end{equation}
We call its blocks $M_R, M_L, M_{\overline{R}}, M_{\overline{L}}$ respectively, with the last two blocks being equal. We thus have:
\begin{equation}
e = 4 \mathrm{tr}(z_R M_R^2) + 4 \mathrm{tr}(z_L M_L^2) + 4 \mathrm{tr}((z_R + z_L) M_{\overline{R}}^2).
\end{equation}
These terms are all straightforward to compute if one uses the fact that all matrices that appear in this expression are block diagonal. The first term is:
\begin{equation}
\begin{aligned}
\mathrm{tr}(z_R M_R^2) =& \mathrm{tr}[z_R (Y^\dagger Y - \frac{C_1}{12N}I_{8N})^2] \\
=& \mathrm{tr}[(m_{\nu} - \frac{C_1}{12N}I_N)^2] \alpha + \mathrm{tr}[(m_e - \frac{C_1}{12N}I_N)^2] \beta + 3 \mathrm{tr}[(m_u - \frac{C_1}{12N}I_N)^2] \gamma \\ &+ 3 \mathrm{tr}[(m_d - \frac{C_1}{12N}I_N)^2] \delta.
\end{aligned}
\end{equation}
For the second term, recall that:
\begin{equation}
YY^\dagger+\frac{j_{Q_i}}{2} = I_2 \otimes (1 \otimes \frac{m_\nu + m_e}{2} \oplus I_3 \otimes \frac{m_u + m_d}{2}),
\end{equation}
from which we deduce that:
\begin{equation}
\begin{aligned}
\mathrm{tr}(z_L M_L^2) =& \mathrm{tr}[(\frac{m_\nu + m_e}{2} - \frac{C_1}{8N}I_N)^2] \mu + \mathrm{tr}[(\frac{m_\nu + m_e}{2} - \frac{C_1}{8N}I_N)^2] \nu \\ &+ 3 \mathrm{tr}[(\frac{m_u + m_d}{2} - \frac{C_1}{8N}I_N)^2] \mu + 3 \mathrm{tr}[(\frac{m_u + m_d}{2} - \frac{C_1}{8N}I_N)^2] \nu \\
=& \left(\mathrm{tr}[(\frac{m_\nu + m_e}{2} - \frac{C_1}{8N}I_N)^2] + 3 \mathrm{tr}[(\frac{m_u + m_d}{2} - \frac{C_1}{8N}I_N)^2]) \right) (\mu+\nu).
\end{aligned}
\end{equation}
We have:
\begin{equation}
\begin{aligned}
\mathrm{tr} & [(\frac{m_\nu + m_e}{2} - \frac{C_1}{8N}I_N)^2] + 3 \mathrm{tr}[(\frac{m_u + m_d}{2} - \frac{C_1}{8N}I_N)^2] \\
&= \mathrm{tr}[(\frac{m_\nu + m_e}{2})^2 + 3(\frac{m_u + m_d}{2})^2 -\frac{C_1}{8N}(m_\nu + m_e + 3 m_u + 3 m_d) + 4 \frac{C_1^2}{64N^2} I_N] \\
&= \frac{1}{4} \mathrm{tr}[m_\nu^2 + m_e^2 + 3 m_u^2 + 3 m_d^2] + \frac{1}{2}\mathrm{tr}[m_\nu m_e + 3 m_u m_d] - \frac{C_1^2}{8N} + \frac{C_1^2}{16N} \\
&= \frac{C_2}{4} + \frac{C_3}{2} - \frac{C_1^2}{16N} \\
&= \frac{1}{16}[4C_2 + 8C_3 - \frac{C_1^2}{N}].
\end{aligned}
\end{equation}
We thus have:
\begin{equation}
\mathrm{tr}(z_L M_L^2) = [4C_2 + 8C_3 - \frac{C_1^2}{N}]\frac{\mu+\nu}{16}.
\end{equation}
Finally, we compute the last term:
\begin{equation}
\begin{aligned}
\mathrm{tr}((z_R + z_L) M_{\overline{R}}^2) &= \frac{C_1^2}{144N^2} \mathrm{tr}[(z_R + z_L) (I_{2} \otimes (1 \oplus 0) \otimes I_N)] \\
&= \frac{C_1^2}{144N^2} (\alpha+\beta+\mu+\nu) N \\
\mathrm{tr}((z_R + z_L) M_{\overline{R}}^2) &= \frac{C_1^2}{144N}(\alpha+\beta+\mu+\nu).
\end{aligned}
\end{equation}
We thus have:
\begin{equation}
\begin{aligned}
e =& 4\left( \mathrm{tr}[(m_\nu - \frac{C_1}{12N}I_N)^2] + \frac{C_1^2}{144N} \right) \alpha + 4\left( \mathrm{tr}[(m_e - \frac{C_1}{12N}I_N)^2] + \frac{C_1^2}{144N} \right) \beta \\
&+ 12 \mathrm{tr}[(m_u - \frac{C_1}{12N}I_N)^2] \gamma + 12 \mathrm{tr}[(m_d - \frac{C_1}{12N}I_N)^2] \delta + [C_2 + 2 C_3 - \frac{2 C_1^2}{9N}] (\mu+\nu).
\end{aligned}
\end{equation}

To summarize:
\begin{equation}
\begin{aligned}
a =& \frac{4N}{3}(3 \alpha+ 5 \beta + 3 \gamma + 5 \delta + 3 \mu + \nu) \\
b =& 2N (\mu + 3 \nu) \\
c =& 2N (\beta + \delta + 2 \nu) \\
d =& \mathrm{tr}(4 m_{\nu}) \alpha + \mathrm{tr}(12 m_u) \beta + \mathrm{tr}(4 m_e) \gamma + \mathrm{tr}(12 m_d) \delta + \mathrm{tr}(4 (m_{\nu} + m_e)) \mu + \mathrm{tr}(12(m_u + m_d)) \nu \\
e =& 4\left( \mathrm{tr}[(m_\nu - \frac{C_1}{12N}I_N)^2] + \frac{C_1^2}{144N} \right) \alpha + 4\left( \mathrm{tr}[(m_e - \frac{C_1}{12N}I_N)^2] + \frac{C_1^2}{144N} \right) \beta \\
&+ 12 \mathrm{tr}[(m_u - \frac{C_1}{12N}I_N)^2] \gamma + 12 \mathrm{tr}[(m_d - \frac{C_1}{12N}I_N)^2] \delta + [C_2 + 2 C_3 - \frac{2 C_1^2}{9N}] (\mu+\nu).
\end{aligned}
\end{equation}
Unless the mass matrices obey very specific constraints, the parameters are all independent. For the next section, we will need these parameters to be all positive. Let us prove that this is possible. We will prove that it is sufficient that $\alpha, \beta, \gamma, \delta, \mu, \nu$ be all positive real numbers. Let us make this assumption. It is clear that $a,b,c$ are positive. The matrices $m_\nu, m_e, m_u, m_d$ are all positive matrices, and so are their traces. The parameter $d$ is thus positive. the matrices $(m_\nu - \frac{C_1}{12N}I_N)^2, \dots$ are all positive matrices, which implies that the first four terms of $e$ are positive. To conclude, we only need to prove that $9 C_2 + 18 C_3 - \frac{2 C_1^2}{N}$ is positive. We will need two useful results for this. Let $A$ be any self-adjoint $N \times N$ matrix. By the Cauchy-Schwartz inequality for the Hilbert-Schmidt norm on matrices, we have:
\begin{equation}
| \mathrm{tr}(I_N A) |^2 \leq \mathrm{tr}(I_N^2) \mathrm{tr}(A^\dagger A),
\end{equation}
from which we deduce that:
\begin{equation}
\mathrm{tr}(A)^2 \leq N \mathrm{tr}(A^2),
\end{equation}
Note also that for any two self-adjoint matrices $A,B$, we have $\mathrm{tr}[(A-B)^2] \geq 0$, from which we deduce that: $2\mathrm{tr}(AB) \leq \mathrm{tr}(A^2 + B^2)$. We thus have:
\begin{equation}
\begin{aligned}
C_1^2 &= \mathrm{tr}[(m_\nu + m_e) + 3(m_u+m_d)]^2 \\
&\leq N \mathrm{tr}[((m_\nu + m_e) + 3(m_u+m_d))^2] \\
&= N \mathrm{tr}[(m_\nu + m_e)^2 + 9 (m_u+m_d)^2] + 6N \mathrm{tr}[(m_\nu + m_e)(m_u+m_d)] \\
&\leq N \mathrm{tr}[(m_\nu + m_e)^2 + 9 (m_u+m_d)^2] + 3N \mathrm{tr}[(m_\nu + m_e)^2 + (m_u+m_d)^2] \\
&= 4N \mathrm{tr}[(m_\nu + m_e)^2 + 3 (m_u+m_d)^2] \\
&= 4N \mathrm{tr}[m_\nu^2 + m_e^2 + 2 m_\nu m_e + 3 m_u^2 + 3 m_d^2 + 6 m_u m_d] \\
C_1^2 &\leq 4N(C_2 + 2C_3).
\end{aligned}
\end{equation}
From this, one infers that:
\begin{equation}
C_2 + 2 C_3 \geq \frac{C_1^2}{4N} \geq \frac{2C_1^2}{9N}.
\end{equation}

\subsection{Normalizing the Fields}

Let us summarize the complete Lagrangian density we obtained:
\begin{equation}
\mathscr{L} = \mathscr{L}_b + \mathscr{L}_f,
\end{equation}
with:
\begin{equation*}
\mathscr{L}_b = - a F^Y_{\mu\nu}F^{Y\mu\nu} - b \mathrm{tr}(F^W_{\mu\nu}F^{W\mu\nu}) - c \mathrm{tr}(F^C_{\mu\nu}F^{C\mu\nu}) + d (D_\mu \Phi)^\dagger D^\mu \Phi - e (|\Phi|^2 - 1)^2,
\end{equation*}
and:
\begin{equation*}
\begin{aligned}
\mathscr{L}_f =& \overline{\Psi_R}(\slashed{D} \otimes 1)\Psi_R + \overline{\Psi_L}(\slashed{D} \otimes 1)\Psi_L + \overline{\Psi_R} \left[\gamma^\mu \otimes (A_\mu^Y T_Y^R - I_2 \otimes \overline{A_\mu^C}) \otimes I_N \right] \Psi_R \\ &+ \overline{\Psi_L} \left[\gamma^\mu \otimes (A_\mu^Y T_Y^L + (A_\mu^W \oplus A_\mu^W \otimes I_3) - I_2 \otimes \overline{A_\mu^C}) \otimes I_N \right] \Psi_L + \overline{\Psi_L} \tilde{q}_\Phi Y \Psi_R \\ &+ \overline{\Psi_R} Y^\dagger \tilde{q}_\Phi^\dagger \Psi_L + \frac{1}{2} \overline{J_{M-} \Psi_R} M \Psi_R + \frac{1}{2} \overline{M \Psi_R} J_{M-} \Psi_R.
\end{aligned}
\end{equation*}

We now need to redefine the fields to give them the correct conventional normalization (\textit{i.e.} kinetic terms with the correct factor \cite{Schwartz, Langacker}). We need to define a hypercharge field $B_\mu$, a weak field $W_\mu$, a gluon field $G_\mu$, and a Higgs field $\phi$ with its quaternion $q_\phi$. From the fermionic action, we see that the fields must be defined the following way:
\begin{equation}
\begin{aligned}
B_\mu =& \frac{1}{g_Y} A^Y_\mu \\
W_\mu =& \frac{1}{g_W} A^W_\mu \\
G_\mu =& -\frac{1}{g_C} \overline{A_\mu^C} \\
\phi =& \frac{1}{g_H} \Phi,
\end{aligned}
\end{equation}
where the constants $g_Y, g_W, g_C, g_H$ are to be determined. We have:
\begin{equation}
\begin{aligned}
F^W_{\mu \nu} =& \partial_\mu A^W_\nu - \partial_\nu A^W_\mu - i[A^W_\mu, A^W_\nu] \\
=& g_W (\partial_\mu W_\nu - \partial_\nu W_\mu - i g_W [W_\mu, W_\nu]),
\end{aligned}
\end{equation}
and similarly for $F^C_{\mu\nu}$ and $F^Y_{\mu\nu}$. This leads to the following definitions for the curvatures of the redefined gauge fields:
\begin{equation}
\begin{aligned}
B_{\mu\nu} =& \partial_\mu B_\nu - \partial_\nu B_\mu \\
W_{\mu\nu} =& \partial_\mu W_\nu - \partial_\nu W_\mu - i g_W [W_\mu, W_\nu] \\
G_{\mu\nu} =& \partial_\mu G_\nu - \partial_\nu G_\mu - i g_C [G_\mu, G_\nu].
\end{aligned}
\end{equation}
This implies that $g_Y, g_W, g_c$ are the gauge couplings! We thus have:
\begin{equation}
\begin{aligned}
F^Y_{\mu\nu} =& g_Y B_{\mu\nu} \\
F^W_{\mu\nu} =& g_W W_{\mu\nu} \\
F^C_{\mu\nu} =& -g_C \overline{G_{\mu\nu}}
\end{aligned}
\end{equation}
(for $G_{\mu\nu}$, the minus sign and complex conjugation in its definition balance each other). We also have:
\begin{equation}
\begin{aligned}
D_\mu \Phi =& (\partial_\mu - iA_\mu^W - iA_\mu^Y) \Phi \\
=& g_H (\partial_\mu - i g_W W_\mu - i g_Y B_\mu) \phi.
\end{aligned}
\end{equation}
We thus define:
\begin{equation}
D_\mu \phi = (\partial_\mu - i g_W W_\mu - i g_Y B_\mu) \phi,
\end{equation}
and we now have:
\begin{equation}
D_\mu \Phi = g_H D_\mu \phi.
\end{equation}
The Bosonic Lagrangian now reads:
\begin{equation}
\mathscr{L}_b = -a g_Y^2 B_{\mu\nu}B^{\mu\nu} - b g_W^2 \mathrm{tr}(W_{\mu\nu}W^{\mu\nu}) - c g_C^2 \mathrm{tr}(G_{\mu\nu}G^{\mu\nu}) + d g_H^2 (D_\mu \phi)^\dagger D^\mu \phi - e (g_H^2 |\phi|^2 - 1)^2.
\end{equation}
We choose bases $(t_a)_a$ and $(\lambda_a)_a$ for the Lie algebras $\mathrm{su}(2)$ and $\mathrm{su}(3)$ respectively, normalized so that:
\begin{equation}
\mathrm{tr}(t_a t_b) = \mathrm{tr}(\lambda_a \lambda_b) = 2 \delta_{ab}.
\end{equation}
In these bases, the curvatures are:
\begin{equation}
\begin{aligned}
W_{\mu\nu} =& W^a_{\mu\nu} t_a \\
G_{\mu\nu} =& G^a_{\mu\nu} \lambda_a,
\end{aligned}
\end{equation}
and the Lagrangian takes the form:
\begin{equation}
\mathscr{L}_b = -a g_Y^2 B_{\mu\nu}B^{\mu\nu} - 2 b g_W^2 W^a_{\mu\nu}W^{a\mu\nu} - 2 c g_C^2 G^a_{\mu\nu}G^{a\mu\nu}+ d g_H^2 (D_\mu \phi)^\dagger D^\mu \phi - e g_H^4 (|\phi|^2 - g_H^{-2})^2,
\end{equation}
where sums over Lie algebra indices are implicit. We can normalize our fields by choosing:
\begin{equation}
\begin{aligned}
g_Y =& \frac{1}{\sqrt{4a}} \\
g_W =& \frac{1}{\sqrt{8b}} \\
g_C =& \frac{1}{\sqrt{8c}} \\
g_H =& \frac{1}{\sqrt{d}}.
\end{aligned}
\end{equation}
With these conventions, The bosonic Lagrangian is:
\begin{equation}
\mathscr{L}_b = -\frac{1}{4} B_{\mu\nu}B^{\mu\nu} - \frac{1}{4} W^a_{\mu\nu}W^{a\mu\nu} - \frac{1}{4} G^a_{\mu\nu}G^{a\mu\nu}+ (D_\mu \phi)^\dagger D^\mu \phi - V_0 (|\phi|^2 - v^2)^2,
\end{equation}
with:
\begin{equation}
\begin{aligned}
V_0 =& e g_H^4 = \frac{e}{d^2} \\
v =& g_H^{-1} = \sqrt{d}.
\end{aligned}
\end{equation}
After the necessary substitutions, the fermionic Lagrangian is given by:
\begin{equation}
\begin{aligned}
\mathscr{L}_f =& \overline{\Psi_R}(\slashed{D} \otimes 1)\Psi_R + \overline{\Psi_L}(\slashed{D} \otimes 1)\Psi_L + \overline{\Psi_R} \left[\gamma^\mu \otimes (g_Y B_\mu T_Y^R + I_2 \otimes g_C G_\mu) \otimes I_N \right] \Psi_R \\ &+ \overline{\Psi_L} \left[\gamma^\mu \otimes (g_Y B_\mu T_Y^L + g_W (W_\mu \oplus W_\mu \otimes I_3) + I_2 \otimes g_C G_\mu) \otimes I_N \right] \Psi_L \\ &+ \overline{\Psi_L} \frac{\tilde{q}_\phi}{v} Y \Psi_R + \overline{\Psi_R} Y^\dagger \frac{\tilde{q}_\phi^\dagger}{v} \Psi_L + \frac{1}{2} \overline{J_{M-} \Psi_R} M \Psi_R + \frac{1}{2} \overline{M \Psi_R} J_{M-} \Psi_R.
\end{aligned}
\end{equation}

\subsection{Comparison with the Standard Model}

We now study the action found above and compare it to the Standard Model (see \cite{Schwartz, Langacker, PDG}). We found the bosonic Lagrangian:
\begin{equation}
\mathscr{L}_b = -\frac{1}{4} B_{\mu\nu}B^{\mu\nu} - \frac{1}{4} W^a_{\mu\nu}W^{a\mu\nu} - \frac{1}{4} G^a_{\mu\nu}G^{a\mu\nu}+ (D_\mu \phi)^\dagger D^\mu \phi - V_0 (|\phi|^2 - v^2)^2,
\end{equation}
where:
\begin{equation}
\begin{aligned}
V_0 =& \frac{e}{d^2} \\
v =& \sqrt{d} \\
g_Y =& \frac{1}{\sqrt{4a}} \\
g_W =& \frac{1}{\sqrt{8b}} \\
g_C =& \frac{1}{\sqrt{8c}}
\end{aligned}
\end{equation}
and:
\begin{equation}
\begin{aligned}
B_{\mu\nu} =& \partial_\mu B_\nu - \partial_\nu B_\mu \\
W_{\mu\nu} =& \partial_\mu W_\nu - \partial_\nu W_\mu - i g_W [W_\mu, W_\nu] \\
G_{\mu\nu} =& \partial_\mu G_\nu - \partial_\nu G_\mu - i g_C [G_\mu, G_\nu] \\
D_\mu \phi =& (\partial_\mu - i g_W W_\mu - i g_Y B_\mu) \phi.
\end{aligned}
\end{equation}
This bosonic Lagrangian clearly matches the Standard Model. 

The fermionic Lagrangian we found is equal to:
\begin{equation}
\begin{aligned}
\mathscr{L}_f =& \overline{\Psi_R}(\slashed{D} \otimes 1)\Psi_R + \overline{\Psi_L}(\slashed{D} \otimes 1)\Psi_L + \overline{\Psi_R} \left[\gamma^\mu \otimes (g_Y B_\mu T_Y^R + I_2 \otimes g_C G_\mu) \otimes I_N \right] \Psi_R \\ &+ \overline{\Psi_L} \left[\gamma^\mu \otimes (g_Y B_\mu T_Y^L + g_W (W_\mu \oplus W_\mu \otimes I_3) + I_2 \otimes g_C G_\mu) \otimes I_N \right] \Psi_L \\ &+ \overline{\Psi_L} \frac{\tilde{q}_\phi}{v} Y \Psi_R + \overline{\Psi_R} Y^\dagger \frac{\tilde{q}_\phi^\dagger}{v} \Psi_L + \frac{1}{2} \overline{J_{M-} \Psi_R} M \Psi_R + \frac{1}{2} \overline{M \Psi_R} J_{M-} \Psi_R.
\end{aligned}
\end{equation}
We need to expand this fermionic Lagrangian and compare it to the Standard Model. To this end, we need to expand the fermionic fields $\Psi_R$ and $\Psi_L$ in terms of the spinors describing each individual fermion. We will need for this the description of the finite space $\mathcal{K}_0$ in section \ref{SectionSMIST}. We use the following notations for right-handed fermions:
\begin{itemize}
	\item The right neutrino of the $i$-th generation is denoted $\nu_R^i \in \mathcal{K}_M$
	\item The right electron of the $i$-th generation is denoted $e_R^i$
	\item The right up quark of the $i$-th generation and color $c \in \{r,g,b\}$ is denoted $u_R^{ic}$
	\item The right down quark of the $i$-th generation and color $c \in \{r,g,b\}$ is denoted $d_R^{ic}$	
	\item We also define quark multi-spinors that describe all three colors at once:
	\begin{equation}
	\begin{aligned}
	u_R^i =& \sum_c u_R^{ic} \otimes c\in \mathcal{K}_M \otimes \mathbb{C}_c^3 \\
	d_R^i =& \sum_c d_R^{ic} \otimes c\in \mathcal{K}_M \otimes \mathbb{C}_c^3.
	\end{aligned}	
	\end{equation}
\end{itemize}
The notation is similar for left-handed fermions, with the index $R$ replaced with an $L$. We additionally need to define $SU(2)$ doublets:
\begin{equation}
\begin{aligned}
L^i =& \nu_L^i \otimes \nu + e_L^i \otimes e \in \mathcal{K}_M \otimes \mathbb{C}_l^2 \\
Q^i =& u_L^i \otimes u + d_L^i \otimes d \in \mathcal{K}_M \otimes \mathbb{C}_c^3 \otimes \mathbb{C}_q^2.
\end{aligned}
\end{equation}

With these notations, the total multi-spinor $\Psi_R$ is equal to:
\begin{equation}
\Psi_R = \sum_i \left[(\nu_R^i \otimes \nu + e_R^i \otimes e) \oplus \sum_c (u_R^{ic} \otimes u + d_R^{ic} \otimes d ) \otimes c \right] \otimes f_i
\end{equation}
and similarly for $\Psi_L$. We will often use the duality relation:
\begin{equation}
\begin{aligned}
S_x \otimes V &\longrightarrow S_x^\ast \otimes V^\ast \\
\varphi(x) \otimes u &\longmapsto \overline{\varphi \otimes u}(x) = H_x (\varphi, \cdot) \otimes u^\dagger,
\end{aligned}
\end{equation}
where $V$ is any of the finite-dimensional vector spaces $\mathbb{C}_l^2, \mathbb{C}_c^3$, etc.

The fermionic Lagrangian the sum of three contributions:
\begin{equation}
\mathscr{L}_f = \mathscr{L}_{\mathrm{Kin}} + \mathscr{L}_g + \mathscr{L}_h.
\end{equation}
The first term is a kinetic term:
\begin{equation}
\mathscr{L}_{\mathrm{Kin}} = \overline{\Psi_R}(\slashed{D} \otimes 1)\Psi_R + \overline{\Psi_L}(\slashed{D} \otimes 1)\Psi_L,
\end{equation}
and can be easily shown to be a sum of kinetic terms for all fermions:
\begin{equation}
\begin{aligned}
\mathscr{L}_{\mathrm{Kin}} =&  \sum_i \left[\overline{\nu_R^i} \slashed{D} \nu_R^i + \text{All other fermions} \right].
\end{aligned}
\end{equation}

The second term contains all coupling to gauge fields:
\begin{equation*}
\begin{aligned}
\mathscr{L}_g =& \overline{\Psi_R} \left[\gamma^\mu \otimes (g_Y B_\mu T_Y^R + I_2 \otimes g_C G_\mu) \otimes I_N \right] \Psi_R \\ &+ \overline{\Psi_L} \left[\gamma^\mu \otimes (g_Y B_\mu T_Y^L + g_W (W_\mu \oplus W_\mu \otimes I_3) + I_2 \otimes g_C G_\mu) \otimes I_N \right] \Psi_L.
\end{aligned}
\end{equation*}
We will expand it into three coupling terms, one for each gauge field:
\begin{equation}
\mathscr{L}_g = g_Y \mathscr{L}_Y + g_W \mathscr{L}_W + g_C \mathscr{L}_C.
\end{equation}
The first is a coupling term to the hypercharge field:
\begin{equation}
\mathscr{L}_Y = \overline{\Psi_R} \left[\gamma^\mu \otimes B_\mu T_Y^R \otimes I_N \right] \Psi_R + \overline{\Psi_L} \left[\gamma^\mu \otimes B_\mu T_Y^L \otimes I_N \right] \Psi_L.
\end{equation}
Now recall that:
\begin{equation*}
\begin{aligned}
T_Y^R =& \begin{pmatrix} 0 & \\ & -2 \end{pmatrix} \oplus \begin{pmatrix} \frac{4}{3} & \\ & -\frac{2}{3} \end{pmatrix} \otimes I_3 \\
T_Y^L =& -I_2 \oplus \frac{1}{3} I_2 \otimes I_3.
\end{aligned}
\end{equation*}
A lengthy but straightforward computation lends the following result:
\begin{equation}
\begin{aligned}
\mathscr{L}_Y =& \sum_i [ -2 \overline{e_R^i} \gamma^\mu B_\mu e_R^i + \frac{4}{3} \overline{u_R^i}(\gamma^\mu B_\mu \otimes I_3) u_R^i - \frac{2}{3} \overline{d_R^i}(\gamma^\mu B_\mu \otimes I_3) d_R^i \\ &- \overline{\nu_L^i} \gamma^\mu B_\mu \nu_L^i - \overline{e_L^i} \gamma^\mu B_\mu e_L^i + \frac{1}{3} \overline{u_L^i}(\gamma^\mu B_\mu \otimes I_3) u_L^i + \frac{1}{3} \overline{d_L^i}(\gamma^\mu B_\mu \otimes I_3) d_L^i ].
\end{aligned}
\end{equation}
To interpret this result and deduce from it the hypercharge of each particle, one needs to compare it to the kinetic terms. For example, we have for the $i$-th right electron:
\begin{equation*}
\mathscr{L}_f = i \overline{e_R^i} \gamma^\mu (\nabla_\mu^S + 2 i g_Y B_\mu) e_R^i + \dots.
\end{equation*}
Our conventions for gauge transformations are such that the gauge covariant derivative of a particle of charge $q$ with respect to some gauge field $A$ is: $D_\mu = \partial_\mu - iqA_\mu + \dots$ (see for example the covariant derivative of the Higgs field $D_\mu \phi = (\partial_\mu - i g_W W_\mu - i g_Y B_\mu) \phi$, which is an $SU(2)$ doublet of hypercharge $+1$). We deduce from this that the hypercharge of the right electron is -2, in accordance with the Standard Model. The correspondence with the Standard Model can be proven similarly for all other particles. The second gauge term is:
\begin{equation}
\mathscr{L}_W = \overline{\Psi_L} \left[\gamma^\mu \otimes (W_\mu \oplus W_\mu \otimes I_3) \otimes I_N \right] \Psi_L.
\end{equation}
A quick computation proves that:
\begin{equation}
\mathscr{L}_W = \sum_i [ \overline{L^i} (\gamma^\mu \otimes W_\mu) L^i + \overline{Q^i} (\gamma^\mu \otimes I_3 \otimes W_\mu) Q^i ],
\end{equation}
in accordance with the Standard Model (recall that for the quark doublets $Q^i$ as defined above, the color indices come before the doublet indices). Finally, the last gauge term is:
\begin{equation}
\mathscr{L}_C = \overline{\Psi_R} \left[\gamma^\mu \otimes (I_2 \otimes G_\mu \otimes I_N) \right] \Psi_R + \overline{\Psi_L} \left[\gamma^\mu \otimes (I_2 \otimes G_\mu \otimes I_N) \right] \Psi_L.
\end{equation}
One can easily prove that:
\begin{equation}
\mathscr{L}_C = \sum_i [ \overline{u_R^i} (\gamma^\mu \otimes G_\mu) u_R^i + \overline{d_R^i} (\gamma^\mu \otimes G_\mu) d_R^i + \overline{u_L^i} (\gamma^\mu \otimes G_\mu) u_L^i + \overline{d_L^i} (\gamma^\mu \otimes G_\mu) d_L^i ],
\end{equation}
which is in accordance with the Standard Model.

To conclude this comparison, we look at the third term of the fermionic Lagrangian, the Higgs (or mass) term:
\begin{equation}
\mathscr{L}_h = \overline{\Psi_L} \frac{\tilde{q}_\phi}{v} Y \Psi_R + \overline{\Psi_R} Y^\dagger \frac{\tilde{q}_\phi^\dagger}{v} \Psi_L + \frac{1}{2} \overline{J_{M-} \Psi_R} M \Psi_R + \frac{1}{2} \overline{M \Psi_R} J_{M-} \Psi_R.
\end{equation}
We start with the simplest terms: the third and fourth ones. Their sum can be rewritten:
\begin{equation*}
\frac{1}{2} \overline{J_{M-} \Psi_R} M \Psi_R + \frac{1}{2} \overline{M \Psi_R} J_{M-} \Psi_R = \frac{1}{2} \sum_{ij} \left( \overline{J_{M-} \nu_R^i} Y_R^{ij} \nu_R^i + \mathrm{h.c.} \right),
\end{equation*}
where h.c. denotes the hermitian conjugate of the previous term, as the two mass terms are conjugate to each other. Here $Y_R^{ij}$ are the components of the matrix $Y_R$. This is a Majorana mass term for the right neutrinos. We now take a look at the first two terms, which are also conjugate to each other (we thus only need to expand the first one). From section \ref{SectionSMIST}, recall that:
\begin{equation*}
\tilde{q}_\phi = (q_\phi \oplus q_\phi \otimes I_3) \otimes I_N,
\end{equation*}
with:
\begin{equation*}
q_\phi = \begin{pmatrix} \alpha & \beta \\ -\overline{\beta} & \overline{\alpha}	\end{pmatrix}.
\end{equation*}
The Higgs field itself is the second column of this matrix:
\begin{equation*}
\phi=\begin{pmatrix} \beta \\ \overline{\alpha} \end{pmatrix}.
\end{equation*}
It is traditional to denote the first column the following way:
\begin{equation}
\phi^\ast = \begin{pmatrix} \alpha \\ -\overline{\beta} \end{pmatrix}.
\end{equation}
It is related to $\phi$ by the following identity:
\begin{equation}
\phi^\ast = i \sigma^2 \overline{\phi},
\end{equation}
with $\overline{\phi}$ being the complex conjugate of $\phi$, and $\sigma^2$ the second Pauli matrix. We can thus write:
\begin{equation}
q_\phi = \begin{pmatrix} \phi^\ast, & \phi \end{pmatrix}.
\end{equation}
Recall also that:
\begin{equation*}
Y = (E_{\nu \nu} \otimes Y_\nu + E_{ee} \otimes Y_e) \oplus (E_{uu} \otimes I_3 \otimes Y_u + E_{dd} \otimes I_3 \otimes Y_d).
\end{equation*}
Let us compute first $Y \Psi_R$. The $Y_p$ matrices act on the generation basis vectors. We find:
\begin{equation*}
Y \Psi_R = \sum_{ij} \left[(Y_\nu^{ij} \nu_R^i \otimes \nu + Y_e^{ij} e_R^i \otimes e) \oplus \sum_c (Y_u^{ij} u_R^{ic} \otimes u + Y_d^{ij} d_R^{ic} \otimes d) \otimes c \right] \otimes f_j.
\end{equation*}
Next, $\tilde{q}_\phi$ acts on doublet degrees of freedom, with $\phi^\ast$ acting on $\nu$ and $u$, and $\phi$ acting on $e$ and $d$. We thus find:
\begin{equation*}
\tilde{q}_\phi Y \Psi_R = \sum_{ij} \left[(Y_\nu^{ij} \nu_R^i \otimes \phi^\ast + Y_e^{ij} e_R^i \otimes \phi) \oplus \sum_c (Y_u^{ij} u_R^{ic} \otimes \phi^\ast + Y_d^{ij} d_R^{ic} \otimes \phi) \otimes c \right] \otimes f_j.
\end{equation*}
Finally, we multiply with $\overline{\Psi_L}$ to find:
\begin{equation*}
\begin{aligned}
\overline{\Psi_L} \frac{\tilde{q}_\phi}{v} Y \Psi_R + \overline{\Psi_R} Y^\dagger \frac{\tilde{q}_\phi^\dagger}{v} \Psi_L = \frac{1}{v} \sum_{ij} & [\overline{L^j} (Y_\nu^{ij} \nu_R^i \otimes \phi^\ast) + \overline{L^j} (Y_e^{ij} e_R^i \otimes \phi) \\ &\oplus \overline{Q^j} (Y_u^{ij} u_R^i \otimes \phi^\ast) + \overline{Q^j} (Y_d^{ij} d_R^i \otimes \phi)] + \mathrm{h.c.},
\end{aligned}
\end{equation*}
and thus that:
\begin{equation}
\begin{aligned}
\mathscr{L}_h =& \frac{1}{2} \sum_{ij} \overline{J_{M-} \nu_R^i} Y_R^{ij} \nu_R^i + \frac{1}{v} \sum_{ij} [\overline{L^j} (Y_\nu^{ij} \nu_R^i \otimes \phi^\ast) + \overline{L^j} (Y_e^{ij} e_R^i \otimes \phi)] \\ & \oplus \frac{1}{v} \sum_{ij} [\overline{Q^j} (Y_u^{ij} u_R^i \otimes \phi^\ast) + \overline{Q^j} (Y_d^{ij} d_R^i \otimes \phi)] + \mathrm{h.c.}
\end{aligned}
\end{equation}
To interpret this correctly, notice that the doublet degrees of freedom of $L^i$ and $Q^i$ couple with those of $\phi$ and $\phi^\ast$, since $\nu_R^i \otimes \phi^\ast \in \mathcal{K}_M \otimes \mathbb{C}_l^2, \dots$, and that the triplet degrees of freedom of $Q^i$ couple with with those of $u_R^i$ and $d_R^i$, since $u_R^i \otimes \phi^\ast \in \mathcal{K}_M \otimes \mathbb{C}_c^3 \otimes \mathbb{C}_q^2, \dots$. With this information, we see that the coupling of fermions to the Higgs field is in accordance with the Standard Model with right neutrino.

We have thus recovered the entire Lagrangian of the Standard Model (with right neutrinos).

\chapter{Conclusion}

The main two reproaches directed at the NCG version of the standard model are: i) it is a Euclidean theory; ii) it is a classical theory. We can consider that the present thesis solves the first problem. The second problem is still wide open. It must also be stressed that the NCG standard model has not reached a completely satisfactory stage as long as the unimodularity condition and the massless photon conditions are not given a clear mathematical meaning.

In this conclusion, we would like to possible extensions of the present work. The first extension is in the direction of Grand Unified Theories. It is worth mentionning that the metric-$KO$ dimensions $(2,0)$ of the standard model are compatible with the Clifford algebra $\mathrm{Cl}(9,9)$ which, by Chevalley's theorem, can be considered as the graded tensor product of the Clifford algebra $\mathrm{Cl}(1,3)$ of spacetime and the Clifford algebra $\mathrm{Cl}(8,6)$. The dimension of the irreducible representation of $\mathrm{Cl}(8,6)$ is 128, which would accomodate a standard model with four generations instead of the three generations experimentally observed. It would be interesting to explore the corresponding gauge theory based on the Lie algebra
$so(8,6)$. Such real Lie algebras have a bad reputation because the corresponding group is not compact, but Margolin and Strazhev proved that they are meaningful and renormalizable \cite{Margolin-90,Margolin-92}. We hope to come back to this interesting subject in the future.

A second extension of the present thesis is towards
possible applications of indefinite spectral triples to solid-state
physics and in particular topological insulators and supraconductors. Indeed, topological insulators also consider systems where two self-adjoint involution (e.g. inversion symmetry and mirror symmetry) and an antilinear operator (e.g. charge conjugation or time-reversal symmetry) are involved.  By our classification theorem \cite{3B-IST}, we can associate a metric-$KO$ pair of dimensions to such a system. We can also make tensor products to describe many-body insulators or
supraconductors.

\end{document}